\documentclass[printer]{tADP2e}
\usepackage{rotating}
\usepackage{cite}
\newcounter{saveeqn}%
\newcommand{\ts}{\textstyle }
\newcommand{\lsim}{ {{}_{\ts <} \atop {\ts \sim}} }
\newcommand{\gsim}{ {{}_{\ts >} \atop {\ts \sim}} }

\renewcommand{\vec}[1]{{\bf #1}}

\begin{document}
\doi{10.1080/00018730600645636}
\issn{1460-6976}
\issnp{0001-8732} \jvol{55} \jnum{1-2} \jyear{2006} \jmonth{January-April}

\markboth{M. Eschrig}{
The effect of collective spin-1 excitations on electronic spectra
in high-$T_c$ superconductors
}

\title{
The effect of collective spin-1 excitations on electronic spectra
in high-$T_c$ superconductors
}

\author{MATTHIAS ESCHRIG$^{\ast }$\thanks{
$^{\ast }$E-mail: eschrig@tfp.physik.uni-karlsruhe.de
\vspace{6pt}\newline\centerline{\tiny{
{\em Advances in Physics}}}\newline\centerline{\tiny{
ISSN 0001-8732 print/ISSN 1460-6976 online
\textcopyright 2006 Taylor \& Francis Ltd}}
\newline\centerline{\tiny{ http://www.tandf.co.uk/journals}}\newline \centerline{\tiny{DOI:
10.1080/00018730600645636}}}\break
Institut f{\"u}r Theoretische Festk{\"o}rperphysik,
Universit{\"a}t Karlsruhe, \\ D-76128 Karlsruhe, Germany} 
\received{Received 11 October 2005; in final form 21 February 2006}

\maketitle

\begin{abstract}
We review recent experimental and theoretical results on the interaction between
single-particle excitations and collective spin excitations
in the superconducting state of high-$T_c$ cuprates.
We concentrate on the traces, that sharp features in the magnetic-excitation
spectrum (measured by inelastic neutron scattering)
imprint in the spectra of single-particle excitations (measured
e.g.  by angle-resolved photoemission spectroscopy, 
tunneling spectroscopy, and indirectly also by optical spectroscopy).
The ideal object to obtain a quantitative picture for these interaction
effects is a spin-1 excitation around 40 meV, termed 'resonance mode'.
Although the total weight of this spin-1 excitation is small,
the confinement of its weight to a rather
narrow momentum region around the antiferromagnetic wavevector makes it 
possible to observe strong self-energy effects in parts of the electronic Brillouin zone.
Notably the sharpness of the magnetic excitation in energy has allowed to 
trace these self-energy effects in the single-particle spectrum 
rather precisely.
Namely, the doping- and temperature dependence together with
the characteristic energy- and momentum behavior of the 
resonance mode has been used as a tool to examine the 
corresponding self-energy effects in the dispersion and in the 
spectral lineshape of the single-particle spectra, 
and to separate them from similar effects due to electron-phonon interaction.
This leads to the unique possibility to single out the self-energy effects 
due to the spin-fermion interaction and to
directly determine the strength of this interaction in high-$T_c$ 
cuprate superconductors. The knowledge of this interaction is important
for the interpretation of other experimental results
as well as for the quest for the still unknown pairing mechanism in these interesting 
superconducting materials.\\

\centerline{\bfseries Contents }\medskip

\noindent
{1.}~Introduction \hfill \pageref{Intro}\\
{2.}~Experimental evidence of a sharp collective spin excitation and its coupling to fermions \hfill \pageref{experiments}\\
\hspace*{10pt}{2.1.}~Inelastic Neutron Scattering\hfill \pageref{secINS}\\
\hspace*{24pt}{2.1.1.}~Magnetic coupling\hfill \pageref{secMagCoup} \\
\hspace*{24pt}{2.1.2.}~The magnetic resonance feature\hfill \pageref{magres}\\
\hspace*{24pt}{2.1.3.}~Bilayer effects\hfill \pageref{secBil}\\
\hspace*{24pt}{2.1.4.}~Temperature dependence\hfill \pageref{temp_res}\\
\hspace*{24pt}{2.1.5.}~Doping dependence\hfill \pageref{Dop_res}\\
\hspace*{24pt}{2.1.6.}~Dependence on disorder\hfill \pageref{Dis_res}\\
\hspace*{24pt}{2.1.7.}~Isotope effect\hfill \pageref{Iso_res}\\
\hspace*{24pt}{2.1.8.}~Dependence on magnetic field\hfill \pageref{Magn_res}\\
\hspace*{24pt}{2.1.9.}~The incommensurate part of the spectrum\hfill \pageref{Incomm}\\
\hspace*{24pt}{2.1.10.}~The spin gap\hfill \pageref{Spingap}\\
\hspace*{24pt}{2.1.11.}~The spin fluctuation continuum\hfill \pageref{SpinCont}\\
\hspace*{24pt}{2.1.12.}~Normal state spin susceptibility\hfill \pageref{Normalsusc}\\
\hspace*{10pt}{2.2.}~Angle resolved photoemission\hfill \pageref{ARPES}\\
\hspace*{24pt}{2.2.1.}~Fermi surface\hfill \pageref{FS}\\
\hspace*{24pt}{2.2.2.}~Normal-state dispersion and the flat-band region\hfill \pageref{FlatBand}\\
\hspace*{24pt}{2.2.3.}~MDC and EDC\hfill \pageref{MDCEDC}\\
\hspace*{24pt}{2.2.4.}~Bilayer splitting\hfill \pageref{Bil_exp}\\
\hspace*{24pt}{2.2.5.}~Superconducting coherence\hfill \pageref{SupCoh}\\
\hspace*{24pt}{2.2.6.}~EDC-derived dispersion anomalies\hfill \pageref{DispAn_ex}\\
\hspace*{24pt}{2.2.7.}~The $S$-shaped MDC-dispersion anomaly \hfill \pageref{S-shape}\\
\hspace*{24pt}{2.2.8.}~The nodal kink\hfill \pageref{NK}\\
\hspace*{24pt}{2.2.9.}~Fermi velocity\hfill \pageref{FV_ex}\\
\hspace*{24pt}{2.2.10.}~Spectral lineshape\hfill \pageref{SpecLS_ex}\\
\hspace*{24pt}{2.2.11.}~The antinodal quasiparticle peak\hfill \pageref{AQP}\\
\hspace*{24pt}{2.2.12.}~The spectral dip feature\hfill \pageref{SDF}\\
\hspace*{24pt}{2.2.13.}~Real part of self energy: renormalization of dispersion\hfill \pageref{RPSE}\\
\hspace*{24pt}{2.2.14.} Imaginary part of self energy: quasiparticle lifetime\hfill \pageref{IPSE}\\
\hspace*{24pt}{2.2.15.} Isotope effect\hfill \pageref{Iso}\\
\hspace*{24pt}{2.2.16.} Relation to pseudogap phase\hfill \pageref{RPP}\\
\hspace*{10pt}{2.3.}  C-axis tunneling spectroscopy\hfill \pageref{CTS}\\
\hspace*{10pt}{2.4.}  Optical spectroscopy\hfill \pageref{OPS}\\
{3.}~The collective mode as spin-1 exciton\hfill \pageref{collectivemode}\\
\hspace*{10pt}{3.1.}~Theoretical models\hfill \pageref{TM}\\
\hspace*{10pt}{3.2.}~Characteristic energies\hfill \pageref{CE}\\
\hspace*{10pt}{3.3.}~The resonance mode\hfill \pageref{RM}\\ 
\hspace*{24pt}{3.3.1.}~Development of spin exciton\hfill \pageref{DSE}\\
\hspace*{24pt}{3.3.2.}~Doping dependence\hfill \pageref{DD}\\
\hspace*{24pt}{3.3.3.}~Dependence on disorder\hfill \pageref{ThDis_res}\\
\hspace*{24pt}{3.3.4.}~Dependence on magnetic field\hfill \pageref{Magfield}\\
\hspace*{24pt}{3.3.5.}~Even and odd mode in bilayer cuprates\hfill \pageref{EvenOdd}\\
\hspace*{10pt}{3.4.}~The incommensurate response\hfill \pageref{IncommRes}\\ 
\hspace*{10pt}{3.5.}~Effective low-energy theories\hfill \pageref{EffLET}\\ 
\hspace*{10pt}{3.6}~Magnetic coherence in La$_{2-x}$Sr$_x$CuO$_4$\hfill \pageref{SpinLSCO}\\
{4.}~Coupling of quasiparticles to the magnetic resonance mode\hfill \pageref{coupling}\\
\hspace*{10pt}{4.1.}~The coupling constant and the weight of the spin resonance\hfill \pageref{CC}\\
\hspace*{10pt}{4.2.}~Theoretical model\hfill \pageref{Theory}\\
\hspace*{24pt}{4.2.1.}~Tight binding fit to normal state dispersion \hfill \pageref{TBF}\\
\hspace*{24pt}{4.2.2.}~Model spectrum and basic equations \hfill \pageref{BEQ}\\
\hspace*{10pt}{4.3.}~Contribution from the spin fluctuation mode\hfill \pageref{CSFM}\\
\hspace*{24pt}{4.3.1.}~Characteristic electronic scattering processes\hfill \pageref{ES}\\
\hspace*{24pt}{4.3.2.}~Electronic self energy\hfill \pageref{RFEL}\\
\hspace*{24pt}{4.3.3.}~Renormalization function and quasiparticle scattering rate\hfill \pageref{QPSR}\\
\hspace*{24pt}{4.3.4.}~Spectral functions at the $M$ point \hfill \pageref{SFMP}\\
\hspace*{10pt}{4.4.}~Contribution of the spin fluctuation continuum\hfill \pageref{CSFC}\\
\hspace*{10pt}{4.5.}~Renormalization of EDC and MDC dispersions\hfill \pageref{RenEDCMDC}\\
\hspace*{10pt}{4.8.}~Bilayer splitting\hfill \pageref{BLS}\\
\hspace*{10pt}{4.6.}~Tunneling spectra\hfill \pageref{Tunnel}\\
\hspace*{10pt}{4.8.}~Doping dependence\hfill \pageref{Doping}\\
{5.}~Discussion of phonon effects\hfill \pageref{Phonon}\\
{6.}~Open problems\hfill \pageref{OP}\\
{7.}~Conclusions\hfill \pageref{conclusions}\\
Acknowledgements\hfill \pageref{Ackn}\\
References\hfill \pageref{Refs}\\ \\
\end{abstract}

\section{Introduction}
\label{Intro}

Cuprate high-$T_c$ superconductivity, discovered in 1986 \cite{Bednorz86},
arises when a sufficient amount of charge carriers (holes or electrons) 
is doped into an antiferromagnetic, Mott-insulating 
parent compound \cite{Tokura89,Kastner98}.
It is one of the fields which continues to inspire both theoretical and experimental research.
The development of new methods and the improvement
of existing ones as a result of the research in cuprate superconductivity
have influenced many other fields in condensed-matter physics.
However, there is no generally accepted agreement about the pairing mechanism
in these materials, and not even the normal state has been described in
a satisfactory way up to date. 

Due to dramatic improvements in the resolution in angle-resolved 
photoemission (ARPES)
experiments during the last years, the properties of single-particle 
electronic excitations throughout the Brillouin zone
have been thoroughly studied.
An agreement has emerged that at least in the superconducting state electronic
quasiparticle excitations are well defined \cite{Dessau91,Hwu91} and are the entities
participating in superconducting pairing \cite{Matsui03}.
However, there are numerous anomalies, caused by self-energy effects, which
complicate the dispersions and spectral lineshapes observed in ARPES experiments.

The recent developments in testing fermionic single-particle excitations 
in high-$T_c$ cuprate superconductors were 
to a large extend driven by a suggestion 
that several dispersion anomalies observed in angle-resolved-photoemission
experiments can be explained in a unified picture
invoking a strong coupling to a resonant magnetic mode at antiferromagnetic
wavevector $(\pi, \pi)$, which is
observed in inelastic-neutron-scattering experiments \cite{Eschrig00}. 
In this scenario, 
the finite momentum width of the resonance mode
plays a crucial role, leading to scattering of quasiparticles 
that is maximal for
points in the Brillouin zone separated by a $(\pi, \pi)$ wavevector, but
to a less extend also present for scattering between points separated
by a wavevector deviating from $(\pi, \pi)$. 
This crucial generalization
of a model by Kampf and Schrieffer \cite{Kampf90} 
allowed to explain the variety of
observed effects in one single model.

The self-energy effects in the single-particle dispersions, 
which are being studied experimentally in great detail,
open a unique possibility to determine the crucial parameters for a successful
theoretical description of the high-$T_c$ phenomenon, namely the strength of
the coupling between the electronic single-particle excitations and the 
collective excitations due to lattice modes (phonons) as well as electronic modes 
present in the spin-, charge- or pairing channel.
The knowledge of these interaction strengths is pivotal for a correct theoretical
description of both the normal and superconducting state of cuprate superconductors.

Numerous experimental techniques have been used to
analyze collective excitations of various types. For example, inelastic
neutron scattering (INS) is a direct probe of both the phonon spectrum
and of the spectrum of electronic collective excitations. 
In particular, it is possible by spin-polarized INS techniques to separate 
electronic excitations of magnetic origin from non-magnetic excitations.
The experimental results obtained in this way are the second crucial ingredient
for the determination of the relevant coupling constants for electronic excitations
in cuprates. Namely, in order to assign the correct collective excitations 
to the various self-energy effects observed by ARPES techniques, it is necessary
to compare the temperature- and doping dependence of the self-energy
effects with that of 
the corresponding collective modes observed by INS techniques.
Only if both the energy range and the magnitude of the observed dispersion anomalies
match the energy and intensity of the corresponding collective excitations, 
is it possible to extract the necessary information for the interaction constants.

Motivated by earlier
work \cite{Kampf90,Dahm96e,Dahm96,Dahm96d,Dahm97a,Dahm98,Shen97,Norman97,Norman98},
a thorough study along these lines has emerged during the last years,
which has found a clear correlation between the spin-fluctuation spectrum measured
in INS experiments and the self-energy effects measured in ARPES experiments
\cite{Abanov99,Abanov00b,Abanov00c,Abanov00d,Abanov01a,Abanov01b,Abanov02a,Eschrig00,Eschrig02,Eschrig03,Li00,Wu01,Manske01,Manske01a,Norman01a,Norman03}.
This theoretical development stimulated great experimental interest. In particular,
it led to doping- and temperature dependent studies of the self-energy effects related
to the magnetic resonance excitation 
by ARPES experiments 
\cite{Campuzano99,Bogdanov00,Kaminski01,Gromko02,Gromko03,Valla00,Johnson01,Sato03,Kordyuk04,Borisenko03,Kim03a,Koitzsch04},
tunneling spectroscopy \cite{Zasadzinski01},
and optical spectroscopy \cite{Hwang04}.

It turned out that it was also necessary to analyze the momentum dependence
of the self-energy effects and to relate them to the momentum-width of the
collective spin excitation 
\cite{Eschrig00,Eschrig03}
in order to be able to distinguish it clearly from
other collective modes like for example phonons.
In high-$T_c$ cuprates
phonons are generally accepted to couple to electrons in a moderate way,
and on theoretical grounds phononic features should be observable in the 
single-particle
spectra as well. Corresponding effects have been found and have been examined
experimentally by INS \cite{McQueeney99,Egami02,Chung03} and ARPES
\cite{Lanzara01,Shen02,Lanzara02,Cuk04a,Zhou04,Gweon04a} as well as 
theoretically (see \cite{Devereaux04,Roesch04,Sandvik04}, and references therein).
We concentrate in this review on the interaction of electronic single-particle excitations with
collective spin excitations in cuprates.
It has been shown e.g. by inelastic neutron scattering and by spatially resolved NMR techniques,
that spin fluctuations play an important role not only above $T_c$
but also
in the superconducting vortex state \cite{Lake01,Lake02c,Mitrovic01,Mitrovic03}.
The results of INS and ARPES experiments as well as other experimental
techniques, as tunneling spectroscopy and optical spectroscopy, and the
correlation between the data obtained from these different techniques, allowed
for the first time
a rather direct and precise determination of the coupling strength
between conduction electrons
and spin collective excitations in cuprate systems.

\begin{figure}
\centerline{
\epsfxsize=0.7\textwidth{\epsfbox{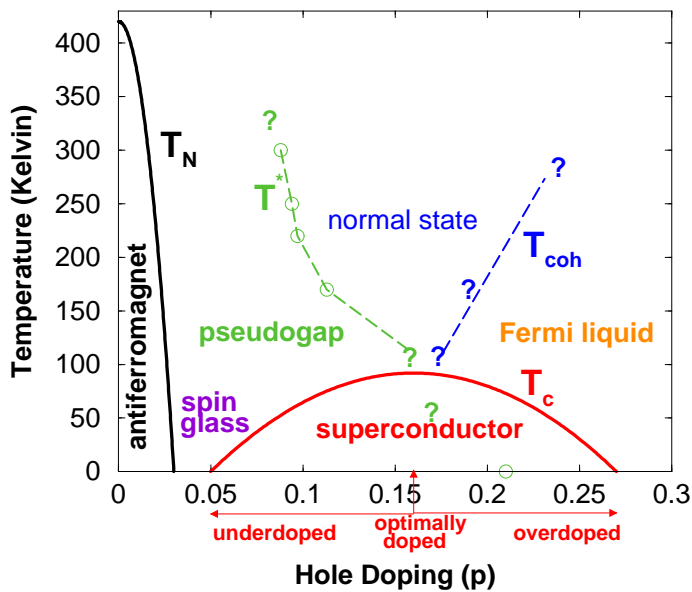}}
}
\caption{
\label{PhaseD}
Typical phase diagram for high-$T_c$ cuprates over the number of
doped holes per Cu ion, $p$. 
Superconductivity arises when the antiferromagnetic parent compound 
($p=0$) is doped with a certain amount of charge carriers (here holes).
Optimal doping corresponds to the doping value with highest $T_c$.
The circles are recent data from Ref. \cite{Fauque05}
of an additional transition observed in neutron scattering experiments.
Question marks denote regions of the phase diagram that are theoretically not
well understood.  (After Refs. \cite{Norman05,Fauque05})
}
\end{figure}

The dominant interaction for single-particle electronic excitations
(quasiparticles) in three-dimensional metals
and superconductors is the electron-phonon interaction.
In contrast, for lower-dimensional systems
the interaction between quasiparticles
and collective electronic excitations becomes relevant.
This is a direct result of the Pauli exclusion principle, which
leads to stronger kinematic phase-space restrictions in higher dimensions.
In quasi-twodimensional materials single particle excitations
are in general modified (but not completely destroyed) 
by interactions with collective modes \cite{Eschrig94}
(this is unlike to quasi-onedimensional materials 
where the interactions between single
particle excitations and collective excitations are dominating the
physics).  It is therefore not surprising that 
in high-temperature cuprate superconductors, which are quasi-twodimensional
materials, such collective excitations have a strong impact on quasiparticles.
Experimentally it was observed, that at least in the
superconducting state quasiparticle-like excitations are well defined,
and to a large extend can be successfully described by a $d$-wave modification
of the Bardeen-Cooper-Schrieffer theory of superconductivity \cite{Schrieffer64}.
The normal state of high-temperature superconductors poses more problems
in this respect. 
For this reason the study of the homogeneous superconducting state might be easier than
that of the normal state, and might give some support for the more difficult tasks
of understanding the pseudogap phase and inhomogeneous superconducting phases.
Thus, we concentrate in this review on the superconducting state and refer the reader for
the interesting questions of the normal-state and pseudogap-state behavior to
other reviews \cite{Timusk99,Randeria98,Norman05}.
In Fig.~\ref{PhaseD} the typical phase diagram for the 
cuprate superconductors as a function of hole-doping is shown.

Superconductivity can also be achieved by electron doping. In this review,
however, we restrict ourselves to the hole-doped materials, as the vast majority
of INS and ARPES experiments were performed for those.
So far in the experimental investigations of
electron-doped cuprates the characteristic self-energy effects as
well as the resonance mode in the spin excitation spectrum that are
the main topic of this review have not been found.

We start in Section \ref{experiments} with a review of the available experimental data, concentrating
on the most recent data referring to self-energy effects observed by ARPES 
experiments and spin-collective modes observed by INS experiments.
Then, in Section \ref{collectivemode} we review theoretical developments concerning the
interpretation of the collective spin-excitation as spin-1 excitonic mode below the spin-fluctuation continuum.
In Section \ref{coupling} we review the methods used to extract the
interaction effects between the single-particle excitations and the collective spin-1 excitonic 
excitations from available experimental data. Using the results of INS experiments and normal state
parameters obtained from ARPES experiments, the various self-energy effects observed in the superconducting
state by ARPES and tunneling experiments are then compared with the theoretical results and are shown to
give a consistent picture. 
Section \ref{Phonon} is devoted to the discussion on self-energy effects due to electron-phonon interaction.
Finally, in Section \ref{OP} we discuss open problems and
in Section \ref{conclusions} we summarize the important
implications of this field of high-$T_c$ research for an understanding of superconductivity
in these systems.

\section{Experimental evidence of a sharp collective spin excitation and its
coupling to fermions}
\label{experiments}

\subsection{Inelastic Neutron Scattering}
\label{secINS}

Neutron scattering experiments have been important in the study of collective
excitations in high-temperature superconductors. 
Inelastic neutron scattering experiments probe both collective excitations
of the lattice (phonons) and collective excitations of the
electronic system. In typical metals such electronic collective excitations
are small perturbations to the liquid of quasiparticles above the
Fermi sea ground state. The reason for this is, that the dynamics of
single-particle excitations for such liquids can be described by
a quantum transport equation for many body excitations which have
signatures of single particles. 
In three dimensional systems, the only collective excitations which
affect the collision terms of the transport equation in leading order
in a controlled approximation are phonons.
Cuprates are quasi-twodimensional, and it was shown that in two-dimensional
systems quasiparticle collision terms are governed by electronic collective
excitations even far from instabilities \cite{Eschrig94,Eschrig00b}.
But still, in these cases, the equilibrium properties of such
quasi-twodimensional systems are expected to be unaffected
by collective electronic modes in leading order in $T/E_F$, 
as long as singular corrections are not important in higher orders. 
For the importance of such singular corrections in two dimensions see
Ref.~\cite{Chubukov05} and references therein.
In cuprates equilibrium properties
show unusual behavior at least in the normal state \cite{Timusk99}.
Thus, it is possible that collective electronic modes play an important role
even for equilibrium properties.
Examples for such electronic modes are spin fluctuations (as a precursor
for spin wave modes), charge fluctuations (precursor for density waves),
combinations of both (`stripes', \cite{Zaanen89,Poilblanc89,Machida89,Schulz90,Emery99,Zaanen01,Kivelson03,Tranquada05,Raczkowski05}), 
pair fluctuations, and combinations of
all of those modes with lattice deformations ('polarons').

There has been an enormous
amount of work in revealing the properties of magnetic excitations in these
materials. The magnetic part of the inelastic neutron scattering (INS)
signal is usually much smaller than the signal from e.g. phonons, and
special techniques had to be applied in order to extract it.
Fortunately the magnetic response is strongly enhanced near the antiferromagnetic
wavevector in cuprate superconductors, which allowed for an experimental analysis
of the magnetic excitation spectrum in cuprates. The main focus of this study
has been a strong peak at the antiferromagnetic wavevector, which is sharp in 
energy and has for optimally doped materials an excitation energy near 40 meV.
We will concentrate in the following on the magnetic excitation spectrum observed
in INS in the superconducting state of high-$T_c$ materials, with particular
weight on the above-mentioned sharp resonant mode.

\subsubsection{Magnetic coupling}
\label{secMagCoup}

In cuprate superconductors the superconducting structural units are either one single copper oxide layer 
or several closely spaced copper oxide layers. These units are separated by much larger distances than the layers within such
units. Correspondingly, there is a hierarchy of magnetic superexchange couplings. 

The in-plane magnetic superexchange coupling $J_\parallel $ is of the order of 120-150 meV.
In the antiferromagnetic insulating state it can be obtained experimentally by fitting
the spin wave velocity to quantum Monte-Carlo calculations. In La$_2$CuO$_4$ this procedure gave
$J_\parallel = 133 $ meV \cite{Aeppli89}.

The superexchange between different superconducting units is more than four orders of magnitude smaller
than the primary coupling within one copper oxygen plane, $J_\parallel $. It is of the order of
0.02 meV \cite{Tranquada89,Rossat93}.
Inelastic neutron scattering experiments show that the neutron scattering signal in the metallic
state of cuprates can be described
by an incoherent superposition of the signals from different superconducting units. Thus, these
units can be considered as magnetically decoupled.
This is expected on theoretical grounds from the fact that the magnetic in-plane correlation lengths 
in cuprates are only a few lattice spacings.

The coupling $J_\perp$ between different planes within a superconducting unit, 
however, is only one order of magnitude less than $J_\parallel $, $J_\perp \sim 10 $ meV \cite{Reznik96}.
Even in the metallic regime a strong magnetic coupling $J_\perp $ 
remains \cite{Rossat93,Bourges98,Regnault98,Tranquada92}.
This coupling is e.g. reflected in a pronounced $Q_z$-dependence of the inelastic neutron scattering
signal for bilayer cuprates \cite{Tranquada92}.
The corresponding signal is proportional to the imaginary part of the susceptibility
\begin{equation}
\chi (\vec{Q},Q_z,\omega ) = \sum_{ij} e^{iQ_z(z_i-z_j)} \chi^{ij} (\vec{Q},\omega )
\end{equation}
where
\begin{equation}
\chi^{ij} (\vec{Q},\omega ) = \langle T_\tau \hat S_z^i(\vec{Q},\tau) \hat S_z^j(-\vec{Q},0)\rangle
\end{equation}
with the spin-density operator $ \hat S_z^i (\vec{Q} )= \sum_{\vec{k},\alpha \beta} c^\dagger_{i,\vec{k}+\vec{Q},\alpha} \; \hat \sigma^z_{\alpha \beta} \; c_{i,\vec{k},\beta }$ (here, $\vec{Q}$ and $\vec{k}$ are in-plane vectors).
The fact that a pronounced $Q_z$-dependence is observed both in the normal and superconducting
state indicates that there is no significant change in coherence between the planes within a bilayer
due to onset of superconductivity \cite{Tranquada92}.

The magnetic part of the spectrum measured in inelastic neutron scattering experiments 
describes the spectrum of spin fluctuations. 
In the superconducting state of cuprates it
typically consists of three parts. The first is a continuum, which is gapped at low energies.
The main feature for cuprates with $T_c$ around 90 K 
is a resonance feature peaked at the antiferromagnetic wave vector, and is present
at energies below the continuum.
Below the resonance energy,
an incommensurate response develops \cite{Dai98,Mook98a,Mook98b}, which however
never extends to zero energy, but instead
the spectrum is limited at low energies
by the so called spin gap $E_{sg}$ \cite{Dai01}.
As also above the resonance an incommensurate response is observed,
the incommensurate response in superconducting cuprates shows a typical
hour-glass shape \cite{Aeppli97,Tranquada04,Pailhes04,Stock05}.

\subsubsection{The magnetic resonance feature}
\label{magres}

The magnetic resonance mode was first observed in inelastic neutron scattering experiments
for bilayer cuprates in the
superconducting state, with energy near 40 meV in optimally doped
compounds \cite{Rossat91,Mook93,Fong95,Fong96,Bourges96,Fong99}.
This resonance is sharp (resolution limited, where the instrumental resolution is typically
less than 10 meV) in energy and magnetic in origin \cite{Mook93}.
It is centered in
momentum around the antiferromagnetic wavevector $\vec{Q}=(\pi,\pi)$.
In contrast to its sharpness in energy, in momentum the resonance has a finite
width of typically $0.25 \AA^{-1}$ (full width half maximum, FWHM).

The total momentum width of the spectrum is minimal at the resonance energy \cite{Bourges95,Bourges96},
where it is (in contrast to the off-resonant momentum width) only weakly doping dependent, with a
full momentum width of about $0.22 \AA^{-1}$ \cite{Bourges95,Balatsky99,Dai01}.
This corresponds to a correlation length $\xi_{sfl}$ of about two lattice spacings.
Note, however, that the spectrum above and below the resonance consists of
incommensurate peaks which strongly overlap, and thus the total momentum
width overestimates the momentum width of the incommensurate spin excitations.

A similar resonance feature is also observed in underdoped
YBa$_2$Cu$_3$O$_{6+x}$, but at reduced energy \cite{Dai96,Fong97,Bourges97,Dai99,Stock04}.
Also in Bi${_2}$Sr${_2}$CaCu${_2}$O${_{8+\delta}}$ the resonance was found
both in the optimally doped \cite{Fong99,He01} and overdoped \cite{He01} regime.
For a comparison with tunneling and ARPES data, which were predominantly performed on
Bi${_2}$Sr${_2}$CaCu${_2}$O${_{8+\delta}}$, it is important to notice
that the characteristic features are very similar to those for YBa$_2$Cu$_3$O$_{6+x}$.
\begin{figure}
\centerline{
\epsfxsize=0.44\textwidth{\epsfbox{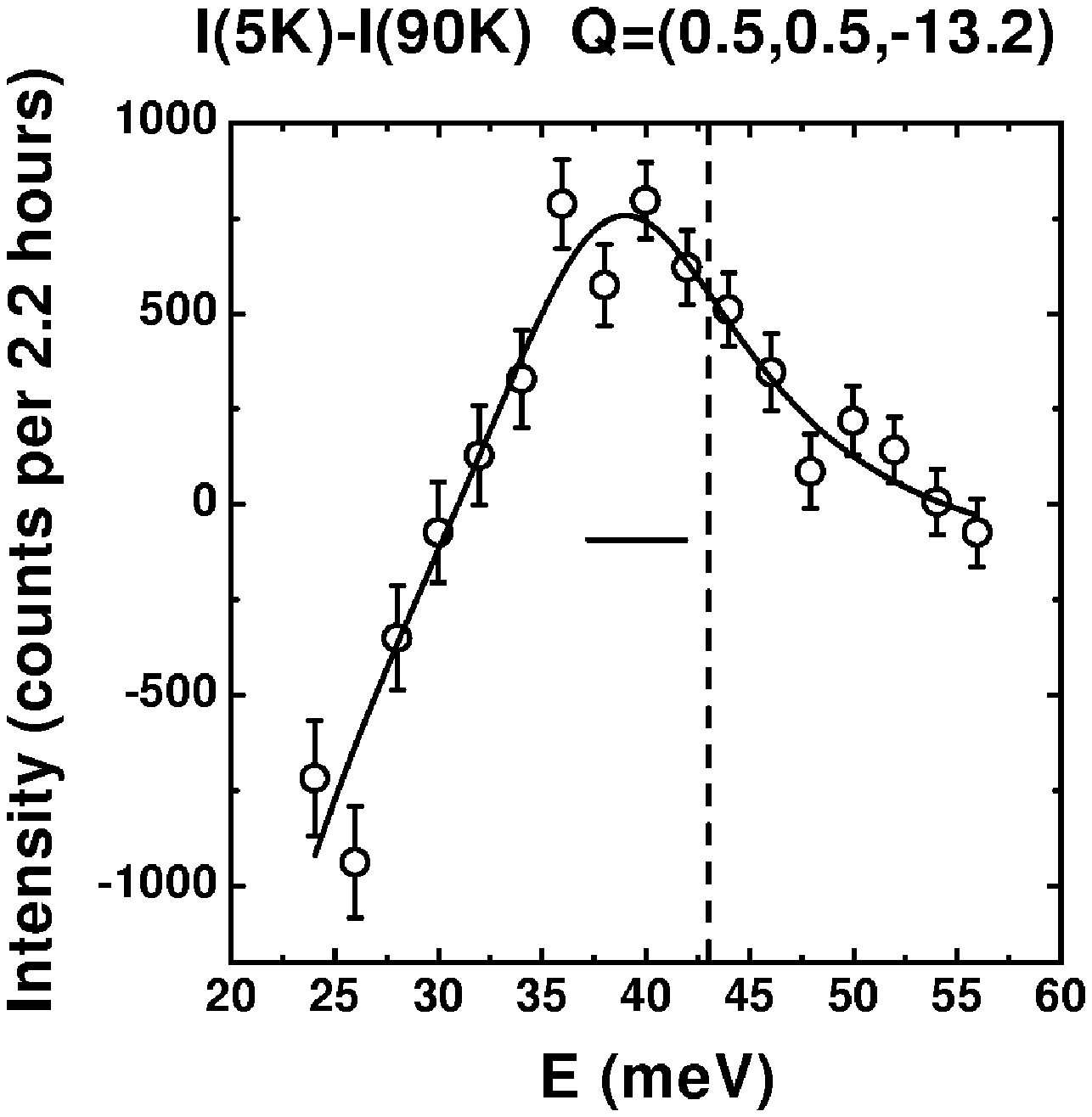}}
\hfill
\epsfxsize=0.44\textwidth{\epsfbox{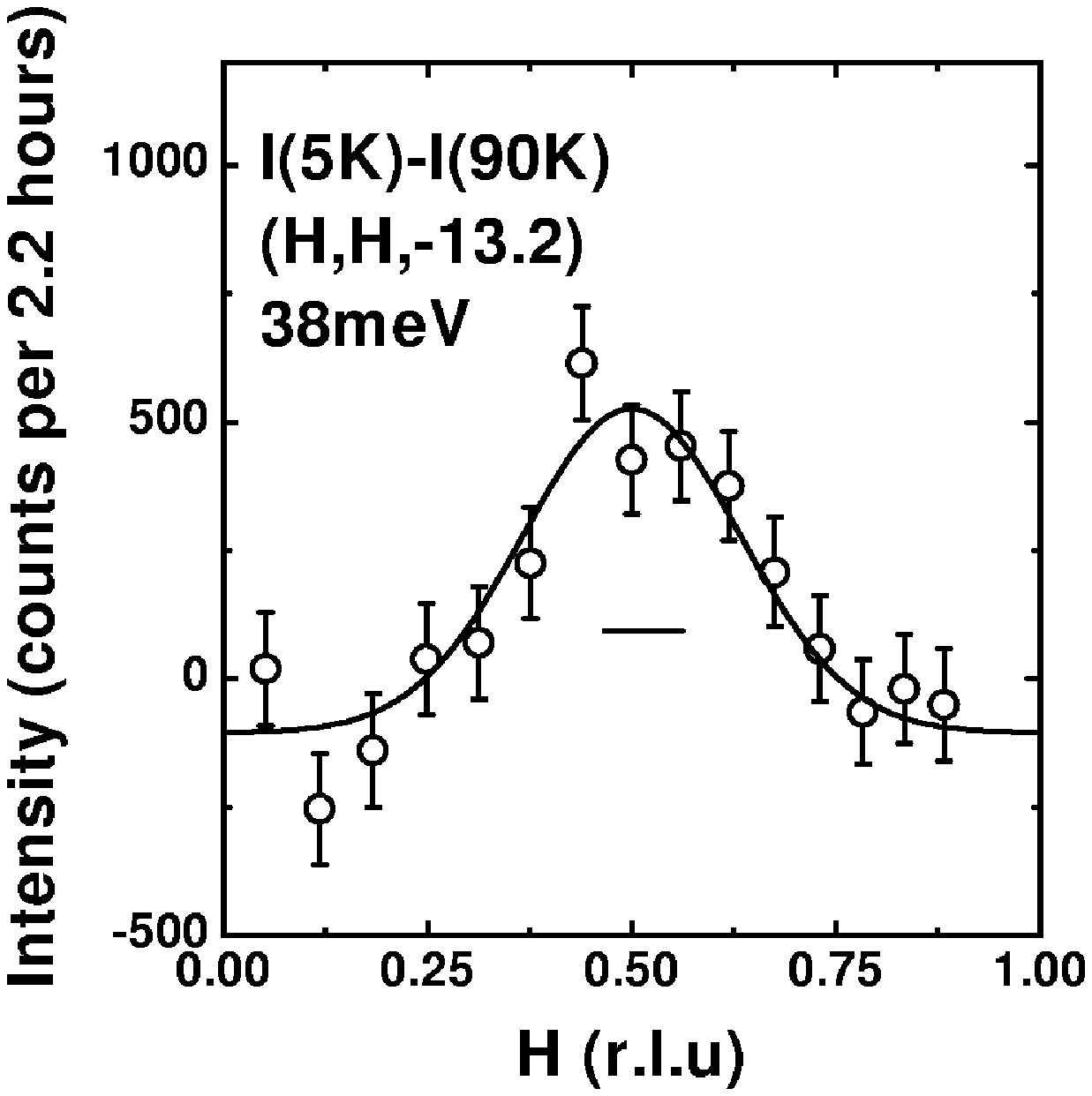}}
}
\caption{
\label{Fig2_He01}
Magnetic resonance in the odd channel for overdoped
Bi$_2$Sr$_2$CaCu$_2$O$_{8+\delta}$ 
($T_c$=83 K).
The difference spectra of the INS intensity measured at 5 K and 90 K at the
antiferromagnetic wavevector as a function of energy
shows a resonance peak at 38 meV (left picture).
The constant energy scans at resonance energy, shown in the right picture,
shows that the resonance is concentrated near the antiferromagnetic 
wavevector, however with a finite momentum width given in relative lattice units (i.e. units of
inverse lattice constant).
(From Ref. \cite{He01},
Copyright \copyright 2001 APS)
}
\end{figure}
In Fig. \ref{Fig2_He01} the INS data for overdoped
Bi${_2}$Sr${_2}$CaCu${_2}$O${_{8+\delta}}$ are reproduced \cite{He01}. The dashed line in the left panel of Fig. 
\ref{Fig2_He01} indicates the resonance energy for an optimally doped
Bi${_2}$Sr${_2}$CaCu${_2}$O${_{8+\delta}}$ sample. Also the instrumental resolution is shown.
The momentum width, shown in the right picture of Fig.
\ref{Fig2_He01} is somewhat broader than that for YBa$_2$Cu$_3$O$_{6+x}$.

Importantly, the resonance has not only been observed in bilayered cuprates, but also in 
the single layered cuprate Tl$_2$Ba$_2$CuO$_{6+\delta }$ \cite{He02}.
Thus, it is not a specific feature of closely spaced layers within a unit cell,
but an intrinsic property of the whole superconductor.

In the normal state these systems show a much weaker response, which is centered
around $\vec{Q}=(\pi,\pi)$ and is broader in momentum than in the superconducting state. In the pseudogap state, some intermediate picture is observed, with
a gradually sharpening response at the antiferromagnetic wavevector, which
can be regarded as a precursor of the magnetic resonance mode below $T_c$ 
\cite{Fong00,Bourges00,Dai01}.

The resonant feature has not been observed in the single-layered system
La$_{2-x}$Sr$_x$CuO$_4$. In contrast, in this compound the magnetic excitations are
strong both in the normal and superconducting state and located at incommensurate planar
wave vectors $\vec{Q}_\delta = (\pi (1\pm \delta )),\pi)$ and $(\pi,\pi(1\pm \delta))$ \cite{Shirane89,Mason96,Aeppli97,Yamada98}.
These incommensurate peaks are enhanced and sharpen in momentum at low energies ($E < 5k_BT_c$)
when entering the superconducting state \cite{Mason96}.
However, it was recently shown \cite{Christensen04} that a
dispersion similar to that in YBa$_2$Cu$_3$O$_{6+x}$ does also exist in
optimally doped La$_{2-x}$Sr$_x$CuO$_4$. In this case, however the maximal
intensity is at the low-energy incommensurate part of the dispersion spectrum.
The recent experiments \cite{Christensen04,Hayden04} support the idea that
the resonance peak in the 90 K cuprates and the incommensurate response in the La$_{2-x}$Sr$_x$CuO$_4$ possibly have a common origin.

The characteristic parameters for the resonance feature are summarized in Tab. \ref{res} for the different studied compounds.
In this table,
the intensity of the mode $W(\vec{Q})$ at the antiferromagnetic wavevector
is defined by $W=\int {\rm d}\omega {\rm Im} \chi(\vec{Q},\omega )$, 
and 
$\langle W(\vec{Q})\rangle $ denotes the momentum average of $W(\vec{q})$
over the entire Brillouin zone.

\begin{table*}
\begin{center}
\tbl{
\label{res}
Characteristic parameters for the resonance feature in different cuprate superconductors
as determined by inelastic neutron scattering experiments at low temperatures ($T\ll T_c$). 
Here, D is the doping (o=overdoped, u=underdoped, op=optimally doped), $S$ denotes the symmetry with respect to the exchange of the layers within the unit cell
(e=even, o=odd), $\Omega_{res}$ is
the resonance frequency, $\Delta Q$ its FWHM momentum width, $W(\vec{Q})$ its
weight at the antiferromagnetic wave vector, and $\langle W(\vec{Q})\rangle $ denotes the momentum averaged resonance intensity.
}
{\begin{tabular}{|c|c|c|c|c|c|c|c|c|}
\hline 
\hline 
Ref.& compound & D & $T_c$ & S & $\Omega_{res} $ & $\Delta Q $ & 
$W(\vec{Q}) $ & $\langle W(\vec{Q})\rangle $\\
& & & [K] & & [meV] & [$\AA^{-1}$] & [$\mu_B^2/{\rm f.u.} $] & 
[$\mu_B^2/{\rm f.u.}$] \\
\hline 
\hline 
\cite{Fong00}&YBa$_2$Cu$_3$O$_7$ & op & 93 & o & 40 & 0.25 &  1.6 & 0.043 \\
\hline 
\cite{Pailhes05a}&Y$_{0.85}$Ca$_{0.15}$Ba$_2$Cu$_3$O$_7$ & o & 75 & o & 34 &   & 0.45 & \\
& & & &e &  35 & & 0.18 & \\
\hline 
\cite{Pailhes03}&Y$_{0.9}$Ca$_{0.1}$Ba$_2$Cu$_3$O$_7$ & o & 85.5 & o & 36 &  0.36 & 1.2$^\ast $ & 0.042 \\
\cite{Pailhes05a}$^\ast $& & & &e &  43 & 0.45 & 0.6$^\ast $ & 0.036\\
\hline
\cite{Pailhes04} &YBa$_2$Cu$_3$O$_{6.85}$ & u & 89 & o & 41 &0.25  &1.8$^\ast $ & 0.07$^\dagger $ \\
\cite{Pailhes05a}$^\ast $& & & &e &  53 & 0.41 & 0.55$^\ast $ &\\
\cite{Fong00}$^\dagger $& & & &&  & & &\\
\hline 
\cite{Pailhes05a}&YBa$_2$Cu$_3$O$_{6.6}$ & u & 63 & o & 37 &   & 2.0 & \\
& & & &e &  55 & & 0.2 & \\
\hline
\cite{Fong00} & YBa$_2$Cu$_3$O$_{6.7}$ &u  & 67 & o & 33  & 0.25 
& 2.1  & 0.056 \\
\hline
\cite{Fong00} & YBa$_2$Cu$_3$O$_{6.5}$ & u & 52 & o & 25 & 0.25 
& 2.6 & 0.069 \\
\hline
\cite{Sidis00} & YBa$_2$(Cu$_{0.97}$Ni$_{0.03}$)$_3$O$_7$ & & 80 & o & 35 & 0.49 
 & 1.6  & 0.2 \\
\hline
\cite{Fong99a} & YBa$_2$(Cu$_{0.995}$Zn$_{0.005}$)$_3$O$_7$ & & 87 & o & 40 & 0.25 
& 2.2 & 0.056 \\
\hline
\cite{Sidis00} & YBa$_2$(Cu$_{0.99}$Zn$_{0.01}$)$_3$O$_7$ & & 78 & o & 38 & 0.44&& \\
\hline
\cite{Fong99} & Bi$_2$Sr$_2$CaCu$_2$O$_{8+\delta}$& op & 91 & o & 43 & 0.52 
& 1.9 & 0.23 \\
\hline 
\cite{He01} & Bi$_2$Sr$_2$CaCu$_2$O$_{8+\delta}$& o & 83 & o & 38 &&&\\
\hline 
\cite{He02}&Tl$_2$Ba$_2$CuO$_{6+\delta }$ & op &92.5 & - & 47 & 0.23  & 0.7 & 0.02 \\
\hline 
\hline 
\end{tabular}}
\end{center}
\end{table*}

\subsubsection{Bilayer effects}
\label{secBil}
In doubly layered materials, under the assumption of coherent coupling
between the planes within a bilayer, the dispersion is classified by the notion of bonding bands (BB)
and antibonding bands (AB). 
In contrast, if the coupling is predominantly incoherent, a classification
according to the layer index is more appropriate.

Because of the symmetry under exchange of the planes within a bilayer, the susceptibility
$\chi_{ij}$ (where $i,j=1,2$ are layer indices) has only two independent components,
$\chi_\parallel \equiv \chi_{11}=\chi_{22}$ and $\chi_\perp \equiv \chi_{12}=\chi_{21}$
\cite{Millis96}.
Using those, the neutron scattering cross section for bilayer cuprates is given by
\begin{equation}
\label{cross1}
\frac{d^2 \sigma }{d\Omega dE} \sim
F^2(\vec{Q} ) \left[ \chi''_\parallel(\vec{Q},\omega )+
\cos(Q_zd) \chi''_\perp (\vec{Q},\omega )\right],
\end{equation}
where $\chi''_{\parallel,\perp }(\vec{Q},\omega )$ is the imaginary part of the dynamical magnetic 
susceptibility within and between the layers, respectively, 
$d\approx 3.3 \AA $ is the distance between the CuO$_2$ planes
within a bilayer, and $F(\vec{Q})$ is the magnetic form factor of the Cu$^{2+}$ ion \cite{Fong00}.

It is common to introduce components for excitations 
even and odd under interchange of planes within a bilayer, 
$\hat S_z^e=(\hat S_z^1+\hat S_z^2)/2$ and $\hat S_z^o=(\hat S_z^1-\hat S_z^2)/2$. 
The corresponding even and odd susceptibilities are,
\begin{eqnarray}
\label{evenodd1}
\chi_e  &\equiv & \langle T_\tau \hat S_z^e(\vec{Q},\tau) \hat S_z^e(-\vec{Q},0)\rangle 
=
\chi_\parallel (\vec{Q},\omega ) +\chi_\perp (\vec{Q},\omega )\\
\label{evenodd1a}
\chi_o (\vec{Q},\omega ) &\equiv & \langle T_\tau \hat S_z^o(\vec{Q},\tau) \hat S_z^o(-\vec{Q},0)\rangle 
=
\chi_\parallel(\vec{Q},\omega )-\chi_\perp(\vec{Q},\omega ).
\end{eqnarray}
The neutron scattering cross section for bilayer cuprates is given in terms of even and odd 
susceptibilities by,
\begin{eqnarray}
\label{cross2}
\frac{d^2 \sigma }{d\Omega dE} \sim &&
F^2(\vec{Q} ) \Big[ 
\sin^2(\frac{Q_zd}{2}) \chi''_o(\vec{Q},\omega ) 
+
\cos^2(\frac{Q_zd}{2}) \chi''_e(\vec{Q},\omega )\Big],
\end{eqnarray}
where $\chi''_{o/e}(\vec{Q},\omega )$ is the imaginary part of the dynamical magnetic susceptibility in
the odd and even channels, respectively \cite{Fong00}.

In the case of coherent coupling of the planes within a bilayer the more appropriate classification
is in terms of susceptibilities within the basis of bonding  ($b$) and antibonding  ($a$) bands.
In this case, the odd susceptibility component describes
scattering between opposite type of bands, and the even susceptibility describes
scattering between same type of bands, according to \cite{Tranquada92}
\begin{eqnarray}
\label{evenodd2}
\chi_e (\vec{Q},\omega )&=& \frac{1}{2}\left[\chi_{aa} (\vec{Q},\omega )+ \chi_{bb}(\vec{Q},\omega )\right]\\
\label{evenodd2a}
\chi_o (\vec{Q},\omega )&=& \frac{1}{2}\left[ \chi_{ab}(\vec{Q},\omega ) + \chi_{ba}(\vec{Q},\omega )\right].
\end{eqnarray}

The spin resonance
was for a long time only observed in the odd channel \cite{Rossat93}, where it
lies below a gapped continuum, the latter having a signal typically
a factor of 30 less than the maximum at $\vec{Q}$ at the mode
energy \cite{Bourges98}.
The continuum is gapped in both the even and odd scattering channels
(the even channel is gapped by $\approx 60 $ meV even in the normal state) \cite{Reznik96}.

\begin{figure}
\centerline{
\epsfxsize=0.55\textwidth{\epsfbox{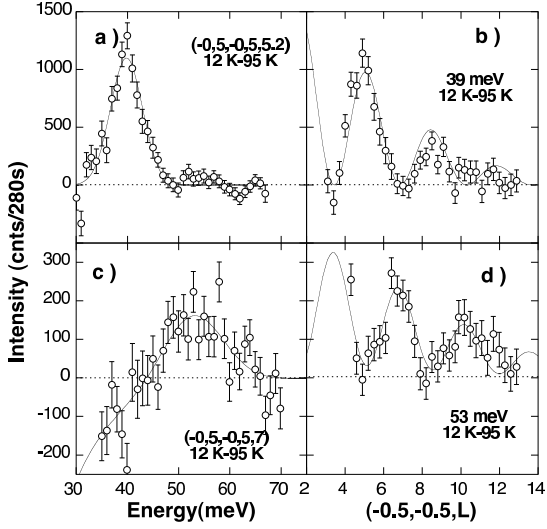}}
}
\caption{
\label{Fig1_Keimer04}
Differences of the INS intensity measured at 12 K and 95 K for
an underdoped YBa$_2$Cu$_3$O$_{6.85}$ sample ($T_c=89$ K, \cite{Pailhes04}).
a) and b) are for the odd channel, c) and d) for the even channel.
The INS intensity as function of energy shows resonance peaks at
39 meV in the odd channel and at 53 meV in the even channel.
The constant energy scans at resonance energy, shown in b) and d),
reveal the typical sin-square modulation for the odd-channel mode and
the cos-square modulation for the even-channel mode (full lines).
(From Ref. \cite{Pailhes04}, 
Copyright \copyright 2004 APS).
}
\end{figure}
Recently the resolution of neutron scattering experiments has increased
considerably to allow for the observation of the resonance mode also
in the channel even with respect to the layer-interchange within
a bilayer. 
The corresponding mode has been observed in both overdoped
and underdoped YBa$_2$Cu$_3$O$_{7-\delta}$ \cite{Pailhes03,Pailhes04,Pailhes05a}.
As shown in Fig. \ref{Fig1_Keimer04} b) and d) for underdoped
YBa$_2$Cu$_3$O$_{6.85}$,
the dependence of the magnetic INS signal as function
of $Q_z$ shows the characteristic sin-squared and cos-squared modulations for the odd and
even channels, respectively \cite{Pailhes04}.
The corresponding resonance energies, as obtained from
Fig. \ref{Fig1_Keimer04} a) and c), are  39 meV in the odd channel, and 53 meV in the even channel.
The intensity in the even channel is much smaller than in the odd channel, which is the reason
why it was for such a long time overlooked.

The peak energy in overdoped Y$_{0.9}$Ca$_{0.1}$Ba$_2$Cu$_3$O$_7$ for the
odd channel is at 36 meV,
lower than the mode energy in optimally doped YBa$_2$Cu$_3$O$_{7-\delta}$.
The corresponding peak width in momentum space around $\vec{Q}=(\pi,\pi)$
is $\Delta Q = 0.36 \pm 0.05 \AA^{-1}$.
The even channel resonance mode is at 43 meV, and has a 
$Q$-width of $\Delta Q = 0.45 \pm 0.05 \AA^{-1}$.
The intensity data of the two modes
are shown in Table \ref{res}.

\subsubsection{Temperature dependence}
\label{temp_res}

\begin{figure*}
\centerline{
\hfill
\epsfxsize=0.35\textwidth{\epsfbox{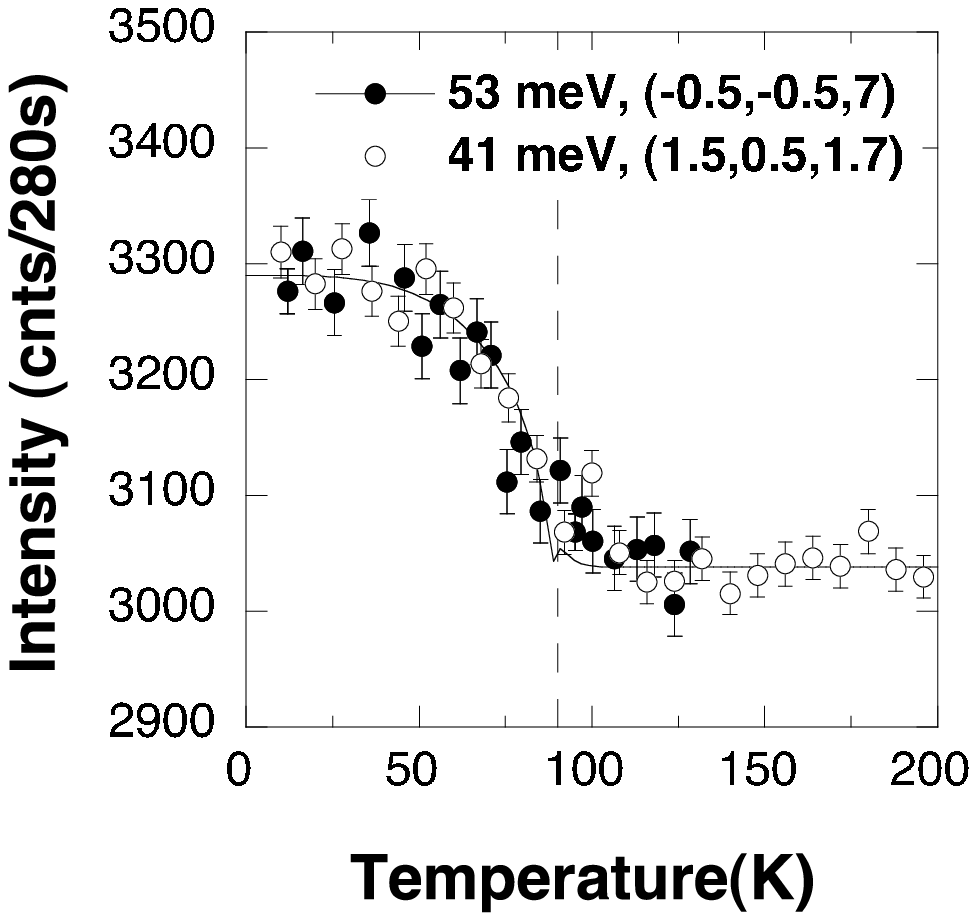}}
\hfill
\epsfxsize=0.45\textwidth{\epsfbox{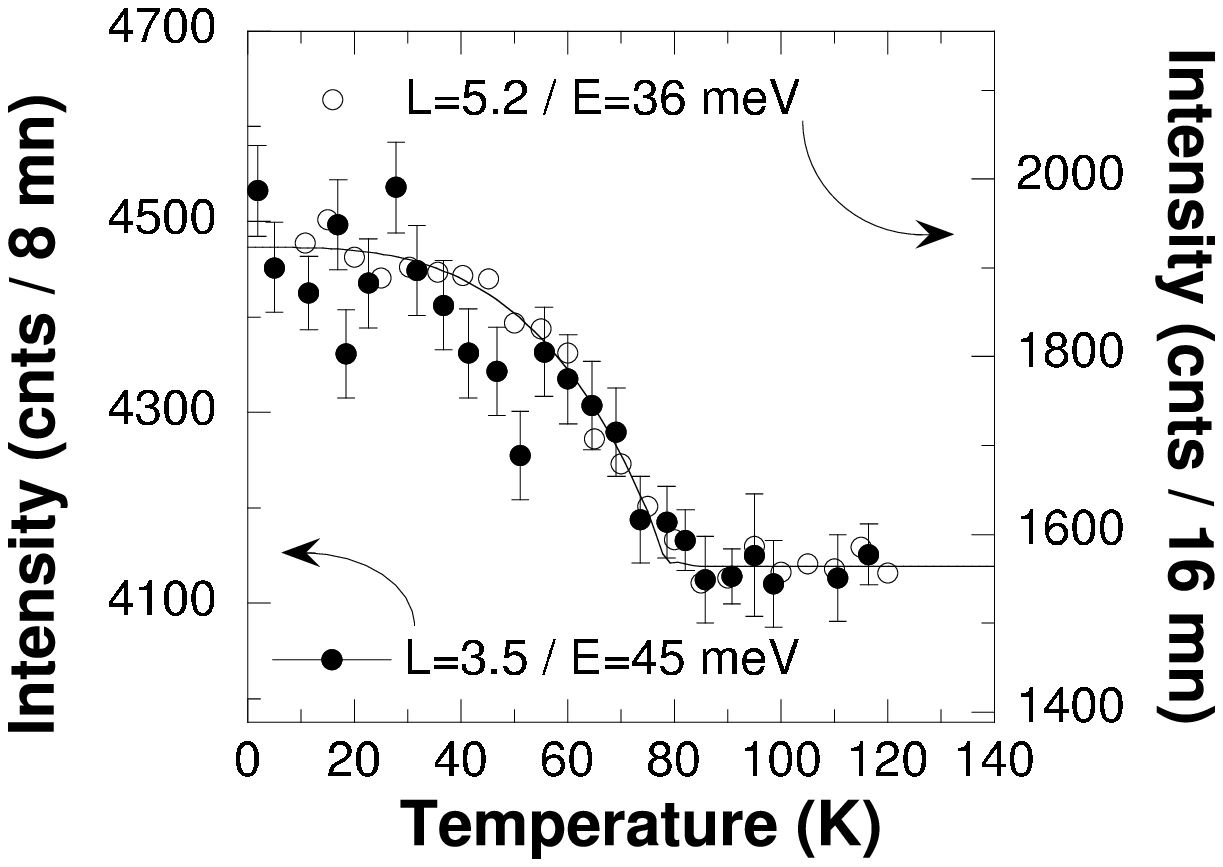}}
\hfill
}
\caption{
\label{Fig3_KeimerBil}
Temperature dependence of the neutron scattering intensity at the
antiferromagnetic wavevector for the resonance mode in the even channel
(full circles) and in the odd channel (open circles). The
even mode intensity has been rescaled to that of the odd mode.
The measurements were performed 
by INS in underdoped 
YBa$_2$Cu$_3$O$_{6.85}$
(left picture, $T_c$=89 K, \cite{Pailhes04}) and overdoped
Y$_{0.9}$Ca$_{0.1}$Ba$_2$Cu$_3$O$_7$
(right picture, $T_c$=85.5 K, \cite{Pailhes03}) YBa$_2$Cu$_3$O$_{7-\delta}$.
(From Refs. \cite{Pailhes03},
Copyright \copyright 2003 APS,
and \cite{Pailhes04},
Copyright \copyright 2004 APS).
}
\end{figure*}

A sharp resonance mode is not observed above $T_c$ \cite{Fong95,Bourges00}.
However, a broadened version is present in the pseudogap state \cite{Dai01}, which can be
regarded as a pre-cursor for the resonance.
On approaching $T_c$ from below the resonance {\it energy} does not change \cite{Bourges96,Fong95,Dai96},
however its {\it intensity} is vanishing toward $T_c$ for optimally doped compounds,
following an order parameter like behavior \cite{Rossat91,Mook93,Bourges96,Dai96,Bourges00}.
The temperature dependence of the even and odd resonance mode intensity is, when
properly rescaled, identical \cite{Pailhes03,Pailhes04}. This is shown in Fig. \ref{Fig3_KeimerBil}
for moderately underdoped YBa$_2$Cu$_3$O$_{6.85}$ and overdoped Y$_{0.9}$Ca$_{0.1}$Ba$_2$Cu$_3$O$_7$.
The peak amplitude of both the even and odd mode vanish at $T_c$.
\begin{figure*}
\centerline{
\hfill
\epsfxsize=0.32\textwidth{\epsfbox{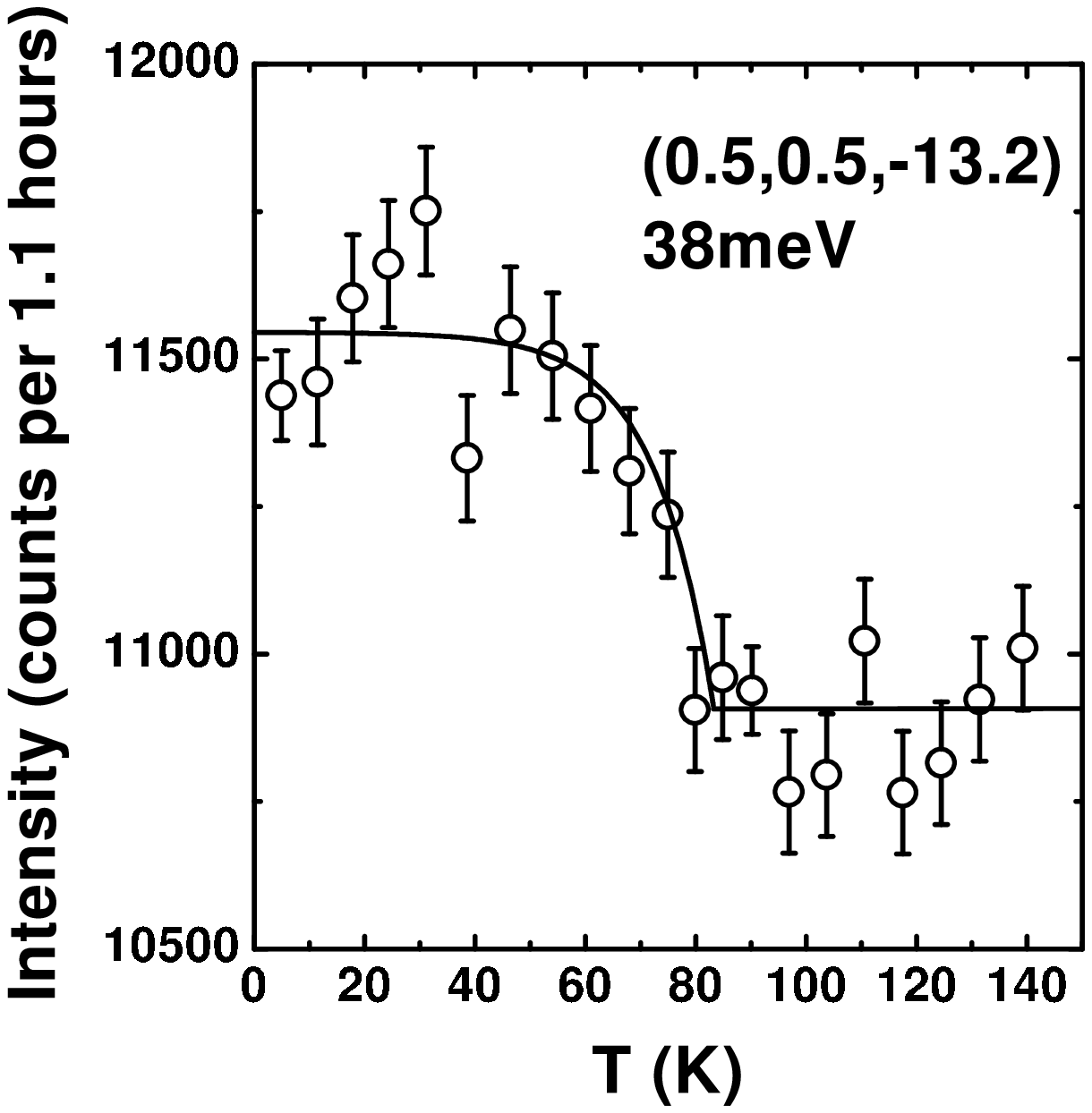}}
\hfill
\epsfxsize=0.43\textwidth{\epsfbox{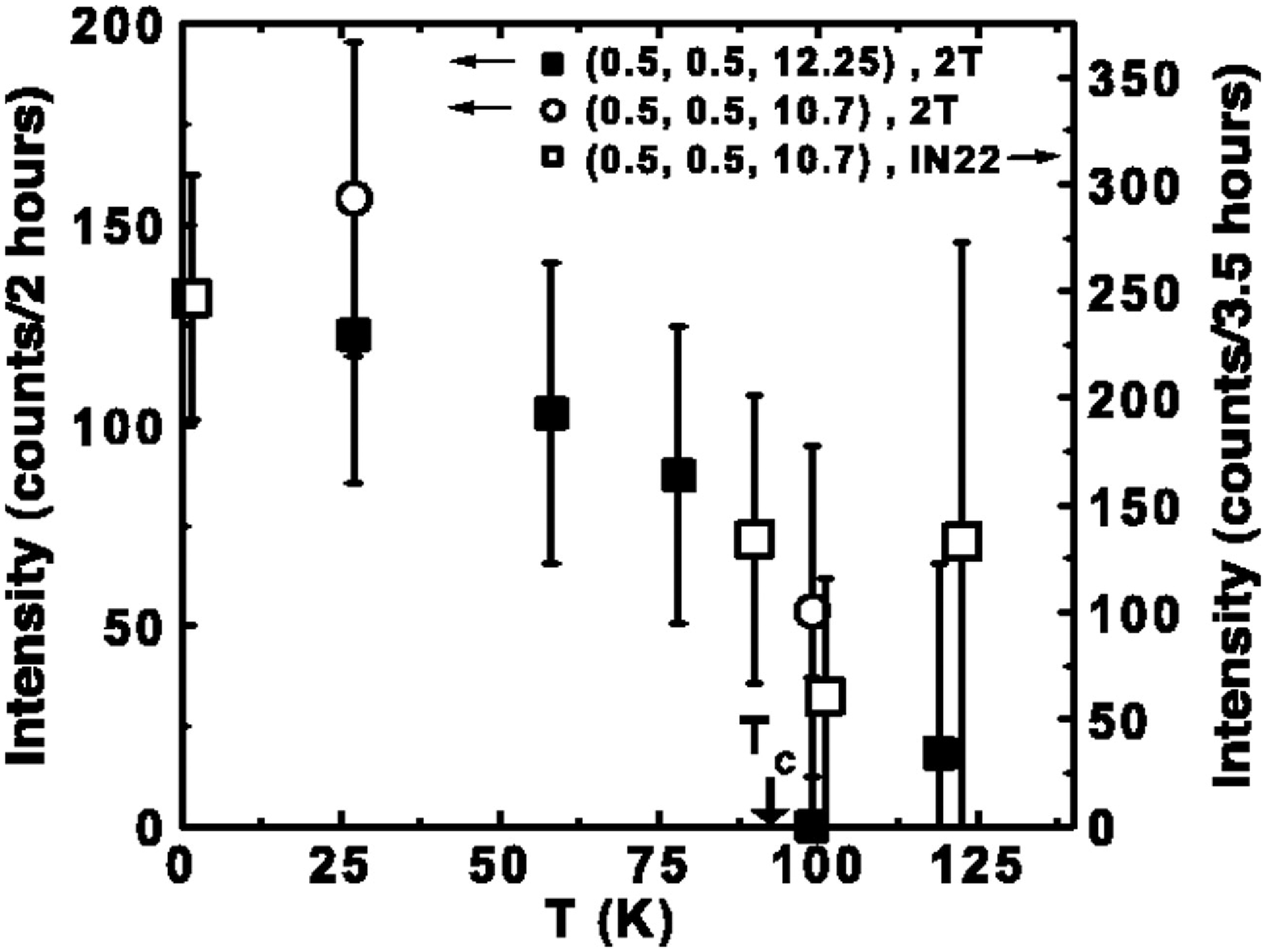}}
\hfill
}
\caption{
\label{Fig4_He}
Temperature dependence of the neutron scattering intensity at the
antiferromagnetic wavevector for the odd-channel resonance mode 
in overdoped Bi$_2$Sr$_2$CaCu$_2$O$_{8+\delta}$ (left picture, $T_c$=83 K, \cite{Fong99,He01}) 
and for the
single layered compound Tl$_2$Ba$_2$CuO$_{6+\delta }$ near optimal doping
(right picture, $T_c$=90 K, \cite{He02}).
(From Ref. \cite{He01}, 
Copyright \copyright 2001 APS,
and Ref. \cite{He02}, 
Reprinted with permission from Science, Ref. \cite{He02}. Copyright \copyright 2002 AAAS).
}
\end{figure*}

When going to stronger underdoped samples, the INS intensity at resonance energy
and antiferromagnetic wavevector is present also above $T_c$ up to a temperature
which characterizes pseudogap phenomena in cuprates \cite{Dai99}.
This correlates with other thermodynamic quantities like the specific heat as
function of temperature for different degrees of doping \cite{Dai99}.
It was argued, that a similar correspondence exists also as a function of
applied magnetic field \cite{Janko99,Dai00}.

The characteristic temperature dependence of the mode intensity was observed also
in overdoped Bi$_2$Sr$_2$CaCu$_2$O$_{8+\delta}$ \cite{Fong99,He01}
and in the single layered compound Tl$_2$Ba$_2$CuO$_{6+\delta }$ near optimal doping,
as shown in Fig. \ref{Fig4_He}.

\subsubsection{Doping dependence}
\label{Dop_res}

The width of the resonance in the odd channel
is smaller than the instrumental resolution
(of typically less than 10 meV) for optimally and moderately underdoped
materials. Strongly underdoped materials show a small broadening of the order
of 10 meV \cite{Bourges98,Fong00}.
The mode frequency decreases with underdoping and has its
maximal value of about 40 meV at optimal doping \cite{Dai96,Fong97,Bourges97,Dai99}.
In both underdoped and overdoped regimes, the resonance energy,
$\Omega_{res}$, is proportional to $T_c$ with $\Omega_{res}\approx
(5\ldots 5.5) k_BT_c$ \cite{Bourges98,Fong00,Dai99,Fong99,He01}.
\begin{figure}
\centerline{
\epsfxsize=0.65\textwidth{\epsfbox{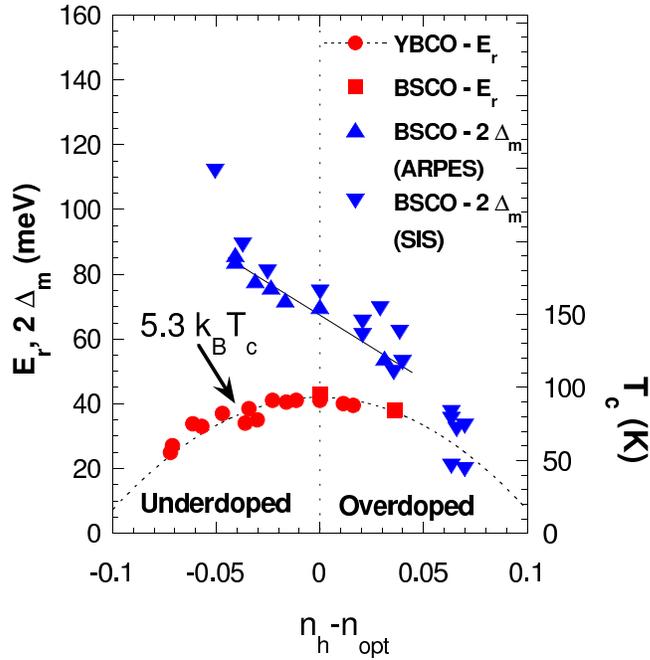}}
}
\caption{
\label{Fig2_KeimerDop}
Doping dependence of magnetic-resonance energy, $\Omega_{res}$ ($\equiv E_r$), as measured by INS in
YBa$_2$Cu$_3$O$_{7-\delta}$
\cite{Rossat91,Mook93,Fong95,Bourges96,Fong00,Dai01},
and Bi$_2$Sr$_2$CaCu$_2$O$_{8+\delta}$ \cite{Fong99,He01}.
For comparison, twice the maximal superconducting gap, $\Delta_m$, 
for Bi$_2$Sr$_2$CaCu$_2$O$_{8+\delta}$,
as determined from ARPES \cite{Mesot99} and from SIS tunneling
\cite{Zasadzinski01}, is shown.
(From Ref. \cite{Sidis04}, Copyright \copyright 2004 WILEY-VCH).
}
\end{figure}
In Fig. \ref{Fig2_KeimerDop} we reproduce the data from Ref. \cite{Sidis04}.
As can be seen the resonance-mode energy tracks very precisely the curve for
5.3 $T_c$. Also shown are the values for twice the maximal superconducting gap as
determined from ARPES \cite{Mesot99} and from SIS break junctions tunneling data
\cite{Zasadzinski01}. The resonance feature stays always below this continuum
edge, indicating an excitonic origin. The doping dependence in
Fig.~\ref{Fig2_KeimerDop} should be compared with the right panel in
Fig.~\ref{Fig3_Zasadsinski}. The similarity is striking.

The total spectral weight related to the resonance peak remains
approximately constant as a function of doping, and amounts to
0.06$\mu_B^2$ per formula unit at low temperatures \cite{Dai99,Fong00}.
This represents about 2\% of the spectral weight contained in the
spin-wave spectra of the undoped materials.
With underdoping 
the intensity of the resonance
at $\vec q=(\pi,\pi)$ increases from about 1.6 $\mu_B^2$
for YBa$_2$Cu$_3$O$_7$ to about 2.6 $\mu_B^2$ per unit volume for
YBa$_2$Cu$_3$O$_{6.5}$ \cite{Bourges98,Fong00}.
With overdoping, the intensity at the antiferromagnetic wavevector decreases,
however there are no data available for strongly overdoped samples.

\subsubsection{Dependence on disorder}
\label{Dis_res}

In cuprates the superconducting transition temperature can be varied also without
changing the carrier concentration by introducing disorder through impurity substitution
in the CuO$_2$-layers. Due to the unconventional energy gap such impurities have a strong
effect on equilibrium properties of the superconducting state, in contrast to conventional
$s$-wave superconductors.
Two types of impurities were used in substitution for Cu$^{2+}$ ions in inelastic neutron scattering
experiments. 
First, non-magnetic Zn$^{2+}$ ions ($3d^{10}, S=0 $ configuration) \cite{Fong99a,Sidis00},
and second, magnetic Ni$^{2+}$ ions ($3d^8, S=1$ configuration) \cite{Sidis96,Sidis00}.
In general, the influence on the superconducting transition temperature
of non-magnetic impurities is stronger than the influence of magnetic impurities in cuprates.
The $T_c$ reduction is 3 times stronger for Zn$^{2+}$ ions than for Ni$^{2+}$ 
ions \cite{Tarascon88,Mendels99}.
It was found that for YBa$_2$(Cu$_{1-y}$Ni$_y$)$_3$O$_7$ with $y=3\% $,
$T_c=80 $K, the resonance shifted to lower energy, preserving the ratio $\Omega_{res}/k_BT_c$,
whereas for YBa$_2$(Cu$_{1-y}$Zn$_y$)$_3$O$_7$ 
the shift of the resonance energy with impurity doping is much smaller \cite{Sidis96,Sidis00}.
The intrinsic energy width of the resonance peak is very sensitive to both types of impurities, being
$\Delta E=11 $meV for YBa$_2$(Cu$_{0.997}$Ni$_{0.003}$)$_3$O$_7$ and
$\Delta E=9 $meV for YBa$_2$(Cu$_{0.999}$Zn$_{0.001}$)$_3$O$_7$ \cite{Sidis00}.
However, there are differences in the temperature dependence of the magnetic response function
for the two types of impurities. Whereas Ni impurities do not measurably enhance the normal state
response, a broad peak with characteristic energy somewhat lower than the resonance energy
of pure YBa$_2$Cu$_3$O$_7$ appears in the normal state for systems containing Zn impurities \cite{Sidis00}.

\subsubsection{Isotope effect}
\label{Iso_res}

Very recently also the influence of an change of the oxygen isotope was
studied in YBa$_2$Cu$_3$O$_{6.89}$ \cite{Pailhes05}. 
It was shown, that there is no shift in the
resonance frequency when exchanging the oxygen isotope  $^{16}$O
by $^{18}$O. This shows the absence of interaction
between the spin-1 excitation and phonons in
high-$T_c$ cuprates near optimal doping.
However, the amplitudes of the
peaks are slightly different, and also the energy widths differ slightly;
the energy integrated magnetic spectral weight, however,
stays unaffected. This modifications could possibly be related to
a certain amount of introduced disorder due to isotope exchange.

\subsubsection{Dependence on magnetic field}
\label{Magn_res}

It was found that a
c-axis magnetic field suppresses the intensity of the magnetic resonance \cite{Dai00},
as predicted from an analysis of specific heat data \cite{Janko99}.
Since the same effect was not observed for in-plane fields \cite{Bourges97a},
this indicates that the resonance is sensitive to the presence of Abrikosov
vortices, and thus intimately connected to the nature of the superconducting
ground state.  This has obvious implications for microscopic theories of
the resonance.
\begin{figure}
\centerline{
\epsfxsize=0.8\textwidth{\epsfbox{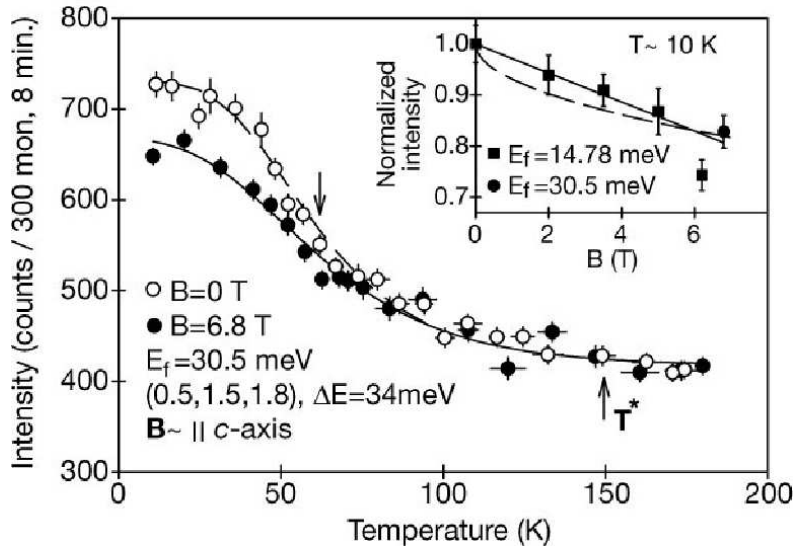}}
}
\caption{
\label{Fig4_DaiNa}
Effect of the magnetic field on the temperature dependence of the
magnetic resonance intensity and field dependence of the resonance intensity
in underdoped YBa$_2$Cu$_3$O$_{6.6}$  ($T_c=$ 62.7 K, odd resonance energy
$\Omega_{res}=$ 34 meV)
at the antiferromagnetic wavevector at $T\sim 10 $K \cite{Dai00}.
The magnetic field points in $c$-axis direction. Open circles are for zero
field, full circles for $B=$6.8 T ($E_f$ denotes the final neutron energy). 
The inset shows the magnetic field 
dependence of the normalized resonance intensity at 10 K, the solid line
corresponds to $I/I_0 = 1-(B/\mbox{36 T})$. The characteristic field of 36 T
is not far from the upper critical field $B_{c2}=$45 T for this sample.
(Reprinted with permission from Mcmillan Publishers Ltd: Nature, Ref. \cite{Dai00}, Copyright \copyright 2000 NPG).
}
\end{figure}
As shown in the inset of Fig. \ref{Fig4_DaiNa},
the experimental suppression goes like $1-H/H^*$, which is highly suggestive
of a vortex core effect, as originally noted by Dai {\it et al.} \cite{Dai00}.
It is interesting to remark that the sample studied experimentally had an
anomalously long magnetic correlation length. Other samples studied by
neutron scattering have a significantly smaller correlation length \cite{Dai01}.
This fact probably lead to a larger effect of the magnetic field on the magnetic resonance
than in other samples, allowing its experimental observation.
The resulting temperature dependence for the resonance intensity with and without applied
magnetic field is reproduced in Fig.  \ref{Fig4_DaiNa}.
As the sample is strongly underdoped, there is a considerable magnetic intensity at the resonance energy
left even above $T_c$, 
as mentioned in Subsection \ref{temp_res}.
The suppression of the mode intensity with magnetic field in $c$-direction is clearly visible below $T_c$.

\subsubsection{The incommensurate part of the spectrum}
\label{Incomm}

\begin{figure}
\centerline{
\epsfxsize=0.78\textwidth{\epsfbox{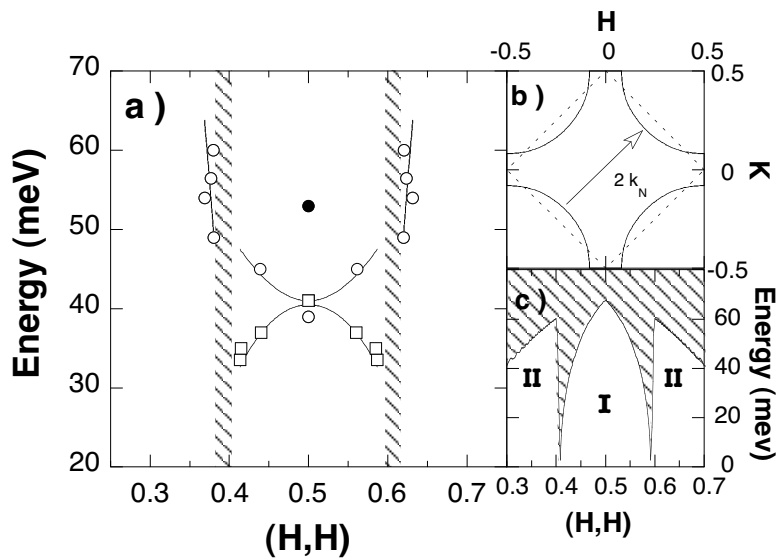}}
}
\caption{
\label{Fig5_Keimer04}
Dispersion of the incommensurate magnetic response 
for underdoped YBa$_2$Cu$_3$O$_{6.85}$ along the (110) direction of the
magnetic Brillouin zone \cite{Pailhes04}.
The open symbols in a) show
the incommensurate spin excitations and the 41 meV resonance mode at the antiferromagnetic
wavevector in the odd channel, the full circle
shows the 53 meV resonance mode in the even channel. Below 30 meV the 
magnetic INS intensity is strongly reduced due to a spin-gap. In the hatched
region no magnetic spin-excitations are observed. The electron-hole continuum
for magnetic spin excitations is shown in c) along the (110) direction for
a $d$-wave superconductor with maximal gap of 35 meV. The inset b) shows the
wavevector within the fermionic Brillouin zone, for which the 
spin continuum goes to zero.
(From Ref. \cite{Pailhes04},
Copyright \copyright 2004 APS).
}
\end{figure}

There has been observed an incommensurate response 
both above and 
(for bilayer materials in the odd channel) below the
magnetic resonance energy. The incommensurate spectrum above the resonance
energy is broad in momentum and shows a dispersion similar to spin waves
\cite{Bourges97,Arai99,Bourges00a,Pailhes04,Stock05}. Below the resonance energy
an incommensurate response was observed in underdoped \cite{Dai01,Mook98b,Arai99,Dai98,Pailhes04}
and optimally doped \cite{Dai01,Bourges00a,Bourges00} 
YBa$_2$Cu$_3$O$_{6+x}$ at the incommensurate wavevectors
$\vec{q}=(\pi \pm \delta,\pi)$ and $(\pi, \pi \pm \delta)$. 
This kind of incommensurability is similar to that observed in 
La$_{2-x}$Sr$_x$CuO$_4$ \cite{Mason93,Yamada95,Christensen04}.
The corresponding four peaks in momentum space disperse away
from the antiferromagnetic wavevector with energy decreasing from the resonance
energy \cite{Arai99,Bourges00a,Pailhes04}. 
In contrast, above the resonance a new type of resonant feature arises,
the so-called `Q$^{\ast}$ mode' \cite{Pailhes04,Eremin05}, 
which shows an incommensurate pattern along
the zone diagonal with maxima at $(\pi\pm\delta^\ast ,\pi\pm\delta^\ast )$
\cite{Hayden04}.
The resulting hour-glass shape dispersion below and 
above the resonance is shown in (110) direction in Fig.~\ref{Fig5_Keimer04}.
As can be seen in a), the magnetic resonance is part of an incommensurate
response which extends in energy down to an energy of $\sim 30 $ meV (for
smaller energies the intensity drops below the background level).
In a certain distance from the antiferromagnetic wavevector on either side,
the dispersion is interrupted by a momentum region, in
which no resonant magnetic spin-excitations are observed. This region, shown as hatched
area in Fig. \ref{Fig5_Keimer04}, can be identified with the region in which
particle-hole-continuum excitations exist. For a $d$-wave superconductor such
a region extends all the way down to zero energy, as shown in Fig. \ref{Fig5_Keimer04} c),
corresponding to
continuum excitations due to node-node scattering. The corresponding wavevector
is given by the node-node wavevector, $2k_N$, which in superconducting cuprates is
slightly displaced from the $(\pi,\pi )$ antiferromagnetic wavevector as can be seen from
Fig. \ref{Fig5_Keimer04} b).

In connection with the fact, that the resonance is part of a dispersive
spin excitation branch,
it is important to realize that the momentum width of the resonance 
is inhomogeneously broadened as a result of a finite
energy window in experiments. Depending on the degree of flatness of the
dispersion near the resonance energy the measured momentum width can differ
considerably. This is a possible reason why in 
Bi$_2$Sr$_2$CaCu$_2$O$_{8+\delta}$ the resonance has a much broader momentum
width than in YBa$_2$Cu$_3$O$_{6+x}$.

\subsubsection{The spin gap}
\label{Spingap}

In fact, the incommensurate excitations are not observed experimentally
down to zero energy. Instead,
in the low energy region of the incommensurate excitations the
INS intensity drops drastically. This `spin-gap' was measured in the
superconducting state of 
La$_{2-x}$Sr$_x$CuO$_4$ \cite{Yamada95,Lake99}, where it is of size 3-6 meV,
as well as in the superconducting state of YBa$_2$Cu$_3$O$_{6+x}$.
In both systems the spin-gap phenomenon was shown to be sensitive to
disorder, with impurities introducing additional states below the spin gap
\cite{Kakurai93,Sidis96}. 
For La$_{2-x}$Sr$_x$CuO$_4$ it was shown in addition that the spin gap
is sensitive to an applied magnetic field in $c$-direction, which also
introduces additional states \cite{Lake01}.

\begin{figure*}
\centerline{
\epsfxsize=0.46\textwidth{\epsfbox{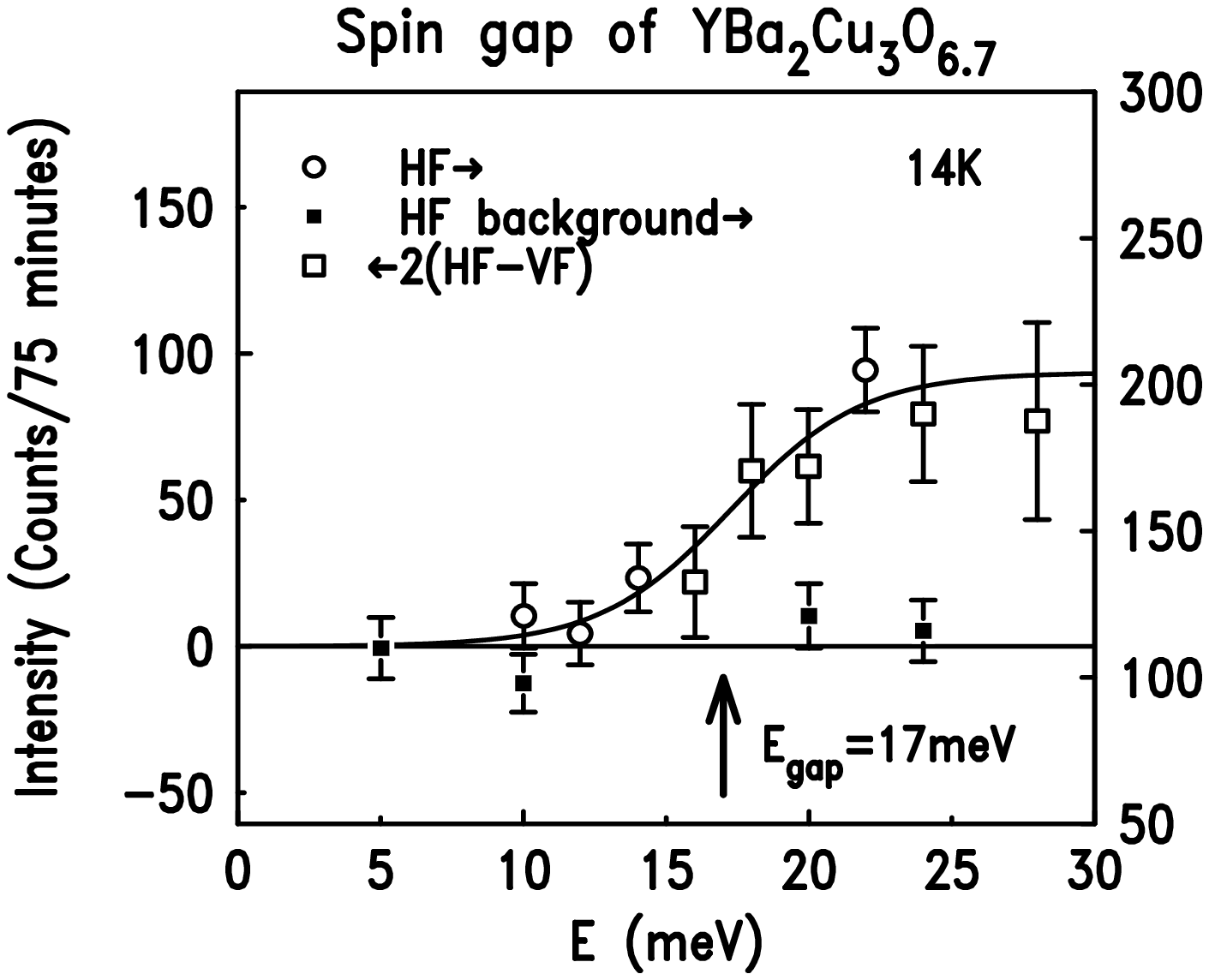}}
\epsfxsize=0.52\textwidth{\epsfbox{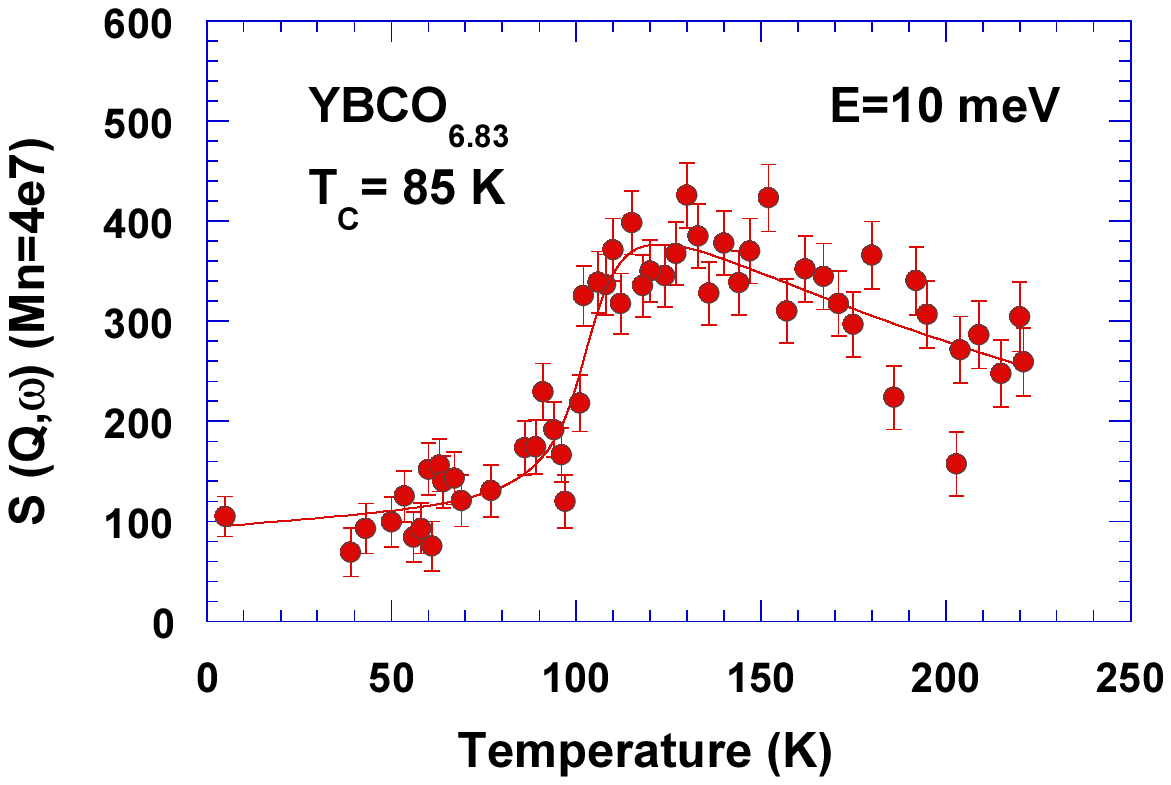}}
}
\caption{
\label{Fig13_Bourges98}
Left:
Spin gap in YBa$_2$Cu$_3$O$_{6.7}$ ($T_c=$67 K, $\Omega_{res}=$ 33 meV) at $T= $14 K 
\cite{Fong00}. 
Right: Temperature dependence of the INS peak intensity at low energy in
YBa$_2$Cu$_3$O$_{6.83}$ \cite{Bourges98}.
(From Refs. \cite{Fong00} 
Copyright \copyright 2000 APS,
and \cite{Bourges98},
Copyright \copyright 1998 Springer).
}
\end{figure*}
In Fig. \ref{Fig13_Bourges98}, left picture, the spin gap is shown for underdoped
YBa$_2$Cu$_3$O$_{6.7}$, in which case it amounts to 17 meV \cite{Fong00}. 
In general the spin-gap magnitude follows closely $T_c$ according
to $E_{sg}=3.8 k_{\rm B}T_c$ \cite{Dai01}. Only at very low doping it
deviates from this linear relation, and shows a smaller spin-gap, e.g.
of 5 meV in YBa$_2$Cu$_3$O$_{6.5}$ \cite{Fong00}.
The temperature dependence of the INS intensity at low energy is shown
in Fig. \ref{Fig13_Bourges98}, right picture. It shows that in underdoped
materials the spin-gap persists to temperatures above $T_c$.

\subsubsection{The spin fluctuation continuum}
\label{SpinCont}

\begin{figure*}
\centerline{
\epsfxsize=0.9\textwidth{\epsfbox{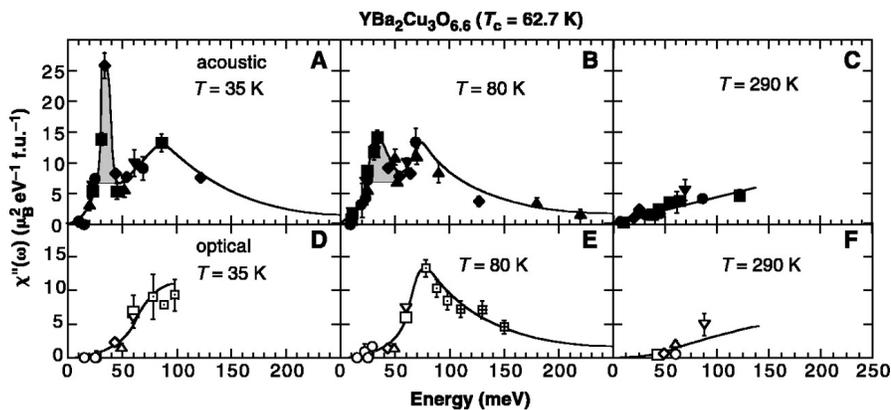}}
}
\caption{
\label{Fig1_Dai99}
Wavevector integrated frequency dependent magnetic susceptibility for
odd (acoustic) and even (optical) modes in underdoped YBa$_2$Cu$_3$O$_{6.6}$ 
($T_c$=62.7  K, \cite{Dai99}), for several temperatures. The resonance is
shadowed. Below the resonance there are contributions from an incommensurate
response, and above the resonance there is a continuum spectrum with
a lower threshold of twice the superconducting gap at the Fermi surface
points which are connected by a $(\pi,\pi)$ wavevector.
(Reprinted with permission from Science, Ref. \cite{Dai99}. Copyright \copyright 1999 AAAS).
}
\end{figure*}

In addition to the resonance and the dispersive features above and below it, there is also
a spin fluctuation continuum, which extends to high energies, see Fig.~\ref{Fig5_Keimer04} c. 
In Fig. \ref{Fig1_Dai99} 
the local (wavevector integrated) susceptibility is shown. The continuum extends well above 200 meV. 
The continuum is more clearly visible
in the local susceptibility, as it has a much broader momentum
width than the resonance feature and the low-energy incommensurate response.
The momentum width of the resonance and of the continuum part of the spectrum
show different doping dependence.
Also, the doping dependence of the
total spectral weight is different for the resonance peak and
the continuum part of the spectrum.
The ratio between the spectral weight of the resonance and the spectral
weight of the continuum actually decreases with underdoping, due to
a stronger increase of the continuum part of the spectrum \cite{Fong00}.

In optimally and overdoped materials the continuum part of the spectrum becomes very
small in (energy resolved) intensity 
and only the resonance part of the spectrum can be observed there.
However, because the continuum is spread over a large energy scale, the
total (energy integrated) intensity can still be considerable.

\subsubsection{Normal state spin susceptibility}
\label{Normalsusc}

\begin{figure*}
\centerline{
\epsfxsize=0.7\textwidth{\epsfbox{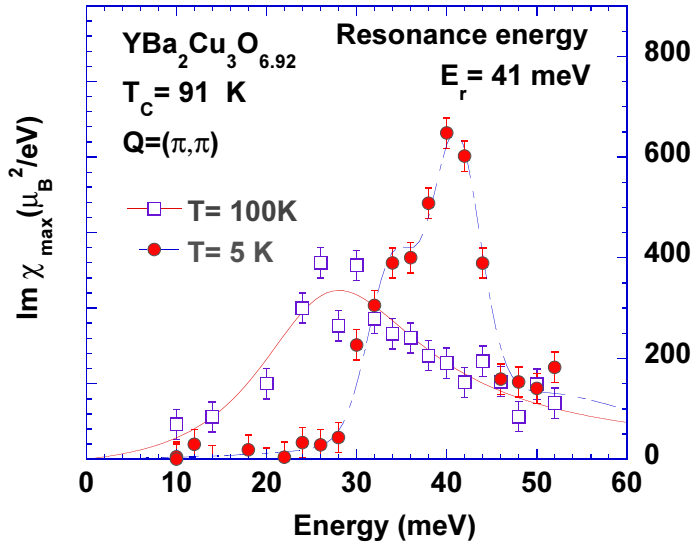}}
}
\caption{
\label{Fig1_Bourges99}
Frequency dependent spin susceptibility at antiferromagnetic wavevector
$\vec{Q}=(\pi,\pi)$ in absolute units, for 
odd (acoustic) excitations in optimally doped YBa$_2$Cu$_3$O$_{6.92}$ 
($T_c$=91  K, \cite{Bourges99}). Shown are data for the normal state (squares) 
and for the superconducting state (circles). 
The normal state susceptibility is characterized by a broad maximum around
$\Omega_{max}\sim 30$~meV, that does not coincide with the resonance
energy in the superconducting state of $\Omega_{res}\approx 41$~meV.
(From Ref. \cite{Bourges99}
Copyright \copyright 1999 AIP, and \cite{Bourges95}
Copyright \copyright 1995 Elsevier).
}
\end{figure*}

In the normal state the spin susceptibility is peaked 
at the commensurate wavevector $\vec{Q}=(\pi,\pi)$ except for
La$_{2-x}$Sr$_x$CuO$_4$ which shows four incommensurate peaks in the
normal state. As function of energy, it is peaked around a characteristic
frequency $\Omega_{max}$, which decreases with underdoping \cite{Rossat93,Bourges95,Regnault95,Bourges98,Regnault98}. 
An example is shown in Fig.~\ref{Fig1_Bourges99} for optimally doped
YBa$_2$Cu$_3$O$_{6.92}$. 
The peak intensity decreases with increasing
doping \cite{Regnault95,Bourges98,Regnault98}.
The overall momentum width of the commensurate response in the normal state
was shown to scale with $T_c$ \cite{Balatsky99}.
For optimally and overdoped materials the normal state susceptibility at
the antiferromagnetic wavevector can be
reasonably well described by an overrelaxational form,
\begin{equation}
\label{Omax}
\chi(\vec{Q},\Omega) = \frac{\chi_{\vec{Q}}}{\displaystyle 1-i \; \Omega/\Omega_{max}}.
\end{equation}
The spin-fluctuation frequency $\Omega_{max}$ characterizes the
normal state response.
In the pseudogap state for underdoped cuprates, Eq.~(\ref{Omax}) is not
a good description at low $\Omega\sim E_{sg}$, because of the persistence of
the spin-gap. It has been shown that the spin excitation spectrum in the pseudogap state
is qualitatively different from that in the superconducting state, showing no 
resonance feature and a steep incommensurate dispersion with a 
strongly anisotropic in-plane geometry \cite{Hinkov06}.

\subsection{Angle resolved photoemission}
\label{ARPES}

Angle resolved photoemission experiments have achieved several 
important goals in characterizing high-$T_c$ materials.
First, they showed that a large and well defined Fermi surface 
exists in these materials. Thus, one can expect that the important
fermionic excitations reside near this Fermi surface and
thus populate only a small fraction of the phase space.
Second, they showed the presence of a shallow extended 
saddle point in the regions of the Brillouin zone, where
the $d$-wave oder parameter is maximal ({\it antinodal} regions).
Third, it turned out that ARPES spectra near the {\it nodal}
directions (where the $d$-wave order parameter vanishes)
and near the antinodal directions of the Brillouin zone
are very different from each other.
Near the antinode spectra are dominated by strong self-energy effects,
with characteristic $S$-shaped regions
in the dispersion and non-trivial line-shapes
of the spectra.
These self-energy effects persist also away from the antinodes, and
continuously evolve into the nodal spectra, which have a simpler
line-shape, and where the dispersion anomalies appear in form of
kinks.
Fourth, the line-widths of the spectra contain important information
about the scattering of quasiparticles.

\subsubsection{Fermi surface}
\label{FS}

\begin{figure}
\centerline{
\epsfxsize=0.7\textwidth{\epsfbox{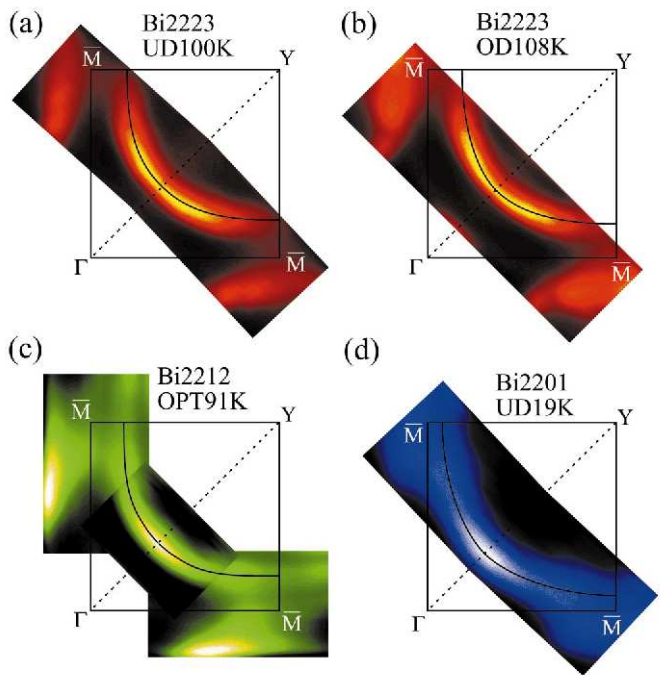}}
}
\caption{
\label{Fig1_Matsui03a}
Experimental normal state Fermi surfaces for 
Bi$_2$Sr$_2$Ca$_{n-1}$Cu$_n$O$_{2n+4}$ with $n=1,2,3$.
(From Ref. \cite{Matsui03a},
Copyright \copyright 2003 APS).
}
\end{figure}

The existence of a well defined normal state Fermi surface was an
object of discussion for some time. 
The matter is settled in the meanwhile, and a consistent picture has emerged
\cite{Damascelli03,Campuzano04}.
The existence of a large, hole-like Fermi surface in the normal state 
was taken as support for the validity of Luttinger's theorem
\cite{Olson90,Campuzano90,Shen95,Ding96b}, a conclusion that was confirmed for
underdoped, optimally doped and moderately overdoped materials \cite{Kordyuk02}.
An example for the quality of experimental Fermi surfaces for 
Bi$_2$Sr$_2$Ca$_{n-1}$Cu$_n$O$_{2n+4}$ 
materials (with $n=1,2,3$ layers per unit cell) is shown in 
Fig. \ref{Fig1_Matsui03a}
for different doping levels. The Fermi surface is large and hole-like, showing only slight variations with doping. 

It was found that on the strongly overdoped side of
(Bi,Pb)$_2$(Sr,La)$_2$CuO$_{6+\delta }$
(Pb-Bi2201) the Fermi surface stays hole-like \cite{Sato01,Sato02} even 
when $T_c$ is reduced to less than 4 K.
For even stronger overdoping ($T_c<2$K)
a transition to an electron-like Fermi surface was suggested \cite{Takeuchi01,kondo04}.
A similar change of the topology of the Fermi surface was observed in L$_{2-x}$Sr$_x$CuO$_4$
\cite{Fujimori98,Ino99,Ino02}, where the transition takes place for 
$x\approx 0.2$, and the Luttinger sum-rule is fulfilled above and below the
transition \cite{Ino99,Ino02}.
Concerning Bi-2212, recent measurements have shown that in
heavily overdoped Bi$_2$Sr$_2$CaCu$_2$O$_{8-\delta }$ the bonding
Fermi-surface sheet stays hole-like, whereas for the antibonding Fermi-surface
sheet a change in the Fermi-surface topology from hole-like to
electron-like takes place with increasing doping \cite{Kaminski05}.
The critical doping value was determined as 0.23, corresponding to
$T_c=55$K. 
Recently also Tl$_2$Ba$_2$CuO$_{6+\delta}$ (Tl2201) 
was studied in the overdoped range, finding a single large hole-like
Fermi-surface for samples with $Tc=63$ K and $T_c=30$ K;
it was concluded that a topological transition will eventually take place
at even higher doping levels for this system as well \cite{Plate05}.
The important observation is, that this change in Fermi-surface topology
is not accompanied by any abrupt changes in $T_c$ as a function of doping.
Also, it takes place at a doping level which does not correspond to the
extrapolation of the pseudogap crossover line to zero temperature (at
doping level 0.19 for Bi2212). 

Whereas for 
overdoped materials the Fermi surface in the normal state is well defined,
in optimally and underdoped materials a pseudogap phase exists above
the superconducting transition temperature, in which the Fermi surface
is present in form of Fermi-surface arcs
near the nodal points, separated by gapped antinodal regions \cite{Norman98n}. 
The length of the arcs increases with temperature, until at a characteristic
temperature $T^{\ast }$ the arcs join each other and the Fermi surface is restored
\cite{Norman98n}.

\subsubsection{Normal-state dispersion and the flat-band region}
\label{FlatBand}

\begin{figure}
\centerline{
\epsfxsize=0.7\textwidth{\epsfbox{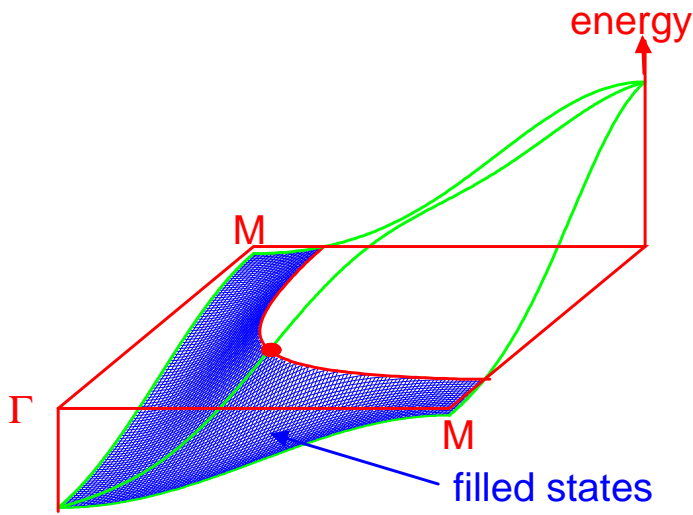}}
}
\caption{
\label{brill0}
Typical normal state dispersion for a single layered high-$T_c$ cuprate
shown in one quarter of the Brillouin zone. The filled states are 
shown as shadowed surface. The Fermi surface is shown as thick curve.
At the $M$-points of the Brillouin zone
the dispersion has Van Hove singularities very close to the chemical potential.
In the superconducting state only the nodal points of the Fermi surface 
remain, shown as a thick dot in the figure.
}
\end{figure}

Typically, for all high-$T_c$ cuprates
the dispersion of electronic states around the Fermi surface is characterized by
the presence of saddle points close to the chemical potential. 
For simple tetragonal symmetry
the corresponding points are the so-called $M$-points in the two-dimensional
Brillouin zone, situated at $(0,\pi) $ and $(\pi,0)$ (in units of the inverse
lattice constant).
To clarify the notation we show in Fig.~\ref{brill0} schematically
the behavior of the dispersion in one quarter of the Brillouin zone.
Such a dispersion is typical for single layered cuprates. For
cuprates with more than one layer per unit cell a splitting of the bands
is expected. The case for bilayer compounds, in which a splitting
of the Fermi surface into two occurs, will be discussed further below. 

The proximity of the van-Hove singularity is seen in Fig.~\ref{brill0}
near the points marked `$M$'. Also seen as thick dot is the position of the
order-parameter node on the Fermi surface, when
the material enters a $d$-wave superconducting state.
The Fermi velocity in cuprates is of the order of eV$\AA $.
This means, that in the vicinity of this node,
excitations with energies within the range of $\pm 100$ meV
are restricted to a very narrow shell around the
Fermi surface. 
The same is not true for the regions around the $M$-points, as the
Van-Hove singularity is within the range of typical excitation energies.

In fact, the dispersion near the $M$-points of the Brillouin zone
shows a surprisingly flat behavior
in the direction parallel to the Fermi surface
\cite{Abrikosov93, Dessau93, Gofron94,Shen95}. The binding energy of that
flat-band region is comparable to the maximal
superconducting $d$-wave gap near optimal doping,
and increases with underdoping. 
In the superconducting state the flatness of the dispersion 
for near-optimally doped materials
is even more pronounced.
It was suggested \cite{Abrikosov93} that these saddle point singularities may be
extended van-Hove singularities in the sense that the quasiparticle mass diverges in one direction, or becomes very large.

\begin{figure}
\centerline{
\epsfxsize=0.78\textwidth{\epsfbox{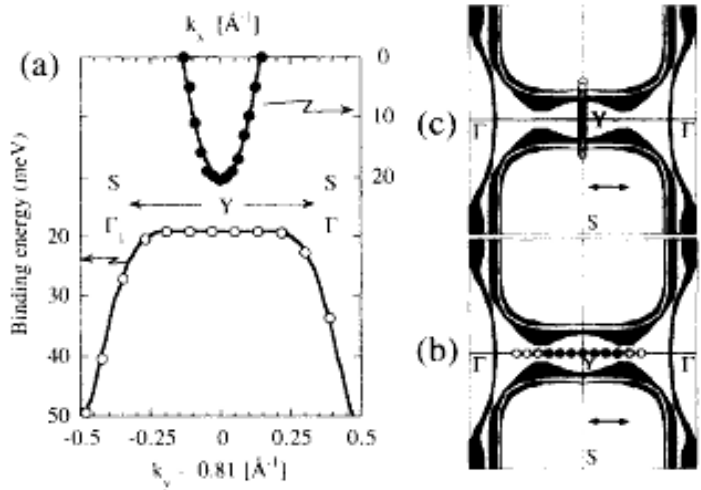}}
}
\caption{
\label{Fig3_Gofron94}
(a) Experimental band dispersion near the van-Hove singularity for untwinned
YBa$_2$Cu$_4$O$_8$ ($T_c=82 $K) at the $Y$-point $(0,\pi)$ of the 
Brillouin zone. Shown in (b) and (c) are the positions in momentum space where the
ARPES spectra were taken. Filled dots correspond to the flat band region.
Superimposed is the theoretically calculated Fermi surface projected on the basal plane
of the Brillouin zone.
(From Ref. \cite{Gofron94},
Copyright \copyright 1994 APS).
}
\end{figure}

As an example, the dispersion near the saddle points for the compound YBa$_2$Cu$_4$O$_8$ is shown in
Fig.~\ref{Fig3_Gofron94}. In YBa$_2$Cu$_4$O$_8$ the symmetry classification is somewhat different
from the case of Bi$_2$Sr$_2$CaCu$_2$O$_{8-\delta }$, and the saddle point corresponds here to the
$Y$ point of the Brillouin zone (see Fig.~\ref{Fig3_Gofron94} b,c). From Fig. \ref{Fig3_Gofron94} (a)
the flat band region in the direction from the saddle points toward 
the center of the Brillouin zone 
is evident. In contrast, the dispersion perpendicular to this is parabolic with Fermi crossings close by.

If not stated otherwise,
we will neglect from now on throughout this paper 
deviations from tetragonal symmetry, which occurs in several high-$T_c$ 
cuprate materials, as these deviations are not important for
the understanding of the physics of superconductivity.
Accordingly, we use a notation adapted for simple tetragonal symmetry, 
and commonly used for the system best studied in ARPES, namely
Bi$_2$Sr$_2$CaCu$_2$O$_{8-\delta }$.

From the line-widths of the excitations near the $M$-points one can conclude
that scattering is strong between the $M$-point regions. As a result,
the flat dispersion at the saddle points is probably a many-body effect,
and not a property of the bare electronic structure \cite{Campuzano04}.
Also, for the same reason the flat-band region does not lead
to any singularity in the density of states, as there are no sharp
quasiparticles present there. However, it can lead to a sizable particle-hole
asymmetry.

Finally, for underdoped materials the $M$ point regions stay gapped above
the superconducting transition temperature, and these gapped regions
are connected by 
Fermi-surface arcs that grow out of the nodal points of the Brillouin zone
\cite{Norman98n}.
This simultaneous presence of gapped and non-gapped regions
leads to the pseudogap-effect.

\subsubsection{MDC and EDC}
\label{MDCEDC}

\begin{figure}
\centerline{
\epsfxsize=0.5\textwidth{\epsfbox{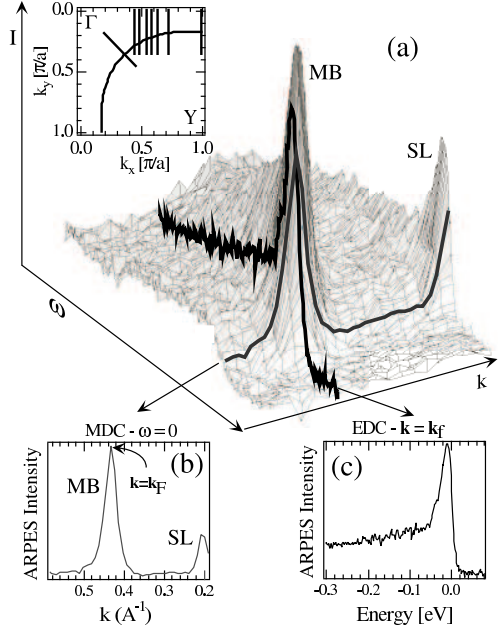}}
}
\caption{
\label{Fig1_Kaminski}
(a) The experimental ARPES intensity as function of binding energy $\omega $ and momentum $\vec{k}$
for optimally doped Bi${_2}$Sr${_2}$CaCu${_2}$O${_{8+\delta}}$ at $T=40 $K (SL is a superlattice
image, MB is the main quasiparticle band). (b) Momentum distribution curves (MDC) from (a).
(c) Energy distribution curves (EDC) from (a). 
The diagonal line in the $\Gamma -Y$ direction of the zone inset,
crossing the Fermi surface, shows the corresponding $\vec{k}$ cut. 
(From Ref. \cite{Kaminski01},
Copyright \copyright 2001 APS).
}
\end{figure}

For the experimental study of cuprates it turned out important to consider
not only spectra for fixed momentum in the Brillouin zone as function
of binding energy (energy distribution curves, EDC), but also
spectra for fixed energy as function of momentum along a certain cut
in the fermionic Brillouin zone (momentum distribution curves, MDC).
The difference is illustrated in Fig. \ref{Fig1_Kaminski} for a typical
set of spectra taken on optimally doped
Bi${_2}$Sr${_2}$CaCu${_2}$O${_{8+\delta}}$ at $T=40 $K.
The corresponding MDC is shown in (b) and the EDC in (c).
It is clearly seen that the EDC poses several problems. First, it has a strongly
asymmetric line-shape, which does not allow for the determination of
the quasiparticle lifetime in a simple way. Instead, the full energy dependent
self energy must be extracted in order to characterize quasiparticles.
In addition, there is an energy-dependent background at higher energies, which
must be subtracted. In contrast, the MDC in panel (b) shows a Lorentzian lineshape
(note that MB is the main band; the additional feature, denoted by SL, is due to a superstructure),
and the background is rather momentum independent and can easily be subtracted.
Also seen in Fig.~\ref{Fig1_Kaminski} is that the maxima of the MDC and EDC dispersions
at $\omega=0$ and $\vec{k}=\vec{k}_F$, respectively,
do not coincide. Accordingly, it is very important to specify which spectra are used
in order to study dispersion anomalies.
It turned out that both types of spectra contain important information. 
In the beginning years of experimental research in the field of cuprates 
almost exclusively EDC spectra were analyzed. 
Only in recent years the resolution of experiments became good enough for
analyzing MDC's as well.
In Refs. \cite{Norman01a,Eschrig02,Eschrig03} the importance of MDC spectra for comparison
with theoretical models was pointed out.

The line shape of the ARPES signal is determined by the spectral function
$A(\epsilon ,\vec{k})$ (multiplied with the Fermi distribution function).
The fact, that the MDC line shape (in contrast to the EDC lineshape) is
approximately Lorentzian,
can be quantified in terms of a self energy $\Sigma (\epsilon, \vec{k})$
with real part $\Sigma'$ and imaginary part $\Sigma''$, that
in the normal state is related to the spectral function by,
\begin{equation}
\label{Specfun}
A(\epsilon ,\vec{k}) = \frac{1}{\pi } \; \frac{\Sigma''(\epsilon ,\vec{k})}{
[\epsilon - \xi_{\vec{k}} - \Sigma'(\epsilon ,\vec{k}) ]^2 + 
[\Sigma''(\epsilon ,\vec{k})]^2 \; ,
}
\end{equation}
where $\xi_{\vec{k}} = \epsilon_{\vec{k}}-\mu$ is the bare band-structure
dispersion.  It is clear from this expression, 
that the experimental findings are consistent with the
notion of a weak momentum dependence 
and a strong energy dependence of the self energy.
Indeed, for momentum independent $\Sigma $ 
and for $\xi_{\vec{k}} \approx \vec{v}_{F0} (\vec{k}-\vec{k}_{F0})$,
the MDC is a Lorentzian 
with a half-width half-maximum
(HWHM) of $W_{MDC}=\Sigma''(\epsilon)/v_{F0}$ 
(assuming for simplicity that the MDC-cut is parallel to $\vec{v}_{F0}$).

\subsubsection{Bilayer splitting}
\label{Bil_exp}

Bilayer splitting for cuprate superconductors with two conducting layers per unit cell
was predicted long ago on theoretical grounds \cite{Massida88,Andersen94}, but only in recent years was found in experiments. 
It is most clearly pronounced in overdoped materials, where it was found first \cite{Feng01a,Chuang01}. 

For dominantly coherent coupling between the planes the appropriate basis
is in terms of bonding and antibonding bands. Their dispersion is given
in terms of the dispersion for a single layer, $\xi_{\vec{k}}$, by 
\begin{eqnarray}
\xi_{\bf k}^{(b)}&=& \xi_{\bf k} - t_\perp ({\bf k}) \nonumber \\
\xi_{\bf k}^{(a)}&=& \xi_{\bf k} + t_\perp ({\bf k})
\end{eqnarray}
with an interlayer hopping term $t_{\perp}({\bf k})$.
The interlayer hopping has the form \cite{Andersen95,Chakravarty93}
\begin{equation}
\label{tperp}
t_\perp({\bf k})=  \frac{1}{4} t_\perp \left[\cos (k_xa)-\cos (k_ya) \right]^2 \; .
\end{equation}
It describes coherent hopping between the CuO$_2$ planes.
Sometimes, a momentum independent incoherent hopping term is added on
the right side of Eq.~(\ref{tperp}).

\begin{figure}
\centerline{
\epsfxsize=0.50\textwidth{\epsfbox{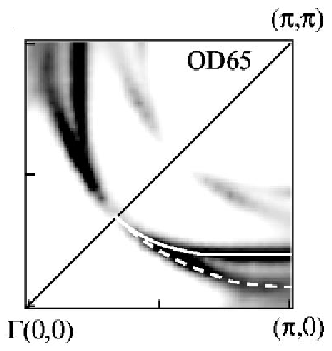}}
\epsfxsize=0.30\textwidth{\epsfbox{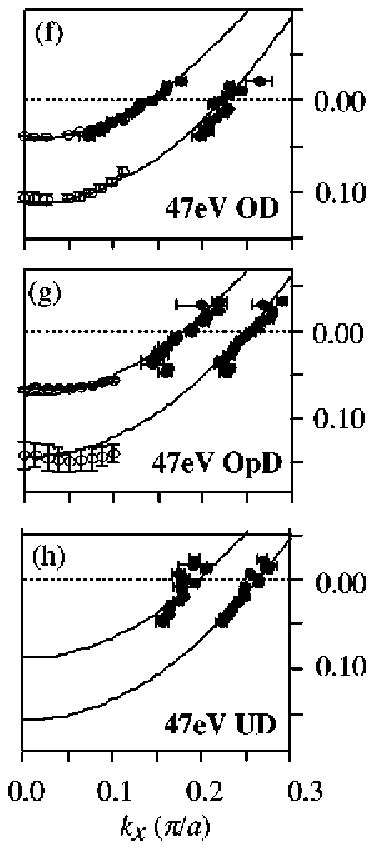}}
}
\caption{
\label{Fig4_Chuang04}
\label{Fig1_Feng02a}
Left:
Bilayer-split normal state Fermi surface of heavily overdoped 
Bi$_{2}$Sr$_{2}$CaCu$_{2}$O$_{8+\delta}$ ($T_c$=65 K).
(The two weaker features are shadow Fermi surfaces 
due to superstructure). Solid and dashed lines represent
the bonding and antibonding Fermi surfaces, respectively.
(From Ref. \cite{Feng02a},
Copyright \copyright 2002 APS).
Right:
Normal state dispersion at $k_y=1.27 \pi $ as a function of $k_x$ 
for Bi$_{2}$Sr$_{2}$CaCu$_{2}$O$_{8+\delta}$. OD refers to
overdoped ($T_c$=55 K) at $T=$80 K, OpD to optimally doped ($T_c$=91 K) at $T=$100 K, and
UD to underdoped ($T_c$=78 K) at $T=$100 K. The open circles are peak centroids
from EDC's and the closed circles from MDC's. The lines are parabolas separated by
70 meV, and shifted with respect to the OD ones by -30 meV for OpD, and by -40 meV for UD.
A fit to $\delta \xi (\vec{k}) = 0.5 t_\perp [\cos (k_xa) - \cos (k_ya)]^2$ gives
$t_\perp=(57\pm 4)$ meV and a maximum splitting of $114\pm 8$ meV.
(From Ref. \cite{Chuang04},
Copyright \copyright 2004 APS).
}
\end{figure}

\begin{figure}
\centerline{
\epsfxsize=0.3\textwidth{\epsfbox{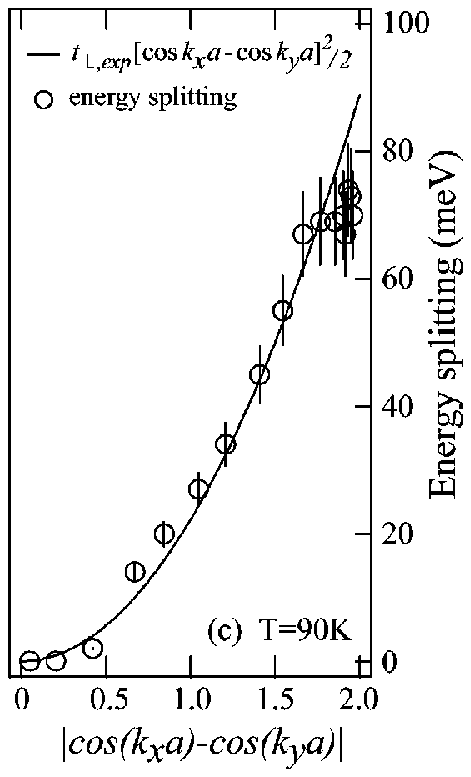}}
\epsfxsize=0.6\textwidth{\epsfbox{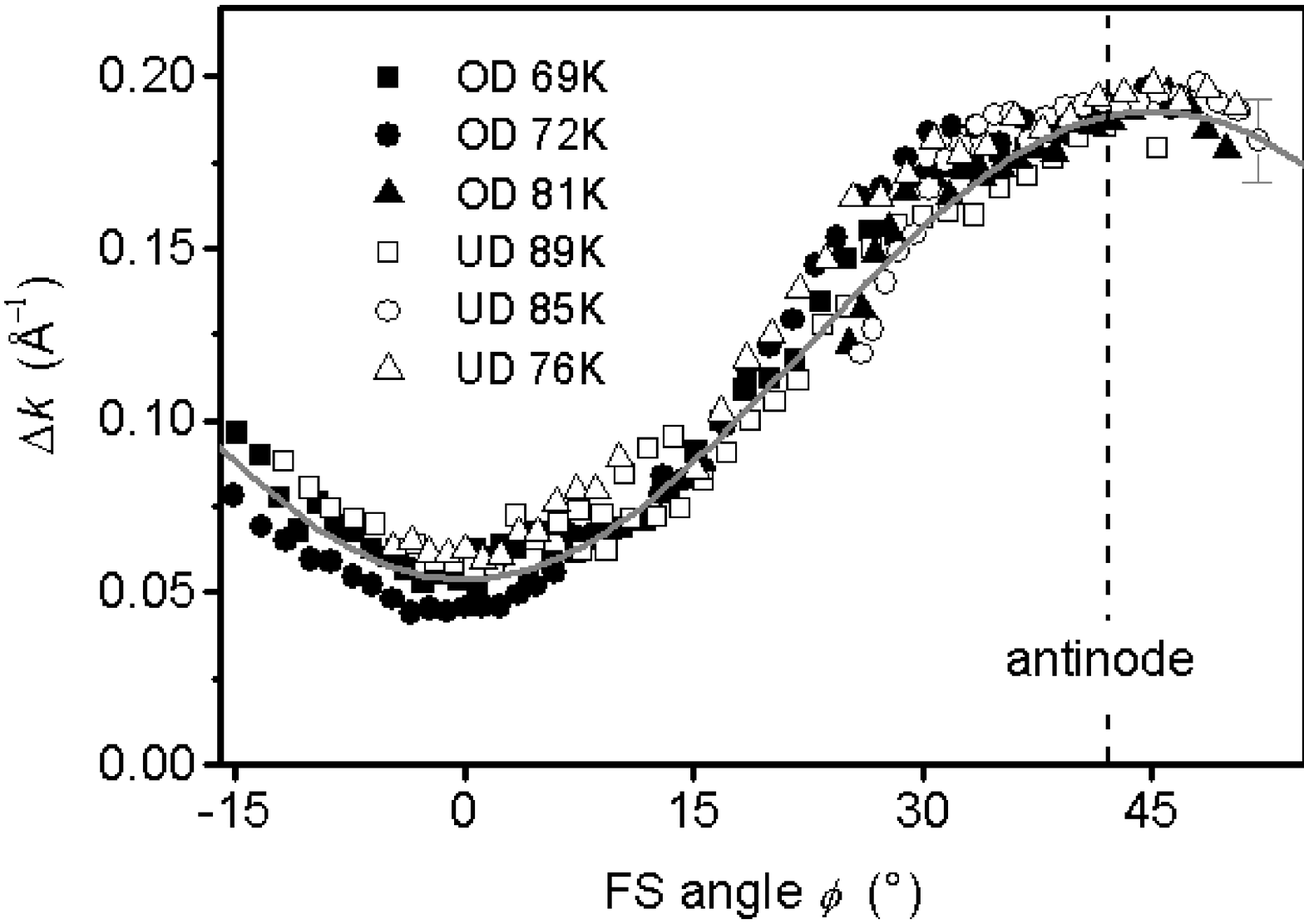}}
}
\caption{
Left:
\label{Fig4_Feng01}
Energy splitting along the antibonding Fermi surface
for an overdoped Bi$_{2}$Sr$_{2}$CaCu$_{2}$O$_{8+\delta}$ sample ($T_c=$65 K).
The curve is 
$0.5 t_\perp [\cos (k_x a)-\cos (k_y a)]^2$ with the experimentally determined fitting parameter
$t_\perp=44\pm 5 $ meV.
(From Ref. \cite{Feng01a},
Copyright \copyright 2001 APS).
Right:
\label{Fig3_Kordyuk02}
Width of the main Fermi surface, $\Delta k$, versus the Fermi surface angle
$\phi $, seen from $(\pi,\pi)$ and measured from the nodal line, for
(Bi,Pb)$_{2}$Sr$_{2}$CaCu$_{2}$O$_{8+\delta}$
at several amounts of doping (the $T_c$ is indicated).  The solid line represents
$\Delta k(\phi) = \Delta k_0 + \Delta k_1 \sin^2 (2\phi)$.
(From Ref. \cite{Kordyuk02},
Copyright \copyright 2002 APS).
}
\end{figure}

In Figs. \ref{Fig1_Feng02a} and \ref{Fig4_Feng01} the main experimental results for the bilayer splitting
are shown. 
First, the bilayer splitting is strongly anisotropic and
follows the 
theoretical predictions \cite{Andersen95,Chakravarty93}. 
The functional dependence of the bilayer splitting on the momentum, Eq.~(\ref{tperp}), is
experimentally verified \cite{Feng01a,Chuang01}. An example for the fit to this
functional form is shown on the left in Fig. \ref{Fig4_Feng01}.
According to Eq.~(\ref{tperp}) the bilayer splitting is zero along the nodal direction
$k_x=k_y$.  
Near the $M$-point of the Brillouin zone the bilayer splitting is maximal. 
Recently, it was suggested that a small bilayer splitting of approximately 23 meV remains for 
Bi$_{2}$Sr$_{2}$CaCu$_{2}$O$_{8+\delta}$
also in nodal direction \cite{Kordyuk03a,Valla05}; furthermore, in YBa$_2$Cu$_3$O$_{6+x}$ a nearly
five times larger nodal bilayer splitting has been observed \cite{Borisenko06}.

The second issue refers to the doping dependence of the bilayer splitting.
In optimally and underdoped compounds the bilayer splitting was also reported 
\cite{Chuang02,Kordyuk02a,Borisenko02,Borisenko03}, and its magnitude was shown to
be independent of doping \cite{Chuang02,Chuang04,Borisenko06}.
As can be seen in Fig. \ref{Fig4_Chuang04}, the maximal bilayer splitting
amounts to $\approx 114 $ meV for all studied values of doping, 
which suggests a value $t_\perp\approx 57$ meV.
Thus, with underdoping the bilayer splitting is not lost,
but the coherence between the bonding and antibonding bands worsens \cite{Kaminski03}.
The idea, that the bilayer splitting stays constant as a function of doping, is also
supported by the observation that the total momentum width of the Fermi surface in
the normal state depends strongly on the position on the Fermi surface, but almost
not on doping, as illustrated on the right in Fig.~\ref{Fig3_Kordyuk02}.
This indicates an unresolved bilayer splitting as source for
the strong anisotropy of the momentum width for all doping levels, which itself
is doping independent \cite{Kordyuk02}. 
The bilayer splitting in optimally and underdoped materials is of the same order as the
linewidth of the quasiparticle excitations,
and strong scattering between the
bonding and antibonding band can lead to the destruction of coherence between
the layers \cite{Kaminski03}. 
In this case, the strong mixing between bonding and antibonding
band often allows to consider both as a single entity. 
Assuming as scattering mechanism a spin-fermion interaction this scattering increases
with underdoping and is weak in overdoped materials.

\subsubsection{Superconducting coherence}
\label{SupCoh}

\begin{figure}
\centerline{
\epsfxsize=0.55\textwidth{\epsfbox{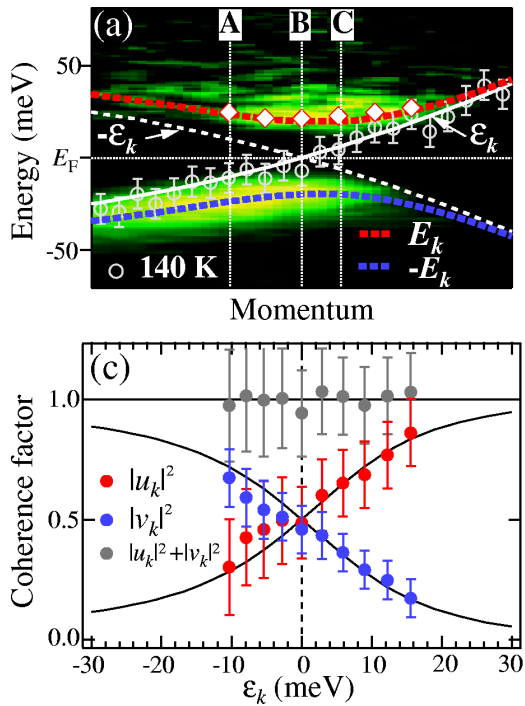}}
}
\caption{
(a) ARPES intensity plot for overdoped Bi$_2$Sr$_2$Ca$_2$Cu$_3$O$_{10+\delta}$ ($T_c$=108 K)
in the superconducting state ($T=$60 K). Shown as 
open white circles is also the dispersion
for the normal state ($T= $140 K). The two thick dashed curves are calculated from the BCS spectrum
$E_k=\sqrt{\epsilon_k^2+|\Delta_k|^2}$, where for $\epsilon_k$ the 
white full line is used, and for $|\Delta_k|$
the experimental peak energy from 60 K spectra is employed. 
The white dashed curve shows $-\epsilon_k$.
(c) shows the coherence factors
as experimentally determined from the ARPES intensity 
(solid circles) compared to
the BCS coherence factors derived from the ARPES dispersion (solid lines).
\label{Fig3_Matsui03}
(From Ref. \cite{Matsui03},
Copyright \copyright 2003 APS).
}
\end{figure}

First experiments showing particle-hole coherence in the superconducting state
were performed by Campuzano {\it et al.} \cite{Campuzano96}.
There it was shown that the hole dispersion branch shows a back-bending effect when crossing
the Fermi momentum, as expected from BCS theory of superconductivity.
Recent improvements in the resolution of ARPES spectroscopy allowed for an impressive
experimental verification of particle-hole coherence in the superconducting state including
the BCS coherence factors. The main results are reproduced in Fig. \ref{Fig3_Matsui03}.

As can be inferred from this figure, the particle-hole mixing is clearly seen in the dispersion
both of the hole as well as of the particle branch. The minimum gap between the particle and hole
branch is at $k_F$, the dispersive features are almost symmetric with respect to $E_F$ and
both the particle and the hole bands show the typical back-bending effect at $k_F$.
Matsui {\it et al.} \cite{Matsui03} also studied the spectral intensity of the two bands as function of $\vec{k}-\vec{k}_F$.
These weights determine the coherence factors in BCS theory, and they are shown in Fig. \ref{Fig3_Matsui03}
(c). The agreement with the BCS theory is striking. The experimental values are very close to
\begin{equation}
|u_{\vec k}|^2 = 1-|v_{\vec k}|^2 = \frac{1}{2} \left( 1+ \frac{\xi_{\vec k}}{E_{\vec k}} \right)
\end{equation}
with $E_{\vec k} = \sqrt{\xi_{\vec k}^2 + |\Delta_{\vec k}|^2}$. Here,
$\Delta_{\vec k} = \Delta_0 [\cos (k_x a) -\cos (k_y  a) ]/2$ is the $d$-wave gap and $\xi_{\vec k}$ is the normal state dispersion
which was obtained from the experimental peak positions at 140 K.
Note, that the sum of the squares of the coherence factors adds up to one. 
As $|u_k|^2$ and $|v_k|^2$ were determined independently,
this condition was not imposed but is an experimental verification of the sum rule $|u_{\vec k}|^2+|v_{\vec k}|^2=1$.

This study unambiguously established the Bogoliubov-quasiparticle nature of the sharp superconducting
quasiparticle peaks near $(\pi,0)$.
It is striking, that in spite of all anomalies observed in high-$T_c$ superconductors the
superconducting coherence of the quasiparticle peaks is described by these simple BCS formulas.

\subsubsection{EDC-derived dispersion anomalies}
\label{DispAn_ex}

Important information about the interaction of quasiparticles 
with collective excitations 
is obtained by studying anomalous behavior of the quasiparticle dispersion. 
Such anomalies are due to self-energy effects which arise when quasiparticles
couple strongly to collective excitations with finite frequency, leading
to inelastic scattering processes.
There are two types of experimental dispersions one can study: EDC-derived dispersions
and MDC-derived dispersions. 

\begin{figure}
\centerline{
\epsfxsize=0.70\textwidth{\epsfbox{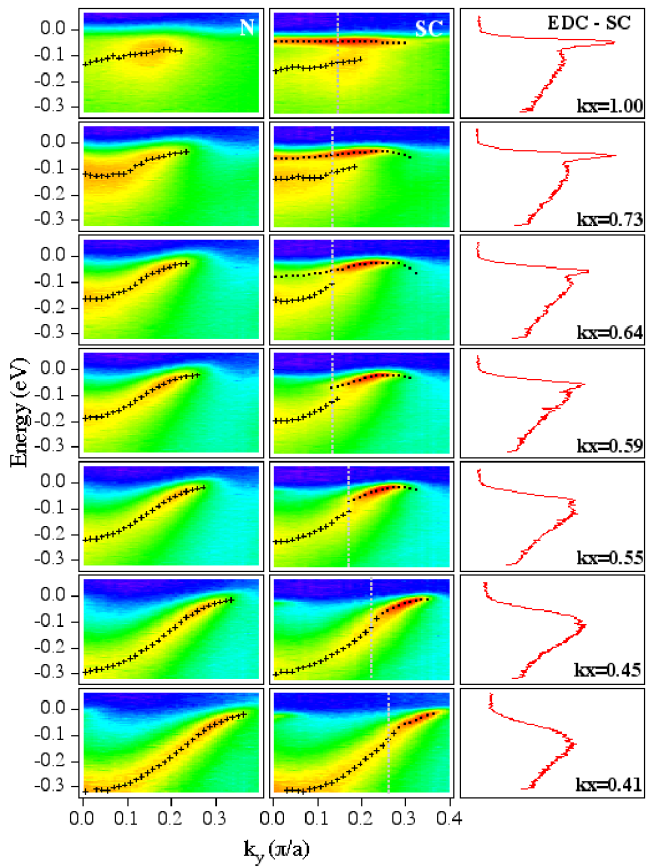}}
}
\caption{
\label{Fig4_Kaminski01}
(Left) Normal ($T=$140 K) and (middle) superconducting state ($T=$40 K) ARPES
intensity measured on optimally  doped
Bi${_2}$Sr${_2}$CaCu${_2}$O${_{8+\delta}}$  ($T_c$=89 K)
throughout the Brillouin zone along cuts indicated in the inset of Fig. \ref{Fig1_Kaminski}.
(Right) superconducting state EDC's from the momenta indicated by the dashed lines in
the middle panels.
(From Ref. \cite{Kaminski01},
Copyright \copyright 2001 APS).
}
\end{figure}

Advances in the momentum resolution of ARPES have led to a
detailed mapping of the spectral function in the high $T_c$ superconductor
Bi${_2}$Sr${_2}$CaCu${_2}$O${_{8+\delta}}$ throughout the Brillouin
zone \cite{Bogdanov00,Kaminski01}. In these first systematic experimental studies of
self-energy effects the emphasis was on the EDC-derived dispersions and
on the spectral line-shapes as function of energy. The main results of
Kaminski {\it et al.} \cite{Kaminski01} are reproduced in Fig. \ref{Fig4_Kaminski01}. 
The link of these data to the finite momentum width of the
magnetic resonance mode \cite{Eschrig00,Eschrig03} led
to a vivid discussion about the fundamental question of what are the relevant low-lying collective
excitations that couple to conduction electrons in cuprate superconductors.
On the right column in Fig.~\ref{Fig4_Kaminski01} superconducting EDC spectra are shown for positions in the
Brillouin zone near the Fermi surface, varying from the region near the $M$-point (top spectrum)
to the region near the zone diagonal (bottom spectrum), where the node of the $d$-wave order parameter is situated. 
The spectra near the $M$-point show
a characteristic low-energy `peak' and a broader high-energy `hump', separated by a `dip' in the spectrum.
The peak and hump-maxima define dispersion branches, that are presented in the middle column
as EDC-derived dispersions in the superconducting state. The corresponding normal state EDC dispersions
are shown in the left column.

The data indicate a seemingly unrelated effect
near the $d$-wave node of the superconducting gap, where the dispersion
shows a characteristic `kink' feature: for binding energies less than the
kink energy, the spectra exhibit sharp peaks with a weaker dispersion;
beyond this, broad peaks with a stronger dispersion \cite{Kaminski00,Bogdanov00,Kaminski01}.
The kink feature near the node is seen both in EDC and MDC derived dispersions.
This MDC-derived kink is present at a particular energy all around the Fermi 
surface \cite{Bogdanov00}, and away from the node the dispersion as determined
from MDC-derived spectra shows an
S-like shape in the vicinity of the kink \cite{Norman01a}.
The similarity between the excitation energy where the kink
is observed and the dip energy at $M$, however, suggests that these effects
are related \cite{Eschrig00}.

As seen in Fig. \ref{Fig4_Kaminski01},
away from the node the
kink in the dispersion as determined from EDC spectra
develops into a `break'; the two resulting
branches are separated by an energy gap, and overlap in momentum space.
Towards $M$, the break evolves into a pronounced spectral `dip' separating
the almost dispersionless quasiparticle branch from the weakly dispersing
high energy branch.
The kink, break, and dip features all occur at roughly
the same energy, independent of position in the zone \cite{Kaminski01},
the kink being at a slightly smaller energy than the break feature \cite{Johnson01}.
Additionally, the observation that the spectral width for binding energies
greater than the kink energy is much broader than that 
for smaller energies \cite{Kaminski00,Bogdanov00,Kaminski01}
is very similar to the difference in the linewidth between the peak and
the hump at the $M$ points.

\begin{figure}
\centerline{
\epsfxsize=0.68\textwidth{\epsfbox{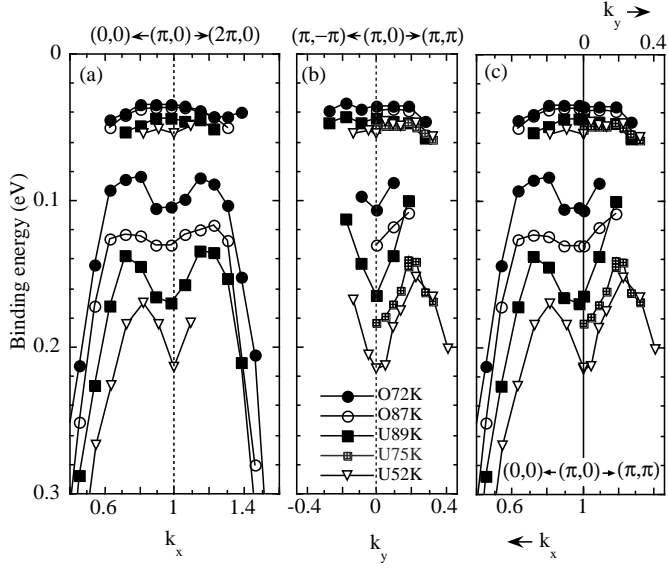}}
}
\caption{
\label{Fig3_Campuzano99}
Doping dependence of the dispersion from (a) $(\pi,0) \to (\pi\pm\pi,0)$,
(b) $(\pi,0) \to (\pi,\pm\pi)$, and (c) both directions, for the
peak and hump in the superconducting state
of Bi$_{2}$Sr$_{2}$CaCu$_{2}$O$_{8+\delta}$.
U is underdoped and O is
overdoped.
(From Ref. \cite{Campuzano99},
Copyright \copyright 1999 APS).
}
\end{figure}

Another important result comes from the comparison of the dispersions 
along the $(0,0) \to (0,\pi)$ direction and the $(0,\pi) \to (\pi,\pi)$ direction,
which is reproduced in Fig. \ref{Fig3_Campuzano99}.
Apart from the increase of the binding energy with underdoping of the
high-binding-energy branch one observes a pronounced dispersion minimum also
in the direction $(0,0) \to (0,\pi)$. This minimum resembles a
mixing between  the $(0,\pi)$ and $(\pi,0)$ regions in the Brillouin zone
due to scattering, which increases with underdoping.

Finally, the dispersion anomalies were observed also in overdoped
materials, taking into account the well resolved bilayer splitting.
In this case dispersion anomalies very similar to the ones discussed above,
are observed for the bonding band in the antinodal region \cite{Feng01a,Gromko03,Gromko04}.
These careful experiments show very clearly that for overdoped materials
self-energy effects are strong in the antinodal region, however become
weak toward the nodal regions. At the nodal point these self-energy effects
are unobservable for strongly overdoped materials, whereas other effects,
presumably due to electron-phonon scattering, remain observable in the nodal
region of the Brillouin zone.

Different experiments focussed on different scattering mechanisms, divided mainly
between coupling to phonons and to antiferromagnetic spin fluctuations.
Whereas certainly both scattering mechanisms are at work in cuprates, it is of
considerable interest to determine the respective coupling constants. For
this goal it is important to differentiate between the separate scattering
channels. Fortunately this became possible through the 
careful studies by several
groups \cite{Johnson01,Gromko03,Gromko04,Sato03,Kordyuk04}.
These studies are important also from the point of view that they show the
intrinsic nature of the dispersion anomalies, which persist even when
the bilayer split bands are resolved.

\subsubsection{The $S$-shaped MDC-dispersion anomaly}
\label{S-shape}

The traditional way of analyzing ARPES data has been that for EDC's,
namely at fixed momentum as a function of binding energy.
A much improved precision in momentum space during recent years, however,
made it possible to analyze in detail also MDC curves in cuprates, taken
at fixed binding energy as a function of momentum
\cite{Kaminski00,Bogdanov00,Valla99}.  
MDC's have been used 
in the high temperature cuprate superconductors to study a variety of 
phenomena, for example as a test for the marginal Fermi liquid 
hypothesis \cite{Valla99,Valla00}, or to elucidate a dispersion kink along 
the nodal direction \cite{Bogdanov00}, the origin of which was subject of a long
debate \cite{Kaminski01,Lanzara01,Johnson01}.

In the normal state, it is relatively straightforward to analyze
MDC's, as there is no energy gap complicating the dispersion; the
same applies for the superconducting state along the nodal direction
\cite{Kaminski01}.  
However, qualitative changes occur in the MDC's due to the energy gap.  By 
analyzing MDC dispersions, one can gain important information on 
many-body effects in the superconducting state.

\begin{figure}
\centerline{
\epsfxsize=0.35\textwidth
\epsfbox{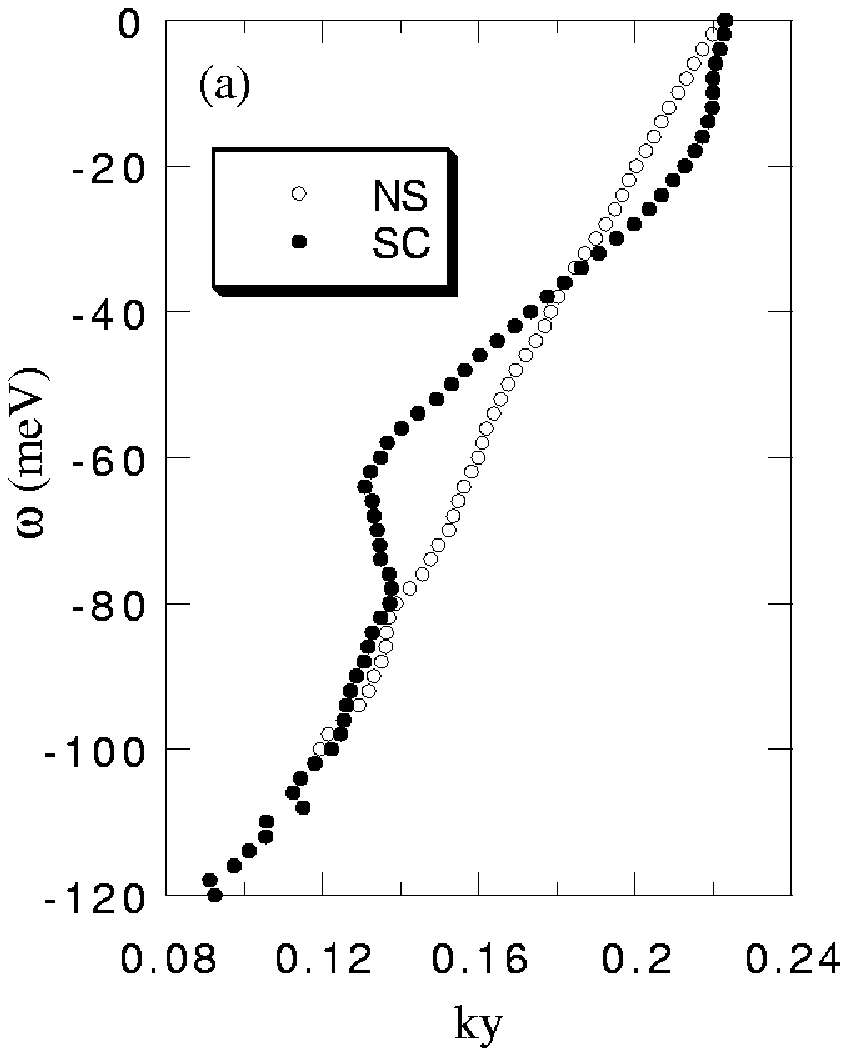}
\hspace{0.1\textwidth}
\epsfxsize=0.27\textwidth
\epsfbox{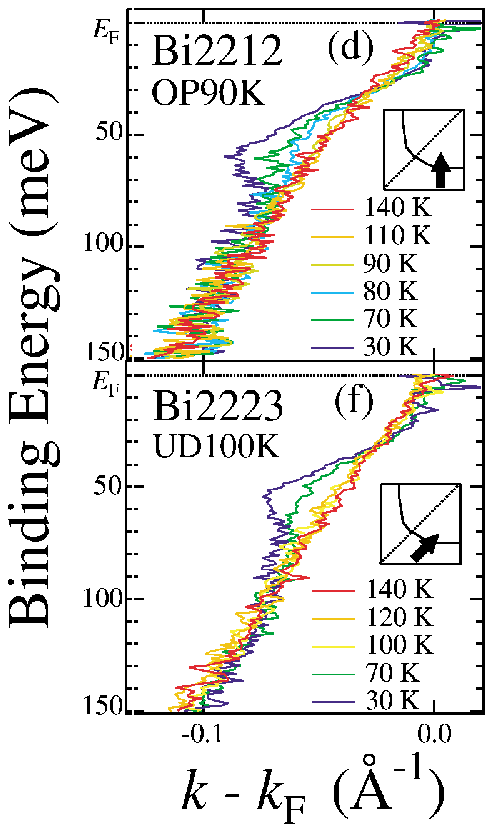}
}
\vspace{0.3cm}
\caption{
Left:
Experimental
MDC dispersion in the superconducting state (SC, T=40K) versus that
in the normal state (NS, T=140K)
of an optimally doped ($T_c$=90K)
Bi$_2$Sr$_2$CaCu$_2$O$_{8+\delta}$ sample.
k$_y$ is in units of $\pi/a$.  For this momentum cut, k$_x$=0.59$\pi/a$.
(From Ref. \cite{Norman01a},
Copyright \copyright 2001 APS).
\label{Fig1_Norman01a}
Right: Experimental MDC peak dispersions for
Bi2212 
(Bi$_2$Sr$_2$CaCu$_2$O$_{8+\delta}$) 
and Bi2223 
(Bi$_2$Sr$_2$Ca$_2$Cu$_3$O$_{10+\delta}$) 
measured along the cuts indicated in the insets as arrow, 
for several temperatures
from above to below $T_c$. The strongest effects correspond to 
lowest temperatures.
(From Ref. \cite{Sato03},
Copyright \copyright 2003 APS).
}
\end{figure}
In Fig. \ref{Fig1_Norman01a}, MDC dispersions are shown in both the normal and 
superconducting states roughly midway between the nodal and 
the antinodal point of the Brillouin zone.  
The normal state dispersion shows roughly a linear behavior in 
${\bf k}$ in the energy range of interest.  In 
the range of 20-60 meV, the superconducting dispersion is also linear, 
but with a slope approximately half that of the normal state, as noted 
earlier by Valla {\it et al. } \cite{Valla00}.  
This implies an additional many-body renormalization of 
the superconducting state dispersion relative to that in the normal 
state.  

Below the gap energy, the MDC derived dispersion shows a completely different 
behavior when compared with the EDC derived dispersions.
The former shows an almost vertical branch toward the chemical potential,
whereas the latter shows a backbending from the chemical potential. This
effect is, however, 
easily explained within a BCS picture when taking into account
finite lifetime effects of the quasiparticles \cite{Norman01a}.

Another, 
more interesting renormalization effect is the $S$-shaped dispersion in
the range between 60 and 80 meV,
before recovering back to the normal state dispersion at higher binding
energies. This $S$-shaped part of the MDC dispersion corresponds to the
`break'-region in the EDC-derived dispersions, or to the `dip' feature
in the EDC's. 
Such effects are typical of electrons interacting with a 
bosonic mode \cite{Scalapino_Parks,Norman98}, and the mode in the current case has been 
identified as a spin exciton by some authors \cite{Norman97,Kaminski01,Johnson01} and a 
phonon by others \cite{Lanzara01}.
However, in explaining the effect, one has to bear in mind that the
$S$-shaped dispersion anomaly is associated with the superconducting state,
which gives an additional restriction for possible interaction mechanisms.

\begin{figure}
\centerline{
\epsfxsize=0.9\textwidth{\epsfbox{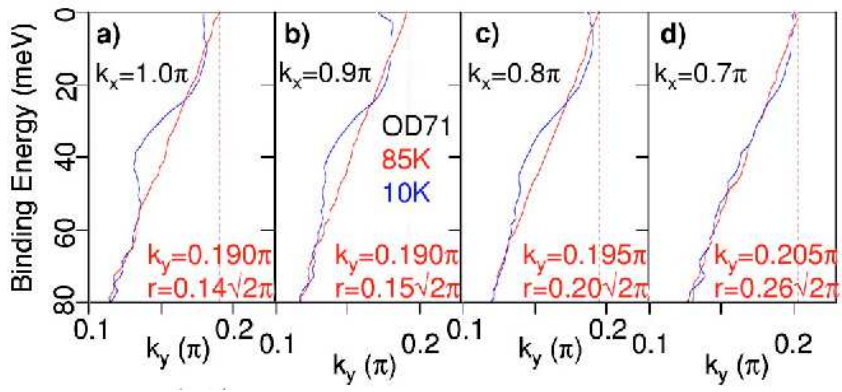}}
}
\caption{
\label{Fig1_Gromko04}
Dependence of the MDC dispersion of the bonding band for an overdoped
($T_c=71$ K)
Bi$_{2}$Sr$_{2}$CaCu$_{2}$O$_{8+\delta}$ sample on the position in 
the Brillouin zone.
Here, $r$ denotes the radial distance  from the $(\pi,0)$ point.
The dashed lines mark the Fermi surface crossings.
(From Ref. \cite{Gromko03},
Copyright \copyright 2003 APS).
}
\end{figure}

The $S$-shaped regions are observed also when bilayer splitting is resolved,
in this case in the bonding band. An example is shown in 
Fig.~\ref{Fig1_Gromko04}. 
In this case, for an overdoped sample, however, the degree to which the corresponding effects spread toward the nodal region is smaller than in
optimally and underdoped materials. 

The $S$-shaped dispersion in the MDC spectra is not observed in nodal
direction. Instead, a kink-like feature is present \cite{Bogdanov00,Kaminski01,Lanzara01,Johnson01,Sato03}, which sharpens 
in the superconducting state, and this extra sharpening has a temperature
dependence similar to that of the antinodal dispersion \cite{Johnson01,Kordyuk04}.

\subsubsection{The nodal kink}
\label{NK}

In nodal direction the dispersion is linear in the high-energy and low-energy
regions, with different slopes respectively. The two regions are
separated by a `kink', which is rather sharp in the superconducting state
\cite{Valla99,Bogdanov00,Kaminski01,Lanzara01}.
This kink is seen both in the EDC and MDC derived dispersion. However,
because the MDC width above the kink is rather large, the
EDC and MDC derived dispersions differ at high energies.
\begin{figure}
\centerline{
\epsfxsize=0.8\textwidth{\epsfbox{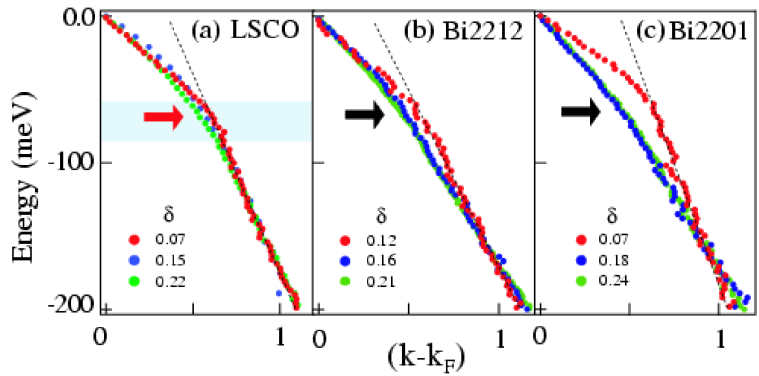}}
}
\caption{
\label{Fig1_Lanzara01}
MDC-derived dispersion for La$_{2-x}$Sr$_x$CuO$_4$ (LSCO),
Bi$_{2}$Sr$_{2}$CaCu$_{2}$O$_{8+\delta}$ (Bi2212), and
Bi$_{2}$Sr$_{2}$CaCuO$_{6+\delta}$ (Bi2201), for several doping levels
as indicated (optimal doping corresponds to $\delta=0.16$),
along the nodal ($\Gamma-Y$) direction. Data are taken at 20 K (a,b) and
30 K (c).
(From Ref. \cite{Damascelli03}, 
Copyright \copyright 2003 APS,
reprinted by permission from Mcmillan Publishers Ltd: Nature, Ref. \cite{Lanzara01} Copyright \copyright 2001 NPG).
}
\end{figure}
The nodal kink is seen in a large number of materials with different 
amounts of doping.
As can be seen in Fig.~ \ref{Fig1_Lanzara01}, it increases with 
underdoping and decreases with overdoping. In strongly overdoped samples
it is almost absent. As a function of temperature, it is sharp below $T_c$
and becomes more rounded in the normal state. 

The derivative of the nodal dispersion curve shows a jump at the kink position,
and constant parts below and above. These constants define a
Fermi velocity $v_F$ and a high-energy velocity $v_{HE}$.

\subsubsection{Fermi velocity}
\label{FV_ex}

\begin{figure}
\centerline{
\epsfxsize=0.40\textwidth{\epsfbox{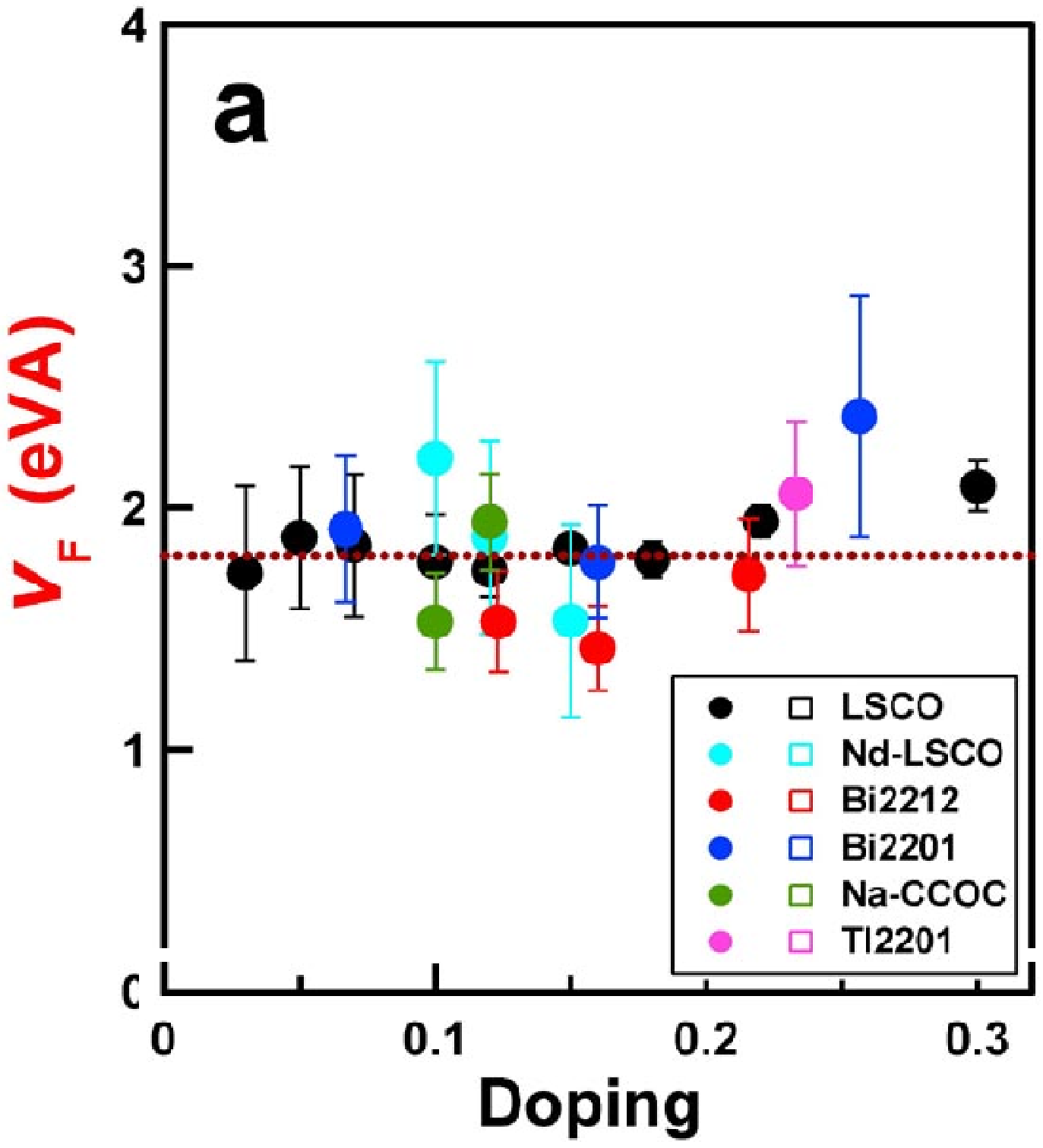}}
\epsfxsize=0.40\textwidth{\epsfbox{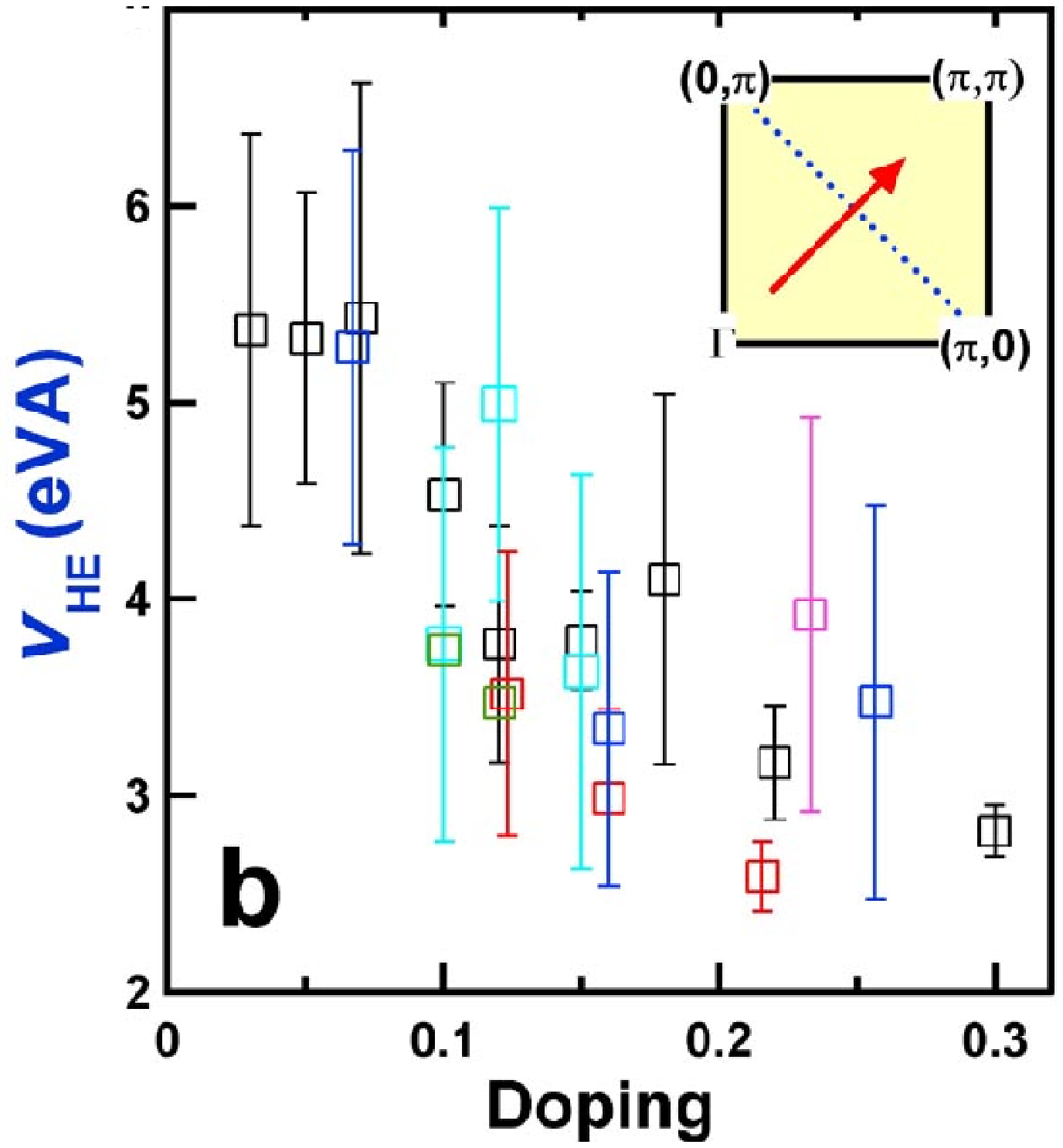}}
}
\caption{
\label{Fig2_Zhou03}
Normal state quasiparticle velocity along the nodal direction, obtained from
momentum-distribution curves, for various cuprates as function of doping.
The high energy dispersion is separated by a kink at about 70 meV from the low energy
dispersion with different slope. Accordingly, the velocities differ for the low-energy region and the
high-energy region.
(a) Low energy (determined from 0-50 meV) Fermi velocity  and (b) high energy (determined from 100-200 meV) velocity.
(Reprinted by permission from Mcmillan Publishers Ltd: Nature, Ref. \cite{Zhou03} Copyright \copyright 2001 NPG).
}
\end{figure}

Experimentally, because the dispersion of the EDC maxima and the MDC maxima differ from each other,
it is important to specify how the Fermi velocity is extracted from the data. 
Clearly, the picture is complicated by a strong energy dependence of self-energy effects.
Here the fact helps, that self-energy effects are weakly momentum dependent. Thus, although
the line shape of the EDC's is highly non-trivial both at
anti-nodal as well as at nodal points, the shape of the MDSs
is very well approximated by a Lorentzian.
From Eq.~(\ref{Specfun}) one can see that the velocity at binding energy
$\epsilon $ is given by,
\begin{equation}
\label{velocity}
\vec{v}(\epsilon )=\frac{\displaystyle \vec{v}_{0}(\epsilon)+
\partial_{\vec{k}} \Sigma'(\epsilon,\hat \vec{k})}{\displaystyle
1-\partial_{\epsilon } \Sigma'(\epsilon,\hat \vec{k})},
\end{equation}
where $\hat \vec{k}$ is the position of the MDC maximum, and it was assumed that
the momentum variation of $\Sigma''$ is negligible.
Assuming that (away from the saddle point)
the bare velocity is a constant in the energy region of interest,
$\vec{v}_0(\epsilon)\equiv \vec{v}_{F0}$, and taking into account that
the momentum dependence of the self energy is weak (this follows from the
fact that the MDC spectra are Lorentzian), then one can neglect the
energy dependence of $\partial_{\vec{k}}\Sigma'(\epsilon,\hat \vec{k})$ and
the main energy dependence comes from the renormalization factor
$Z(\epsilon,\vec{k})=1-\partial_{\epsilon } \Sigma'(\epsilon,\vec{k})$.
It is clear that
the ratio between the velocities $v_{HE}$ and $v_{F}$ in Fig.~\ref{Fig2_Zhou03}
gives directly the ratio between the quasiparticle renormalization factor
$Z=Z(0,\vec{k}_F)$ and a high energy renormalization $Z_{HE}(\epsilon )$, which
is only weakly energy dependent \cite{Randeria04}.

The Fermi velocity near the nodal point is large and of the order of 1.8 eV$\AA $ \cite{Kaminski01,Zhou03}
and virtually doping independent \cite{Zhou03}; weak systematic changes with doping are
e.g. in YBa$_2$Cu$_3$O$_{6+x}$ within 0.2 eV$\AA $ \cite{Borisenko06}.
This is in contrast to the slope of the dispersion above roughly
70 meV, which changes strongly with doping and amounts to 2.5-5.5 eV$\AA $ \cite{Johnson01,Zhou03,Borisenko06}.
In Fig. \ref{Fig2_Zhou03} the two velocities are shown for several cuprates as function of doping.

\begin{figure}
\centerline{
\epsfxsize=0.50\textwidth{\epsfbox{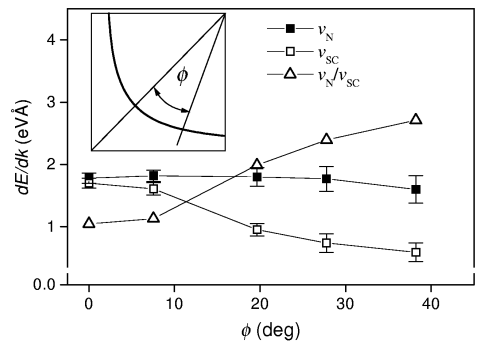}}
}
\caption{
\label{Fig3_Valla00}
Fermi velocities for optimally doped 
Bi$_{2}$Sr$_{2}$CaCu$_{2}$O$_{8+\delta}$ ($T_c=$91 K)
in normal (solid squares) and superconducting
(open squares) state as function of Fermi surface angle $\phi $ defined in the
inset. The ratio between normal state and superconducting state
Fermi velocities is also shown as open triangles.
(From Ref. \cite{Valla00},
Copyright \copyright 2000 APS).
}
\end{figure}

The angular dependence of the Fermi velocity $v_F$ along the Fermi surface is shown in Fig. \ref{Fig3_Valla00}. It is
rather isotropic in the normal state, however is renormalized differently in the superconducting state, leading
to an anisotropy along the Fermi surface. 
From Fig. \ref{Fig3_Valla00} one can see that
the Fermi velocity is reduced near the $(\pi,0) $ points of the Brillouin zone, however only slightly so near the nodes.

It is interesting to note, that the higher energy part of the nodal dispersion is linear to highest measured energies
\cite{Ronning03}, and does not extrapolate to the Fermi crossing
\cite{Bogdanov00,Lanzara01}. This suggests, that the high energy dispersion is 
strongly renormalized and cannot be described by simple models assuming a fixed bandwidth as function of doping.
Certainly, the whole band structure changes with doping, and it is the low energy part which stays surprisingly
stable from the overdoped to underdoped materials. This is reflected in the constance of the Fermi velocity shown
in Fig. \ref{Fig2_Zhou03} (a), in the weak change of the Fermi surface, and in the pinning of the saddle point 
singularity at the $M$ points to the low-energy region, within about 100 meV near the chemical potential.
The strong renormalization of the high-energy part of the spectrum 
lets us conclude, that if the continuum part of a bosonic spectrum
that couples to electrons is 
responsible for these renormalizations, then this bosonic spectrum
must extend to high ($\sim $eV) energies.

Finally, an estimate of the {\it bare} Fermi velocity can be obtained
by comparing the MDC and EDC widths of the spectra (assuming a definite
energy dependence of the self energy in the vicinity of the Fermi surface,
and a Lorentzian MDC lineshape).
In Ref. \cite{Kaminski05}
the bare Fermi velocity was determined for optimally doped Bi$_2$Sr$_2$CaCu$_2$O$_{8+\delta}$,
and was shown to vary from 4 eV$\AA $ at the node to 2 eV$\AA $ at the antinode.
This is consistent with Ref. \cite{Kordyuk05}, which found a nodal
bare Fermi velocity of 3.4 eV$\AA $ in optimally doped Bi(La)-2201 
($T_c=32 $K), of 3.8 eV$\AA $ in underdoped Bi(Pb)-2212 
($T_c=77 $K), and of 3.9 eV$\AA $ in overdoped doped Bi(Pb)-2212 ($T_c=75 $K).

\subsubsection{Spectral lineshape}
\label{SpecLS_ex}

It has been known for some time that near the
$(\pi,0)$ point of the zone, the spectral function in the superconducting
state of Bi$_2$Sr$_2$CaCu$_2$O$_{8+\delta}$
shows an anomalous lineshape, the so called `peak-dip-hump'
structure \cite{Dessau91,Randeria95,Ding96,Norman97}.
This structure was also found in
YBa$_2$Cu$_3$O$_{7-\delta}$ \cite{Lu01}, and in
Bi$_2$Sr$_2$Ca$_2$Cu$_3$O$_{10+\delta}$ \cite{Feng02,Sato02}.

Extensive studies on Bi$_2$Sr$_2$CaCu$_2$O$_{8+\delta}$ 
as a function of temperature revealed
that this characteristic shape of the spectral function is closely related to
the superconducting state.
In the normal state, the ARPES spectral function is broadened
strongly in energy, the broadening increasing with underdoping \cite{Ding96}.
When lowering the temperature below $T_c$, a coherent quasiparticle
peak grows at the position of the leading edge gap, and the incoherent
spectral weight is redistributed to higher energy, giving rise to a dip and hump
structure \cite{Dessau91,Randeria95,Norman97}.
This peak-dip-hump structure is most
strongly developed near the $M$-point of the Brillouin zone.
Below $T_c$, the spectral peak quickly narrows with decreasing
temperature \cite{Norman01}, and sharp
quasiparticle peaks were identified
well below $T_c$ along the entire Fermi surface \cite{Kaminski00}.
\begin{figure}
\centerline{
\epsfxsize=0.75\textwidth{\epsfbox{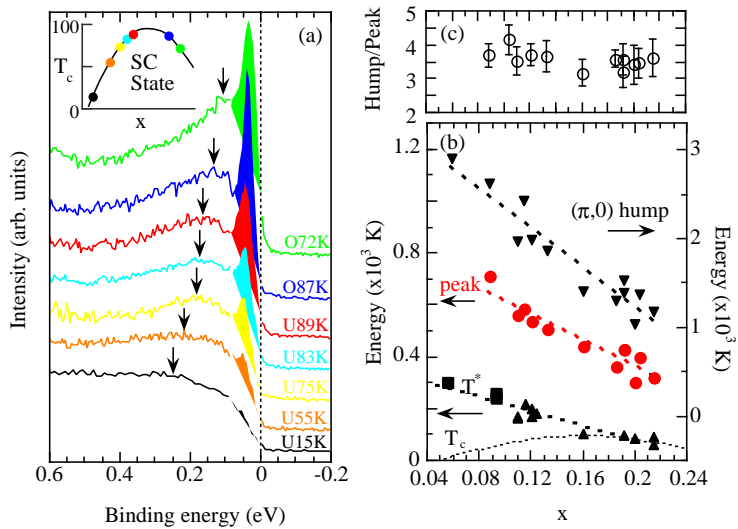}}
}
\caption{
\label{Fig4_Campuzano}
(a) Doping dependence of experimental ARPES spectra at the ($\pi, 0$) point for
Bi$_2$Sr$_2$CaCu$_2$O$_{8+\delta}$ from overdoped (O) 
to underdoped (U) at $T=15 $K. The inset shows the corresponding 
values for $T_c$. The quasiparticle coherent weight is indicated by shadowing and
decreases with underdoping.
(b) Doping dependence of the pseudogap temperature scale $T^\ast$, the
quasiparticle peak binding energy, the binding energy of the hump
feature, denoted by an arrow in (a), and the ratio of the latter two in (c).
(From Ref. \cite{Campuzano99},
Copyright \copyright 1999 APS).
}
\end{figure}
The doping dependence of the spectral lineshape was carefully studied by Campuzano {\it et al.} \cite{Campuzano99}.
In Fig. \ref{Fig4_Campuzano} it is seen that the peak loses weight with underdoping. 
The peak, dip, and hump feature all move to higher binding energy with underdoping.

The well defined quasiparticle peaks at low energies 
contrast to the high energy 
spectra, which show a broad linewidth which grows linearly
in energy \cite{Valla99,Yusof02}.
This implies that a scattering channel present in the normal
state becomes gapped in the superconducting state \cite{Kuroda90}.
The high energy excitations then stay broadened, since they involve
scattering events above the threshold energy.
While this explains the existence of sharp quasiparticle peaks, a gap in the 
bosonic spectrum which mediates electron interactions leads only to a weak 
dip-like feature \cite{Littlewood92}.
This suggests that the dip feature is instead due to the 
interaction of electrons with a sharp (in energy)
bosonic mode. The sharpness implies a strong self-energy
effect at an energy equal to the mode energy plus the quasiparticle peak
energy, giving rise to a spectral dip \cite{Norman98}. The fact that
the effects are strongest at the $M$ points implies a mode momentum
close to the $(\pi,\pi)$ wavevector \cite{Shen97}.

\begin{figure}
\centerline{
\epsfxsize=0.35\textwidth{\epsfbox{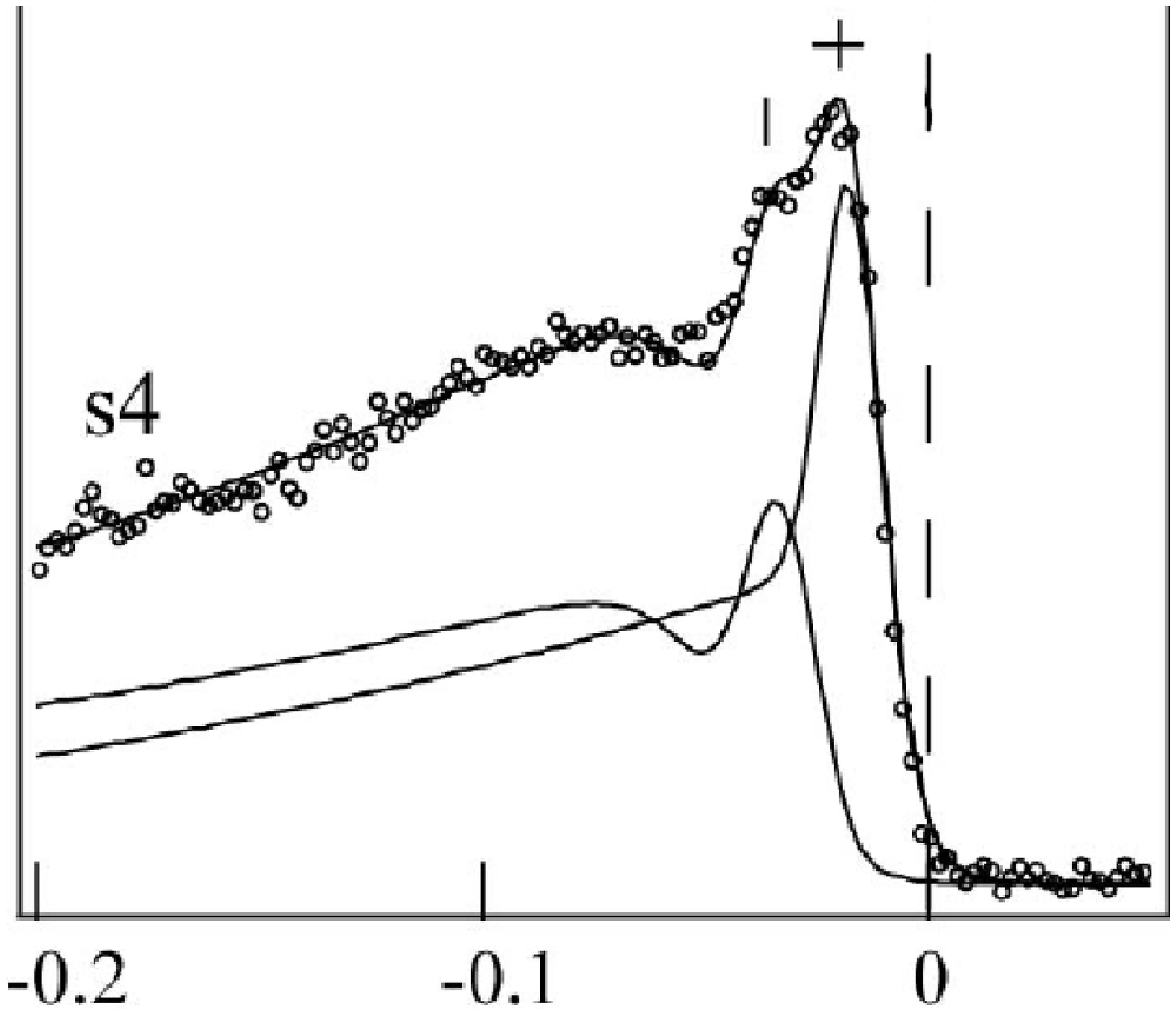}}
\epsfxsize=0.42\textwidth{\epsfbox{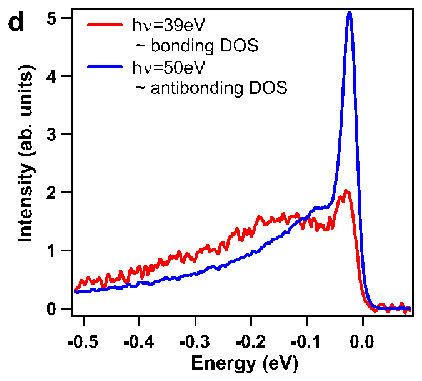}}
}
\caption{
\label{Fig3_Feng01}
Left:
ARPES spectrum for an overdoped
Bi$_{2}$Sr$_{2}$CaCu$_{2}$O$_{8+\delta}$ sample ($T_c=$ 65 K) at
the point $(0.11,1) \pi $ of the Brillouin zone as function of energy
(in eV), measured at $T= 10$ K. 
The cross refers to the antibonding peak and the bar to the bonding peak.
An experimental fitting procedure assigns the high-energy hump mainly to
the bonding spectral function, as seen from the three fitting curves shown
for the bonding, the antibonding, and the total spectrum.
(From Ref. \cite{Feng01a},
Copyright \copyright 2001 APS).
Right:
Bonding band and antibonding band ARPES spectrum for a
(Bi,Pb)$_{2}$Sr$_{2}$CaCu$_{2}$O$_{8+\delta}$ sample at
the $(\pi,0)$ point of the Brillouin zone
measured at $T= 20$ K. Here, the bonding and antibonding contributions have
been separated by varying the photon energy $h\nu$.
(From Ref. \cite{Borisenko05},
Copyright \copyright 2006 APS).
}
\end{figure}

The non-trivial spectral line-shape is further complicated due to the presence
of bilayer splitting. In this case the different behavior of matrix elements
as function of the photon energy for bonding and antibonding bands has to
be used in order to separate the effects.
It was found, that the peak-dip-hump structure is also present when the
bilayer splitting is resolved. Two examples 
are shown in Fig. \ref{Fig3_Feng01}.
In Fig.~\ref{Fig3_Feng01} (left),
the bonding band shows a pronounced dip in the spectrum
near the bonding band Fermi momentum. Consequently, there are strong self-energy
effects present even in such overdoped materials.
The same Figure also reveals that the self-energy effect in the antibonding band
is much weaker. 
In Fig.~\ref{Fig3_Feng01} (right),
the difference in the spectral line-shapes between bonding
and antibonding band spectra is shown at the 
$(\pi,0)$ point of the Brillouin zone, revealing again a stronger
dip feature in the bonding spectrum. This characteristic asymmetry
between bonding and antibonding line-shapes has been an important
information for the assignment of the effect to the interaction of
quasiparticles
with the spin-1 resonance mode \cite{Eschrig02,Borisenko05}.

\subsubsection{The antinodal quasiparticle peak}
\label{AQP}

The antinodal quasiparticle peak (or coherence peak) in the superconducting state 
determines the spectral gap. 
With underdoping, the sharp quasiparticle peak moves
to higher binding energy, indicating that the gap increases \cite{Campuzano99}.
The quasiparticle peak has also been traced as a function of Fermi surface angle, and
has been found consistent with a $d$-wave symmetry of the order parameter 
\cite{Dessau91,Wells92,Shen93}. 
The $d$-wave symmetry of the superconducting order parameter was unambiguously
demonstrated by phase sensitive tests (see \cite{Tsuei00}).

\begin{figure}
\centerline{
\epsfxsize=0.40\textwidth{\epsfbox{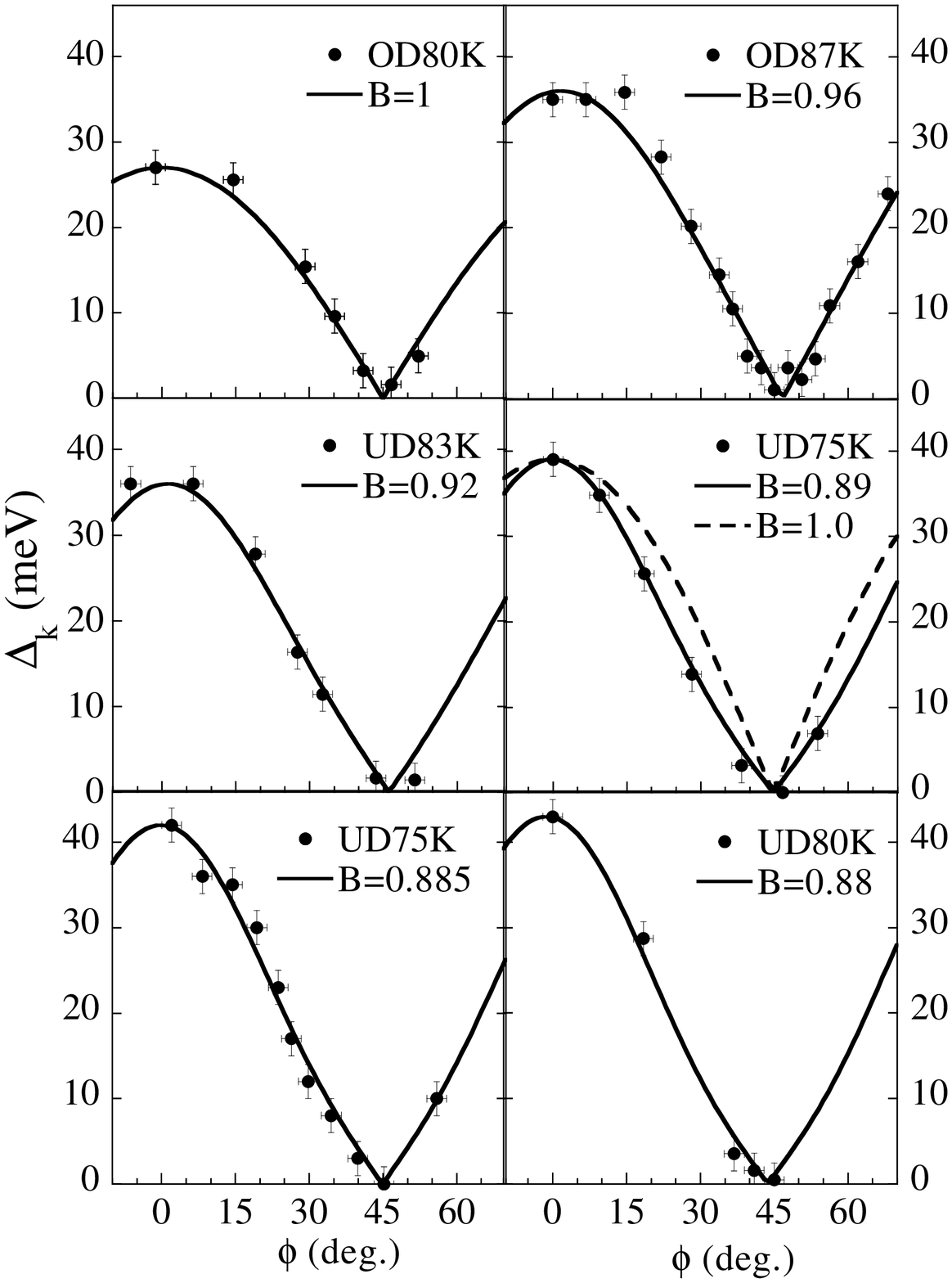}}
\epsfxsize=0.30\textwidth{\epsfbox{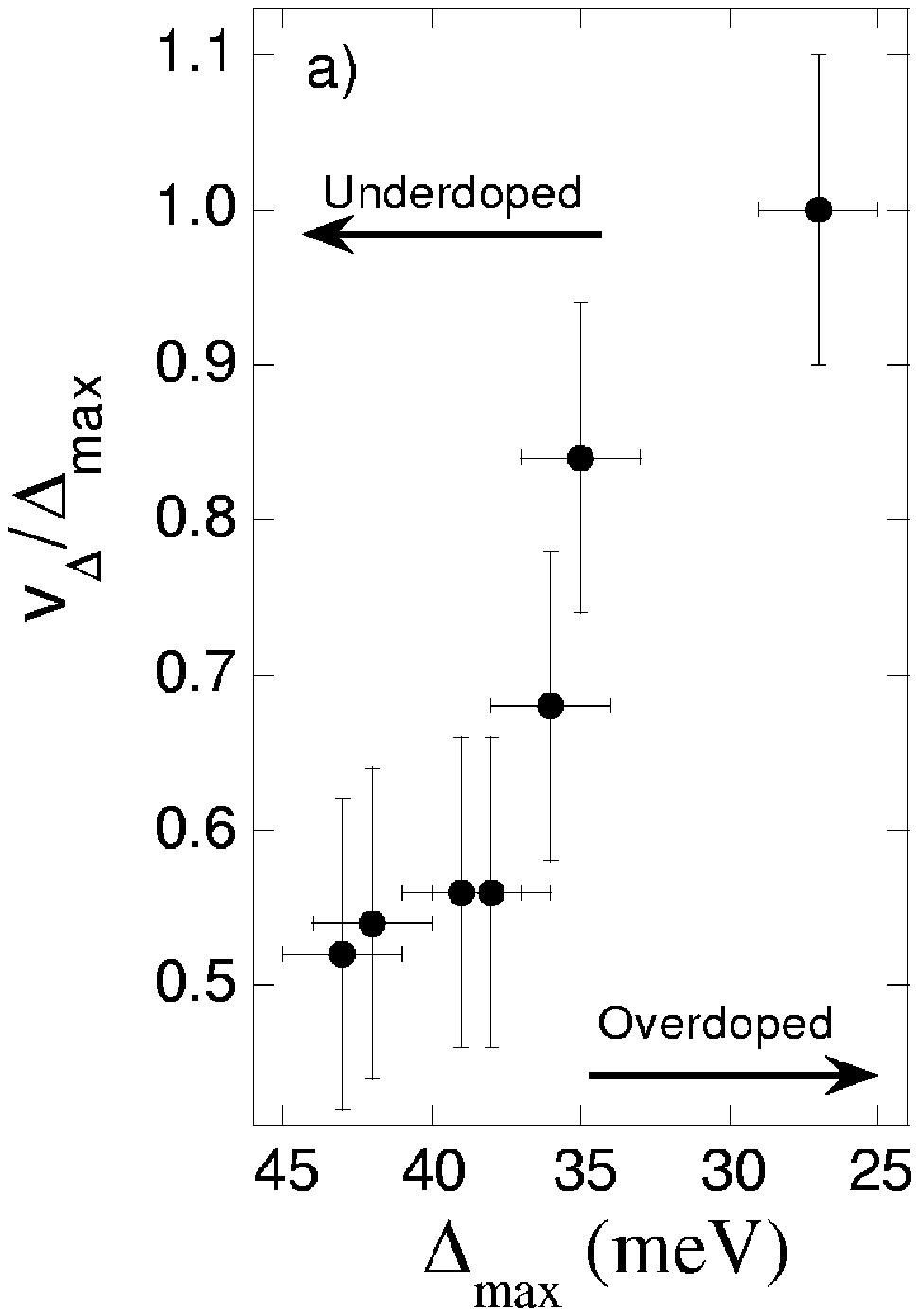}}
\epsfxsize=0.25\textwidth{\epsfbox{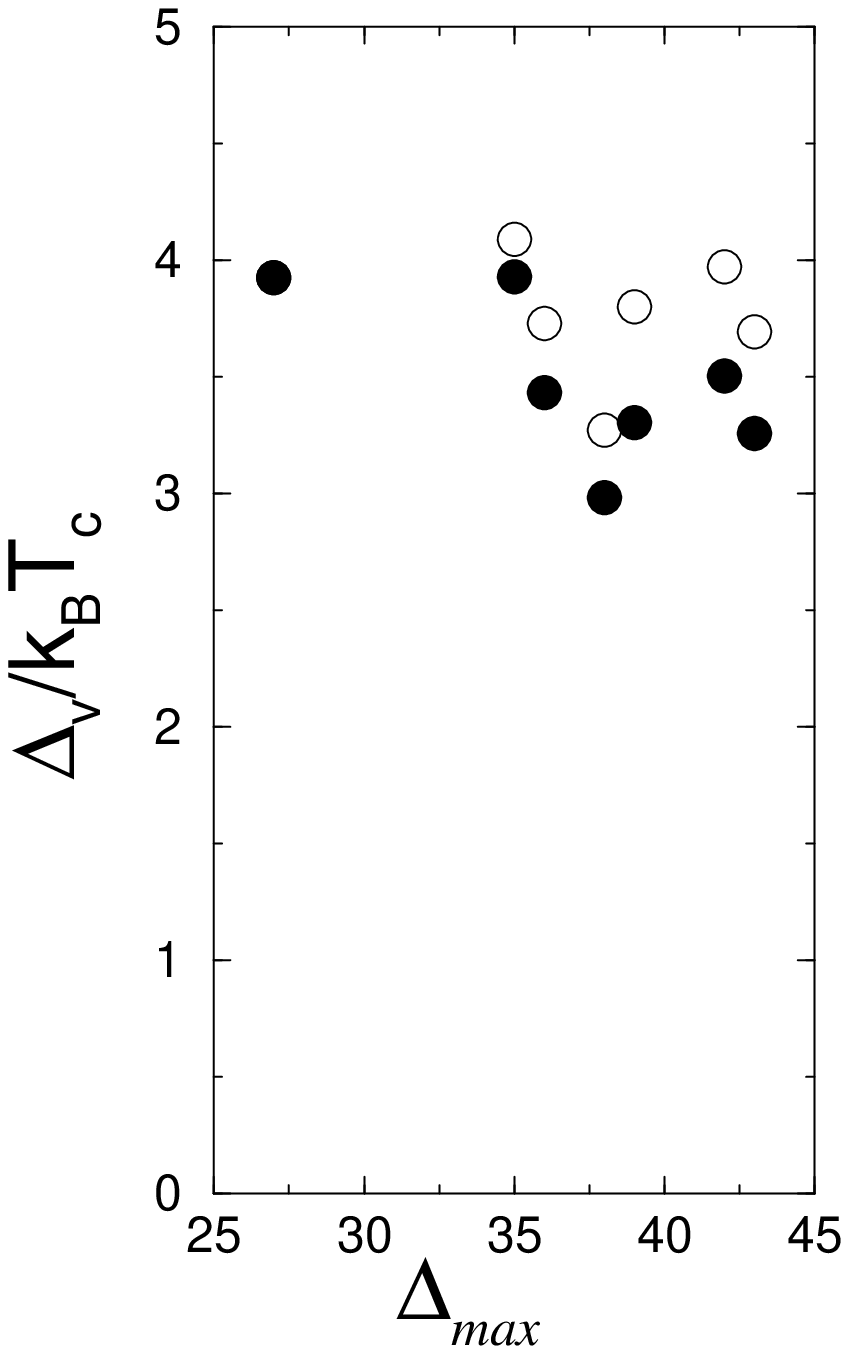}}
}
\caption{
\label{Fig2_Mesot99}
Left:
Superconducting gap as function of Fermi surface angle $\phi$ for a series
of (Bi,Pb)$_{2}$Sr$_{2}$CaCu$_{2}$O$_{8+\delta}$ samples with varying doping.
The fits use a functional form
$\Delta_{k}=\Delta_{\rm max}[B\cos(2\phi)+(1-B)\cos(6\phi)]$. The dashed line
shows for comparison the function $\Delta_{\rm max}\cos(2\phi)$.
(From Ref. \cite{Mesot99}).
Middle: Normalized slope of the gap at the node
($v_{\Delta}/\Delta_{\rm max}$) vs gap maximum $\Delta_{\rm max}$.
(From Ref. \cite{Mesot99},
Copyright \copyright 1999 APS).
Right: The ratio between the nodal order parameter, defined by the
formula $\Delta_{k}=\Delta_{\rm v}\cos(2\phi)$ in such a way that the
slopes $v_\Delta $ coincides with the measured one. Open circles are
directly from experimental $v_\Delta $ (middle panel), full circles are obtained
by using the $v_\Delta $ from the fitted  curves in the left picture.
}
\end{figure}

In Fig.~\ref{Fig2_Mesot99}, data for the gap-anisotropy as determined from ARPES
experiments are shown for several doping levels. 
The magnitude of the gap is 
clearly consistent with a $d_{x^2-y^2}$-wave symmetry of the order parameter. 
In optimally doped materials the magnitude of the gap follows very closely a
cos$(2\phi )$-dependence on the Fermi surface angle \cite{Ding96}, 
\begin{equation}
\label{DWOP1}
\Delta_{\vec k}=\frac{1}{2} \Delta_M \left[ \cos (k_x a) -\cos (k_y a) \right]
\end{equation}
which takes its maximal value $\Delta_M$ closest to the $M$ point of the 
Brillouin zone. 
Whereas the same
holds also for overdoped materials, in underdoped materials
deviations from this simple behavior have been detected \cite{Mesot99,Borisenko02}.
Interestingly, toward underdoping the slope of
the order parameter along the Fermi surface at the nodal point decreases,
although the maximal gap at the antinodal point increases (this is in
contrast to what is deduced from, e.g., thermal conductivity measurements
\cite{Sutherland03}). 
As shown in Fig. \ref{Fig2_Mesot99} (right), the
ratio between this slope and the transition temperature stays roughly constant.
This is what one would expect for a BCS superconductor. In strong contrast,
the ratio of the maximal gap to the transition temperature sharply rises
with underdoping.
This might be an indication
that in underdoped cuprates an additional order parameter is present near
the antinodes. This idea is supported by the experimental 
fact that in the pseudogap phase the antinodal regions stay gapped, whereas
the nodal regions show well defined Fermi surface pieces.

Borisenko {\it et al.} \cite{Borisenko02} have measured the gap separately for the bonding and
antibonding bands and have found, that for underdoped 
Bi(Pb)$_2$Sr$_2$CaCu$_2$O$_{8-\delta }$ ($T_c=$ 77 K) each of the bilayer split
bands supports in the superconducting state a gap of the same magnitude. 
The consequences become clear when transforming the order parameters from
plane representation to bonding-antibonding representation, 
\begin{eqnarray}
\Delta_\parallel (\vec{k}) &=& \frac{\Delta^{(a)}_{\vec{k}}+\Delta^{(b)}_{\vec{k}}}{2} , \\
\Delta_\perp  (\vec{k}) &=&  \frac{\Delta^{(a)}_{\vec{k}}-\Delta^{(b)}_{\vec{k}})}{2} .
\end{eqnarray}
The apparent experimental finding, that
\begin{equation}
\Delta^{(a)}_{\vec{k}}=\Delta^{(b)}_{\vec{k}},
\end{equation}
means that the interplane pairing interaction $V_\perp$ vanishes.

\begin{figure}
\centerline{
\begin{minipage}{0.46\textwidth}
\epsfxsize=1.\textwidth{\epsfbox{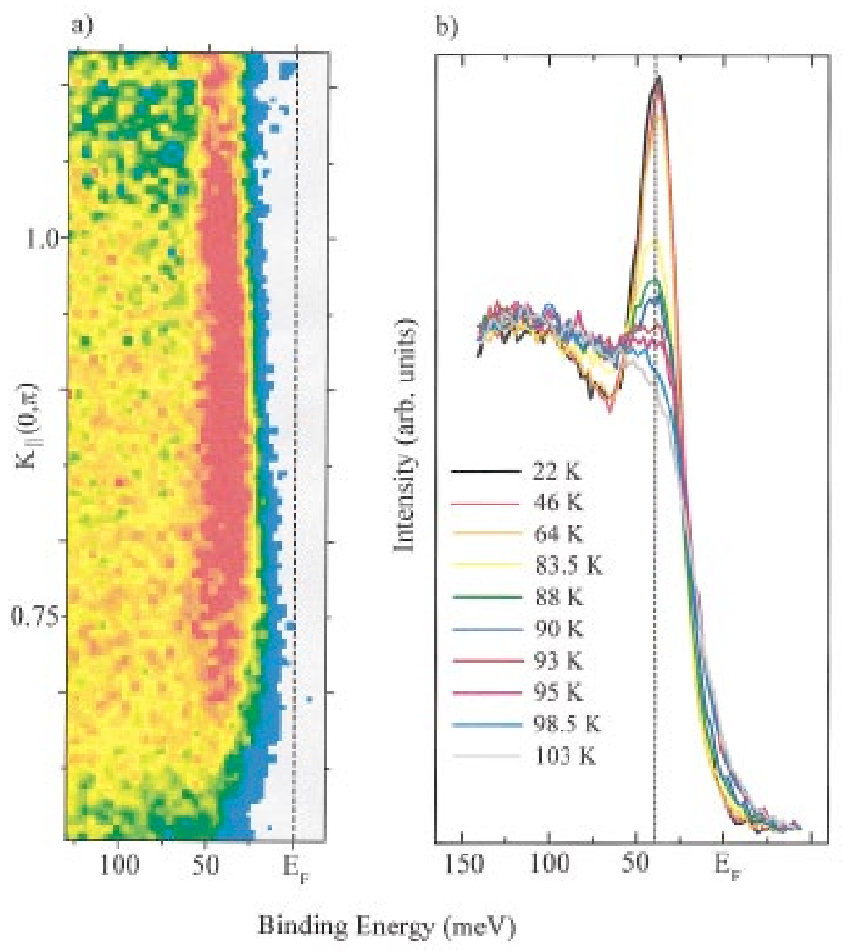}}
\end{minipage}
\begin{minipage}{0.23\textwidth}
\begin{minipage}{1\textwidth}
\epsfxsize=1.\textwidth{\epsfbox{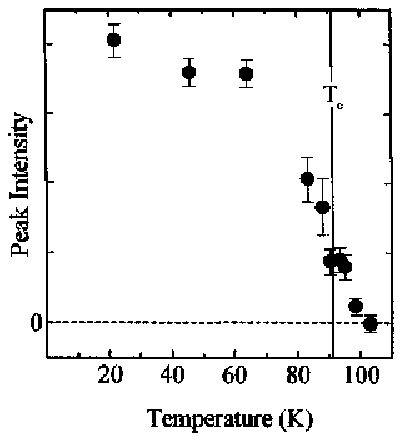}}
\end{minipage}\\
\begin{minipage}{1\textwidth}
\epsfxsize=1.\textwidth{\epsfbox{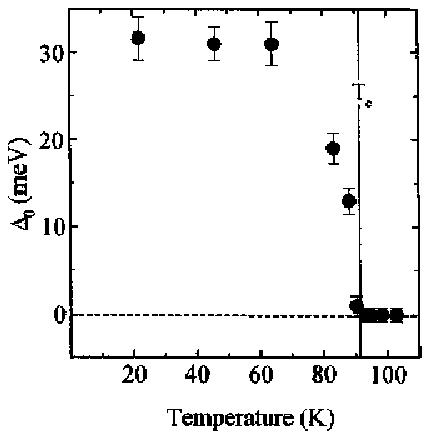}}
\end{minipage}
\end{minipage}
}
\caption{
\label{Fig1_Fedorov99}
(a) Spectral density plot in $\Gamma M$ direction  for
an optimally doped Bi$_2$Sr$_2$CaCu$_2$O$_{8+\delta}$ sample ($T_c$=91 K)
at $T=46$K. The superconducting peak intensity is seen as dispersionless
feature of high intensity between $\sim 25\ldots 50$ meV binding energy. (b) Spectra integrated over the region in (a) as function of temperature.
The strongest peak intensity is observed at lowest temperatures.
On the right the peak intensity and the gap $\Delta_0$ are shown as function
of temperature.
(From Ref. \cite{Fedorov99},
Copyright \copyright 1999 APS).
}
\end{figure}

Apart from the binding energy of the quasiparticle peak its spectral weight is of interest.
In Fig. \ref{Fig1_Fedorov99} the temperature evolution of the spectral weight is
reproduced for optimally doped Bi$_2$Sr$_2$CaCu$_2$O$_{8+\delta}$.
The weight follows an order parameter like behavior, and becomes very small in the
normal state. It was argued \cite{Norman01} that using a different modeling of the
spectral lineshape, the peak broadens drastically when entering the normal state,
instead of a reduction of the peak weight. 
The analysis in Fig. \ref{Fig1_Fedorov99} uses a phenomenological model to fit
the peak and the remaining part of the spectrum separately, in order to extract
the peak weight.

\begin{figure}
\centerline{
\epsfxsize=0.76\textwidth{\epsfbox{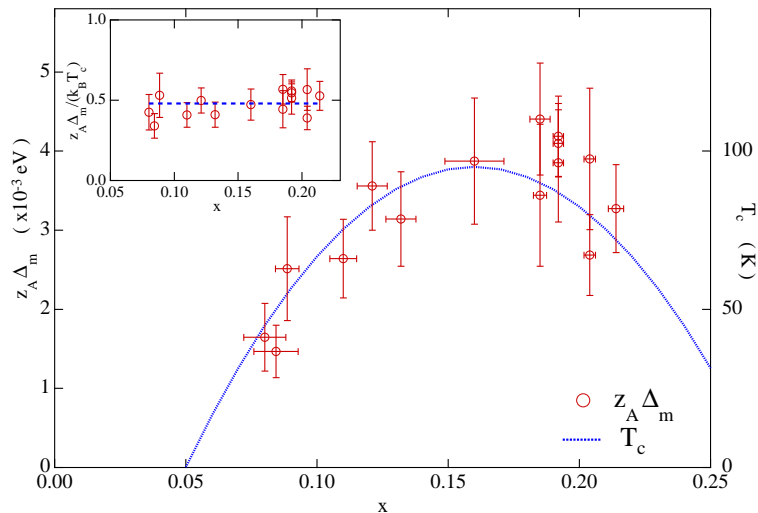}}
}
\caption{
\label{Fig3_Ding01}
Doping dependence
of $z_A\Delta_m$ (open circles) at $(\pi,0)$ in comparison with
$T_c$. 
Here, $z_A$ is the low-temperature coherent weight
and $\Delta_m$ the maximum gap obtained from the
position of the quasiparticle peak. 
The inset shows the ratio $z_A\Delta_m/(k_BT_c)$.
All data are at $T=$14 K.
(From Ref. \cite{Ding01},
Copyright \copyright 2001 APS).
}
\end{figure}

The spectral weight $z_A$ of the peak as a function of doping is 
discussed in Fig. \ref{Fig3_Ding01}.
A drop of the peak weight with underdoping was also analyzed in Refs.
\cite{Campuzano99,Feng00} and can be seen directly in Fig. \ref{Fig4_Campuzano}.
The quantity 
$z\Delta_{M}/k_BT_c$ stays roughly constant as function of doping, as seen
in the inset in Fig. \ref{Fig3_Ding01} \cite{Ding01}.
Also, the hump moves to higher binding energy and loses
weight with underdoping \cite{Campuzano99}.
This doping evolution of the quasiparticle peak points to an increasing
mode intensity at the $(\pi,\pi)$ wavevector with underdoping.
Again, there is a similarity to the nodal direction:
the low energy renormalization of the dispersion 
below the kink energy increases with underdoping \cite{Johnson01},
consistent with a common origin of both effects.

\subsubsection{The spectral dip feature}
\label{SDF}
\begin{figure}
\centerline{
\epsfxsize=0.68\textwidth{\epsfbox{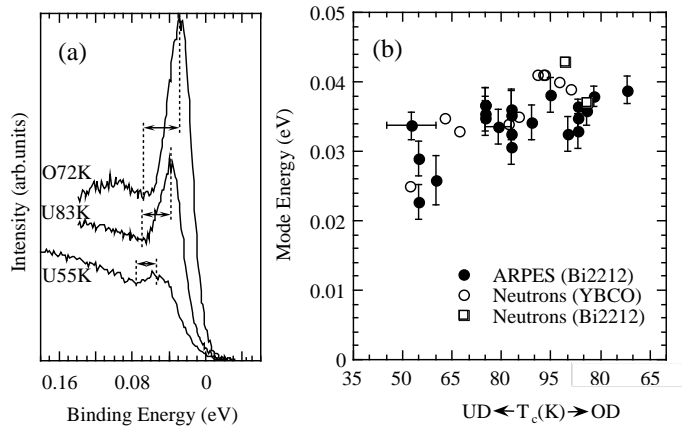}}
}
\caption{
\label{Fig5_Campuzano99}
(a) ARPES-spectra  for Bi$_{2}$Sr$_{2}$CaCu$_{2}$O$_{8+\delta}$
at $(\pi,0)$ for different amounts of doping, showing the determination of the
peak-dip separation. (b) Doping dependence of the mode energy
inferred from ARPES, and from INS \cite{Bourges98,Mook98c,Fong99}.
(From Ref. \cite{Campuzano99},
Copyright \copyright 1999 APS).
}
\end{figure}

If one assumes that the spectral-dip feature is due to coupling of
quasiparticles to a sharp bosonic mode, then one can determine the
mode energy from the ARPES spectra. 
The energy of the bosonic mode, as inferred from
the energy separation $\Omega_0$ between the peak and the dip,
was shown to decrease with underdoping \cite{Campuzano99}.
As was shown in theoretical studies \cite{Norman98,Eschrig00,Eschrig03} it is
the position of the spectral dip with respect to the quasiparticle peak which
determines the characteristic frequency of self-energy effects. The position of the hump
feature does not contain as much reliable informations about the self energy.
The doping variation of the peak-dip separation 
is shown in Fig.\ref{Fig5_Campuzano99}, where for comparison
also the mode frequencies of the magnetic resonance mode as inferred from INS
experiments are plotted. The agreement is striking.

In fact, as inferred from tunneling experiments, the peak-dip separation also
decreases with overdoping, so that it follows roughly the superconducting 
transition temperature \cite{Zasadzinski01}.
Similarly, the kink energy is maximal at optimal doping and decreases both with
underdoping and overdoping \cite{Johnson01}, indicating some relationship
between the kink at the nodal $N$ point 
and the peak-dip-hump structure at the $M$ point.

\subsubsection{Real part of self energy: renormalization of dispersion}
\label{RPSE}

The most direct test for the interaction of quasiparticles with bosonic modes
comes from the recent achievement of the direct determination of the
real part of the self energy. 
This determination is based on the fact that the dispersion of the MDC maxima
$\hat \vec{k}$
is determined from Eq.~(\ref{Specfun}) as
\begin{equation}
\label{ReSig}
\epsilon - \xi_{\hat\vec{k}} - \Sigma'(\epsilon ,\hat\vec{k}) =0,
\end{equation}
where 
$\xi_{\vec{k}} $ is the bare dispersion, and 
$\Sigma'(\epsilon ,\hat \vec{k}) $ the real part of the self energy at
the MDC maximum. It was assumed here that 
$\Sigma''(\epsilon ) $ is not momentum dependent in the region where the
MDC-Lorentzian is peaked (see later).
There are two ways to proceed from here.
One either determines the bare dispersion independently 
(see e.g. Ref. \cite{Kordyuk05}),
or one subtracts from Eq.~(\ref{ReSig}) the corresponding equation for the
normal state. 
Assuming a linear bare dispersion and neglecting the momentum variation of
$\Sigma'(\epsilon, \hat {\bf k})$ for $\hat {\bf k}$ varying between 
$\hat {\bf k}_N$ and $\hat {\bf k}_{SC}$, where (N) refers to the
normal and (SC) to the superconducting state, one obtains
\begin{equation}
{\bf v}_{F0} ( \hat {\bf k}_N -\hat {\bf k}_{SC})\approx
\Sigma_{SC}'(\epsilon ,\hat\vec{k}) - \Sigma_N'(\epsilon ,\hat\vec{k}) .
\end{equation}

\begin{figure}
\centerline{
\begin{minipage}{0.44\textwidth}
\epsfxsize=1.0\textwidth{\epsfbox{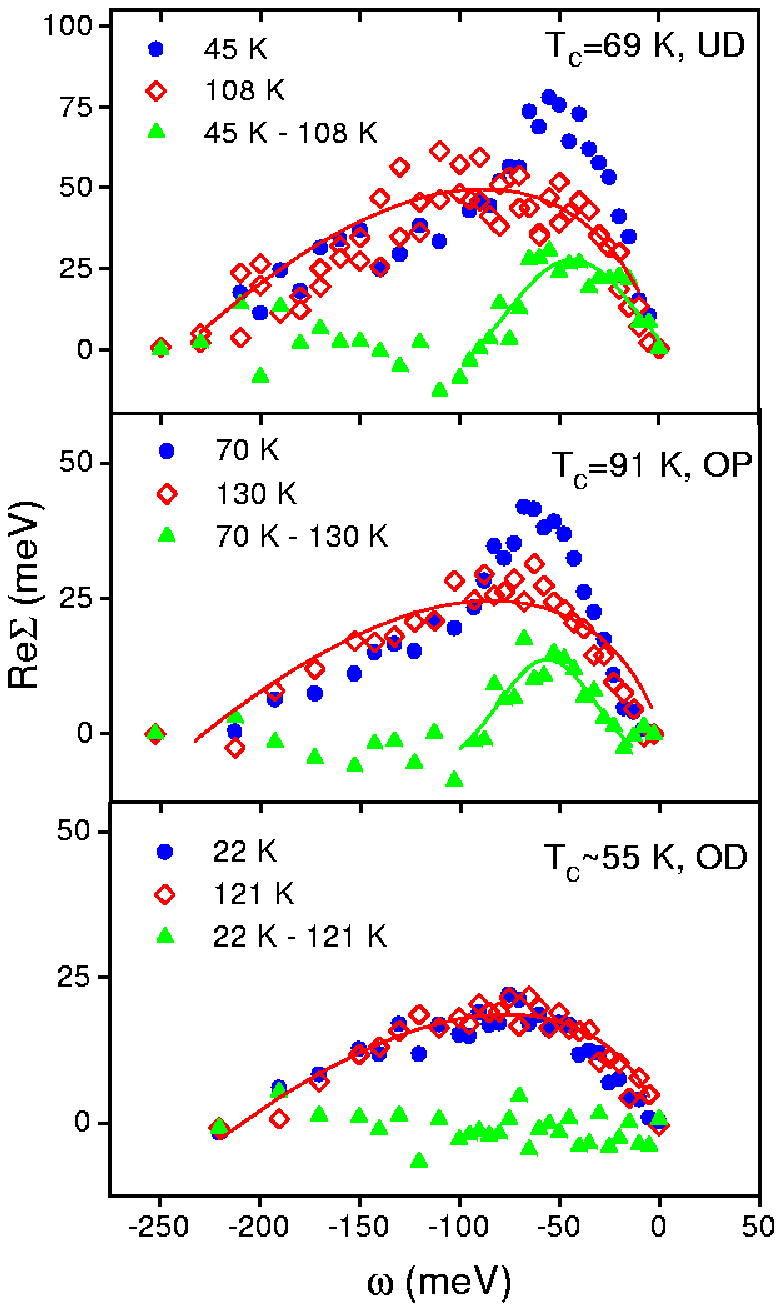}}
\end{minipage}
\begin{minipage}{0.44\textwidth}
\begin{minipage}{1.0\textwidth}
\epsfxsize=0.7\textwidth{\epsfbox{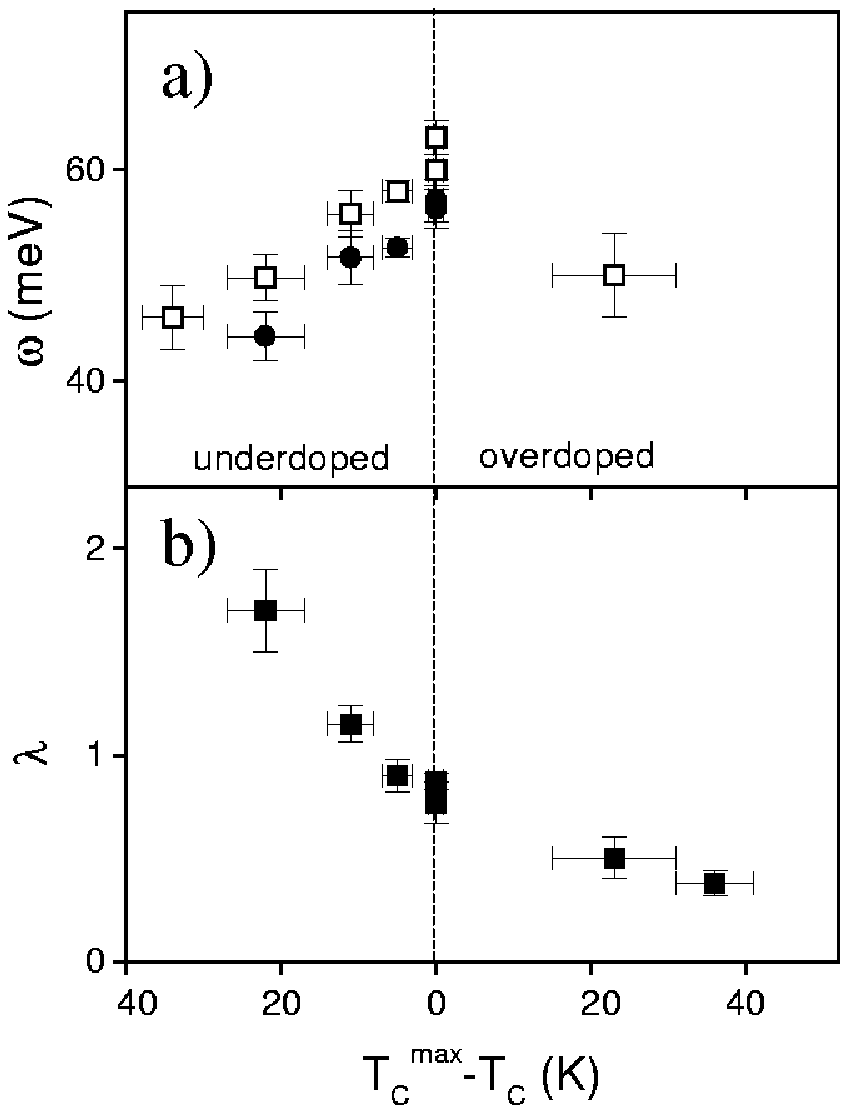}}
\end{minipage}\\
\begin{minipage}{1.0\textwidth}
\epsfxsize=0.9\textwidth{\epsfbox{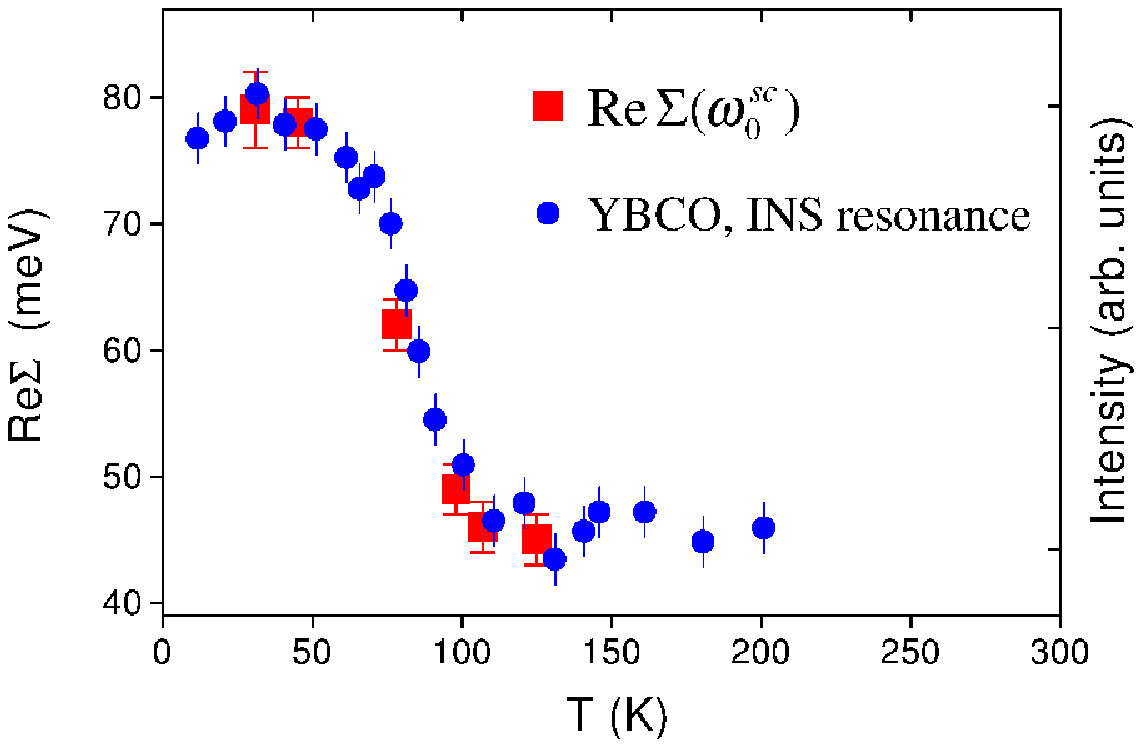}}
\end{minipage}
\end{minipage}
}
\caption{
\label{Fig1_Johnson}
Left:
Re $\Sigma (\omega )$ as function of binding energy for superconducting
(full dots)
and normal state (open diamonds) for an overdoped ($T_c$=55 K),
optimally doped ($T_c$=91 K), and underdoped ($T_c$=69 K) 
Bi$_{2}$Sr$_{2}$CaCu$_{2}$O$_{8+\delta}$  sample
taken in nodal direction. The  difference
between superconducting and normal Re $\Sigma (\omega )$ is shown as
triangles. The peak in this quantity defines the peak energy
$\omega_0^{sc}$. Right: at the top the energy $\omega_0$ of the maximum value
of Re $\Sigma (\omega )$ in the superconducting state, and $\omega_0^{sc}$ are
shown as function of doping. The middle panel shows the coupling constant
$\lambda =(\partial \mbox{Re} \Sigma /\partial \omega)_{E_F}$.
In the bottom panel the temperature dependence of Re $\Sigma (\omega )$ in nodal 
direction is shown for the underdoped sample, together with the intensity
of the resonance mode observed in INS studies of underdoped YBa$_2$Cu$_3$O$_{6+x}$
($T_c=74$ K, \cite{Dai99}).
(From Ref. \cite{Johnson01},
Copyright \copyright 2001 APS).
}
\end{figure}

Using the former technique Johnson {\it et al.} 
\cite{Johnson01} have studied the self-energy
effects in the nodal direction for several doping values.
As seen in Fig. \ref{Fig1_Johnson}, 
in addition to some interactions present already in the normal state
(most probably due to the spin-fluctuation continuum and due to
electron-phonon interaction),
there are clearly self-energy effects
which set in at the superconducting transition temperature.
These self-energy effects show an order-parameter-like temperature dependence,
very similar to the intensity of the resonance mode observed in INS experiments.
This effect can be assigned to the coupling with the magnetic resonance mode.
Note, that by no means that exhausts all self-energy effects in the nodal direction.
It is well established that electron-phonon interaction is present in cuprate
systems, and they lead to additional contributions to the real part of the
self energy \cite{Lanzara01,Lanzara02}. However, these contributions develop smoothly through $T_c$, 
in contrast to the effects discussed in Ref. \cite{Johnson01}.

It is clear from this study, that the self-energy effects in the nodal region
due to the resonance mode play no role for the overdoped materials. In Fig. \ref{Fig1_Johnson}
this can be seen from the lower left panel. The nodal coupling constant, shown in
Fig. \ref{Fig1_Johnson} (b), is in the overdoped region due to other effects.
However, the steep rise as a function of temperature when entering
the superconducting state, that is present 
in the optimally doped and underdoped region,
is consistent with an
interaction with the magnetic resonance mode, 
which sets in at the superconducting transition temperature.

\begin{figure}
\centerline{
\epsfxsize=0.68\textwidth{\epsfbox{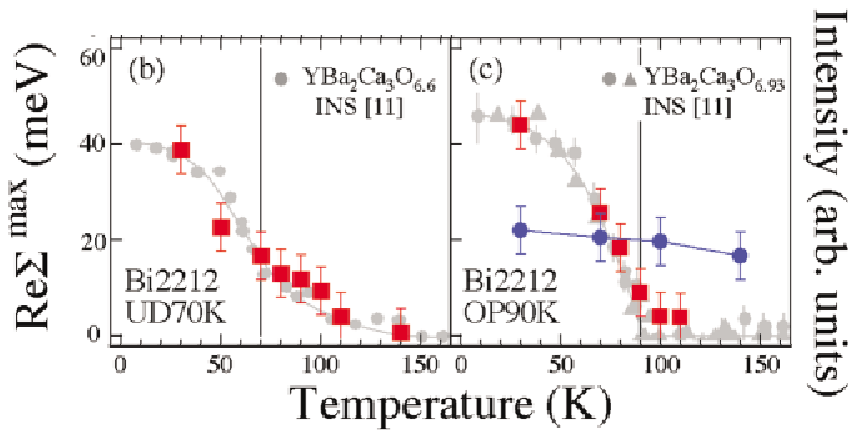}}
\epsfxsize=0.3\textwidth{\epsfbox{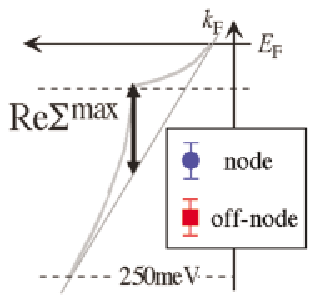}}
}
\caption{
\label{Fig4_Sato03}
Maximum value of Re $\Sigma (\omega )$ as function of temperature measured along
an off-nodal cut for 
underdoped ($T_c=$ 70 K) and optimally doped ($T_c=$ 90 K)
Bi$_{2}$Sr$_{2}$CaCu$_{2}$O$_{8+\delta}$. 
The off-nodal cut is performed about midway between the nodal and antinodal points.
For comparison, also points for a nodal cut are shown for the optimally doped case (circles).
The intensity of a resonance peak of YBa$_2$Cu$_3$O$_{6+x}$ with similar $T_c$ is superposed.
(From Ref. \cite{Sato03},
Copyright \copyright 2003 APS).
}
\end{figure}
A similar picture is obtained for off-nodal cuts, as shown in Fig. \ref{Fig4_Sato03} \cite{Sato03}.
The magnitude of Re $\Sigma $, determined as shown schematically on the right in
Fig. \ref{Fig4_Sato03}, is shown as function of temperature superimposed with the INS data.
The magnitude of the self-energy effects here even more clearly traces the intensity
of the resonance mode, both for underdoped and optimally doped materials. 
It was also shown, that
corresponding effects in single layered compounds are weak \cite{Sato03}.

\begin{figure}
\centerline{
\epsfxsize=0.78\textwidth{\epsfbox{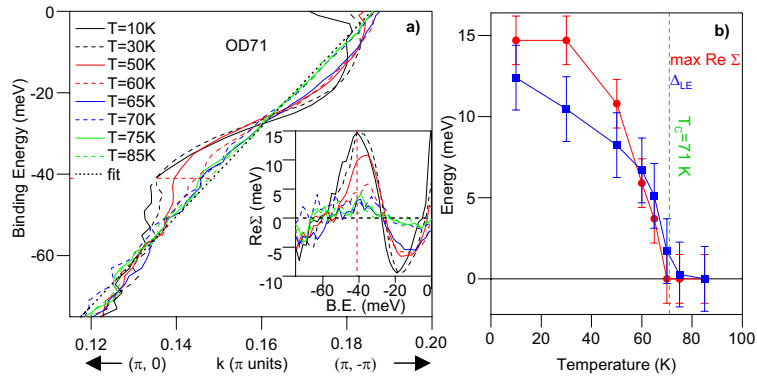}}
}
\caption{
\label{Fig2_Gromko03}
(a) MDC dispersion for an overdoped
Bi$_{2}$Sr$_{2}$CaCu$_{2}$O$_{8+\delta}$ ($T_c=$71 K) sample, along $(\pi,0)-(\pi,\pi)$ for
several temperatures. The inset shows the
temperature dependence of Re $\Sigma $. The strongest effect is observed at
lowest temperatures.
(b) Temperature dependence of the maximum in Re $\Sigma $ (circles) and the
superconducting leading edge gap $\Delta_{LE}$ (squares).
(From Ref. \cite{Gromko03},
Copyright \copyright 2003 APS).
}
\end{figure}
\begin{figure}
\centerline{
\epsfxsize=0.8\textwidth{\epsfbox{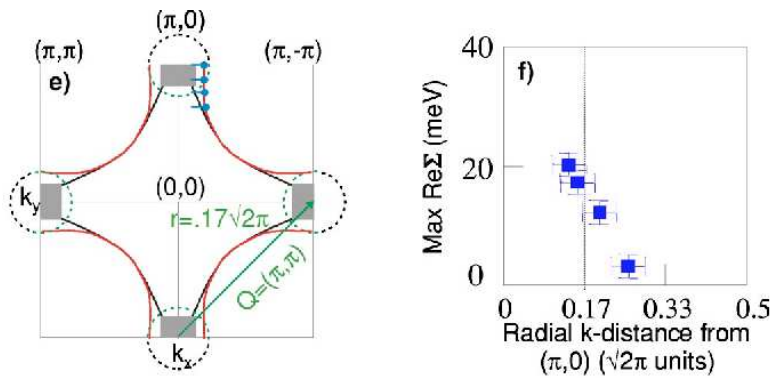}}
}
\caption{
\label{Fig1_Cuk04}
Experimentally deduced self-energy effects for an overdoped
Bi$_{2}$Sr$_{2}$CaCu$_{2}$O$_{8+\delta}$ ($T_c=$71 K) sample, 
analyzed for the bonding band. The locations of the MDC cuts are 
indicated in the Brillouin zone on the left hand side.
The circles denote the locus, where Re$\Sigma $ decays to half its maximal
value at $(\pi,0)$. The real part of the self energy Re$\Sigma $ as a function
of distance from the $(\pi,0)$ point, $r$, is shown in the right panel.
The order parameter node corresponds to $r=0.5$.
(From Ref. \cite{Gromko03},
Copyright \copyright 2003 APS).
}
\end{figure}

Gromko {\it et al.} \cite{Gromko03} have refined the studies of self-energy effects near the antinodal point
of the Brillouin zone by resolving separately bonding and antibonding bands.
An example is shown in Fig. \ref{Fig2_Gromko03} (left), where
for an overdoped Bi$_{2}$Sr$_{2}$CaCu$_{2}$O$_{8+\delta}$ sample 
clearly an $S$-shaped MDC dispersion has been observed near
the $(\pi,0)$ point
in the bonding band when entering the superconducting state. 
In contrast, in the normal state this self-energy effect is absent, as can
be seen from the temperature variation of the effect shown in
Fig~\ref{Fig2_Gromko03}.
This is a convincing prove that self-energy effects near the $(\pi,0) $ point of the
Brillouin zone, 
that are due to interactions of quasiparticles with some sharp mode, are
clearly observed {\it in addition} to bilayer splitting effects.
The temperature behavior of the real part of the self energy 
is shown on the right hand side of Fig~\ref{Fig2_Gromko03}.
As can be seen, the 
temperature behavior follows an order parameter like behavior. 
Furthermore, as the experiment
was performed on an overdoped sample, this shows that although for overdoped materials the
nodal renormalization due to the resonance mode vanishes, this is not so for the antinodal point.
The range for which the interaction with the  resonance mode is relevant is shown in Fig. \ref{Fig1_Cuk04}.
Clearly, for overdoped compounds there are renormalizations present in the antinodal region,
however these do not extend to the nodal point. It is for optimally and underdoped materials, that the
range of interaction also includes the nodal point.

In a study by Kim {\it et al.} \cite{Kim03a} the coupling strength parameter
for quasiparticles with binding energy $\Delta_M$ at the antinodal point, 
defined as
\begin{equation}
\lambda_{\Delta } = \frac{\Sigma' (-\Delta_M, \vec{k}_M )}{\Delta_M },
\end{equation}
was studied both in the bonding and antibonding bands of
(Pb,Bi)$_{2}$Sr$_{2}$CaCu$_{2}$O$_{8+\delta}$ with various doping levels
(the momentum dependence of $\Sigma'$ can be neglected in the region between
the Fermi crossings nearest to the $M$ point).
It was found, that $\lambda_{\Delta }$ increases with underdoping \cite{Borisenko03}. It reaches
very large values of approximately 8 for $p\approx 0.12$ ($T_c=77 $ K),
and for overdoped samples
with $p\approx 0.21$ ($T_c=69 $ K) still is approximately 3 \cite{Kim03a}.
Note in this regard, 
that the gap $\Delta_M$ itself increases with underdoping, and
consequently $\lambda_{\Delta}$ tests the real part of the self energy at
an increasing binding energy with decreasing doping level.
A similar analysis in the normal state yielded a
doping independent value for the coupling strength of around 1.
In the nodal direction, where the gap vanishes, the analogous quantity
\begin{equation}
\lambda_N = -\lim_{\epsilon \to 0}\frac{\Sigma' (\epsilon, \vec{k}_N )}{\epsilon },
\end{equation}
(where $\vec{k}_N$ is the nodal wavevector)
was found to be about 1 for the normal state, 
with additional self-energy effects in the superconducting
state \cite{Koitzsch04}.

\subsubsection{Imaginary part of self energy: quasiparticles lifetime}
\label{IPSE}

Interesting information can be obtained also from the linewidth of the MDC spectra,
which is directly connected to the imaginary part of the self energy, or to the scattering rate
of the quasiparticles.
From Eq.~(\ref{Specfun}) it is seen, that 
under the assumption that the momentum variation of the imaginary part of the
self energy $\Sigma''(\epsilon,\vec{k})$ can be neglected in the region
in which the Lorentzian is peaked,
the MDC half-width half-maximum in direction of $\vec{v}_{F0}$, $W_{MDC}$, is given by,
\begin{equation}
W_{MDC}=\frac{\Sigma''(\epsilon,\vec{k}) }{v_{F0}}.
\end{equation}

\begin{figure}
\centerline{
\epsfxsize=0.60\textwidth{\epsfbox{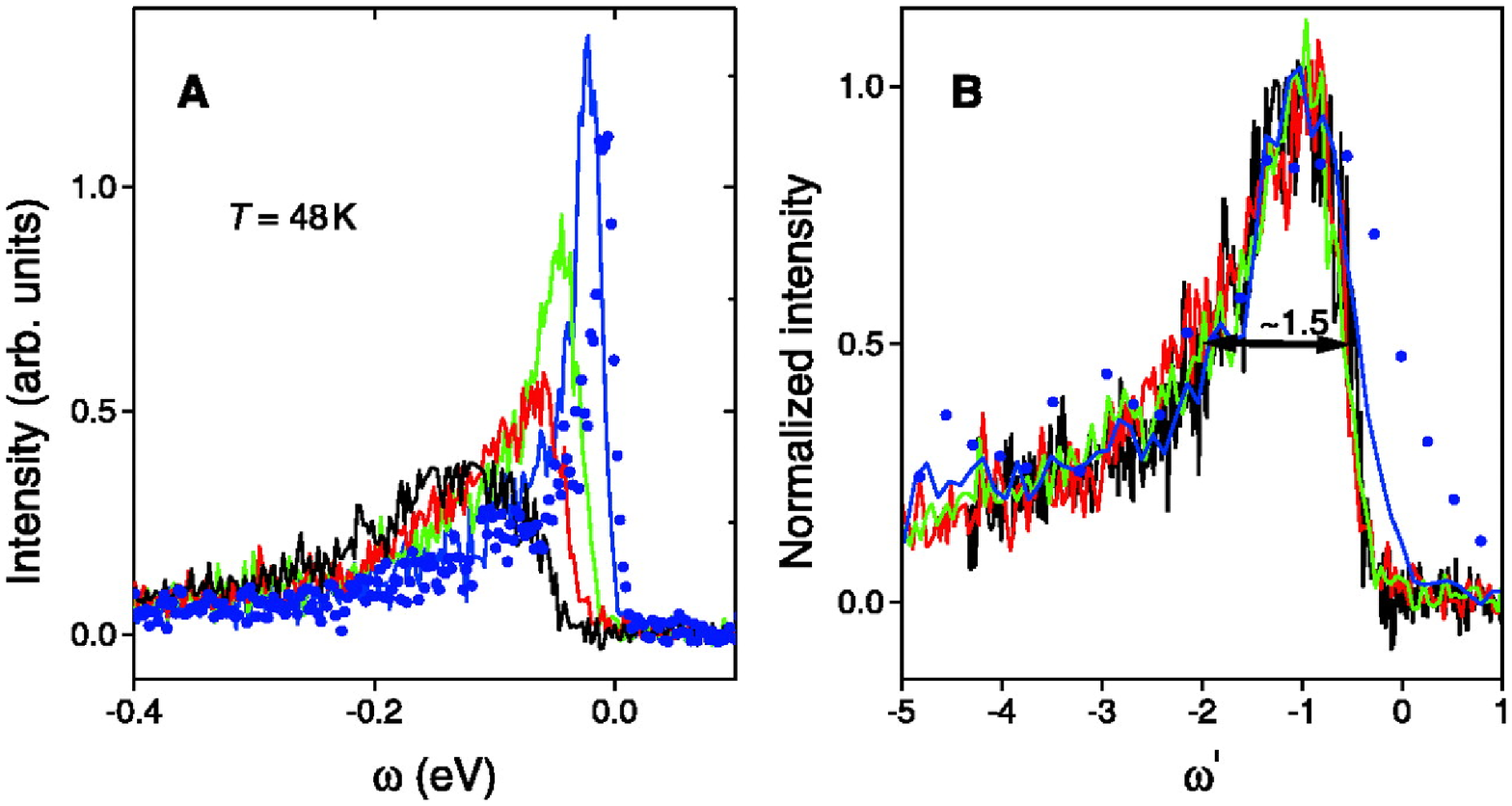}}
\epsfxsize=0.30\textwidth{\epsfbox{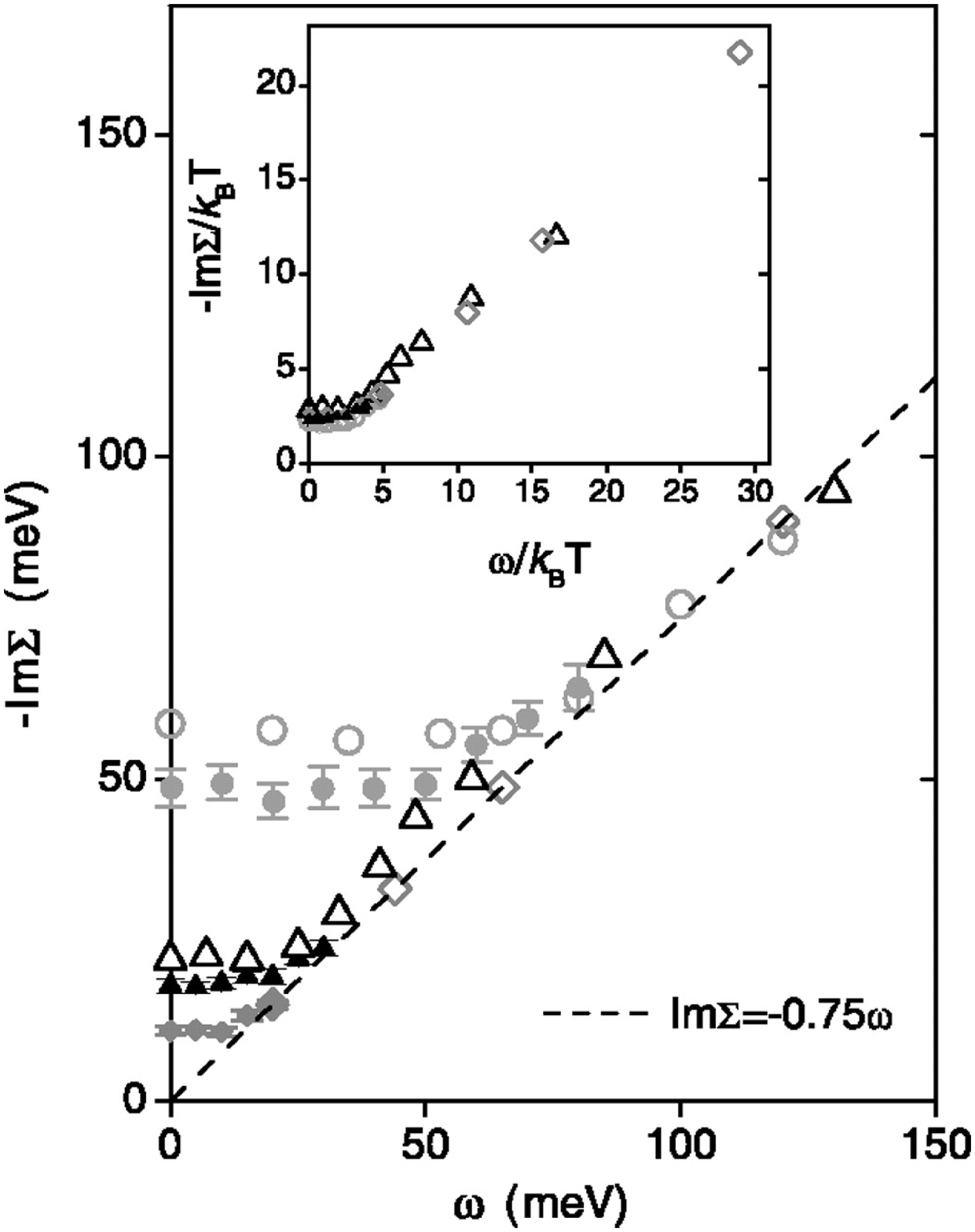}}
}
\caption{
\label{Fig4_Valla99}
(A) EDC's for optimally doped Bi$_{2}$Sr$_{2}$CaCu$_{2}$O$_{8+\delta}$,
obtained in the $(0,0) \to  (\pi,\pi)$ direction (after subtracting a background). 
(B) EDC's scaled to the same peak position showing that the overall shape scales linearly with binding energy.
The peak width is approximately 1.5 times the binding energy.
Right: Im $\Sigma $ obtained from MDC peak widths (solid symbols) and from EDC peak widths (open symbols) 
as a function of binding energy for temperatures of 48 K (diamonds), 90 K (triangles), and 300 K (circles). 
Inset: same data plotted in dimensionless units confirming scaling behavior.
(Reprinted with permission from Science, Ref. \cite{Valla99}. Copyright \copyright 1999 AAAS).
}
\end{figure}

An analysis of the momentum width of the quasiparticles moving in nodal direction as function of binding energy
was performed by Valla {\it et al.} \cite{Valla99}. In Fig. \ref{Fig4_Valla99} the EDC's along the nodal direction
are shown on the left. It was found that the EDC spectra, scaled to the same peak position, coincide.
This corresponds to a scaling relation 
$A(\omega ) = A(E_p) f(\omega/E_p)$
with $\omega $ the electronic energy, and $E_p$ the peak position of the EDC.
The {\it energy} width of the quasiparticle peak, $W_{EDC}$
is proportional to the binding energy, $W_{EDC} (E_p) \approx 1.5 E_p $
\cite{Olson90,Valla99}.
The corresponding data for Im $\Sigma $ extracted from the EDC spectra is shown in Fig.~\ref{Fig4_Valla99}
on the right for several temperatures.
All data approach for high energies the asymptotics Im $\Sigma \approx -0.75 \omega $ independent
of temperature.  For a given
temperature, Im $\Sigma $ is constant up to energy $\omega \approx 2.5 k_BT$.
The inset demonstrates a scaling behavior for the nodal direction,
\begin{equation}
\mbox{Im}\Sigma (\omega,T)  = k_B T\cdot F(\frac{\omega}{k_BT}) \;,
\end{equation}
where $F$ is a scaling function.
This behavior resembles that of the marginal Fermi liquid hypothesis 
of Varma {\it et al.} \cite{Varma89,Littlewood92}.

\begin{figure}
\centerline{
\epsfxsize=0.80\textwidth{\epsfbox{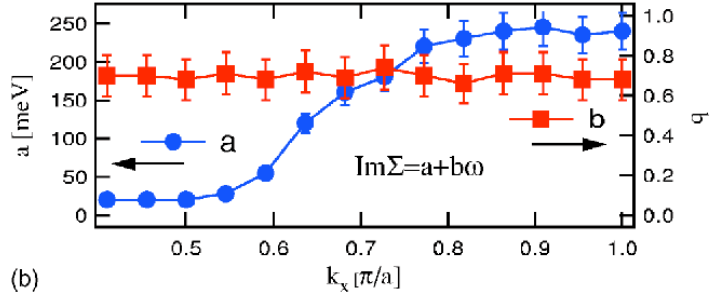}}
}
\caption{
\label{Fig4_Kaminski05}
Variation of the coefficients $a$ and $b$ in the equation
Im$\Sigma(\vec{k}_F,\omega )=a(\vec{k}_F)+b\omega $
around the Fermi surface for optimally doped
Bi$_{2}$Sr$_{2}$CaCu$_{2}$O$_{8+\delta}$ above $T_c$.
(From Ref. \cite{Kaminski05},
Copyright \copyright 2005 APS).
}
\end{figure}

Recently Kaminski {\it et al.} scrutinized the momentum dependence of
this effect in the normal state. When going away from the nodal direction, there is an additional
temperature independent term.
They found that the imaginary part of the self energy above $T_c$
in the intermediate energy region ($2.5 k_BT\lsim \omega \lsim 200$ meV)
can be written as
\begin{equation}
\mbox{Im}\Sigma (\vec{k}_F,\omega)  =  a(\vec{k}_F) + b \omega,
\end{equation}
where the coefficient $b$ is {\it isotropic} for both optimally doped and
overdoped materials. In contrast, the coefficient $a$ is strongly
momentum dependent for underdoped and optimally doped compounds, 
its anisotropy following
the behavior of the pseudogap along the Fermi surface. That means, it is 
zero on a Fermi surface arc around the node, and increases to about 10 times
the pseudogap at the antinodes. For strongly overdoped samples without
a pseudogap the coefficient $a$ is isotropic as well.
In Fig.~\ref{Fig4_Kaminski05} the variation of the coefficients
$a$ and $b$ around the Fermi surface are shown for optimally doped
Bi$_{2}$Sr$_{2}$CaCu$_{2}$O$_{8+\delta}$.

In the superconducting state, it was argued recently that the MDC peak width in near-nodal
direction, when resolving nodal bilayer splitting,
shows at low binding energies ($|\omega |\lsim 75$ meV) a superposition
of an $\omega $ and an $\omega^3$ behavior \cite{Valla05}. 
This would be consistent with theoretical predictions of Refs. \cite{Zhu04,Dahm05}.
Above $\sim 75 $ meV a
nearly linear dependence of the MDC peak width
on binding energy is recovered, that is almost unaffected by
the superconducting transition \cite{Valla05}.

\begin{figure}
\centerline{
\begin{minipage}{0.45\textwidth}
\epsfxsize=1.0\textwidth{\epsfbox{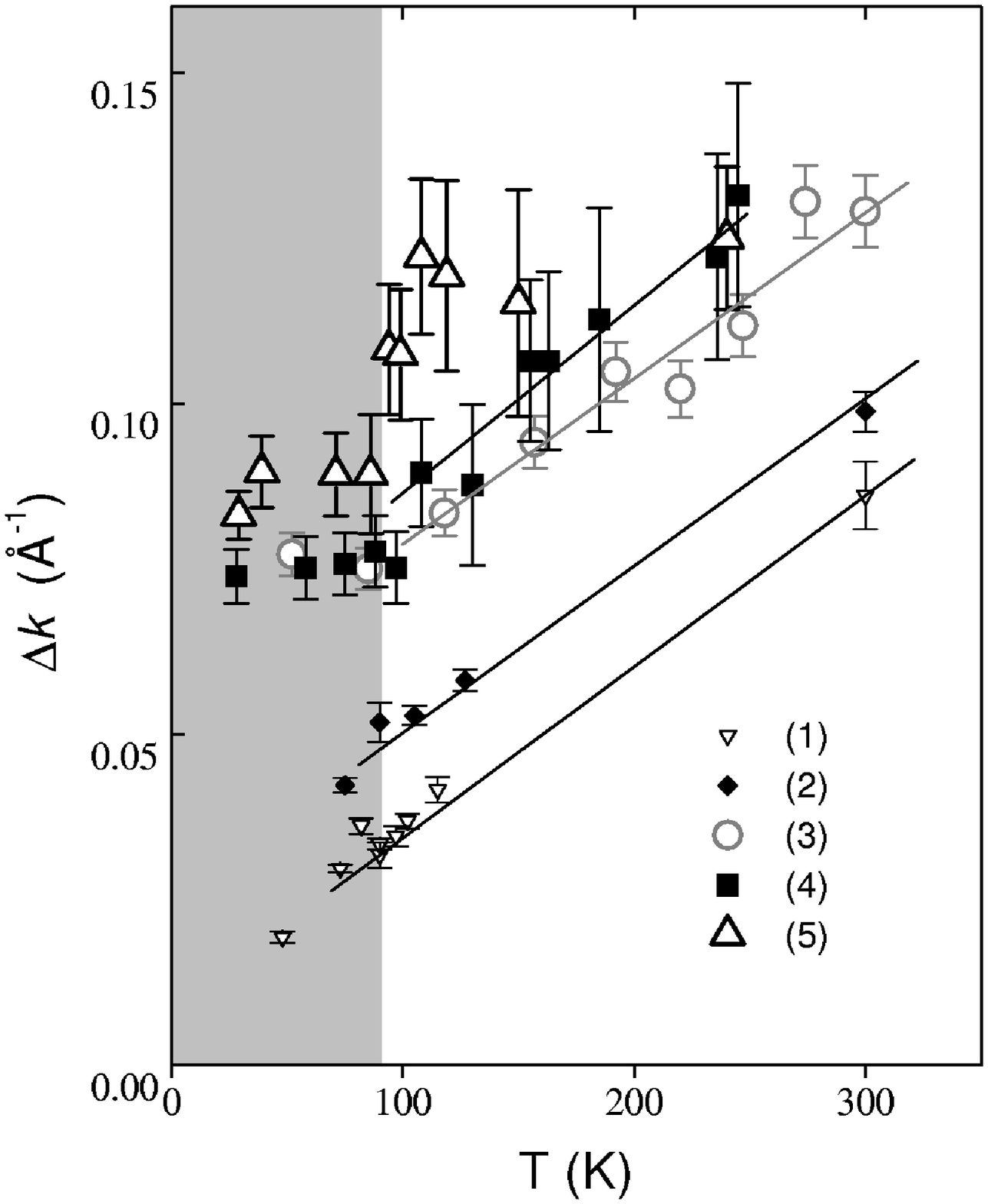}}
\end{minipage}
\begin{minipage}{0.45\textwidth}
\begin{minipage}{1.\textwidth}
\epsfxsize=0.5\textwidth{\epsfbox{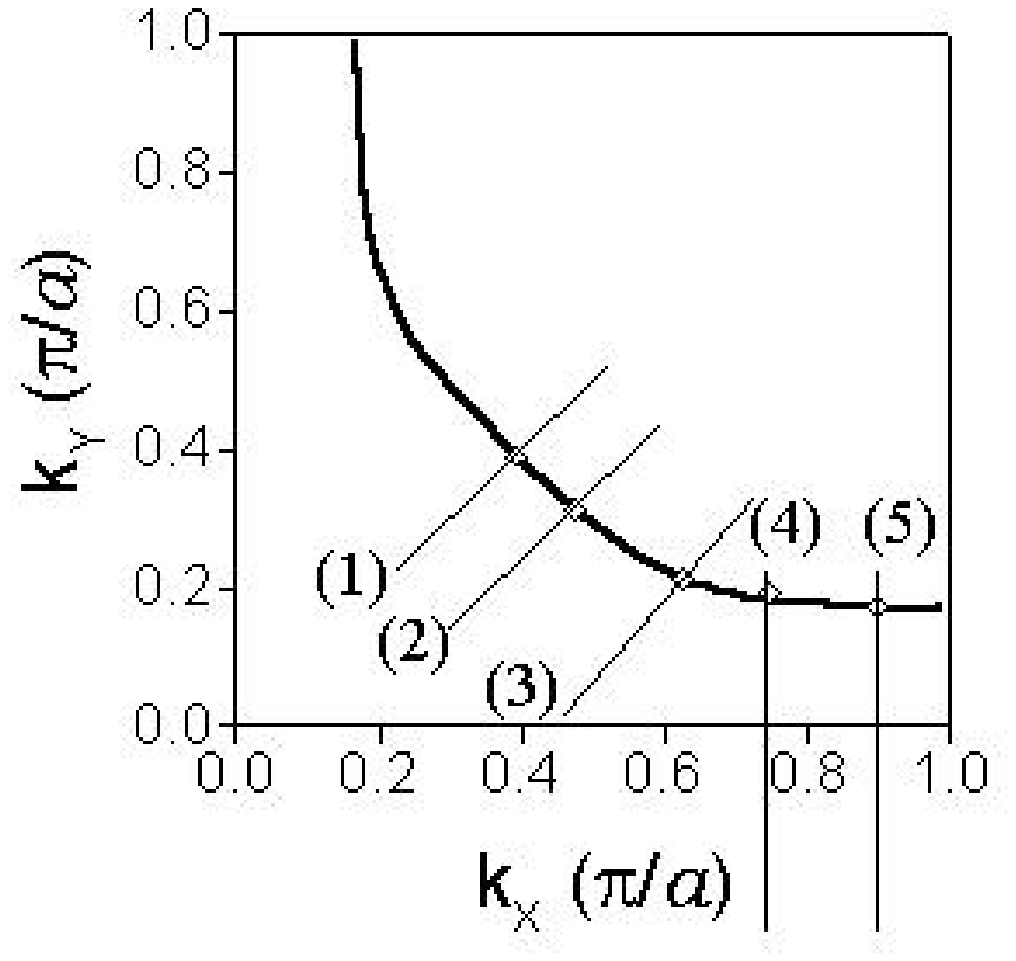}}
\end{minipage}\\
\begin{minipage}{1.\textwidth}
\epsfxsize=0.8\textwidth{\epsfbox{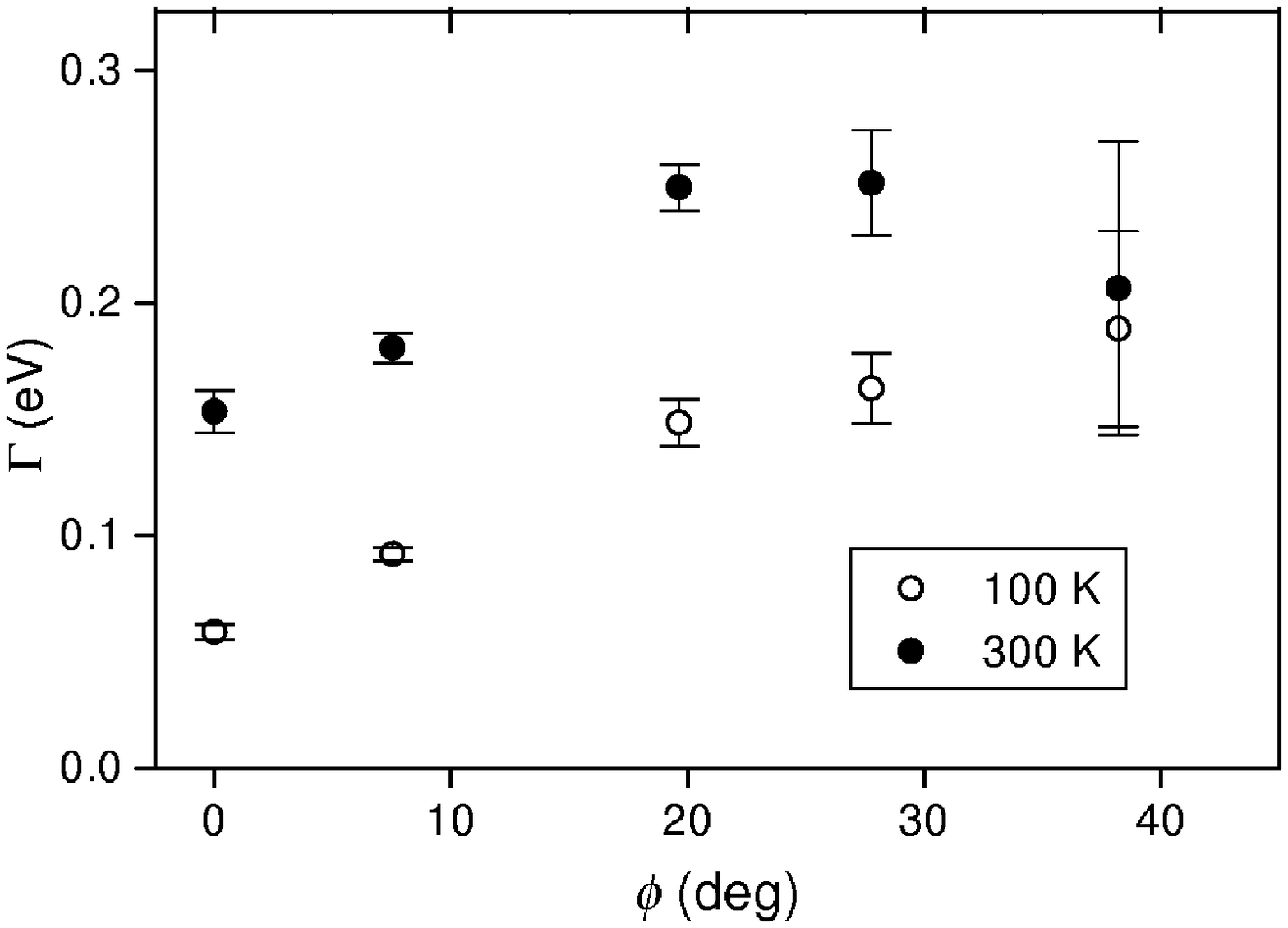}}
\end{minipage}
\end{minipage}
}
\caption{
Normal state
momentum widths as a function of temperature for different positions on the
Fermi surface (indicated in the top right picture) obtained by fitting the MDC's with Lorentzian lineshapes. Widths
are measured at the Fermi level and at the leading edge, in the normal and in
superconducting (gray region) state, respectively. The dependence as function of
position on the Fermi surface is shown in the right lower picture ($\phi $ is measured from the nodal direction).
\label{Fig4_Valla00}
(From Ref. \cite{Valla00},
Copyright \copyright 2000 APS).
}
\end{figure}

Turning to temperature dependence, Valla {\it et al.}
\cite{Valla99} have reported an MDC linewidth growing linearly with temperature in the
range $T_c< T\lsim 300$ K.
In Fig. \ref{Fig4_Valla00}
the momentum widths are shown as function of temperature for different positions on the
Fermi surface. 
The temperature dependence is linear for most of the Fermi surface
with a slope independent on the $k_F$ position. However, the temperature independent offset
varies along the Fermi surface. 
As a consequence, the imaginary part of the self energy (or the `scattering
rate') at the peak 
positions around the Fermi surface follows the behavior
\begin{equation}
\mbox{Im}\Sigma (\vec{k}_F,T)  =  a'(\vec{k}_F) + b' k_BT
\end{equation}
with a momentum independent constant $b'$ and a temperature-independent constant $a'(\vec{k}_F)$. 
As can be inferred from the extrapolation $T\to 0$ in Fig.~\ref{Fig4_Valla00},
for the nodal direction, $a'(\vec{k}_N)\approx 0$ holds \cite{Valla00}.
The scattering rate, given by the product of the momentum width and the normal state Fermi velocity,
is shown as a function of the Fermi surface angle in the lower right panel of Fig. \ref{Fig4_Valla00}.
It can be seen that the scattering rate for fixed temperature increases away from the nodal point of the Fermi surface, revealing the variation
of $a'(\vec{k}_F)$ along the Fermi surface.

In the superconducting state the functional dependence on the binding energy is modified in
the low-energy region, and the MDC peak width in the nodal Fermi surface point ($\omega=0$)
of the bonding state in optimally doped Bi$_{2}$Sr$_{2}$CaCu$_{2}$O$_{8+\delta}$
was fitted to a $T^3$ functional dependence \cite{Valla05}.

In summary, the normal state scattering rate depends for $\omega \ll k_BT$ linearly 
on temperature, but is independent of frequency, and for $\omega \gg  k_B T$
is linearly depending on frequency, but temperature independent.
In both cases the linearity coefficients are isotropic around the Fermi surface,
and there is an energy- and temperature independent but anisotropic term which
vanishes in nodal direction. 
Again both these findings are consistent with the
marginal Fermi liquid hypothesis \cite{Varma89}. For intermediate energies,
however, $\omega \sim k_BT, \Delta_{\vec{k}}$ there are additional effects
beyond the marginal-Fermi-liquid phenomenology present.

\begin{figure}
\centerline{
\epsfxsize=0.44\textwidth{\epsfbox{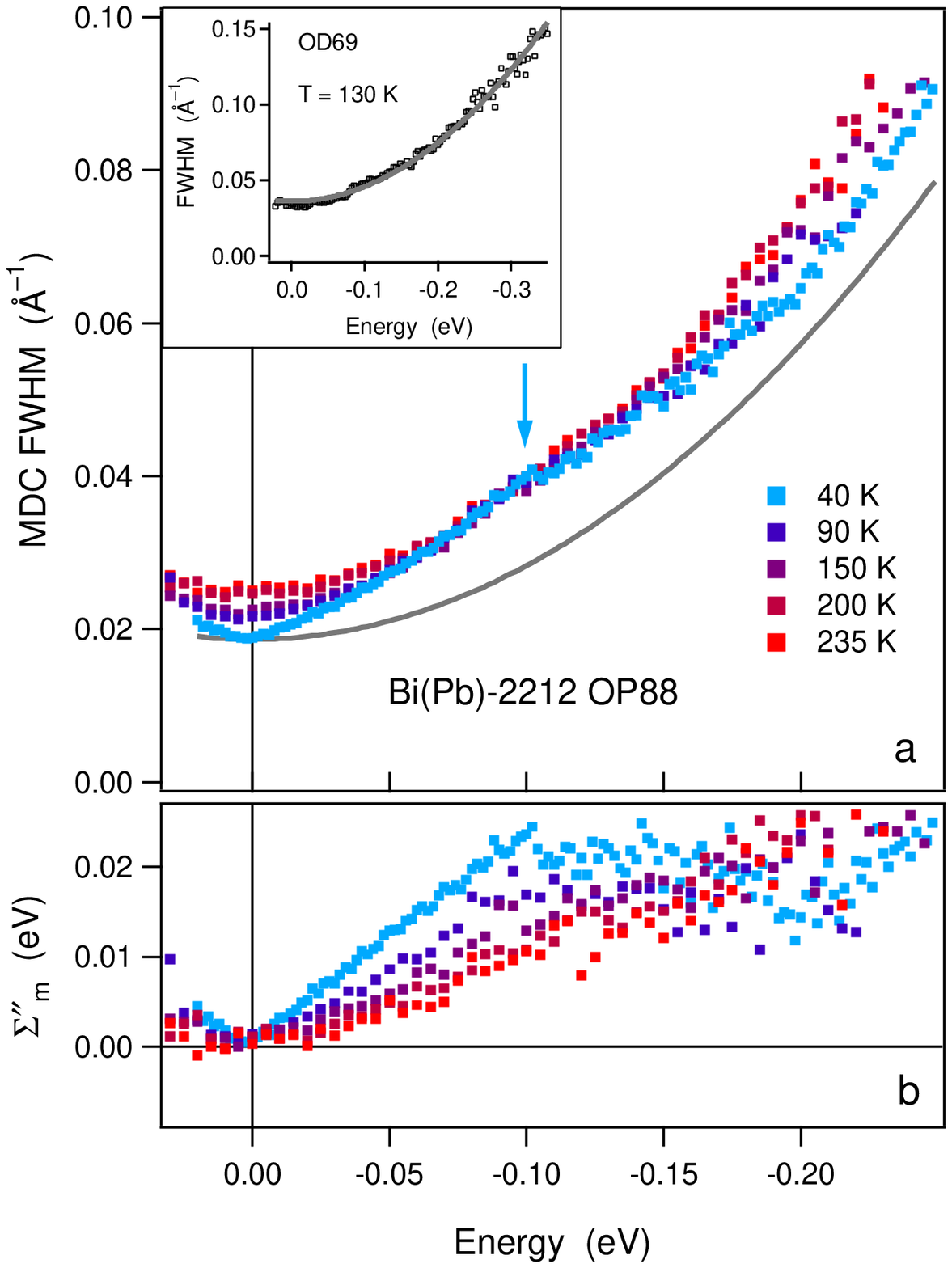}}
\epsfxsize=0.44\textwidth{\epsfbox{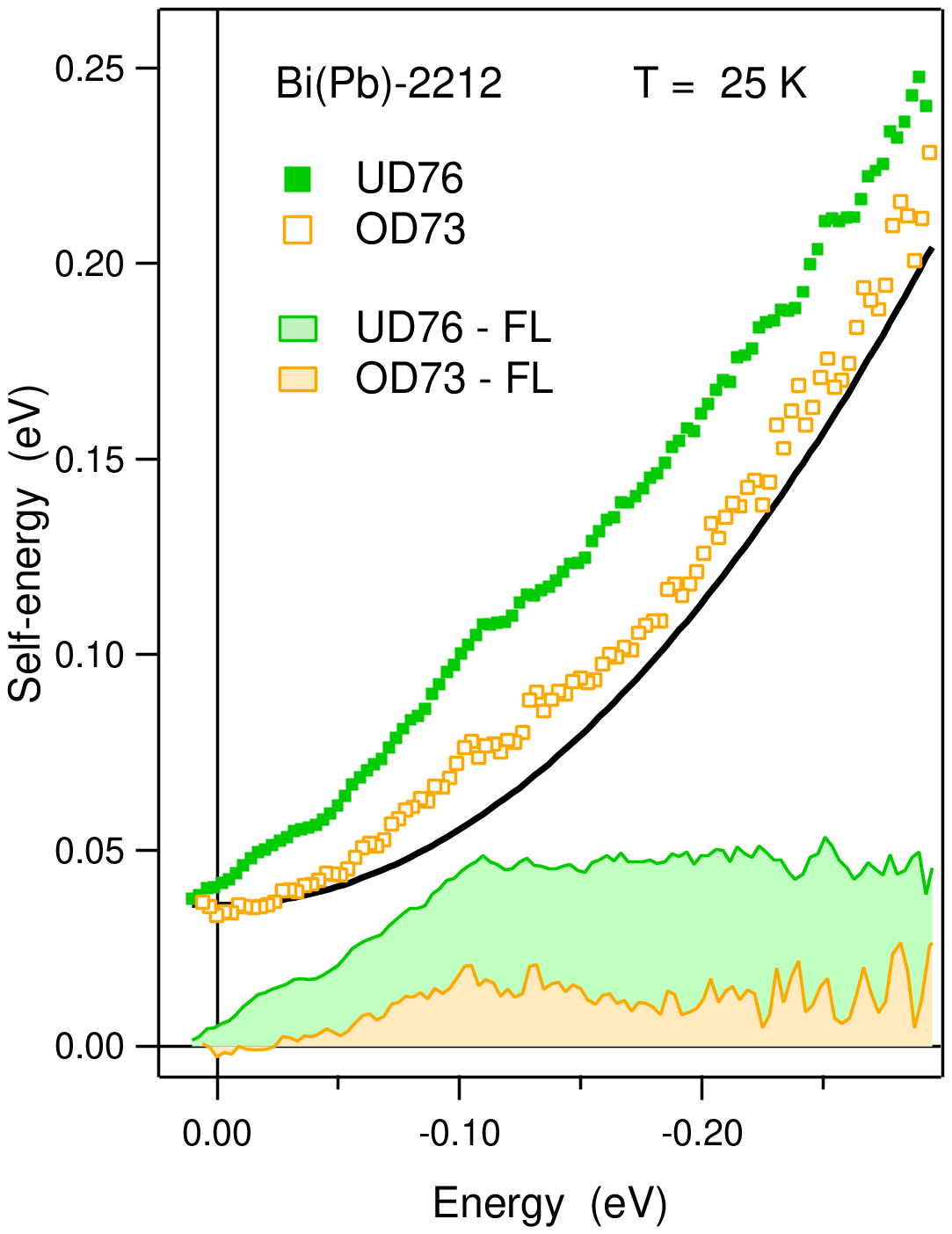}}
}
\caption{
\label{Fig1_Kordyuk04}
Temperature and doping dependence of self-energy effects for nodal quasiparticles
determined from the full width at half maximum of the momentum distribution curves.
Left: temperature dependence for optimally doped
Bi(Pb)$_2$Sr$_2$CaCu$_2$O$_{8+\delta}$; the contribution from a highly overdoped sample ($T_c=69 $K) at
130 K (inset and gray line) is subtracted to obtain the imaginary part of the self energy (using
FWHM/2 times $v_F=4$ eV $\AA $). Right: doping dependence of the corresponding curves; shown are data for
an underdoped ($T_c=76 $K) and an overdoped ($T_c=73 $K) sample at 25 K.
(From Ref. \cite{Kordyuk04},
Copyright \copyright 2004 APS).
}
\end{figure}

For example, in the superconducting state there are additional features and clear
deviations from the linearity of the scattering rate as a function of
binding energy. Generally, for large binding energies the behavior
linear in $\omega $ still is present. 
However, recent studies of the nodal linewidths as function of 
energy \cite{Kordyuk04,Valla05} revealed that at moderate energies there is
a `kink effect' in Im $\Sigma $, in accordance with the finding in the real part of
the self energy by Johnson {\it et al.} \cite{Johnson01,Valla05}. 
This is illustrated in Fig. \ref{Fig1_Kordyuk04}.
The effect is absent for highly overdoped ($T_c=$69 K) compounds, however clearly visible
for overdoped Bi$_{2}$Sr$_{2}$CaCu$_{2}$O$_{8+\delta}$ with $T_c=$73 K,
optimally doped Bi$_{2}$Sr$_{2}$CaCu$_{2}$O$_{8+\delta}$ 
and for underdoped
Bi$_{2}$Sr$_{2}$CaCu$_{2}$O$_{8+\delta}$ ($T_c=$76 K).
As seen from Fig. \ref{Fig1_Kordyuk04} (b), the kink in Im $\Sigma $
vanishes above $T_c$. Also shown in this figure is a comparison between an
overdoped and an underdoped sample, showing that underdoped materials
exhibit a much stronger kink effect that overdoped materials.
This is consistent with the behavior of the real part of the self energy 
in nodal direction, shown in Fig.~\ref{Fig1_Johnson} \cite{Johnson01}.
The doping and temperature dependence of the sudden change in the MDC linewidths shown in Fig.~\ref{Fig1_Kordyuk04}
is consistent with the doping and temperature
dependence of the intensity of the magnetic resonance mode observed in INS.

In summary, for the {\it nodal} direction, the combined information from the
experimental analysis of effects both in Re$\Sigma $ and Im$\Sigma $ 
is, that the resonance mode has sizable contributions
to nodal quasiparticle scattering for underdoped, optimally doped,
and slightly overdoped materials. This is consistent with the scenario for the nodal
self-energy effects proposed in \cite{Eschrig00}.
For strongly overdoped materials its contribution to nodal quasiparticle scattering is negligible.
Furthermore, for a momentum 
region around the {\it antinodal} direction the data are
consistent with a coupling to the spin-fluctuation spectrum
even when the overdoping is so strong that
the nodal effects due to the magnetic resonance mode are unobservable.
This finding is consistent with the apparent decrease in intensity of
the magnetic resonance mode with overdoping. The already weaker nodal effect
disappears earlier than the antinodal one when overdoping the material.

\subsubsection{Isotope effect}
\label{Iso}
\begin{figure}
\centerline{
\epsfxsize=0.6\textwidth{\epsfbox{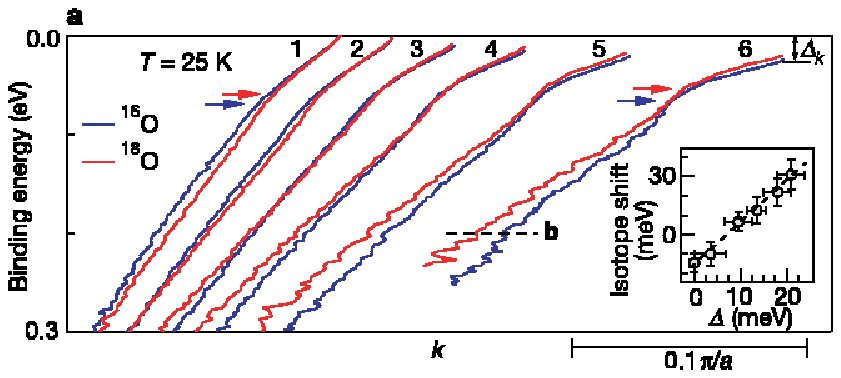}}
\epsfxsize=0.26\textwidth{\epsfbox{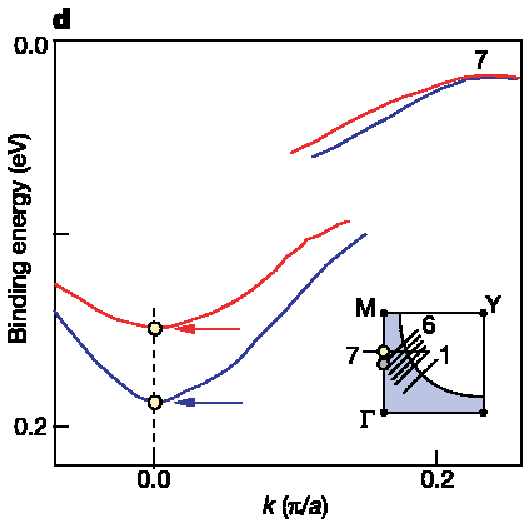}}
}
\caption{
\label{Fig2_Lanzara04}
Isotope induced changes of the dispersion along the cuts shown in the inset of
(d), for optimally doped Bi$_2$Sr$_2$CaCu$_2$O$_{8-\delta}$ with different
oxygen isotopes at $T=25 $K. In (a) the MDC-derived dispersions are shown
for cuts starting from nodal direction 1. The dispersions are shifted in
k-direction for convenience. The inset shows the isotope shift near 220 meV binding
energy as a function of the superconducting gap of the same MDC.
In (d) the EDC-derived dispersion for cut 7 is shown.
(Reprinted by permission from Mcmillan Publishers Ltd: Nature, Ref. \cite{Gweon04a} Copyright \copyright 2001 NPG).
}
\end{figure}

Recently the isotope effect on the dispersion spectra has been studied \cite{Gweon04a}
and it was found that it shows a complicated behavior, shown in
Fig.~\ref{Fig2_Lanzara04}. The shift is
small in the low-energy region, with the isotope effect on the gap
value being small and random between different samples 
in both magnitude and sign, regardless of the isotope mass \cite{Gweon04a}.
Thus, it can be assumed that the low-energy region of the dispersion
is only weakly affected by the isotope exchange.
On the other hand, the high-energy dispersion shows a sample-independent,
reproducible and reversible isotope effect that is strongly anisotropic and
becomes small in the normal state.
The isotope shift changes sign when going from the nodal direction to the
antinodal direction, as is shown in the inset of Fig.~\ref{Fig2_Lanzara04} (a).

However, the strength of the self-energy effects at low binding energies
seems to be independent on the isotope mass. 
This can be seen also from Fig.~\ref{Fig2_Lanzara04}, where 
neither the low binding energy slopes nor the size
of the break in panel (d)
change considerably, although the overall dispersion 
is modified by a renormalization factor.
This renormalization affects the dispersion, but does not seem to affect
the value of the gap. This is also consistent with the negligible
change in $T_c$ from 92 K to 91 K upon isotope substitution $^{16}$O $\to $
$^{18}$O. It seems, that the effect of phonons is to contribute to an overall
renormalization of the dispersion, although the superconducting properties
are only weakly affected.

\subsubsection{Relation to pseudogap phase}
\label{RPP}
Finally, there is important information contained in the doping dependence
of the self-energy effects. 
In underdoped compounds, there is a  pseudogap 
between $T_c$ and $T^{\ast}$ \cite{Marshall96,Ding96};
the pseudogap is maximal near the $M$-point of the Brillouin zone
and is zero at arcs centered at the $N$-points which
increase with temperature \cite{Norman98n}.
In the pseudogap state above $T_c$, there are low energy renormalizations
in the dispersion, and some trace of the kink feature persists.
But in the the work by Johnson {\it et al.} \cite{Johnson01}, it
was clearly shown that an additional renormalization of the dispersion
sets in just at $T_c$. This indicates that the bosonic spectrum redistributes
its spectral weight when entering the superconducting state.
The additional low energy renormalization of the dispersion
below the kink energy follows an order parameter like behavior as a
function of temperature \cite{Johnson01}.
Arguing that the renormalization near the
nodal region for underdoped materials is influenced by the coupling to the same bosonic mode 
which causes the strong self-energy effects at the $M$ point of the Brillouin
zone,
the above implies that some mode intensity may be present in the pseudogap
state already, but there is an abrupt increase in the mode intensity when going
from the pseudogap state into the superconducting state, and this
increase shows an order parameter like behavior as a function of
temperature below $T_c$.

\subsection{C-axis tunneling spectroscopy}
\label{CTS}
Unusual spectral dip features in tunneling data of 
Bi$_2$Sr$_2$CaCu$_2$O$_{8-\delta}$ 
are found in point contact junctions \cite{Huang89},
in scanning tunneling spectroscopy (STM) \cite{Renner95,DeWilde98,Hudson99},
in break junctions \cite{Mandrus91,DeWilde98,Zasadzinski01,Zasadzinski02},
and in intrinsic c-axis tunnel junctions \cite{Yurgens99}.
A consistent picture emerged from all these different tunneling
techniques \cite{Zasadzinski02}.
These data show a peak feature, 
usually assigned to the maximal $d$-wave superconducting gap,
and a hump feature at higher bias, separated from the peak by a
pronounced dip feature. 
SIN junctions usually show a spectrum strongly asymmetric around the
chemical potential. In many SIN data, also the self-energy effects appear
stronger on the occupied side of the 
spectrum \cite{Huang89,Renner95,DeWilde98}.
Recently, however, Zasadzinski {\it et al.} \cite{Zasadzinski03} reanalyzed STM data by 
Hudson {\it et al.} \cite{Hudson99}
and argued that after subtracting a background the resulting tunneling spectra
are rather symmetric. The issue is unsolved and needs further investigation.
The dip feature has been observed in tunneling spectra of
the single Cu-O$_2$ layer compound Tl$_2$Ba$_2$CuO$_6$ as 
well \cite{Zasadzinski00}, where it appears weaker. Interestingly,
also INS data show that a weak resonant magnetic excitation exists in that
material \cite{He02}.

\begin{figure}
\centerline{
\epsfxsize=0.45\textwidth{\epsfbox{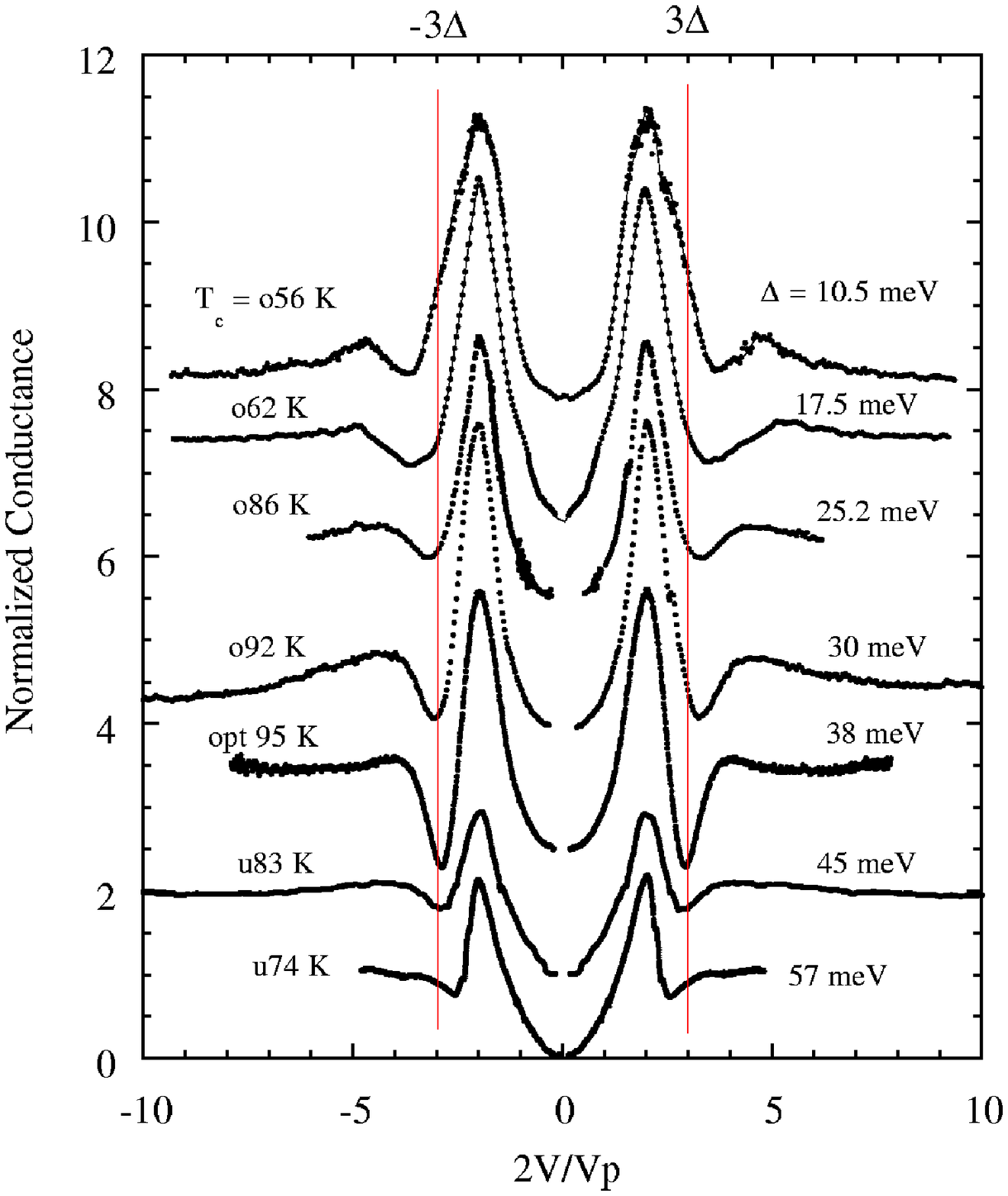}}
\epsfxsize=0.50\textwidth{\epsfbox{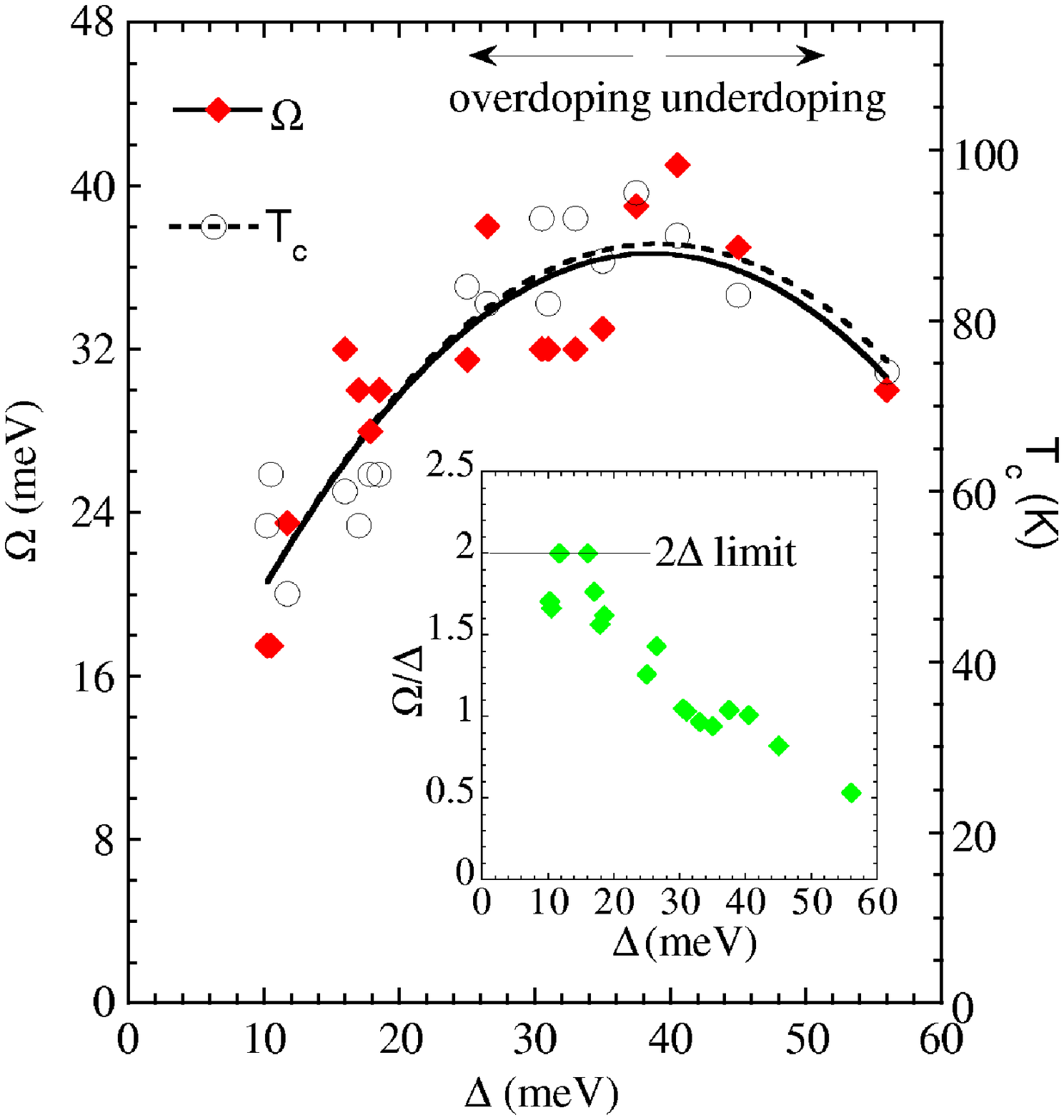}}
}
\caption{
\label{Fig2_Zasadsinski}
Left:
SIS tunneling conductances of
Bi$_2$Sr$_2$CaCu$_2$O$_{8-\delta}$ for various amounts of hole-doping from
overdoped (o) via optimally doped (opt) to underdoped (u).
The voltage has been scaled in units of $\Delta $. The curves
have been shifted for clarity.
\label{Fig3_Zasadsinski}
Right: Dependence of the dip-peak energy separation $\Omega $ as function of
the spectral gap energy $\Delta $ determined from the c-axis tunneling spectra on
Bi$_2$Sr$_2$CaCu$_2$O$_{8-\delta}$.
$\Omega $ follows $T_c$, not $\Delta $.
The inset shows that the ratio $\Omega /\Delta $ never exceeds 2.
(From Ref. \cite{Zasadzinski01},
Copyright \copyright 2001 APS).
}
\end{figure}
In order to extract information about the bosonic mode which 
would produce a dip feature in the tunneling conductance, a systematic
study as a function of doping was performed in break junction
tunneling spectroscopy by Zasadzinski {\it et al. } \cite{Zasadzinski01}.
There, the doping dependence in Bi$_2$Sr$_2$CaCu$_2$O$_{8-\delta}$ of
the peak-dip-hump structure was determined over a wide range of doping. 
Corresponding spectra are reproduced in
Fig. \ref{Fig2_Zasadsinski} (left panel).
It was found that the dip-peak energy separation,
$\Omega $, follows $T_c$ as $\Omega = 4.9 k_BT_c$, as demonstrated in
the right panel of
Fig. \ref{Fig3_Zasadsinski}.
In the inset one can see that,
as expected for an excitonic mode, $\Omega $ approaches
but never exceeds $2\Delta $ in the overdoped region, and $\Omega /\Delta$
monotonically decreases as doping decreases and the superconducting gap
increases. The dip feature is found to be strongest near optimal doping.
Similar shifts of the dip position with overdoping
were reported previously by STM \cite{Renner96}.
Together with the ARPES results, these studies give a detailed picture
about the doping dependence of the mode energy involved in 
electron interactions in the superconducting state.

\subsection{Optical spectroscopy}
\label{OPS}

Self energy effects can also be studied by optical spectroscopy assuming
some model for the optical response. A common way is to write the
complex optical conductivity in terms of an optical single-particle self-energy
$\Sigma^{op}$ as
\begin{equation}
\sigma (\omega ) = \frac{i}{4\pi } \; \frac{\omega_{pl}^2}{\omega - 2 \Sigma^{op} (\omega ) }
\end{equation}
with the plasma frequency $\omega_{pl}$, that can be determined experimentally from the
absorption spectra in the near-infrared region \cite{Hwang04}.
The imaginary part of the self energy determines an energy dependent carrier
scattering rate via
\begin{equation}
\mbox{Im} \Sigma^{op} (\omega ) = -\frac{1}{2\tau (\omega) }.
\end{equation}

Using the above model, Hwang {\it et al.} \cite{Hwang04} studied the self-energy effects
for different doping levels and temperatures in
Bi$_2$Sr$_2$CaCu$_2$O$_{8-\delta}$.
They found that the optical single-particle self-energy shows a sharp feature which
weakens with doping and cannot be resolved anymore beyond a doping level of 0.23 holes
per copper atom.
\begin{figure*}
\centerline{
\epsfxsize=0.65\textwidth{\epsfbox{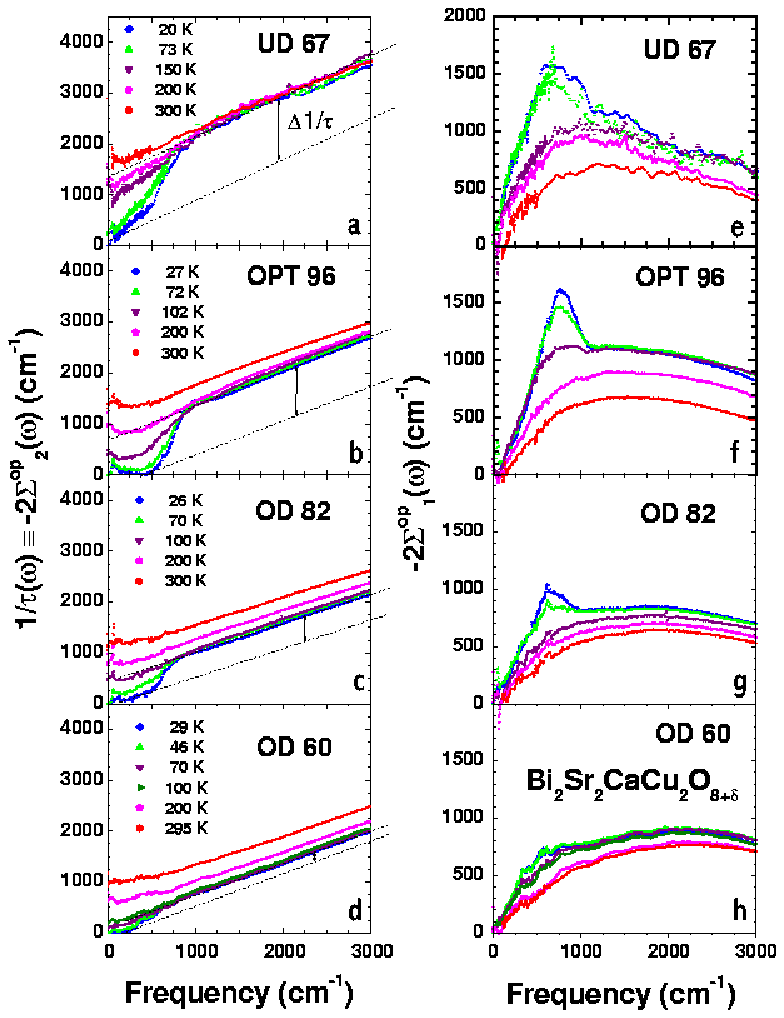}}
\epsfxsize=0.31\textwidth{\epsfbox{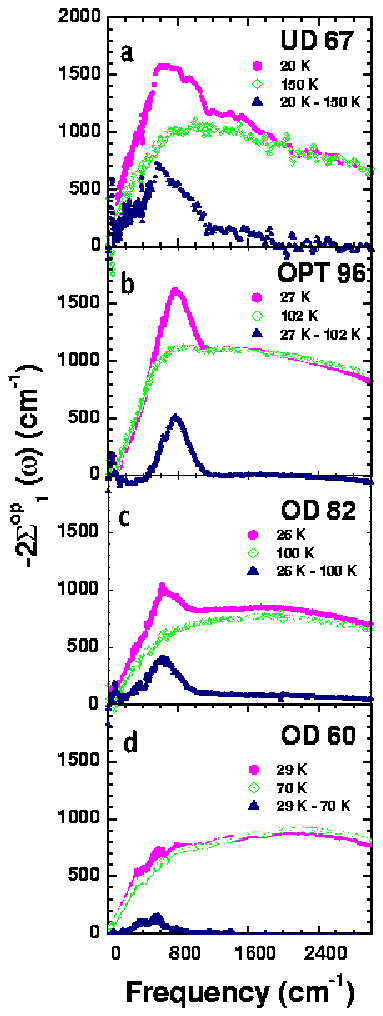}}
}
\caption{
\label{Fig1_Hwang}
The optical self energy for different doping levels of
Bi$_2$Sr$_2$CaCu$_2$O$_{8-\delta}$, with the imaginary part in the
left column, and the real part in the middle column. 
The feature due to the magnetic resonance
can be seen as peak in the difference spectrum
between superconducting state and normal state
of the real part of the optical self energy, shown in the right column.
(Reprinted by permission from Mcmillan Publishers Ltd: Nature, Ref. \cite{Hwang04} Copyright \copyright 2001 NPG).
}
\end{figure*}
As can be seen in Fig. \ref{Fig1_Hwang},
the imaginary part of $\Sigma^{op}$ shows a linear in $\omega $ behavior at high
frequencies, consistent with what is obtained by ARPES spectroscopy. 
This high-frequency scattering rate generally decreases with increasing doping.
In the superconducting state there is a sharp onset of scattering at a finite frequency, seen on the left in Fig. \ref{Fig1_Hwang},
which in the real part of the self energy leads to a sharp peak,
shown in the middle row of Fig. \ref{Fig1_Hwang}. This sharp peak
is restricted to the superconducting state, and a study of the doping and temperature dependence
allowed the authors of Ref.~\cite{Hwang04} to assign this ``optical resonance mode effect'' to the interaction of
quasiparticles with the magnetic resonance mode.
This is demonstrated on the right in Fig. \ref{Fig1_Hwang}, where the difference
of the real part of the self energy between the superconducting and the
normal state is shown.
The frequency position of the peak in the real part of the optical self energy 
is consistent with the frequency position of the peak in the real part of
the self energy obtained from ARPES spectroscopy along the nodal direction,
shown in Fig.~\ref{Fig1_Johnson}. Also consistent with these latter data,
the amplitude of the peak on Re$\Sigma^{op}$ decreases with doping.
The study of the optical self energy also shows that  a weak
effect of the magnetic resonance mode is still observed for overdoped 
Bi$_2$Sr$_2$CaCu$_2$O$_{8-\delta}$ with $T_c=$ 82 K and 60 K. 
The optical studies of \cite{Hwang04} add support to the picture emerging from the
ARPES studies.

Optical data along the $a$-axis of a detwinned single
crystal of YBa$_2$Cu$_3$O$_{6.50}$ (underdoped, $T_c=59$ K) confirmed this
picture and
added further support for a correlation between the effects observed in these
data and the spin-1 magnetic resonance observed in inelastic neutron scattering
\cite{Hwang05}.
As in the case for Bi$_2$Sr$_2$CaCu$_2$O$_{8-\delta}$
a peak in the data for $-2\Sigma^{op}_1 (\omega ,T)$ is present that sharpens in
the superconducting state.
The temperature dependence of the peak magnitude follows closely
the intensity of the magnetic resonance mode \cite{Hwang05}.

An issue of discussion \cite{Cuk04,Valla04} referred to the non-observability
of the optical self-energy effect due to coupling to the magnetic resonance
mode for doping values beyond $p=0.23$ \cite{Hwang04}. 
This claim was based on the 
extrapolation of the measured data to higher doping levels.
However, one has to bear in mind that the optical self energy is a quantity
which is not resolved with respect to the position on the Fermi surface, in
contrast to ARPES. Thus, it is well possible that self-energy effects
persist to the highest doping levels, but are too weak to be observed
by optical spectroscopy. This is also consistent with the finding,
that the region around the $M$ point in which strong self-energy
effects due to coupling to the resonance mode occurs, shrinks with
doping, and consequently contributes less and less to the averaged
self energy.

\begin{figure}
\centerline{
\begin{minipage}{0.45\textwidth}
\centerline{
\begin{minipage}{\textwidth}
\epsfxsize=\textwidth{\epsfbox{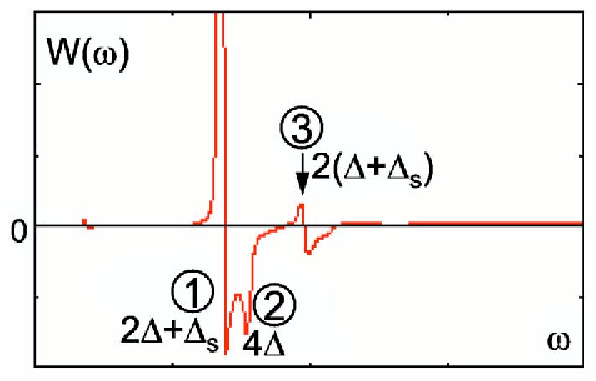}}
\end{minipage}
}
\begin{minipage}{\textwidth}
\epsfxsize=\textwidth{\epsfbox{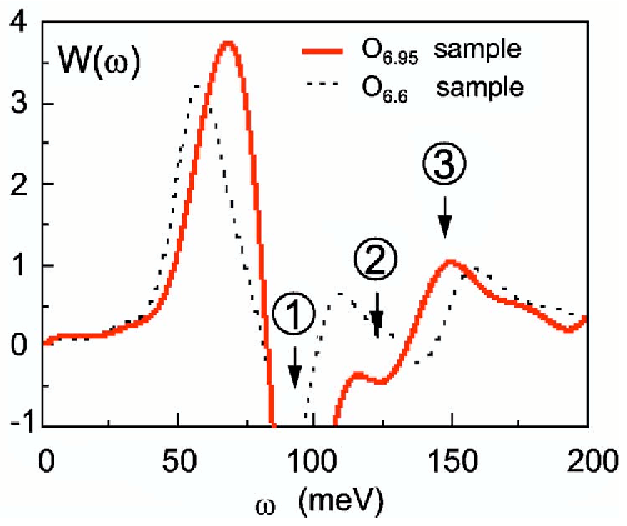}}
\end{minipage}
\end{minipage}
\begin{minipage}{0.45\textwidth}
\epsfxsize=\textwidth{\epsfbox{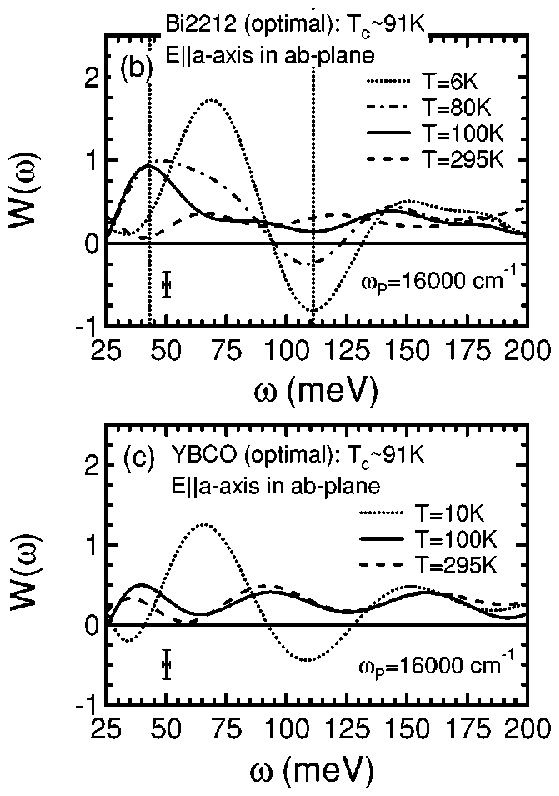}}
\end{minipage}
}
\caption{
\label{Fig4_Tu02}
Left: The lower part of the picture shows
the function $W(\omega )=(2\pi)^{-1}\partial^2_{\omega} [\omega/\tau(\omega)]$
obtained from experimental $a$-axis 
optical scattering rate data for an optimally doped
YBa$_2$Cu$_3$O$_{6.95}$ untwinned single crystal (full line) and for an underdoped
untwinned YBa$_2$Cu$_3$O$_{6.6}$ single crystal (dashed line). 
(Reprinted by permission from Mcmillan Publishers Ltd: Nature, Ref.~\cite{Carbotte99} Copyright \copyright 2001 NPG).
In the upper part the assignment
of the structures to characteristic frequencies obtained from the theory
of Abanov {\it et al.} \cite{Abanov01a} are shown for comparison.
$\Delta $  is the maximal superconducting $d$-wave gap, $\Delta_s$ is the frequency of a sharp bosonic peak coupled to electrons 
(from Ref.~\cite{Abanov01a},
Copyright \copyright 2001 APS).
Right:
The same experimental functions $W(\omega )$ obtained for optimally doped
Bi$_2$Sr$_2$CaCu$_2$O$_{8+\delta}$ (top) and for untwinned optimally doped
YBa$_2$Cu$_3$O$_{6.95}$  for several temperatures
(from Ref.~\cite{Tu02},
Copyright \copyright 2002 APS).
}
\end{figure}

Another method which proofed to be very useful to extract information about
the interaction of the electrons with a bosonic mode from the optical data was
suggested by Marsiglio {\it et al.} \cite{Marsiglio98}. It is based on the
function 
\begin{equation}
W(\omega) = \frac{1}{2\pi} \frac{{\rm d}^2}{{\rm d}\omega^2} \frac{\omega}{\tau(\omega )}.
\end{equation}
In Fig.~\ref{Fig4_Tu02} results of this function extracted from experimental
optical scattering rates are shown for optimally doped materials.
Carbotte {\it et al.}~\cite{Carbotte99} analyzed an optimally doped 
untwinned YBa$_2$Cu$_3$O$_{6.6}$ single crystal. The strong maximum in
$W(\omega )$ was theoretically  
shown to be at $\Delta+\Delta_s$, where $\Delta$  is 
the superconducting gap and $\Delta_s$ the frequency of a 
sharp bosonic mode assumed to be coupled to electrons.
Furthermore, Abanov {\it et al.}~\cite{Abanov01a} assigned additional features, denoted
in Fig.~\ref{Fig4_Tu02} (left) with 1,2 and 3, to the three characteristic
frequencies $2\Delta +\Delta_s$, $4\Delta $ and $2(\Delta + \Delta_s)$.
In a detailled recent study by C{\'a}sek {\it et al.} \cite{Casek05}
the assignments of the maximum at $\Delta+\Delta_s$ (at low temperatures) 
and the minimum at $2\Delta+\Delta_s$ have been
confirmed, however noticeable contributions of quasiparticles away from
the antinodal regions of the Brillouin zone have been found to complicate the
interpretation, in particular for the remaining extrema.

As can be seen in the lower left picture of Fig.~\ref{Fig4_Tu02}, 
the $\Delta+\Delta_s$-peak corresponds to 69 meV, the 
$2\Delta + \Delta_s$-minimum to 100 meV, the $4\Delta$-minimum to 130 meV,
and the $2\Delta + 2\Delta_s$-peak to 150 meV. This is consistent with
$\Delta\approx 30$ meV and $\Delta_s\approx 40$ meV \cite{Abanov01a}.
Tu {\it et al.} found from optical data for optimally doped
Bi$_2$Sr$_2$CaCu$_2$O$_{8+\delta}$ the values 
$\Delta =34 \pm3$ meV and
$\Delta_s=40\pm3$ meV \cite{Tu02}. For optimally doped 
YBa$_2$Cu$_3$O$_{6.95}$ the same procedure gave $\Delta=30\pm4$ meV and
$\Delta_s=40 \pm4 $ meV \cite{Tu02,Homes99}.
The corresponding curves $W(\omega )$ are shown in Fig.~\ref{Fig4_Tu02} (right)
(the feature at $4\Delta $ is not resolved here).
These values are in excellent agreement with the gap values obtained by
ARPES and tunneling experiments, and with the mode frequency of the magnetic
resonance mode observed in INS.
Finally, as can be seen from Fig.~\ref{Fig4_Tu02} (right), 
in the normal state this characteristic structure is not present.
This is consistent with the assumption that in the normal state the mode
contribution is not present.

\section{The collective mode as spin-1 exciton}
\label{collectivemode}

\subsection{Theoretical models}
\label{TM}

There are several theoretical interpretations of the resonance mode
observed in the magnetic INS signal. 
In all theories the resonance mode is an excited pair of quasiparticles
with total lattice momentum $\vec{Q}=(\pi/a,\pi/a)$ and spin $S=1$
\cite{Brinckmann98,Brinckmann99,Demler98,Greiter97}.
Lavagna and Stemman \cite{Lavagna94}, and Abrikosov \cite{Abrikosov98} give an interpretation in terms
of a van Hove singularity in the Stoner continuum.
Demler and Zhang \cite{Demler95} proposed within the SO(5) approach an interpretation in
terms of an antibound state in the particle-particle channel 
($\pi $-resonance or $\pi$-particle, \cite{Demler95,Zhang97,Demler98}).
This interpretation was, however, shown to be in conflict with the experimental finding
that the resonance always appears below twice the maximum $d$-wave gap in
the superconducting state \cite{Brinckmann98,Tchernyshyov01}.
Vojta {\it et al.} \cite{Vojta00} gave an interpretation as a soft mode 
directly related to the nearby antiferromagnetically ordered state.
Herbut and Lee \cite{Herbut03} considered within QED$_3$-theory the effect that vortex fluctuations
have on the spin dynamics and give an interpretation of the resonance as
four overlapping collective particle-hole modes of the phase fluctuating
$d$-wave superconducting state, centered at the node-node wavevectors.
An interpretation in terms of a roton-like excitation, appearing as a coupled
mode in the spin- and charge response (`hybrid spin-charge roton') was
suggested by Uemura \cite{Uemura04,Uemura05}.

Studies of the spin-excitation spectrum for a striped ground state 
(exhibiting stripe like modulations of charge and spin)
have been performed in Refs. 
\cite{Kruger03,Carlson04,Vojta04,Uhrig04,Uhrig05,Andersen05,Seibold05,Seibold05a}.
Striped states have been employed to describe in particular the spin
excitation spectrum in La$_{2-x}$Sr$_x$CuO$_4$.
However, so far this picture has not been as successful in describing
YBa$_2$Cu$_3$O$_{7-\delta}$ and Bi$_2$Sr$_2$CaCu$_2$O$_{8+\delta}$.

Finally, a promising candidate is the interpretation as 
a bound state (spin exciton) in the particle-hole channel.
Here, a number of techniques have been applied, including
slave-particle approaches within the $t-J$ and Hubbard models
\cite{Tanamoto91,Tanamoto94,Zha93,Stemmann94,Liu95,Brinckmann98,Brinckmann99,Brinckmann01,Kao00,Li02,Chen05}, 
Hubbard-operator techniques for the $t-t'-J$ model
\cite{Onufrieva95,Onufrieva99x,Onufrieva02}, 
memory-function approaches within the $t-J$ model \cite{Sega03,Sega05}
BCS models in random phase approximation (RPA)
\cite{Maki94,Mazin95,Bulut96,Salkola98,Norman01b,Eremin05},
FLEX-approximations in the Hubbard model \cite{Pao95,Dahm98,Takimoto98,Manske01a},
or spin-fermion models \cite{Morr98,Abanov99,Chubukov01,Chubukov04a}.
Theories based on such `spin-exciton models' have been especially successful
in reproducing in detail the experimentally determined spin-excitation spectra
in YBa$_2$Cu$_3$O$_{7-\delta}$ and Bi$_2$Sr$_2$CaCu$_2$O$_{8+\delta}$.
With some modifications of the model also the low-energy spectrum in 
La$_{2-x}$Sr$_x$CuO$_4$ has been described \cite{Morr00}.
Detailled quantitative predictions for the energy- and momentum-dependent
spectra determined experimentally have been obtained to date only within 
the spin-exciton models.
The following discussion will concentrate on a review of 
results obtained with those.

Although differing in the doping dependence of the band structure and
exchange coupling, most of the above-mentioned models for the spin-1 exciton 
ultimately determine the magnetic susceptibility for a given fixed doping value
by a relation of the form
\begin{equation}
\chi (\omega, {\bf q})= \frac{\chi_0(\omega ,{\bf q})}{1-J_{{\bf q}}
\chi_0(\omega ,{\bf q})} \; .
\label{chi}
\end{equation}
The parameter $J_{\vec{q}}$ can for example be interpreted as a Hubbard $U$, an
exchange constant $J$, or a spin-fermion coupling $\bar g$, depending on
the model and on the form of the susceptibility $\chi_0$ which is used.
A common form for $J_{\vec{q}}$ is
(from now on we use units in which $a=1$),
\begin{equation}
J_{\bf q}= J_0-2 J \left[\cos (q_x)+\cos (q_y) \right],
\end{equation}
which includes an onsite term $J_0$ and 
an effective exchange interaction $J$
that is restricted to nearest neighbors. This restriction
is generally assumed to be a very good approximation in cuprates.
The `bare' irreducible susceptibility 
$\chi_0(\omega ,{\bf Q})$, is determined as \cite{Schrieffer64}
\begin{eqnarray}
\chi_0(\omega ,{\bf q}) &=& -\sum_{{\bf k}} \sum_{\mu,\nu =\{\pm \}}
\frac{A_{{\bf k}}^\mu A_{{\bf k}+{\bf q}}^\nu + \alpha_{\bf q} C_{{\bf k}}^\mu C_{{\bf k}+{\bf q}}^\nu }{\omega + 
E_{{\bf k}}^\mu-E_{{\bf k}+{\bf q}}^\nu+i\Gamma } 
\left[f(E_{{\bf k}}^\mu ) - f(E_{{\bf k}+{\bf q}}^\nu )\right], \quad
\label{chi0}
\end{eqnarray}
where the excitation spectrum is given by,
\begin{eqnarray}
E_{{\bf k}}^\pm &=& \pm \sqrt{ \xi_{{\bf k}}^2
+|\Delta_{{\bf k}} |^2},
\end{eqnarray}
and the coherence factors are
\begin{eqnarray}
A_{{\bf k}}^\pm = \frac{1}{2}\pm \frac{\xi_{{\bf k}} }{E_{{\bf k}}^+-E_{{\bf k}}^-}  
\qquad C_{{\bf k}}^\pm = \pm\frac{ \Delta_{{\bf k}} }{E_{{\bf k}}^+-E_{{\bf k}}^-}.
\end{eqnarray}
The factor $\alpha_{\bf q} $ is equal to 1 in BCS theory, however was introduced
by Ioffe and Millis \cite{Ioffe99} in order to account for a suppression of the
$\langle \Delta_{{\bf k}}\Delta_{{\bf k}+{\bf q}} \rangle $ correlator in the 
$C_{{\bf k}}C_{{\bf k}+{\bf q}}$ coherence factors (by reducing $\alpha $ to
less than 1). 
The superconducting $d$-wave order parameter to a good approximation is given by
$\Delta_{\bf k}= \Delta \left[\cos (k_x)-\cos (k_y) \right]/2$,
and the effective quasiparticle dispersion $\xi_{{\bf k}}$
is usually parameterized by employing a tight-binding form,
\begin{eqnarray}
\label{tbdisp}
\xi_{\bf k}&=& -2\tilde t \left[\cos (k_x)+\cos (k_y) \right] 
-4\tilde t' \cos (k_x) \cos (k_y) - (\ldots ) -\mu ,
\end{eqnarray}
where $(\ldots )$ stands for possible further then next 
nearest neighbor hopping terms.

\subsection{Characteristic energies}
\label{CE}

\begin{figure}
\centerline{
\epsfxsize=0.5\textwidth{\epsfbox{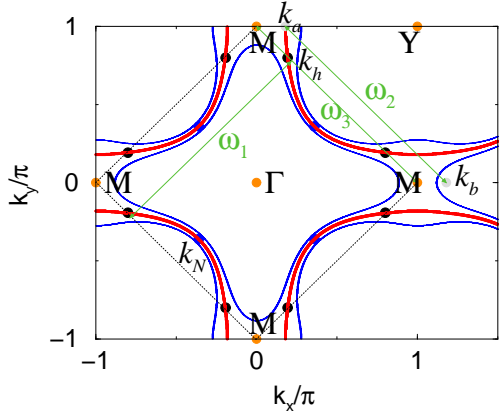}}
}
\caption{
$(\pi,\pi)$-scattering geometry 
with definition of characteristic wavevectors and characteristic
scattering energies discussed in the text. The thick curve is the Fermi surface. Lines of equal energy
at $\pm 50 $ meV around the Fermi energy are shown as thin curves.
\label{Brill_Onf}
}
\end{figure}

We discuss in this section singularities in the energy dependence of the
bare susceptibility at the antiferromagnetic wavevector $\vec{Q}=(\pi,\pi)$.
For a $d$-wave superconductor the bare irreducible susceptibility 
near the antiferromagnetic wavevector has a characteristic shape.
It is connected to the Fermi surface crossing along the $(0,\pi )-(\pi,\pi)$ direction,
or, equivalently,  along the $(\pi,0)-(\pi,\pi) $ direction. The corresponding Fermi
surface wave vector at ${\bf k}_a=(k_a,\pi )$ is the `{\it antinodal }' point of
the Fermi surface.
The scattering geometry for near optimal doping is sketched in 
Fig.~\ref{Brill_Onf}.

The singular behavior of the susceptibility at certain energies $\omega $ 
can be studied analytically by expanding the normal state quasiparticle
dispersion around the $(0,\pi)$ point and the $(\pi,0)$ point of the
Brillouin zone up to quadratic order, 
\begin{eqnarray}
\label{spdisp}
\xi (k_x,\pi+k_y) &=& \xi_M+\alpha k_x^2 - \beta k_y^2 \\
\label{spdisp1}
\xi (\pi+k_x,k_y) &=& \xi_M-\beta k_x^2 + \alpha k_y^2 ,
\end{eqnarray}
where the positive constants $\alpha $ and $\beta $ determine the
effective masses of the dispersion around the $M$ point, and
$\xi_M<0$ determines the position below the chemical potential
of the saddle point singularity in the electronic dispersion.

\begin{figure}
\centerline{
\epsfxsize=0.78\textwidth{\epsfbox{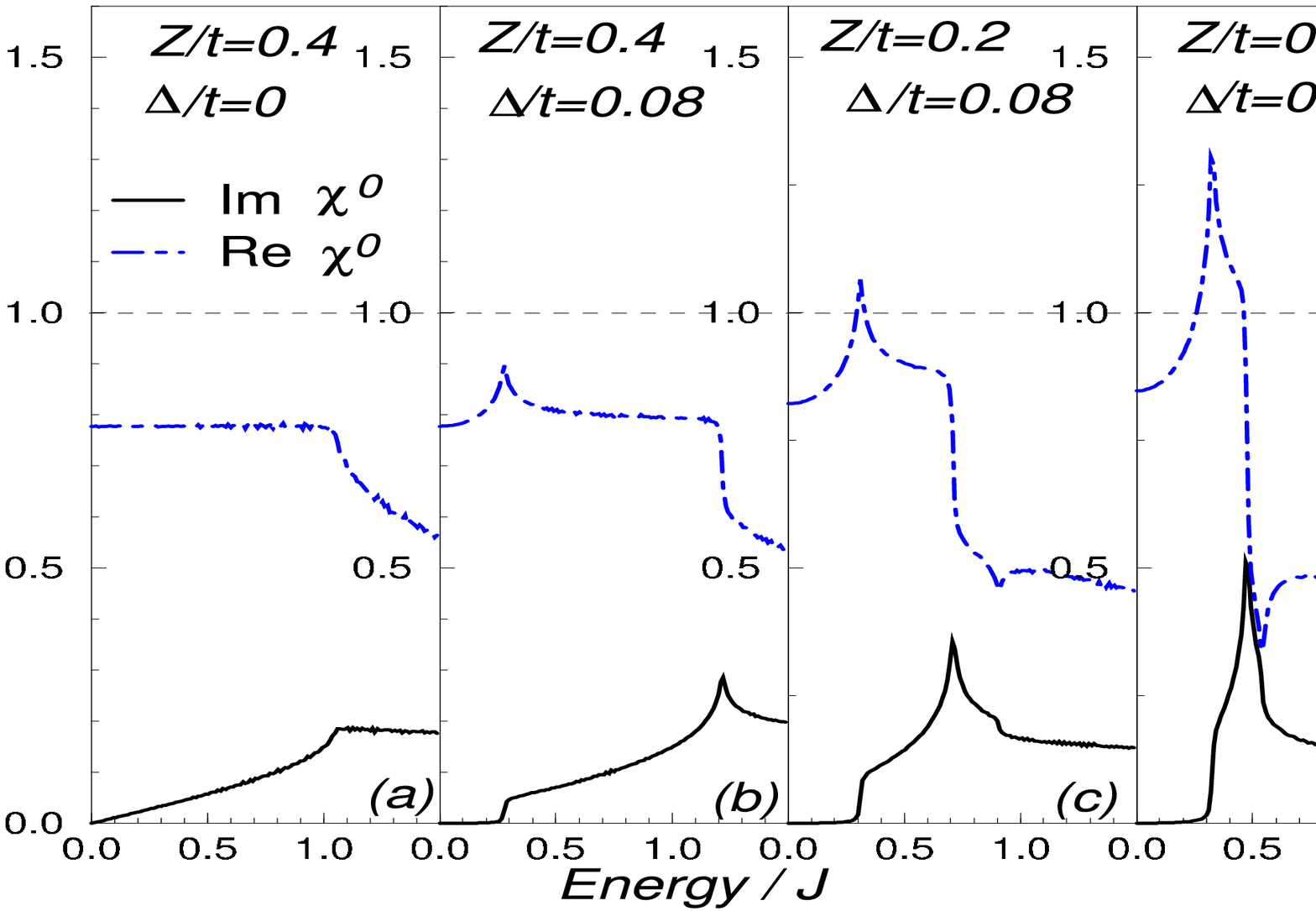}}
}
\caption{
\label{Fig4_Onufrieva02}
The irreducible susceptibility from Eq.~(\ref{chi0}) (in units of 1/4J)
as function of energy at the antiferromagnetic wavevector. $Z$ is the distance
of the saddle point singularity at the $(\pi,0)$ point in the fermionic 
Brillouin zone from the chemical potential, $Z=\mu -\epsilon_M=-\xi_M$. (a) is for
the normal state, (b) for $Z\gg \Delta $, corresponding to underdoping,
(c) corresponding to slight underdoping, and (d)
for $Z\sim \Delta $ corresponding to optimal doping.
The intersection of the dashed  line with Re $\chi_0$ determines the
resonance energy.
(From Refs. \cite{Onufrieva99x,Onufrieva02},
Copyright \copyright 2002 APS).
}
\end{figure}

In the normal state there is only one singularity, which is a (dynamic) Kohn 
singularity of square-root type.
It corresponds to the wavevector which connects the Fermi surface point
at ${\bf k}_a=(k_a,\pi )$ (see Fig.~\ref{Brill_Onf}) with the wavevector 
$ {\bf k}_b= {\bf k}_a+\vec{Q}= (\pi+k_a,0)$ (or equivalent wavevectors). The corresponding
energy is given by $\omega_2=|\xi_b |\approx |\xi_M| (1+\beta/\alpha )  $,
using that $k_a\approx \sqrt{-\xi_M/\alpha }$.
The leading singular term of the susceptibility near $\omega_2$ can be
readily calculated using the approximate dispersion 
Eqs.~(\ref{spdisp})-(\ref{spdisp1})
\cite{Onufrieva99,Onufrieva00},
and is of the form Const$-\frac{i}{\pi (\alpha + \beta) }
\sqrt{\frac{1-\omega /\omega_2}{1-\beta /\alpha }}$.
The behavior of the bare susceptibility is shown in Fig.
\ref{Fig4_Onufrieva02} (a). In this figure the energy is measured in $J$, 
as this is the energy which determines the superconducting gap within the
$t-t'-J$ model used in these calculations \cite{Onufrieva99x,Onufrieva02}.

In the superconducting state, shown in Fig.~\ref{Fig4_Onufrieva02} b-d, this singularity turns into a 
logarithmic singularity due to the qualitative change of the dispersion at the
point ${\bf k}_a$. It shows a jump in the real part and a logarithmic divergence
in the imaginary part of $\chi_0$, situated between the
two absorption edges in Im$\chi_0$ (see figure). The corresponding energy is given
by $\Delta_a+E_b$, 

\begin{equation}
\omega_2=\Delta_a+\sqrt{\xi_b^2+\Delta_b^2} \; . 
\end{equation}

Above this singularity there is an absorption edge within the particle-hole 
continuum which corresponds to the wavevector which connects
the $(\pi,0)$ point with the $(0,\pi )$ point. This singularity only
exists in the superconducting state when there are states available at $(\pi,0)$
and $(0,\pi)$ above the chemical potential due to particle hole mixing.
Calculations show, that the saddle point singularities near 
the $(\pi , 0)$  and $(0,\pi)$ points ($M$ points)
of the fermionic Brillouin zone play an important role.
The distance of this saddle point singularity from the chemical potential, 
$Z=-\xi_M=-(\epsilon_M-\mu)$ 
introduces a parameter which quantifies its importance \cite{Onufrieva95,Onufrieva99x}.
The corresponding energy of the absorption edge is given in terms of
this parameter by $2E_M$, or,
\begin{equation}
\omega_3=2\sqrt{\xi_M^2+\Delta^2} \; . 
\end{equation}
This absorption edge is seen in 
\ref{Fig4_Onufrieva02} (c) and (d).
Note, that near optimal doping $\omega_2\approx \Delta + E_M$, and thus
is close to $\omega_3=2E_M$. 
This can be seen in \ref{Fig4_Onufrieva02} (d).

Finally, in the superconducting state there is a 
continuum threshold toward low energies, which is given
by twice the excitation energy at a momentum, which is slightly displaced
from the `{\it hot spots}' in direction toward the nodes of the order parameter.  
The hot spots are commonly defined
as the crossing points between the Fermi surface and the lines
$k_x\pm k_y=\pi$ ($k_h$ in Fig.~\ref{Brill_Onf}). 
They are connected pairwise by antiferromagnetic 
wavevectors. In order to give an expression for the continuum threshold
we consider the line $k_x+k_y=\pi$, and measure k along this line
from $(0,\pi )$ toward $(\pi,0)$. It crosses the Fermi surface twice
with angle $\alpha $ at the so called hot spot wavevectors.
The order parameter at the first crossing point at $k_h$ will be
denoted by $\Delta_h$ (the order parameter at the other hot spot
along this line is then $-\Delta_h$), and we introduce velocities
$v_{\Delta,h} = |\partial \Delta (k) /\partial k |_{k_h}$ and
$v_{F,h} \sin \alpha = |\partial \xi (k) /\partial k |_{k_h}$.
The continuum threshold is given by twice the excitation energy
at a certain point along this line, which we determine by
minimizing $2\sqrt{\xi (k)^2+\Delta (k)^2}$ with respect to $k$.
Linearizing the $k$-dependence of $\xi_k$ and $\Delta_k$ around $k_h$
leads to
\begin{equation}
k-k_h = - \; \frac{\Delta_h v_{\Delta,h}}{v_{F,h}^2\sin^2\alpha +v_{\Delta,h}^2} \; .
\end{equation}
The corresponding wavevector is displaced toward the nodal line
$k_x=k_y$. From this the continuum threshold energy follows as
\begin{equation}
\omega_1 = 2\Delta_h \sqrt{1-\frac{2\beta_h^2}{\sin^2\alpha +\beta_h^2} },\quad
\beta_h=\frac{v_{\Delta,h}}{v_{F,h}} \; .
\end{equation}
Around optimal doing the correction term in the square-root is small and $k\approx k_h$,
$\omega_1\approx 2\Delta_h$.
The continuum threshold shows a jump in the imaginary part of the bare susceptibility,
and a logarithmic singularity in the real part.
It is seen in Fig.
\ref{Fig4_Onufrieva02} (b)-(d) as the lowest-energy singularity.

The strength of this singularity at the continuum threshold $\omega_1$ 
depends on the position of the hot spots on the Fermi surface.
When the hot spots move closer to the nodal points, which happens toward
underdoping, the threshold energy 
decreases and the magnitude of the jump on the continuum threshold decreases,
and vanishes completely at some critical doping. In the latter case the singularity is replaced
by  a square-root like singularity \cite{Brinckmann01}.

In the 
real part of the bare susceptibility the singularity shows up as a peak at the continuum edge,
which leads to the development of a spin-excitonic mode below
the continuum threshold, as we discuss next.

\subsection{The resonance mode}
\label{RM}

\subsubsection{Development of spin exciton}
\label{DSE}

As can be seen from Eq.~(\ref{chi}), for fixed momentum ${\bf q}$ the condition
\begin{equation}
1- J_{{\bf q}} \mbox{Re} \chi_0(\omega ,{\bf q})=0 ,\; \; 
\mbox{Im} \chi_0(\omega ,{\bf q}) =0
\label{modecond}
\end{equation}
gives sharp collective excitations at certain energies $\omega=\Omega_{res}$.
This condition can be satisfied below the particle-hole
continuum threshold $\Omega_0$, where it is of excitonic type.
The weight of this collective excitation is given by
\begin{equation}
w_{\bf q}= \frac{1}{J_{\bf q}^2} \left( \frac{\partial 
\mbox{Re} \chi_0(\omega ,{\bf q}) }{\partial \omega } \Big|_{\omega=\Omega_{res}}
\right)^{-1}.
\label{modeweight}
\end{equation}

As the real part of the bare susceptibility for a $d$-wave superconductor has
a logarithmic singularity at the continuum threshold,
the weight of the resonance goes linearly to zero when $\Omega_{res}$ approaches the continuum
threshold $\Omega_0$. On the other hand, the weight diverges for $\Omega_{res}\to 0$.

At the antiferromagnetic wavevector, ${\bf q}={\bf Q}$, the corresponding exciton is the
resonance mode observed in INS experiments.
The susceptibility can then be written as the sum of a 
continuum part and a resonant part, and the latter is given through its imaginary part
\begin{equation}
\mbox{Im} \chi_{res} (\omega,{\bf Q}) = \pi w_{\bf Q} \left(\delta(\omega-\Omega_{res})
- \delta(\omega+\Omega_{res}) \right) \; .
\label{resonance}
\end{equation}

Collective excitations stay well defined also for 
$\mbox{Im} \chi_0(\omega ,{\bf q}) \ne 0$ as long as 
\begin{eqnarray}
|J_{{\bf q}} \mbox{Im} \chi_0(\Omega_{res} ,{\bf q})| &\ll & 1 \nonumber \\
|J_{{\bf q}}^2w_{\bf q} \mbox{Im} \chi_0(\Omega_{res} ,{\bf q})| &\ll & \Omega_{res}
\label{modeim}
\end{eqnarray}
holds.
As we do not consider a spin-density wave instability, the condition
\begin{equation}
1- J_{{\bf q}} \mbox{Re} \chi_0(0 ,{\bf q})>0 
\label{SDWinst}
\end{equation}
is necessary. In the following we discuss the resulting 
resonance feature as a function of various parameters.

\subsubsection{Doping dependence}
\label{DD}

Calculations within a $t-t'-J$ model in mean-field approximation lead to
doping and temperature dependent effective hopping parameters $\tilde t$,
$\tilde t'$ and $\tilde t_\perp$. Calculations using a slave-boson
technique were performed by Brinckmann and Lee \cite{Brinckmann99,Brinckmann01}, calculations in
a Hubbard operator technique by Onufrieva and Pfeuty \cite{Onufrieva99x,Onufrieva02}. The Mori memory-function
formalism has been applied by Sega, Prelov{\v s}ek and Bon{\v c}a \cite{Sega03,Sega05,Prelovsek06}.

\begin{figure}
\centerline{
\epsfxsize=0.44\textwidth{\epsfbox{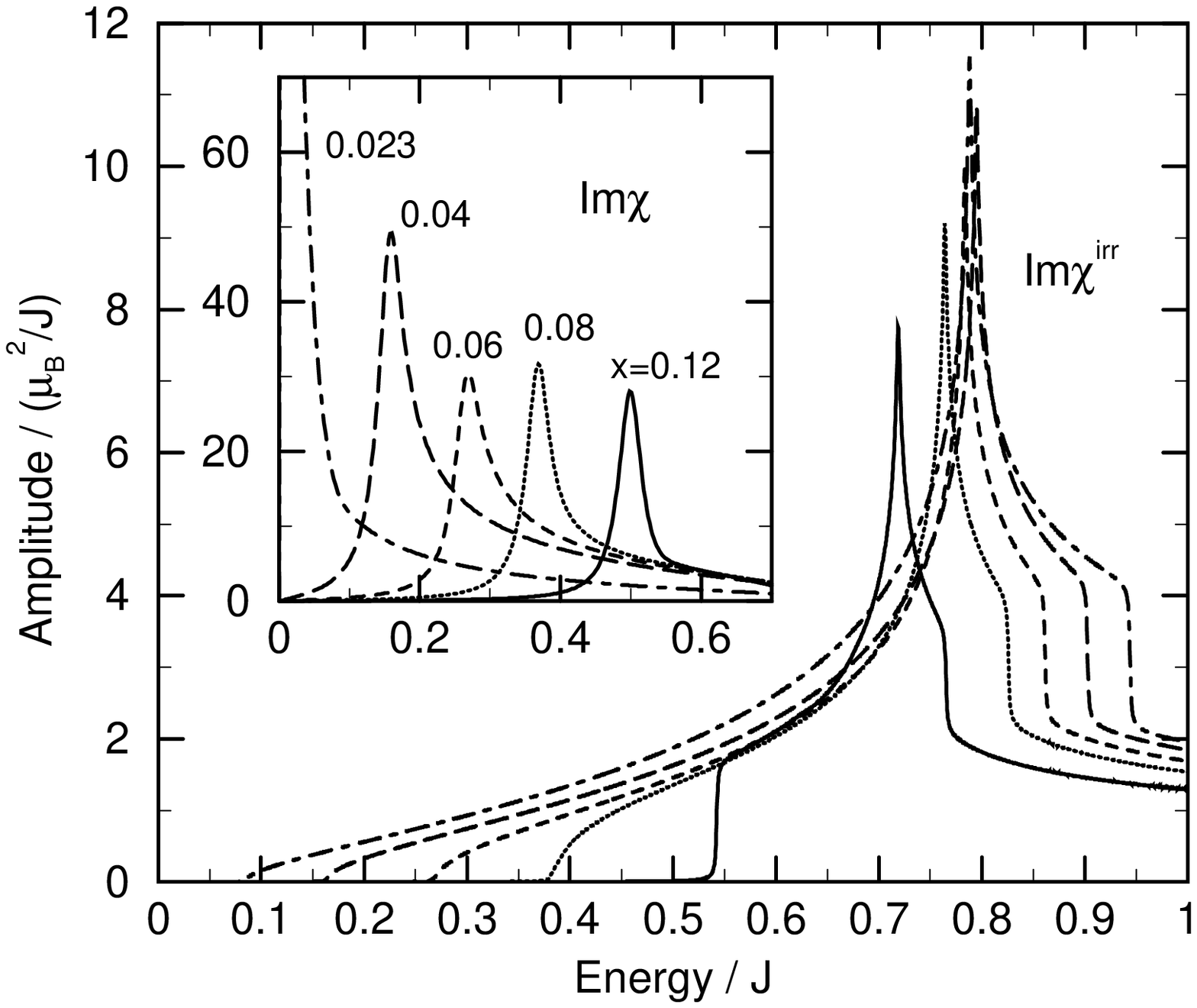}}
\epsfxsize=0.44\textwidth{\epsfbox{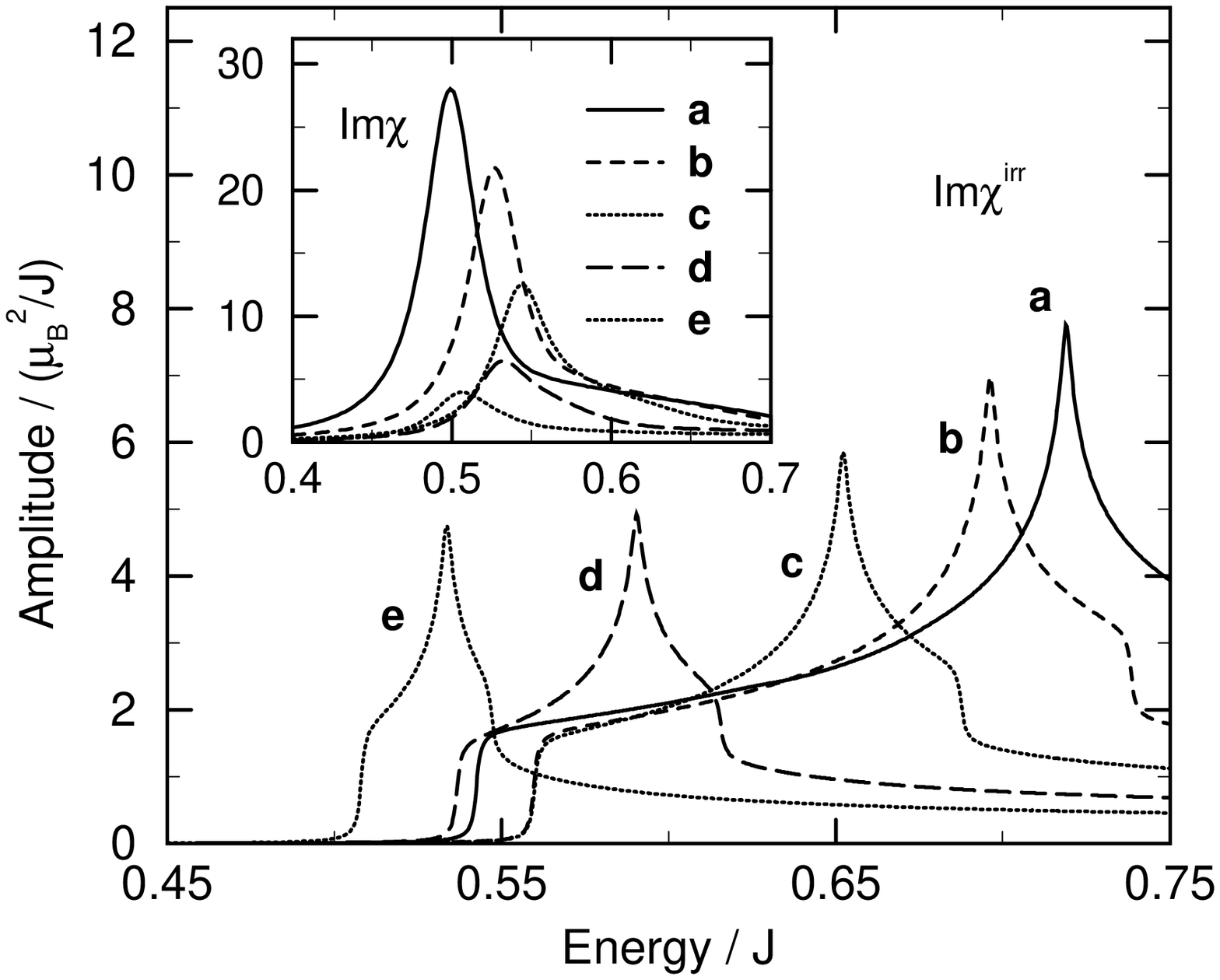}}
}
\caption{
\label{Fig5_Brinck01}
Magnetic response at wave vector $(\pi, \pi)$ for a single CuO$_2$ layer for
optimal to underdoped (left) and for optimal to overdoped (right) hole filling
$x$. Optimal doping corresponds to $x=0.12$. The calculations are for $T=0$.
The hole fillings for the right picture are $x=0.12 (a), 0.14 (b), 0.18 (c),
0.24 (d), 0.30 (e)$. 
The main figures show Im $\chi_0$ calculated from Eq.~(\ref{chi0}).
The insets show the corresponding Im $\chi $  calculated from Eq.~(\ref{chi}),
and using a damping term $\Gamma = $2.5 meV in Eq.~(\ref{chi0}).
(From Ref. \cite{Brinckmann01},
Copyright \copyright 2001 APS).
}
\end{figure}

In Fig. \ref{Fig5_Brinck01} the doping dependence of both Im$\chi_0$ and Im$\chi $ (in the 
insets), obtained with a slave-boson technique, is shown. With underdoping the continuum threshold $\omega_1$ moves to lower energies,
and at the same time the magnitude of the jump decreases. At a critical doping value the
jump vanishes and the continuum threshold shows a weak square-root singularity.
The resonant mode, shown in the inset, moves to lower energies and its weight increases.
With overdoping the continuum threshold first increases, then decreases again. A similar
trend shows up for the resonance frequency \cite{Brinckmann01}.
The qualitative behavior of the experimentally observed doping dependence thus is reproduced by these $t-t'-J$ model
calculations.

In Ref. \cite{Manske01a} the doping dependence of the resonance mode was
studied in a self-consistent FLEX approximation 
combined with random phase approximation. It was found 
that in the overdoped range
the resonance is close to twice the maximal superconducting $d$-wave gap $2\Delta $, 
at optimal doping is given by $\Omega_{res}=2\Delta_0 - \Omega_{max}$ 
(with the normal state spin-fluctuation spectrum given by
Im$\chi(Q,\Omega) = $Im$[\chi_Q/(1-i\Omega/\Omega_{max})]$)
and in the underdoped regime follows the normal state spin-fluctuation 
frequency $\Omega_{max}$. 

In Ref. \cite{Abanov99} the two limiting cases of weak and strong spin-fermion
coupling were studied in an effective low-energy theory.
There it was found that for the weak coupling limit, where
$\Delta\ll\Omega_{max}$
(corresponding to strongly overdoped materials), the resonance position
$\Omega_{res}=2\Delta_h  (1-w_\vec{Q})$ is related to the
weight of the resonance, $w_{\vec{Q}}=e^{-\Omega_{max}/(2\Delta_h)}$ (here,
$\Delta_h$ is the gap at the `hot-spot' wavevector $\vec{k}_h$, see
Fig.~\ref{Brill_Onf}).
Thus, the binding energy of the resonance,
$2\Delta_h-\Omega_{res}$, is
proportional to the weight of the resonance $w_{\vec{Q}}$.
In the strong coupling limit, appropriate for the underdoped region,
they obtain $\Omega_{res}\sim \sqrt{\Omega_{max}\Delta_h}$. This is valid
when $\Omega_{res}< \Delta$, $\Omega_{max}< \Delta $ holds. Both inequalities
are fulfilled for optimally and underdoped materials.

\subsubsection{Dependence on disorder}
\label{ThDis_res}

There have been few theoretical efforts to study the influence of 
non-magnetic impurities on the magnetic resonance. 
In Ref. \cite{Bulut00} the effects of dilute Zn impurities on the 
spin susceptibility in YBa$_2$Cu$_3$O$_{7-\delta}$
was studied using a two-dimensional Hubbard model. Coulomb correlations
were included via the random-phase approximation, before averaging
over the positions of the impurities was performed.
An enhancement of the static susceptibility at the antiferromagnetic 
wavevector with Zn impurity concentration was found.

A different route was chosen in Ref. \cite{Sachdev99}, by studying
a quantum impurity in a system
of coupled spin-ladders, which exhibits a paramagnetic 
ground state for small inter-ladder coupling. It was argued there that
the magnon damping mechanism due to the presence of impurities
in such systems has also relevance for the
broadening of the resonance peak in Zn-doped YBa$_2$Cu$_3$O$_{7}$.

Further theoretical work in studying the interplay between non-magnetic
impurities and antiferromagnetic correlations in superconducting
cuprates is required for a complete understanding of the experimental results
of Section \ref{Dis_res}.

\subsubsection{Dependence on magnetic field}
\label{Magfield}

In order to explain the suppression of the weight of the magnetic resonance with
an applied magnetic field in $c$-axis direction \cite{Dai00}, it is necessary to
consider the influence vortices have on it. 
This problem was treated in Ref. \cite{Eschrig01}.
There are several possible effects of vortices. 
The first effect is associated with 
the influence of the supercurrents
circulating around the vortices on the resonance.  
Second, there might be an effect of a spatially uniform suppression of the 
$\langle \Delta_{{\bf k}}\Delta_{{\bf k}+{\bf Q}} \rangle $ correlator
which enters the coherence factors of the spin susceptibility
(where ${\bf Q}$ is the antiferromagnetic wave vector at which the resonance is
peaked).  Such a suppression could be a result of
dephasing of the pairing in a c-axis field due to the vortices, as observed
in Josephson plasma resonance experiments \cite{Matsuda95}.  
It can be taken into account by
reducing $\alpha$ to less than 1 in Eq.~(\ref{chi0}).
A related model was considered to explain a sum rule violation in c-axis 
infrared conductivity, see \cite{Ioffe99}.
A third possibility is an assumed (field induced) spatially uniform suppression
of the gap magnitude.
Finally, the effect of the vortex cores will have an impact. In this case the
resonance might be locally suppressed in the core regions and recovers only
outside. This assumes of course that the spin correlation length is short,
so that a local suppression of the resonance feature makes sense.
Indeed, in cuprates the correlation length associated with the resonance
is of the same order as the superconducting coherence length.
The fact that the experimental suppression goes like $1-H/H^*$ 
(where $H^*$ is a number of the order of the upper critical field)
is highly suggestive of a vortex core effect \cite{Dai00}.

A calculation of the 
influence of the supercurrents around the vortices on the
resonance in the spin-spin correlation function has shown that
the supercurrent has three effects:
it shifts the position of the resonance to slightly lower energy, it broadens the
resonance, and it reduces the magnitude of the resonance at the peak
energy \cite{Eschrig01}.  
However, the integrated weight between 0 and $\approx 2\Delta $ is conserved.
These findings are in apparent contradiction with the
experimental facts, which are that the resonance does not shift, nor
broaden, and that the integrated weight is reduced by about 15\% at 7
T (~\cite{Dai00}).  

\begin{figure}
\centerline{
\epsfxsize=0.45\textwidth{\epsfbox{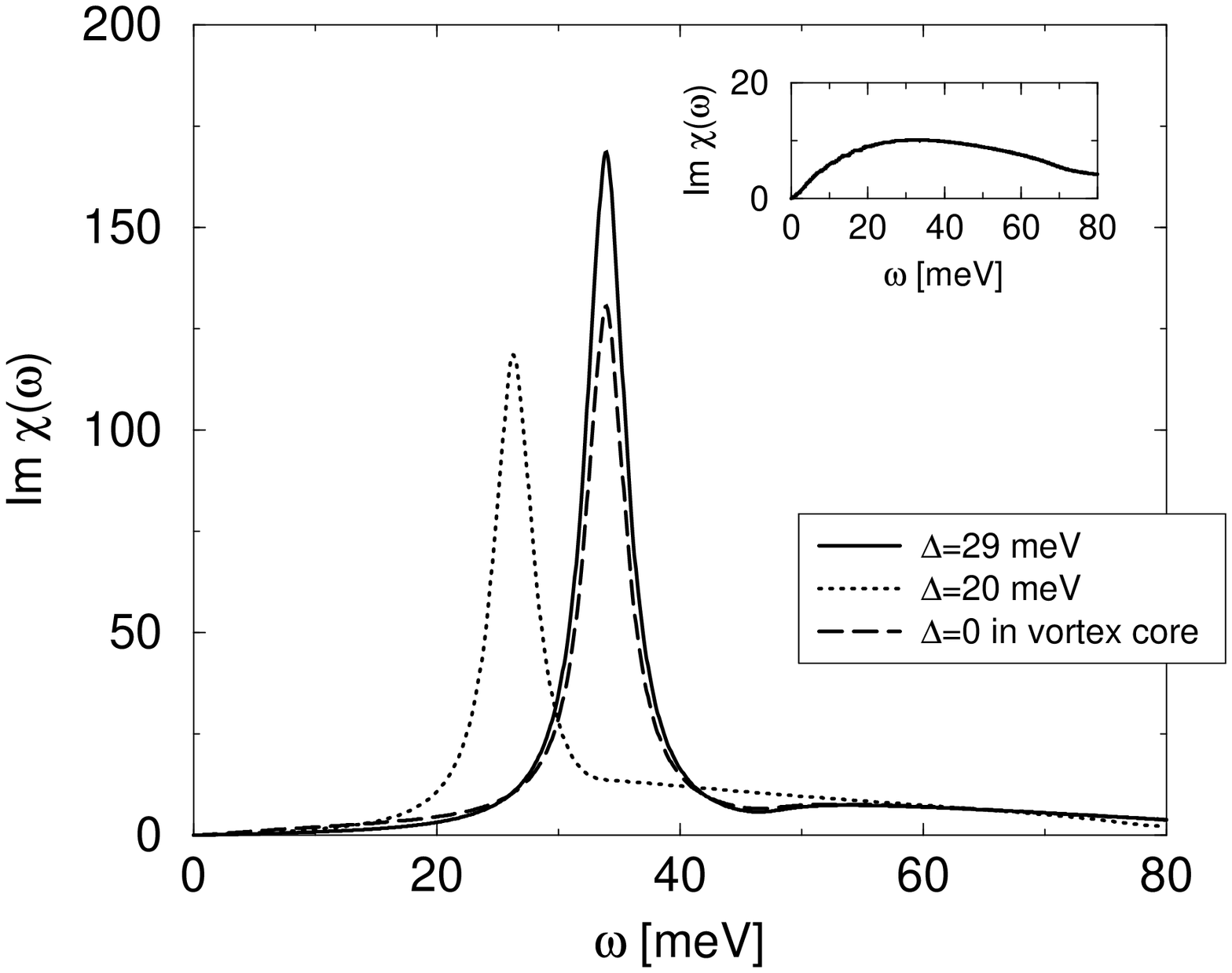}}
\epsfxsize=0.45\textwidth{\epsfbox{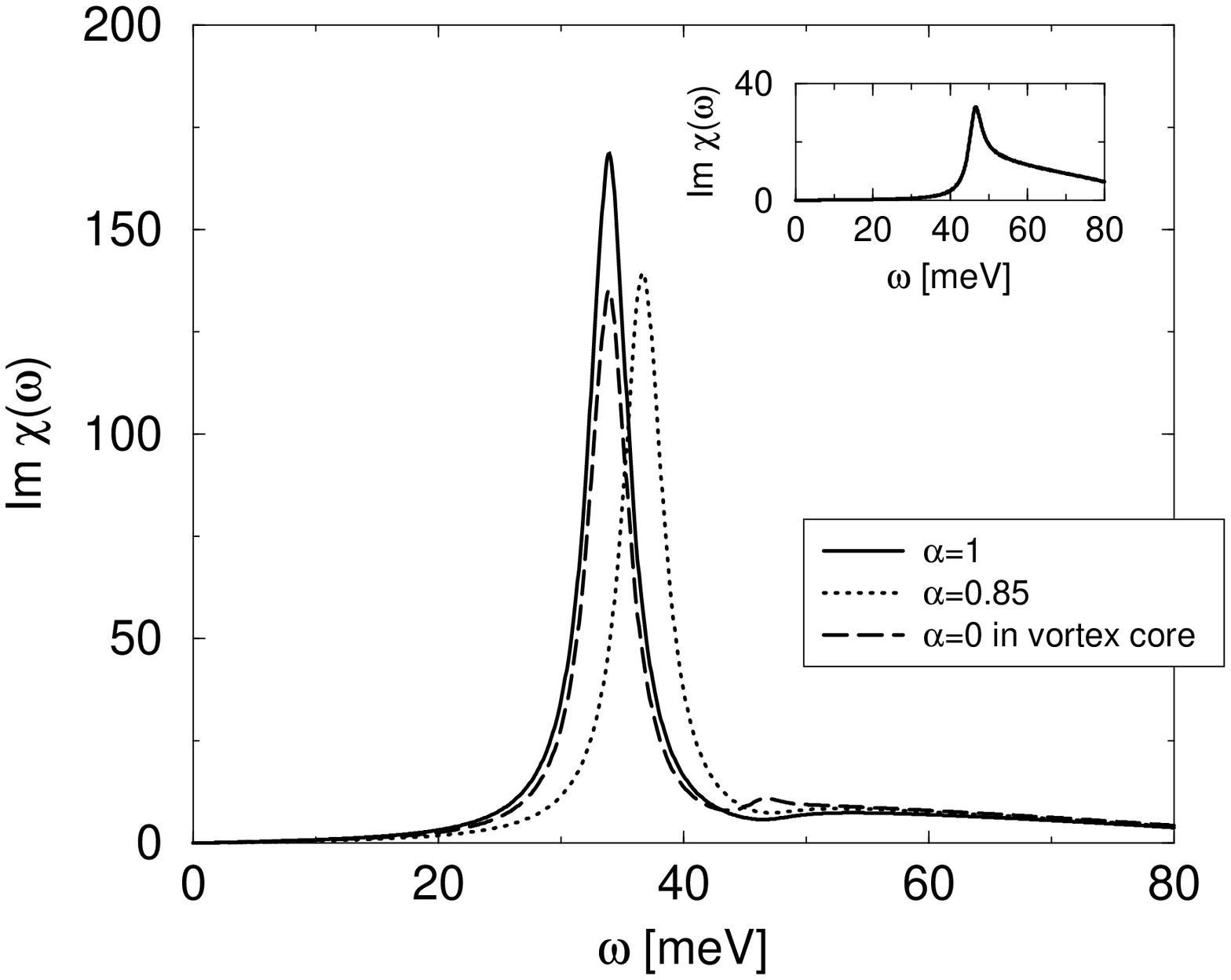}}
}
\caption{
\label{fig2a}
Left:
Comparison of zero field susceptibility with the same susceptibility, but
with reduced gap magnitude, $\Delta$.
Dotted line: assuming a spatially uniform reduction from 29 meV to 20 meV;
dashed line: assuming a reduction to zero in 24\% of the
vortex unit cell area representing the cores (the uniform zero gap
response is shown in the inset).
In both cases the weight of the resonance is reduced, but
for the spatially uniform case, the resonance is shifted considerably
downwards in energy.
Right:
\label{fig2b}
Comparison of zero field susceptibility with the same susceptibility, but
with the $\langle \Delta \Delta \rangle $ correlator in the numerator
of Eq.~(\ref{chi0}) reduced ($\alpha < 1$). Dotted line:
assuming a spatially uniform reduction by 15\%;
dashed line: assuming a reduction to zero in 20\% of the
vortex unit cell area representing the cores (the uniform
zero $\alpha $ response is shown in the inset).
In both cases the weight of the resonance is reduced, but
for the spatially uniform case, the resonance is shifted considerably
upwards in energy.
(From Ref. \cite{Eschrig01},
Copyright \copyright 2001 APS).
}
\end{figure}
In Fig.~\ref{fig2a} the results for the influence of a magnetic field
on the resonance are summarized. 
First, any spatially homogeneous effect is in disagreement with
the experimental data:
(A) a homogeneous reduction of the gap magnitude, shown as the dotted line in
in the left panel of Fig.~\ref{fig2a}, leads to a shift of the resonance to considerably lower
energy compared to the zero field result (full line); 
and (B) a homogeneous reduction of the 
$\langle \Delta \Delta \rangle $ correlator by 15\% ($\alpha=0.85$), shown
as dotted line in the right panel of Fig.~\ref{fig2a}, 
leads to a shift of the resonance to higher energy,
compared to the zero field result (full line). 

The vortex-core effect is illustrated as dashed lines in 
Fig.~\ref{fig2a}.  In the insets the susceptibilities
for zero $\Delta $ and for zero $\langle \Delta \Delta \rangle $ correlator,
used in the vortex core area, are shown, whereas the full curves
of the main panels are used for the inter-vortex regions.
The latter is justified since the Doppler shift has a
negligible effect on the integrated intensity.
In both cases, the resulting curves, calculated for a 15\% reduction in
total integrated weight,
reproduce very well the experimental finding of no shift or broadening of
the resonance. 

The conclusion that the resonance 
is not supported in the region of the vortex core is corroborated by
the following five additional facts \cite{Eschrig01}:
a) the considerable momentum
width of the resonance shows that the corresponding spin excitations 
have a decay length of only two lattice constants, which is smaller than
the coherence length; thus the resonance will be sensitive to variations of the
order parameter on the coherence length scale; 
b) the resonance at zero field only exists in the
superconducting state, and disappears in the normal state;
c) coherence peaks in the single particle density of states at the gap edge
were not found in the core region in STM measurements \cite{Maggio95,Renner98,Pan00a};
this would modify the $2\Delta $-edge in $\chi''_0$ [Eq.~(\ref{chi0})]
and suppress the resonance;
d) in underdoped materials, missing subgap states 
point toward a loss of quasiparticle 
weight due to a pseudogap in the vortex core \cite{Maggio95,Renner98,Pan00a};
e) the dip feature in the tunneling density 
of states, thought to be due to the coupling of 
quasiparticles to the resonance \cite{Eschrig00}, is not observed in the
vortex core region \cite{Maggio95,Renner98,Pan00a}.

As documented by ARPES measurements \cite{Loeser97,Norman98a},
quasiparticle-like peaks in the spectral functions are present
only below $T_c$, the onset temperature of
phase coherence.  
Motivated by this observation it was suggested in Ref.~\cite{Eschrig01} 
that this may lead to a destruction
of quasiparticle excitations in the vortex core region
similar to what happens in the pseudogap state.
The absence of quasiparticle peaks as well as of 
the neutron resonance in the core
region is consistent with the notion that {\em both} these spectral
features require substantial local phase correlations \cite{Janko99}.
Thus, we can assume that the magnetic resonance is simply absent
in the vortex core regions because of the absence of quasiparticle excitations
\cite{Eschrig01}.
The question to whether the vortex core is closely related to the
pseudogap phase, or if even a different type of order exists in the
core region, is not settled up to now and subject of ongoing research.

\subsubsection{Even and odd mode in bilayer cuprates}
\label{EvenOdd}

For bilayer materials the susceptibility is classified into even and odd
components according to Eqs.~(\ref{evenodd1})-(\ref{evenodd1a}) and (\ref{evenodd2})-(\ref{evenodd2a}).
It can be inferred experimentally via Eqs.~(\ref{cross1}) and (\ref{cross2}).

If the coupling between the two planes in the bilayer is coherent, then
the bands split into
bonding and antibonding bands according to Eq.~(\ref{tperp}).
The $d$-wave order parameter has the form Eq.~(\ref{DWOP1}). In agreement
with experiments \cite{Borisenko02} it is assumed to be the same on the bonding and antibonding
bands.

\begin{figure}
\centerline{
\epsfxsize=0.78\textwidth{\epsfbox{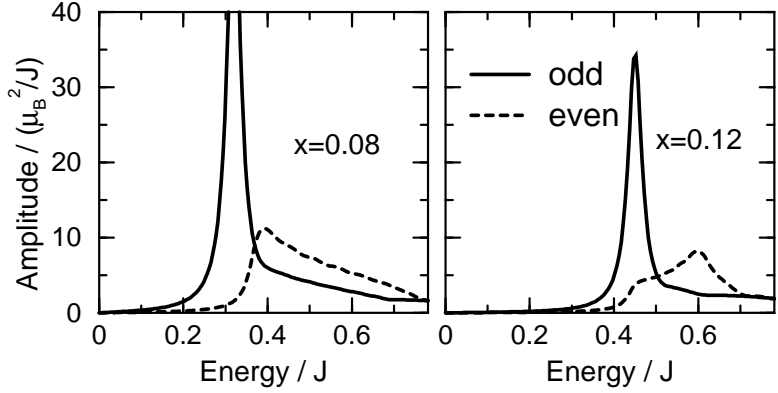}}
}
\caption{
\label{Fig13_Brinck01}
Imaginary part of odd- and even-mode susceptibility of a bilayer system at
wavevector $(\pi ,\pi)$. (From Ref. \cite{Brinckmann01},
Copyright \copyright 2001 APS).
}
\end{figure}

The irreducible susceptibility, Eq.~(\ref{chi0}), is generalized to a matrix in bonding and antibonding
indices, and for example the component $\chi_0^{(ab)}$ reads,
\begin{eqnarray}
\chi_0^{(ab)}(\omega ,{\bf q}) &= &
-\sum_{{\bf k}} \sum_{\mu,\nu =\{\pm \}}
\frac{A^{(a),\mu}_{{\bf k}} A^{(b),\nu}_{{\bf k}+{\bf q}} + \alpha^{(ab)}_{\bf q} C^{(a),\mu}_{{\bf k}} C^{(b),\nu}_{{\bf k}+{\bf q}} }{\omega + 
E^{(a),\mu}_{{\bf k}}-E^{(b),\nu}_{{\bf k}+{\bf q}}+i\Gamma^{(ab)} } 
\nonumber \\ &&\times 
\left(f(E^{(a),\mu}_{{\bf k}} ) - f(E^{(b),\nu}_{{\bf k}+{\bf q}} )\right) \; .
\label{chi0bil}
\end{eqnarray}
with coherence factors and excitation energies calculated with the corresponding bonding and antibonding dispersions and order parameters.
One then builds even and odd components of the irreducible susceptibility according to Eq.~(\ref{evenodd2}).
Taking into account an exchange coupling between the planes within a bilayer,
the exchange coupling also has even and odd components
\begin{eqnarray}
J^{e}_{\bf q}= J_{\bf q} - J_\perp ,\nonumber \\
J^{o}_{\bf q}= J_{\bf q} + J_\perp .
\end{eqnarray}
The susceptibility then can be written as
\begin{eqnarray}
\chi_e (\omega, {\bf q})= \frac{\chi^{(e)}_0(\omega ,{\bf q},p_z)}{1-J^{(e)}_{{\bf q}}
\chi^{(e)}_0(\omega ,{\bf q})} \; ,\nonumber \\
\chi_o (\omega, {\bf q})= \frac{\chi^{(o)}_0(\omega ,{\bf q},p_z)}{1-J^{(o)}_{{\bf q}}
\chi^{(o)}_0(\omega ,{\bf q})} \; .
\label{chibil}
\end{eqnarray}
Results of calculations for the spin susceptibility taking into account
bilayer splitting are reproduced in Fig. \ref{Fig13_Brinck01} for 
two doping levels. 
The even-mode susceptibility shows a considerably
weaker resonance feature, which is placed closer to the
continuum edge, than the odd-mode susceptibility. This is consistent
with the experimental observation of \cite{Pailhes03}.

\subsection{The incommensurate response}
\label{IncommRes}

There are numerous theoretical works explaining the incommensurate response
in the spin susceptibility 
\cite{Lavagna94,Abrikosov98,Onufrieva99x,Abanov99,Brinckmann99,Norman00,Kao00,Manske01a,Onufrieva02,Norman01b,Chubukov01,Eremin05}.

\begin{figure}
\centerline{
\epsfxsize=0.40\textwidth{\epsfbox{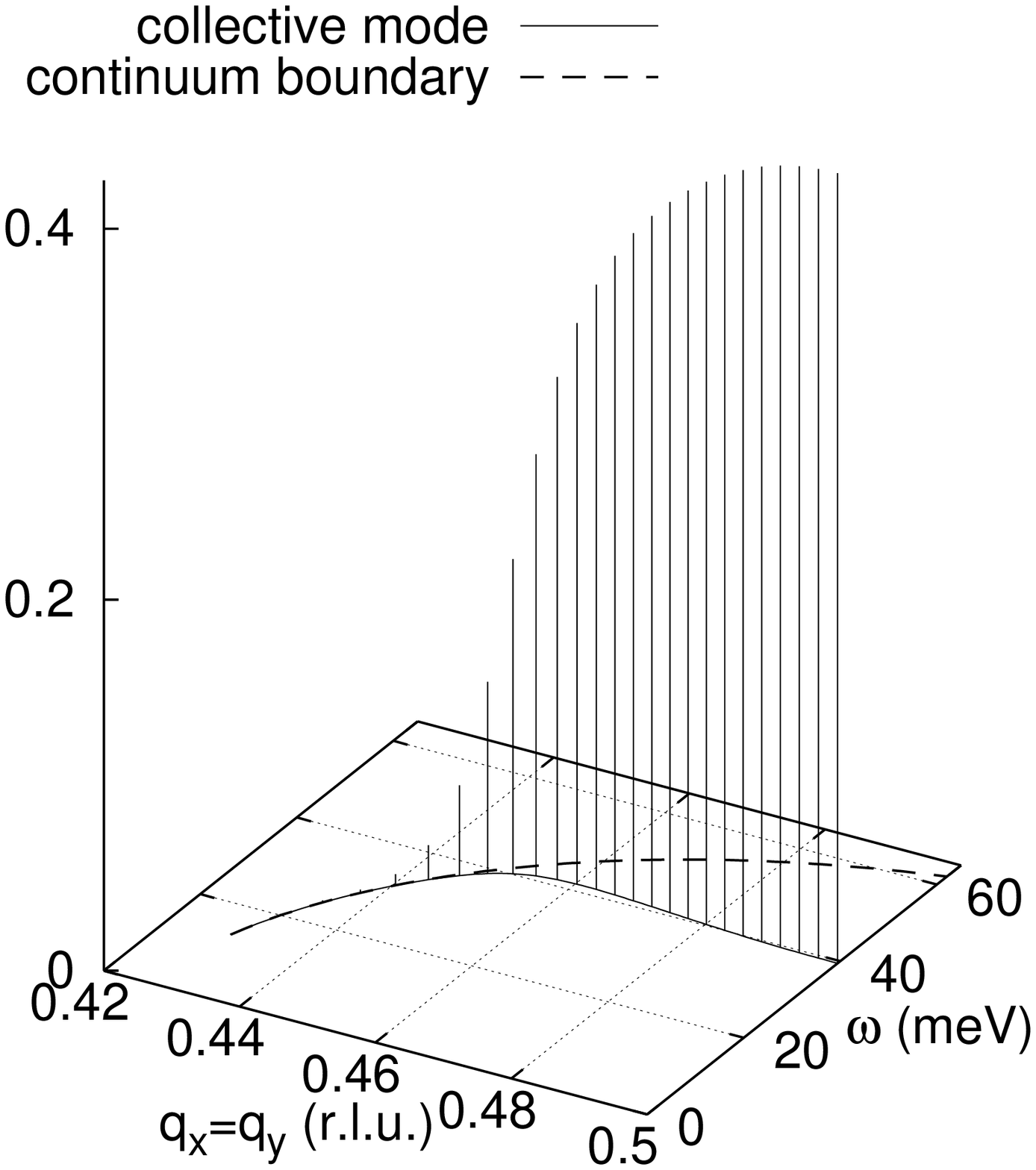}}
\epsfxsize=0.55\textwidth{\epsfbox{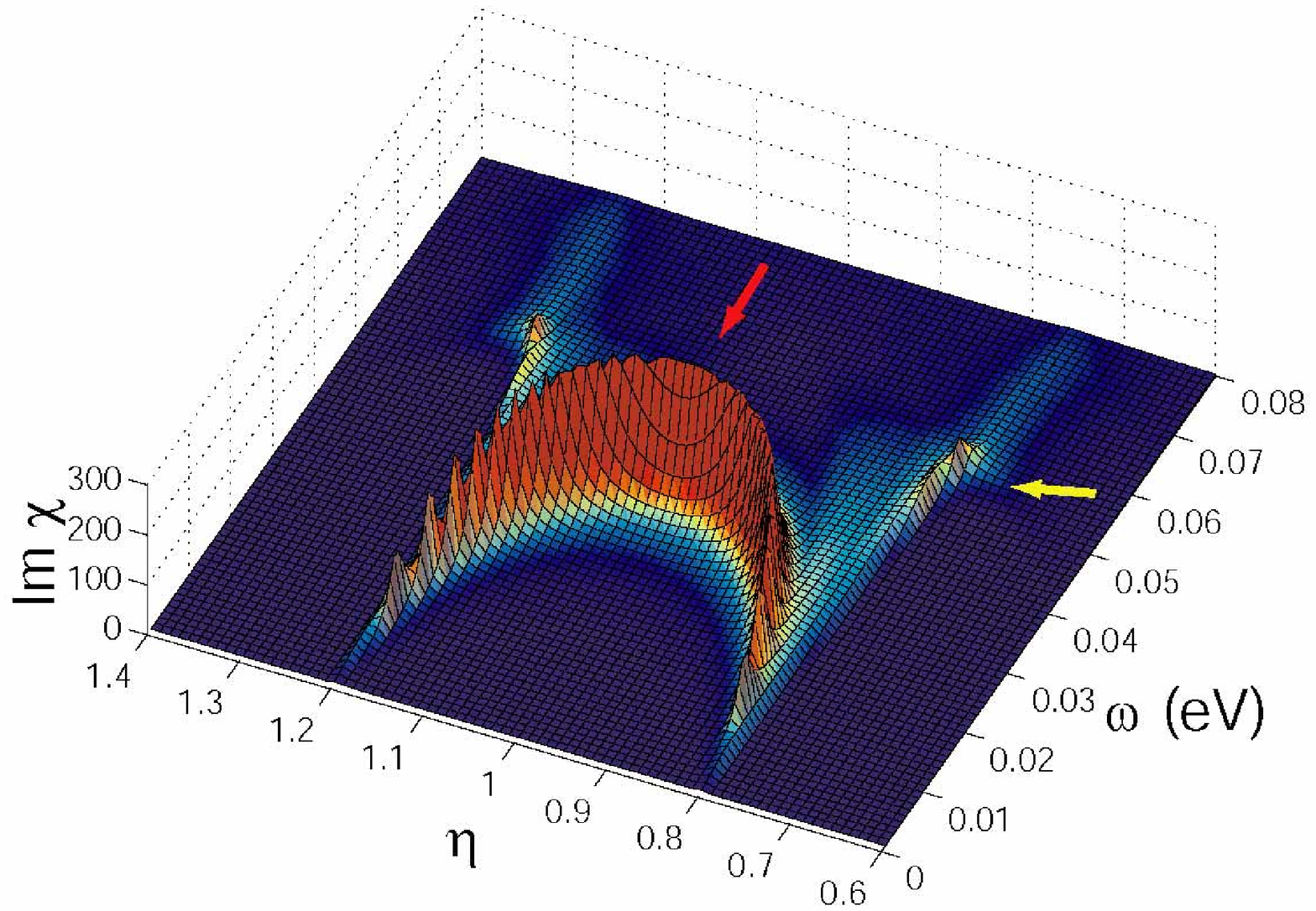}}
}
\caption{
\label{Fig3_Chubukov01}
Left:
Dispersion along the (110) direction and intensity of the spin-1 resonance from 
Eqs.~(\ref{chi})-(\ref{chi0}) using a tight-binding quasiparticle dispersion
inferred from ARPES experiments, corresponding to optimally doped
Bi$_2$Sr$_2$CaCu$_2$O$_{8+\delta}$.  The parameter
$J_{\vec{q}}\equiv J_0$ was chosen to be independent of the wavevector $\vec{q}$.
(From Ref. \cite{Chubukov01},
Copyright \copyright 2001 APS).
Right: The same for parameters suitable to reproduce the experimental 
dispersion 
along the (110) direction of an underdoped YBa$_2$Cu$_3$O$_{6.85}$ sample,
as shown in Fig.~\ref{Fig5_Keimer04}. Here, $\Delta = 42$ meV, and
$J_{\vec{q}}=573 $meV$ [1-0.1(\cos q_x-\cos q_y)]$ was used, to reproduce
details in the experimental dispersion of experiment \cite{Pailhes04}
(indicated by arrows).
(From Ref. \cite{Eremin05},
Copyright \copyright 2005 APS).
}
\end{figure}

In a picture where the magnetic resonance mode is treated as an excitonic
bound state below the particle-hole continuum,
the incommensurate response arises from the resonance condition 
for wavevectors slightly displaced from the antiferromagnetic wavevector.
For this case, the continuum threshold for a $d$-wave superconductor shows a characteristic
dependence on $\delta_x=q_x-\pi$, with a minimum in the continuum threshold $\omega_1(\delta_x)$
at a certain value $\pm \delta_{x,min}$.
As a consequence, the resonance criterion is fulfilled also for wavevectors with non-zero $\delta_x$.
For $\delta_x\ne 0$ the step in Im $\chi_0$ splits into two steps, the lower of which corresponds now
the the continuum threshold. 
The stepsize is however much smaller than the total stepsize at $\delta_x=0$,
leading also the a weaker structure in Re $\chi_0$. 
Thus, the weight of the resonance feature drops
quickly for increasing $|\delta_x|$. This behavior is shown in Fig. \ref{Fig3_Chubukov01} (left), where a
tight-binding dispersion obtained from ARPES data for
Bi$_2$Sr$_2$CaCu$_2$O$_{8+\delta}$ was used as input \cite{Chubukov01}. Interestingly, for this case the dispersion of the
collective mode is rather weak near the resonance energy $\Omega_{res}$, which leads to a large momentum
width of the resonance mode as observed in experiments.
Also shown in Fig. \ref{Fig3_Chubukov01}, on the right, is an example for
the theoretical results appropriate
for an underdoped YBa$_2$Cu$_3$O$_{6.85}$ sample \cite{Eremin05}, reproducing details of the
incommensurate dispersion which are shown in Fig.~\ref{Fig5_Keimer04}.
In particular, it explains a new feature, the so-called $Q^\ast $ mode
(light arrow),
which resides above the resonance mode (dark arrow) 
at incommensurate wavevectors and was observed in 
the experiment \cite{Pailhes04}.

\begin{figure}
\centerline{
\epsfxsize=0.78\textwidth{\epsfbox{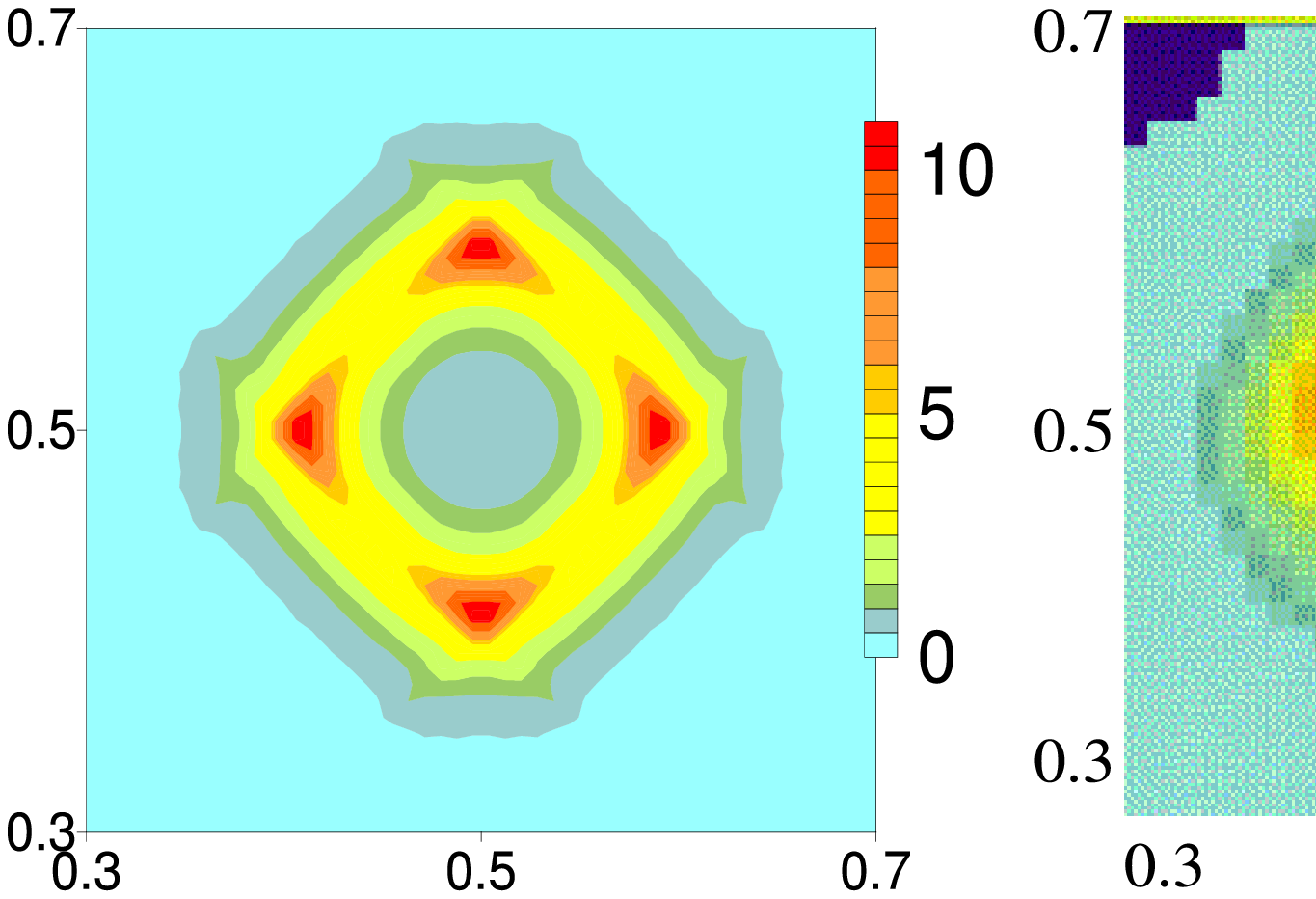}}
}
\caption{
\label{Fig6_Onufrieva02}
$p_x$, $p_y$ dependence of Im $\chi (\vec{p},\omega)$ showing the
incommensurate response around the $(\pi,\pi)$ wavevector (corresponding
to $p_x=p_y=0.5$). 
Left the theoretical calculation 
\cite{Onufrieva99x,Onufrieva02} according to
Eqs.~(\ref{chi}), (\ref{chi0}), for a $t-t'-J$ model with $t'/t=-0.3,
Z/t=0.2, \Delta/J=0.2, \Omega/J=0.23$. Right, experimental results 
from \cite{Mook98b}, measured on  YBa$_2$Cu$_3$O$_{6.6}$ at
$\omega = 24 $meV.
(From Ref. \cite{Onufrieva99x,Onufrieva02},
Copyright \copyright 2002 APS).
}
\end{figure}

The incommensurability pattern below the resonance is characteristic in the sense that
it has four maxima at positions $(\pi \pm \delta_x,\pi)$, and $(\pi ,\pi\pm \delta_x)$.
This is reproduced quantitatively by theory using the formulas 
Eqs.~(\ref{chi})-(\ref{chi0}) \cite{Brinckmann99,Norman00,Kao00,Onufrieva02,Sega05,Eremin05}.
An example is shown in Fig. \ref{Fig6_Onufrieva02}.
Corresponding experimental patterns above the resonance have their maxima at
positions $(\pi \pm \delta^\ast ,\pi\pm \delta^\ast )$ \cite{Hayden04},
consistent with theory \cite{Eremin05,Sega05,Prelovsek06}.

\subsection{Effective low-energy theories}
\label{EffLET}

A different approach is to start from a separation of energy scales into high energy and
low energy. On a formal level this allows for a perturbation expansion in Feynman diagrams,
where all propagators are written as a sum of a high-energy propagator and a low-energy
propagator. Then, all high-energy propagators are combined with their attached interaction
vertices to new, renormalized high-energy interactions. The resulting diagrams contain
low-energy propagators and high-energy effective interaction vertices. In a second
renormalization step the quasiparticle renormalization factor $z_0$ is absorbed by the
attached vertices to built quasiparticle propagators and effective quasiparticle interaction
vertices. For theoretical treatments along these lines see e.g. 
Abanov {\it et al.} \cite{Abanov03} and citations there.

Using this procedure for the susceptibility $\chi $, one usually neglects terms containing
one low-energy propagator and one high-energy propagator in $\chi_0$, so that one has
a high-energy bare susceptibility $\chi_{h}$ and a low-energy bare susceptibility $\chi_l$.
For the full susceptibility one obtains (we omit for simplicity all arguments and we allow for
a more general notation, where multiplication of two quantities can for example 
also symbolize convolutions)
\begin{eqnarray}
\label{renorm}
g_1\chi g_2 &=& g_1\frac{(\chi_h+z_0\chi_lz_0)}{1-J(\chi_h+z_0 \chi_l z_0)}g_2  \\
\label{renorm1}
&=& (g_1\chi g_2)_{\rm inc} + \bar g_1 \frac{\chi_l}{1-\bar J \chi_l} \bar g_2
\end{eqnarray}
with the renormalized quantities
\begin{eqnarray}
(g_1\chi g_2)_{\rm inc}&=& g_1 \chi_h\frac{1}{1-J \chi_h} g_2 \\
\bar g_1 &= &g_1\frac{1}{1-\chi_h J}z_0 \\
\bar g_2 &= &z_0\frac{1}{1-J\chi_h}g_2\\
\bar J &=&z_0\frac{1}{1-J\chi_h} Jz_0 \; .
\end{eqnarray}
The renormalization factors $z_0$ entering $\chi_l$ are in general anisotropic, and are
defined as $\sqrt{z_{\vec{k}} z_{\vec{k}+\vec{q}}}$. Thus, the vertices for the effective
low-energy susceptibility depend in general both on the quasiparticle wavevector
$\vec{k}$ and the external wavevector $\vec{q}$.

We assume that these renormalized high-energy quantities vary on a high-energy scale,
and can be considered in first approximation as constant in energy when considering 
the low energy region. If one chooses to assign some energy-dependence to these quantities,
then it is very weak in the low-energy region.
Only $\chi_l $ is affected by the superconducting transition, $(g_1\chi g_2)_{\rm inc}$, $\bar J$
and $\bar g_{1,2}$ stay unaffected. If one wants to allow for a change of the 
quasiparticle renormalization factors at $T_c$ one needs to keep them explicitly.
However, up to date it is not clear yet, if for high-$T_c$ cuprates the
loss of the coherence peaks is due to a change in $z_0$ or due to a dramatic decrease
of the lifetime \cite{Norman01}. Note in this regard, that the experimental definition
of the coherent quasiparticle weight in ARPES is purely phenomenological and might
be different from the above introduced quasiparticle renormalization factors.

We are interested in strong correlations which lead to
\begin{eqnarray}
\frac{1}{1-J\chi_h} \gg 1 \; 
\end{eqnarray}
at the antiferromagnetic wavevector. This means, all 
`external' effective coupling constants (e.g. $\bar g_1$,$\bar g_2$) as well as the
effective exchange coupling $\bar J$ 
are strongly enhanced at the antiferromagnetic wavevector.
The same is true for the incoherent part $(g_1\chi g_2)_{\rm inc}$, which
determines the {\it real part} of the susceptibility at the antiferromagnetic
wavevector at $\omega=0$.
For the calculation of $\chi_l$ 
the high-energy part of the quasiparticle dispersion is 
not essential, as any re-definition of the separation
procedure in high- and low-energy quantities goes along with a corresponding 
change in the interaction
constants exactly in a way that physical measurable 
quantities are not affected. 
Note that {\it low-energy} quantities are insensitive to the details
of this renormalization procedure.

As an example, when using tight-binding dispersion fits, 
the high-energy part of the dispersion
can be chosen arbitrarily without changing the low-energy physics. One has to bear in
mind, however, that the interaction constants for the quasiparticles 
do depend on the choice of the high-energy dispersion 
(bandwidth, high-energy cut-off etc.) as this determines their
`coherent quasiparticle weight'. Thus, only if it is 
specified if the {\it high-energy} 
quasiparticle renormalization factor is included in both the coupling constants
and high-energy dispersions or not, it makes sense to
compare coupling constants in different theories.

With this in mind, the formulas (\ref{chi}) and (\ref{chi0}) can be used with a dispersion
obtained from tight-binding fits to experimental data near the Fermi energy, and choosing any
reasonable behavior for high energies, assuming that all coupling and interaction constants are
renormalized quantities assigned to the particular choice of the high-energy part of the dispersion.

If one is interested in the behavior near the antiferromagnetic wavevector only,
a common procedure is
to expand the quantity $1-J\chi_h$ according to
\begin{eqnarray}
\label{chi_high}
1-J_{\bf q} \chi_h (\omega,{\bf q}) \approx \kappa^2 + \xi_0^2 ({\bf q}-{\bf Q})^2 -\frac{\xi_0^2 \omega^2}{v^2_{s}} \; ,
\end{eqnarray}
where $v_s$ is a spin-wave velocity, $\xi_0$ is an antiferromagnetic correlation length, and $\kappa^{-2} $ determines
the enhancement of the normal state static susceptibility at the antiferromagnetic wavevector.
The $\omega $ dependence is relevant only for the case that $\kappa $ becomes small. 
This introduces a new low-energy scale, and the separation between high energies and low energies is
not anymore clear cut. 
This case corresponds to strongly underdoped cuprate superconductors, where the pseudogap phenomenon
is dominating the physics. For overdoped, optimally doped and slightly underdoped 
materials the spin-wave term can be neglected.

The condition for a resonance in the superconducting state follows now from 
Eq.~(\ref{renorm1}) as
\begin{eqnarray}
J_{\bf Q}\mbox{Re} \chi_l (\omega,{\bf Q}) = \kappa^2 
\end{eqnarray}
which replaces the condition $J_{\bf Q}\mbox{Re} \chi_0 (\omega,{\bf Q})=1$.

A general low-energy theory for small deviations from the antiferromagnetic wavevector and
small energies would start from Eq.~(\ref{renorm}), using Eq.~(\ref{chi_high}) in
the denominator, and neglecting $\chi_l$ in comparison with $\chi_h$ in the numerator
(in the denominator $\chi_l $ cannot be neglected against $\chi_h$ because $1-J\chi_h$ is
comparable to $Jz_0^2\chi_l$ even though $Jz_0^2\chi_l \ll J\chi_h$).
The high-energy quantities $\chi_h$ and $\xi_0^2$ are renormalized by $\kappa^{-2}$ according to
$\xi=\xi_0/\kappa $, $\chi_Q=\chi_h/\kappa^2$.
It is common to introduce a self energy $\Pi $ via $Jz_0^2\chi_l /\kappa^2= \bar g^2 \chi_Q \Pi $, 
which enters the Ornstein-Zernicke form for the susceptibility in the following way,
\begin{eqnarray}
\label{chi_low}
\chi (\omega,{\bf q}) \approx \frac{\chi_Q}{1 + \xi^2 ({\bf q}-{\bf Q})^2 - \bar g^2\chi_Q \Pi(\vec{q},\omega )} \; .
\end{eqnarray}
The imaginary part of $\chi (\omega,{\bf q})$ 
is not a high-energy quantity, but enters even 
in the normal state only via the low-energy quantity $\Pi(\vec{q},\omega )$.

The use of the above procedures is a powerful way to discuss dispersion
features in the spin-spin response function \cite{Morr98,Abanov99,Chubukov01,Eremin05} and to describe the low-energy physics near the Fermi surface.
It is however semi-phenomenological in the sense that 
(in general momentum dependent) interaction parameters
and high-energy contributions to the real part of the spin-susceptibility
have to be obtained from either experiment or more general theories.

\subsection{Magnetic coherence in La$_{2-x}$Sr$_x$CuO$_4$}
\label{SpinLSCO}

Since the material La$_{2-x}$Sr$_x$CuO$_4$ shows a 
spin excitation spectrum different from the one discussed in the 
previous sections, we briefly summarize here the modifications 
of the theory that are necessary to describe the spin-response of 
this system.

The spectrum of the single layered compound La$_{2-x}$Sr$_x$CuO$_4$ 
has been scrutinized experimentally both in the normal and superconducting
state. The following important modifications in comparison with
the cuprates having optimal temperatures around 90 K exist:
(A) In the normal state the spectrum for $x>0.04$ is characterized by
peaks at incommensurate planar wavevectors
$\vec{Q}_\delta = (\pi (1\pm \delta )),\pi)$ and $(\pi,\pi(1\pm \delta))$ 
\cite{Shirane89,Mason96,Aeppli97,Yamada98}, where $\delta $ increases
with increasing doping.
(B) When entering the superconducting state $\chi'' (\omega,\vec{Q}_\delta )$
is reduced from its normal state value for $\omega < 7$ meV, and
increases for $\omega \gsim 7$ meV; $\chi'(\omega, \vec{Q}_\delta )$ decreases
in the superconducting state \cite{Lake99}.
(C) For $\omega \approx 7$ meV the incommensurate peaks sharpen in
the superconducting state \cite{Mason96,Lake99}.

These effects, termed ``magnetic coherence effects'' \cite{Mason96},
have been explained theoretically \cite{Morr00,Morr00a},
based on a modified form of Eq.~(\ref{chi_low}). The main difference between
the behavior in the spin susceptibility for La$_{2-x}$Sr$_x$CuO$_4$ 
and that for the systems YBa$_2$Cu$_3$O$_{6+x}$, 
Bi$_2$Sr$_2$CaCu$_2$O$_{8+\delta}$, and Tl$_2$Ba$_2$CuO$_{6+\delta }$
is, that for La$_{2-x}$Sr$_x$CuO$_4$ the momentum dependence of the
high-energy contributions to the spin susceptibility in the vicinity of
the antiferromagnetic wavevector cannot be neglected. Thus,
the detailled momentum dependence on the left hand side of Eq.~(\ref{chi_high})
is important, 
and its presence leads to the incommensurate spin response in
the normal state. 
This can be described by an expansion similar to Eq.~(\ref{chi_high}),
however around the incommensurate wavevectors $\vec{Q}_\delta $.
The third term in the modified Eq.~(\ref{chi_high}), though, can be
neglected for not too strongly underdoped compounds,
as experimentally one finds that the normal state low-energy spin excitation
spectrum is almost dispersionless in these cases.

In the normal state the real part of $\Pi $ in the denominator of
Eq.~(\ref{chi_low})
is proportional to the particle-hole asymmetry of the fermionic dispersion
around the chemical potential, which is small near the Fermi surface points
connected by the incommensurate magnetic wavevectors.
Thus, the {\it position} of the incommensurate peaks 
is determined to a good approximation by {\it high-energy fermions} only.
In contrast, the {\it momentum width} of 
these incommensurate peaks is determined by damping due to coupling to
{\it low-energy quasiparticles},
as it comes from the bosonic self-energy $\Pi $ in the denominator
of Eq.~(\ref{chi_low}). 

In the superconducting state the quantity $\Pi(\vec{q}, \omega )$ changes
in two ways:
it develops a non-negligible real part, and its imaginary part is
strongly modified at low $\omega $ due to the presence of the
anisotropic superconducting gap. The latter fact causes restrictions
in the damping of spin-excitations, that lead to a sharpening of
the incommensurate peaks.

The calculations of Refs. \cite{Morr00,Morr00a} along these lines 
give a good account of the experimental facts, and in particular 
reproduces (A) the frequency dependence of $\chi''$ at the incommensurate
wavevectors in the normal and superconducting states, (B) the momentum 
dependence of the spin gap in the superconducting state, (C) the 
changes of the momentum width of the incommensurate peaks when entering
the superconducting state, and (D) the slight dispersion towards the 
antiferromagnetic wavevector with increasing frequency in the 
superconducting state \cite{Morr00}.
It was argued by the same authors that the low-energy incommensurate
dispersion in YBa$_2$Cu$_3$O$_{6+x}$ can be explained in a similar way.

Finally, it is important to note that the magnetic {\it resonance mode}
as described in previous sections arises from the dependence of
$\Pi (\vec{q},\omega )$ on frequency $\omega $ in Eq.~(\ref{chi_low}).
From Eq.~(\ref{chi_high}) it follows that by adding the spin-wave term
(third term on the right hand side) to the high-energy contribution
of the spin susceptibility, one can expect a similar resonance 
effect. This was theoretically studied in Ref. \cite{Morr98}.
Note that these two descriptions are conceptually very different, as
in one case high-energy fermions are responsible for the formation
of the resonance mode, whereas in the other case it is the low-energy
quasiparticles that are responsible. For optimally and overdoped
materials one would, however, expect that one can safely neglect any
spin-wave term in the normal state response. For the underdoped
region the presence or absence of such a term is far from being settled
(note that the strong damping by low-energy quasiparticles in the normal
state makes the experimental observation of such a term non-trivial).

\section{Coupling of quasiparticles to the magnetic resonance mode}
\label{coupling}

The key question for any understanding of the importance of magnetic excitations
for high-$T_c$ superconductivity is how strongly do quasiparticles couple to spin fluctuations.
The observation of clear correlations between the magnetic spin-1 mode and
the self-energy effects in the single-particle spectrum opens the possibility to
determine {\it experimentally} the strength of this coupling. This is extremely
valuable for theoretical treatments, which up to now do not even agree on the
order of magnitude of the coupling constant\cite{Kee02,Abanov02a}. 
In order to achieve this goal, several experimental difficulties had to be
overcome, and it is only very recently that the precision of ARPES experiments
became sufficient to allow for the separation of
the effects of the magnetic resonance
from other effects like bilayer splitting and features due to
electron-phonon coupling \cite{Lanzara01}.
With regard to the latter point, a sharp, weakly dispersing phonon mode, 
if dominating in the phonon spectrum, 
would in a similar way allow for the determination
of the coupling constants between electronic quasiparticles and phonons. 
However here we concentrate on the more controverse coupling between  
electronic quasiparticles and magnetic excitations.

An assignment of the anomalous
ARPES lineshape near the $M$ point of the Brillouin zone
to the coupling between spin fluctuations and
electrons has been made in a number of theoretical treatments
\cite{Kampf90,Dahm96,Dahm96d,Manske01,Shen97,Norman97,Norman98,Abanov99,Eschrig00}.
In particular, the idea that a dispersionless collective 
mode is coupled strongly to electronic quasiparticles was suggested by 
Norman {\it et al.} \cite{Norman97,Norman98} and
by Shen and Schrieffer \cite{Shen97}.
By analogy with the  Holstein effect, the coupling of the magnetic resonance
to electronic quasiparticles should lead to spectral anomalies (``dips'')
in the fermionic spectral function, most prominent near
the $(\pi,0)$ points of the fermionic Brillouin zone \cite{Dessau91}, 
because these are connected by the characteristic wavevector of the resonance.
The separation between the quasiparticle peak and the dip should equal the
resonance energy \cite{Norman98,Abanov99}.
The EDC dispersion shows a break at roughly the same energy and the MDC
dispersion an $S$-shape anomaly. The universality of this energy scale
can be explained in a unified picture \cite{Eschrig00,Eschrig02,Eschrig03}.
At the same time, for not too strongly overdoped materials this coupling
leads to a kink in the quasiparticle dispersion
along the $(\pi,\pi)$ direction \cite{Bogdanov00,Kaminski01,Johnson01}, 
with the kink energy near $\Delta + \Omega_{res}$ \cite{Eschrig00}.
A kink-like feature is also present above $T_c$ in underdoped and optimally
doped materials, due to a peak in the spin-fluctuation spectrum.
However, it sharpens when entering the superconducting state.
For overdoped materials, the effect at the nodes due to the spin resonance
becomes weak and is overshadowed by additional effects, most probably
phonons, which add to the modification of the nodal dispersion.
Similarly, effects of the magnetic resonance
are expected for the density of states, as measured by tunneling
spectroscopy, and were found both in the SIN and SIS tunneling conductance 
\cite{Zasadzinski01}. Finally, also in
optical conductivity \cite{Basov96,Carbotte99,Hwang04} such
self-energy effects are present at 
$2\Delta + \Omega_{res}$ \cite{Abanov99,Abanov01a,Casek05,Abanov01c}
and at $\Delta + \Omega_{\rm res}$ \cite{Carbotte99,Casek05}.
The magnetic resonance mode
can also cause subgap peaks in SNS junctions \cite{Auerbach00}.

Experimentally, as it was discussed in section \ref{experiments},
the features that could be interpreted as due to 
scattering from the resonance have been observed in ARPES
spectra, SIN and SIS tunneling spectra, and
optical conductivity measurements on Bi$_2$Sr$_2$CaCu$_2$O$_{8-\delta }$
at various doping 
concentrations \cite{Abanov01b}. Furthermore, the resonance energies 
inferred from ARPES \cite{Campuzano99} and SIS \cite{Zasadzinski01} measurements as a
function of doping match $\Omega_{res}$ as measured directly by inelastic neutron
scattering.  
The mode extracted from SIS experiments \cite{Zasadzinski01} 
is located very near $2\Delta$ in overdoped materials,
but progressively deviates to lower energies with 
underdoping, as would be expected of a collective 
excitation inside a continuum gap \cite{Zasadzinski01}.
In addition, the real part of the fermionic self-energy at the node as a
function of temperature has been shown to scale with the resonance
intensity \cite{Johnson01}.  
In recent years, 
such self-energy effects have been also separated from bilayer splitting
effects and have been shown to be present independently \cite{Feng01a,Gromko04}.

The minimal set of
characteristic features for the collective mode which is involved follows from
the experimental results from ARPES and tunneling. The mode is characterized by
its energy and its intensity at the $(\pi,\pi)$ wavevector (the wavevector
being suggested by the momentum dependence of the strength of the ARPES
anomalies).
Its properties from ARPES and SIS tunneling are as follows.  The energy
should be weakly dependent on momentum,
roughly 40 meV in optimally doped
cuprates, follow $T_c$ with doping, and
be constant with increasing temperature up to $T_c$.
The intensity should be maximal at the $(\pi,\pi)$ wavevector, where it
should increase with underdoping and follow an order parameter-like
behavior as a function of temperature below $T_{c}$.
The mode should be absent in the normal state; a remnant can be present
in the pseudogap state, but an abrupt increase in intensity should occur
at $T_c$ with lowering temperature.
It is clear that these characteristics extracted from ARPES and tunneling
spectra fit perfectly to the magnetic-resonance mode.

The first question in order is, if the coupling between the
magnetic resonance mode and the electrons is sufficient to lead to the correct
order of magnitude of the self energy effects in the electronic dispersions
as observed in experiment.

\subsection{The coupling constant and the weight of the
spin resonance}
\label{CC}

The issue we will discuss in this section is whether the magnetic
resonance can account
for the measured changes in the fermionic properties of the
cuprates below $T_c$, via a feedback effect similar to the 
Holstein effect in phonon mediated superconductors. 
It was pointed out by Kee {\it et al.} \cite{Kee02} that this is not obvious, since 
the total experimental spectral weight of the resonance peak, 
\begin{eqnarray}
I_0 = \int S({\bf q},\Omega) \frac{d^2q d \Omega}{8\pi^3}, 
\end{eqnarray}
is only a few percent of the local moment sum rule, $S(S+1)/3 = 1/4$. 
Here, we set $\mu_B =1$ and for $S=1/2$  define the spin structure factor as  
\begin{eqnarray}
S({\bf q}, \Omega) &=& \int dt \int d^2 r e^{i(\Omega t - 
{\bf q}{\bf r})} <S_x ({\bf r},t) S_x ({\bf 0},0)>  \nonumber \\
&=& \frac{0.5\hbar}{1 - e^{-\hbar \Omega/k_BT}} \chi^{\prime \prime} ({\bf q}, \Omega). 
\end{eqnarray}
This local moment sum rule is, however, only valid in in the Heisenberg limit,  and
the total integrated weight of the structure factor
should be reduced in the metallic regime.

Motivated by the raised doubts of Ref.~\cite{Kee02}, Abanov {\it et al.}
\cite{Abanov02a} addressed the issue whether the smallness of the
integrated intensity of the peak precludes strong effects on the fermions. Their
main result is that the fermionic self-energy due to scattering 
from the resonance is strong and unrelated to the small 
integrated intensity of the peak. 

The question can be divided into two key points: first, what the order of magnitude
for the values of the spin-fermion coupling $g$ 
and the dimensionless coupling constant $\lambda$
is, and second,
the dependence of the self-energy on the integrated intensity of the peak. 

The second question was dealt with already in Refs.~\cite{Abanov99,Eschrig00,Abanov02a,Eschrig02,Eschrig03},
where detailed numerical calculations have shown that
under the assumption that the coupling constant is of the order of several
hundred meV, the small but finite weight is enough to produce the observed
self energy effects in the fermionic dispersions. 
There it was pointed out
that the effects are enhanced by a factor equal to the volume of the magnetic
Brillouin zone divided by the volume which is populated by the magnetic resonance.
The resonance only exists in a restricted momentum range, which constitutes only
about 6\% of the area of the Brillouin zone. 
This enhancement factor counter-acts the small integrated intensity of the
resonance, leading to large self energy effects in the fermionic dispersion, as
discussed in the following sections.
In other words, 
in the regions where self-energy effects are strong, the entire weight of the
magnetic resonance contributes. Note in this regard that experimentally
$\int d \Omega S({\bf Q}, \Omega) \sim 1.5$ is indeed not small
\cite{Rossat91,Fong00,Dai01}. 

\begin{figure}
\centerline{
\epsfxsize=0.5\textwidth{\epsfbox{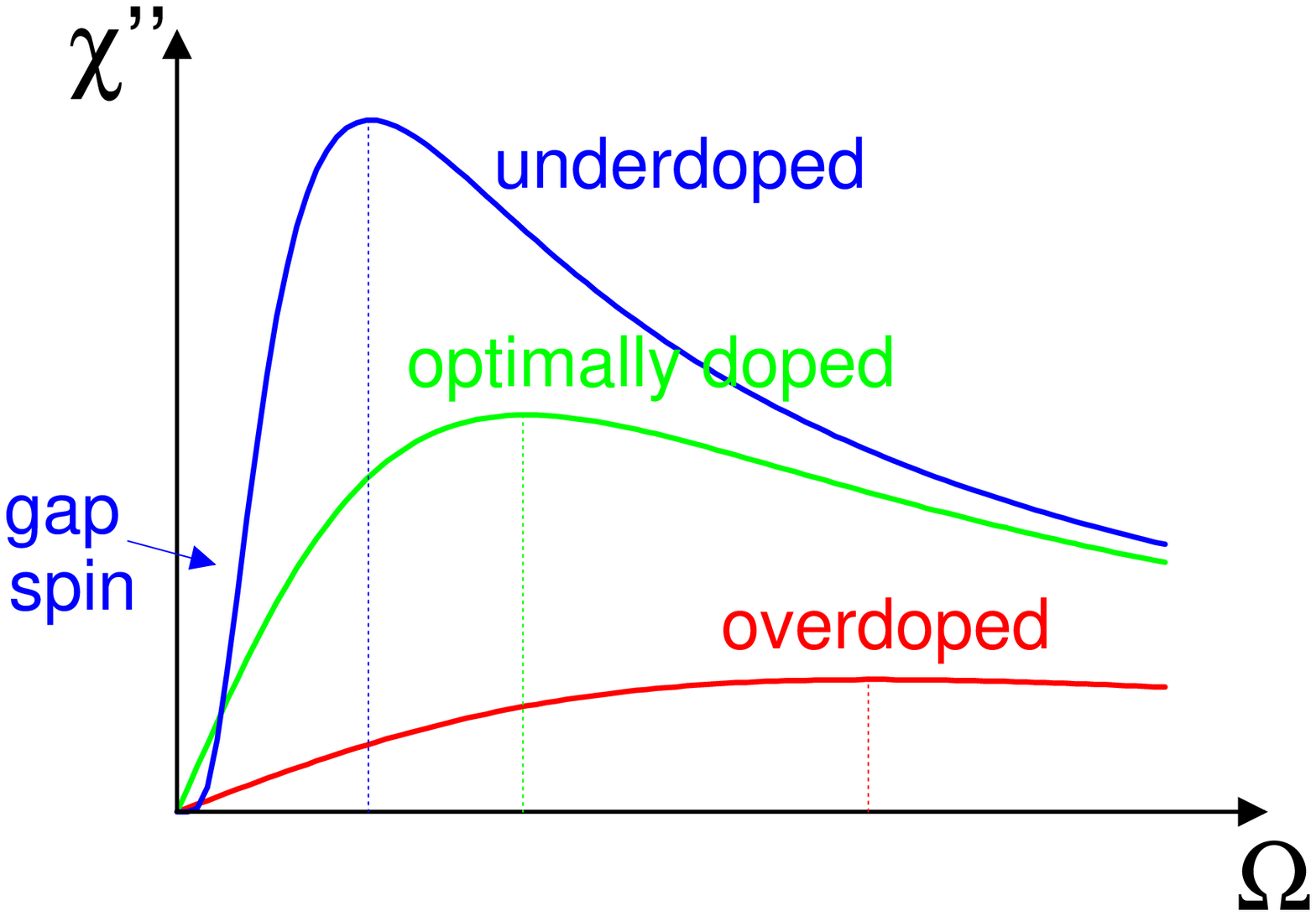}}
\epsfxsize=0.35\textwidth{\epsfbox{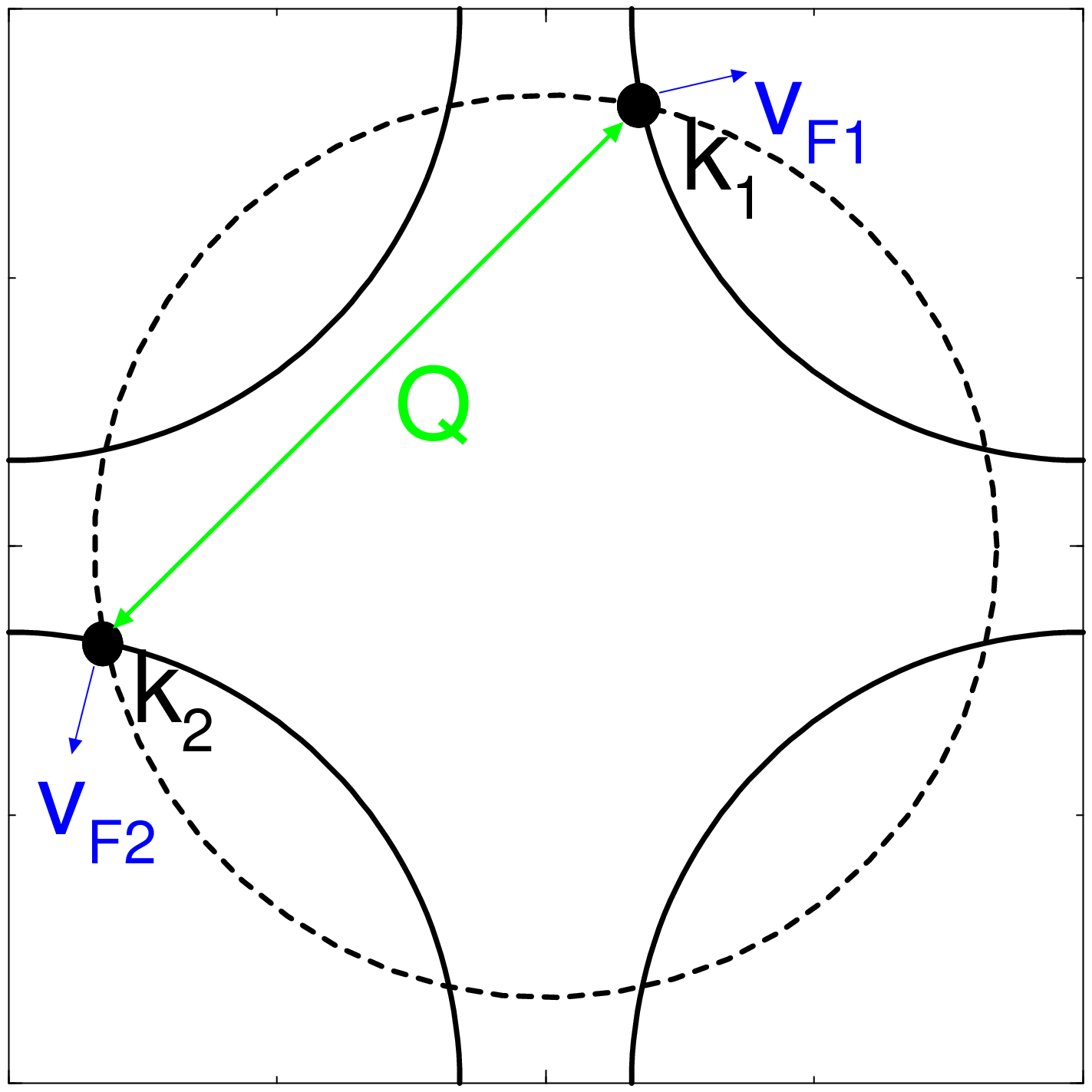}}
}
\caption{
\label{chi_normal}
Left: Schematic behavior of the normal state spin susceptibility in cuprates. Normal state
spin excitations in cuprates are overdamped. In underdoped cuprates a spin-gap effect
additionally comes into play.
Right: The hot spots are connected by a wavevector $\vec{Q}=(\pi,\pi)$. The fermionic
dispersions near the hot spots is determined by the corresponding Fermi velocities $\vec{v}_{F1}$
and $\vec{v}_{F2}$.
}
\end{figure}

The main issue, then, is the first question, namely that of the spin-fermion coupling, $g$.
We show in the following that it is indeed of the order of several hundred
meV.
The most straightforward way to extract 
$g$ is to fit the position of the maximum of the 
normal-state spin susceptibility 
$\chi^{\prime \prime} ({\bf Q}, \Omega)$ 
to experimental data.
Experimentally, this maximum is located 
at $20-25 {\rm meV}$ in optimally doped YBa$_2$Cu$_3$O$_{7-\delta}$ 
\cite{Rossat91,Fong00,Dai01}.
The data are consistent with a relaxational form for the susceptibility, as shown on the left in
Fig. \ref{chi_normal}, which is given by
\begin{equation}
\label{chi_n}
\chi ({\bf Q}, \Omega) = \frac{1}{\chi_Q^{-1} - i \Gamma (\Omega) } = \frac{\chi_Q}{1-i\frac{\Omega }{\Omega_{max}}} ,
\end{equation}
and whose imaginary part has a maximum given by 
$\Gamma (\Omega_{max}) =  \chi_Q^{-1}$.
Writing the susceptibility near the wavevector $Q=(\pi,\pi)$ as in 
Eq.~(\ref{chi_low}),
\begin{equation}
\label{chi_Pi}
\chi (\vec{q},\Omega )= \frac{ \chi_Q}{1+\xi^2(\vec{q}-\vec{Q})^2 - \bar{g}^2\chi_Q \Pi(\vec{q}, \Omega )}
\end{equation}
the above form is obtained by evaluating the fermionic bubble. 
$\Gamma(\Omega)$ is then the imaginary part of the fermionic bubble times
$\bar g^2$.
The coupling constant $\bar g$ is related to $g$ by $3\bar g^2 = g^2$.
It is most easy to calculate the fermionic bubble $\Pi(\vec{q},\Omega )$ at the hot spots, defined
as the points on the Fermi surface connected by $\vec{Q}$.
. The corresponding geometry is shown
in Fig. \ref{chi_normal} on the right, where the Fermi surface shifted by $\vec{Q}$
is shown as dashed line on top of the unshifted Fermi surface.
Linearizing the dispersion around the hot spots
and summing over all 8 hot spots results in \cite{Abanov99,Abanov01d,Abanov03}
\begin{equation}
\Gamma (\Omega)  = \frac{4 \bar g^2 \Omega }{\pi v_F^2 \sin 2\alpha } \; ,
\label{damping}
\end{equation} 
or, equivalently,
\begin{equation}
\Omega_{max} = \frac{\pi v_F^2 \sin 2\alpha }{4 \bar g^2 \chi_Q} \; ,
\label{wmax}
\end{equation} 
where $\alpha $ is the angle between the Fermi velocity at the hot spot and the wavevector $\vec{Q}$,
and $v_F$ is the magnitude of the Fermi velocity at the hot spots.
Near optimal doping in cuprates $\alpha \approx \pi/4$ so that $\sin 2\alpha \approx 1$.
We obtain consequently the coupling constant via the formula
\begin{equation}
\bar g= \sqrt{\frac{\pi v_F^2  \sin 2\alpha }{4 \chi_Q \Omega_{max} }}
\end{equation} 
The parameters $v_F$, $\alpha $, $\Omega_{max}$ are all known from the normal state
data from INS and ARPES. The value for $\chi_Q$ is not so well known for the normal
state because the magnetic INS signal is very small. Thus, we follow 
Ref. \cite{Abanov02a}
and extract the value from the superconducting state data.

\begin{figure}
\centerline{
\epsfxsize=0.4\textwidth{\epsfbox{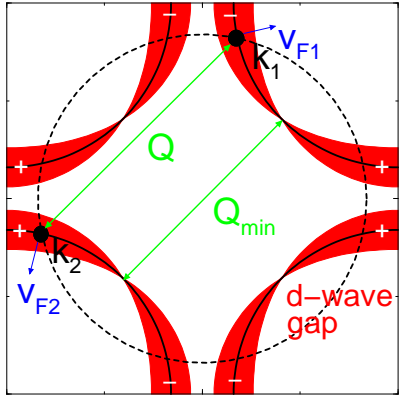}}
}
\caption{
\label{chi_sc}
Fermi surface and scattering geometry in the superconducting state.
The hot spots are connected by a wavevector $\vec{Q}=(\pi,\pi)$. The fermionic
dispersions near the hot spots is determined by the corresponding Fermi velocities $\vec{v}_{F1}$
and $\vec{v}_{F2}$. The $d$-wave gap is indicated. The low-energy incommensurate wavevector $\vec{Q}_{min}$
is determined by node-node scattering.
}
\end{figure}

In the superconducting state we can obtain a semi-phenomenological description of
the resonance by using the same form Eq.~(\ref{chi_Pi}) for the spin susceptibility
in the superconducting state, with the superconducting state fermionic bubble $\Pi (\vec{q},\Omega )$ now
replacing the normal state one.
Using the same approximation as above, as indicated in Fig. \ref{chi_sc}, one obtains
a range in which the imaginary part of $\Pi $ is zero and the real part is given for
small $\Omega $ in an expansion up to order $\Omega^2$ by, 
\begin{equation}
\mbox{Re} \Pi (\vec{Q},\Omega )= \frac{\Omega^2 }{2 v_F^2 \Delta_Q } + \cdots
\end{equation} 
where $\Delta_Q$ is the gap magnitude on the hot-spot positions, and we assumed $\alpha=\pi/4$, which
is a good approximations near the hot spots (however not near the nodes).
Introducing this into the susceptibility, one obtains
in this range $\Omega < 2\Delta_Q$  the form
\begin{equation}
\chi ({\bf Q}, \Omega) \approx \frac{\chi_Q}{1 - (\Omega/\Omega_{res})^2}, 
\end{equation} 
where in this approximation 
\begin{equation}
\Omega_{res}=\sqrt{\frac{2v_F^2\Delta_Q }{\bar g^2 \chi_Q}}=\sqrt{\frac{8\Omega_{max}\Delta_Q}{\pi}} \; .
\end{equation} 
In order to obtain the coupling  constant separately from $\chi_Q$ we need the
energy integrated weight of the resonance.
The resonance contribution to the susceptibility is
\begin{equation}
\chi_{res}=\chi_Q \frac{\Omega_{res}^2}{\Omega_{res}^2-(\Omega+i0^{+} )^2} \; ,
\end{equation} 
and the weight of the resonance is then,
\begin{equation}
W_{res}=\int d \Omega \chi_{res}^{\prime \prime} ({\bf Q}, \Omega) = \frac{\pi}{2} \chi_Q \Omega_{res} 
\end{equation} 
or solving for $\chi_Q$,
\begin{equation}
\chi_Q = \frac{2W_{res}}{\pi \Omega_{res}} \; .
\end{equation} 
The weight $W_{res}$ amounts to about 1.6
\cite{Rossat91,Fong00,Dai01}. Knowing $\Omega_{res}$ very accurately from INS experiments,
this allows to determine $\chi_Q$ as
$\chi_Q \approx $13 eV$^{-1}/$plane.
Using this  experimental $\chi_{Q} \sim 13 $ eV$^{-1}$,
$\Omega_{max}=20 {\rm meV}$,
and $v_F \sim 0.4 {\rm eV}$ (in units where the lattice constant is 1) 
it follows that $\bar g \sim 1.75 v_F \sim 0.7 {\rm eV}$ \cite{Abanov02a}.

The dimensionless coupling $\lambda$ can be extracted from 
the low-energy fermionic self-energy:
$Re\Sigma (\epsilon ) = -\lambda \epsilon $. 
$\Sigma ({\bf k},\epsilon )$ is determined as 
3$\bar g^2$ times a convolution of $\chi({\bf q},\Omega)$ with 
$G_0({\bf k}+{\bf q},\epsilon +\Omega)$.
Again, linearizing the fermionic dispersion about the hot
spots, and expanding $\chi$ quadratically around ${\bf Q}$
with a correlation length, $\xi$, it follows for the normal state
that \cite{Abanov03}  
\begin{equation}
\lambda = \frac{3 \bar g^2 \chi_{Q}}{4 \pi v_F \xi} = \frac{3 v_F}{16 \Omega_{max} \xi}
\end{equation}
Substituting the above numbers and $\xi \sim 2$, we find $\lambda \sim 2$.
Note that $\lambda$ refers here to fermions near the hot spots, 
and is obtained by coupling to the entire spin fluctuation spectrum.
In the superconducting state, 
the part of $\lambda $ which is due to purely the spin resonance has
been numerically obtained in Ref.~\cite{Eschrig03}, which was found to
be $\sim 0.9$ at the $M$ point if the fermionic Brillouin zone for
optimally doped materials. They also gave an approximate analytical formula
for this weight,
\begin{eqnarray}
\lambda_{M}  &\approx &
\frac{g^2I_0}{\pi} \cdot 
\frac{1}{(\Omega_{res}+E_M)^2}  
+\frac{g^2w_{MN}}{\pi v_Nv_{\Delta}}
\frac{2}{\pi}\ln\left(1+\frac{\Delta_A}{\Omega_{res}}\right),
\end{eqnarray}
where $E_M=\sqrt{\xi_M^2+\Delta_M^2}$ is the quasiparticle 
binding energy at the $M$ point of the Brillouin zone, $w_{MN}$ the
resonance intensity at the wavevector $\vec{q}_{MN}=\vec{k}_M-\vec{k}_N$,
$v_N=\partial_k\xi_k|_{k_N}$, $v_\Delta=\partial_k\Delta_k|_{k_N}$, and $\Delta_A$ is
the superconducting gap at the antinodal Fermi surface point.
For optimally doped 
Bi$_2$Sr$_2$CaCu$_2$O$_{8-\delta }$ they found
the first term in this sum equal to $17.44 I_0$ with $I_0=0.035$, which amounts to
about 0.61, and the second term $\approx 0.21$, leading to a total
of $\lambda_M \approx 0.82$.
The spin-fluctuation continuum, not contained in this estimate,
contributes the remaining part to the 
total coupling constant $\lambda \sim 2$ \cite{Eschrig03}.

The such obtained value of $g$ is consistent with
fitting resistivity data to spin fluctuation 
scattering \cite{Monthoux94a,Zha96} and with Eliashberg calculations of $\Delta$ and
$\Omega_{res}$ \cite{Abanov01b}.  Such a large value of $g$ is also expected on 
microscopic grounds:  in the Hubbard model, the effective $g$
is expected to be of the order of the fermionic bandwidth $W$ \cite{Scalapino95,Vilk97} 
which is $1 {\rm eV}$ for the cuprates \cite{Abanov02a}.
The estimate  $\lambda \sim 2$ is consistent with the
velocity renormalization estimated from normal state ARPES 
experiments \cite{Olson90}, and with the bare density of states
extracted from the specific-heat data of 
Ref.~\cite{Loram97}. 
It was shown by Abanov {\it et al.} \cite{Abanov03} that spin fluctuations are slow
compared to fermions, and for that reason an ``effective'' Migdal theorem exists
for fermions near the hot spots, 
which justifies the use of the above perturbation theory.

The above picture of the spin resonance and its effect on fermions 
has been challenged by a number of authors.  
Kee {et al.} \cite{Kee02} argued that $g \sim 14 {\rm meV}$ and $\lambda \sim 10^{-3}$,
two and three orders of magnitude smaller than the values above, respectively.
However, the discrepancy can be attributed to different definitions of the
effective coupling constants in the different theories.
In the low-doping regime, where the
Fermi surface evolves toward small hole pockets, fermions couple 
to antiferromagnetic magnons \cite{Abanov02a}.
In this case, the spin mode couples to fermions only through gradients,
i.e., the renormalized coupling $g_{eff}$ is much smaller than $g$. 
This reduction from $g$ to $g_{eff}$ is the result
of strong vertex corrections if antiferromagnetic magnons are present 
in the normal state \cite{Shraiman90,Schrieffer95,Sachdev95b}, and occurs because antiferromagnetic magnons
are only compatible with a small Fermi surface (hole pockets), in which 
case $g$ has been absorbed into the definition of renormalized   
fermions with a spin density wave energy gap \cite{Kampf90,Chubukov97}.
Here we are dealing, however, with the metallic phase where a large
Fermi surface exists, and where 
the normal state spin dynamics is purely relaxational.
In this case, $g$ is the appropriate coupling to use, not $g_{eff}$.

To summarize, the 
large intensity of the resonance at ${\bf Q}= (\pi,\pi)$ is
consistent with the small value of the total momentum and frequency
integrated intensity of the resonance peak, and also with the fact that the
magnetic part of the condensation energy is only a small fraction of $J$ \cite{Abanov02a}.
The spin-fermion coupling $g$ is of the order of $1 {\rm eV}$
and this value of $g$ is consistent with experiment.
This $g$ is sufficiently large that scattering from the resonance can
substantially affect the electronic properties of the cuprates below $T_c$.

\subsection{Theoretical model}
\label{Theory}

As we are interested in the renormalization of the fermionic dispersion
as a result of the coupling of electrons to the sharp magnetic-fluctuation
mode at low energies, we search for a minimal model which captures the
low-energy physics correctly. Such a model was suggested
in Refs. \cite{Eschrig02,Eschrig03}. 
The idea there is to concentrate on the superconducting state assuming that
superconducting order is already
established without coupling to this resonant feature in the
spin-fluctuation spectrum. Thus, the superconducting
state is described by an independent order parameter $\Delta_k$. 
The origin of the superconducting pairing instability can be for example
the spin fluctuations continuum, which extends to high energies,
or other sources. The pairing problem is assumed to be of BCS type.
The spin fluctuation resonance supports pairing, but does not cause
superconductivity in and of itself \cite{Eschrig00,Eschrig03,Abanov02a}.
Although semi-phenomenological, this approach has the advantage that 
the conclusions drawn are independent on the specific microscopic
pairing mechanism.

The order parameter
is chosen to have $d$-wave symmetry 
(here and in the following the
unit of length is the lattice constant $a$),
\begin{equation}
\label{DWOP}
\Delta_k=\frac{\Delta_M}{2} (\cos k_x-\cos k_y).
\end{equation}
In agreement with the experiments of \cite{Borisenko02} it is assumed that
in the case of bilayer materials the pairing interaction has no interplane contribution.
The magnitude $\Delta_M$ is determined by the peak position 
of the ARPES spectrum
at the $M$ point after including self-energy effects due to coupling to the
spin fluctuations. The coupling to the resonant spin-fluctuation mode leads in general
to a renormalization of the gap, and it is this renormalized gap which is observed
in experiments.

The model can be formulated in terms of retarded Green functions, $G^R_{\epsilon,k}$, 
for fermionic excitations in the superconducting state. These
Green functions are functionals of the 
normal state electronic dispersion $\xi_k$, the order parameter $\Delta_k$, and
the self energies due to coupling to spin fluctuations, 
$\Sigma^R_{\epsilon,k},\Phi^R_{\epsilon,k}$. 
The `normal state' refers here to 
the state at the same temperature, but with zero order parameter.

\subsubsection{Tight binding fit to normal state dispersion}
\label{TBF}
In order to model the normal-state properties as realistic as possible 
the electronic dispersion $\xi_k$ is obtained from 
ARPES experiments in the normal state, and parameterized by using
a six parameter tight binding fit of the form
\begin{eqnarray}
\xi_k&=&t_0+t_1\frac{\cos k_x+\cos k_y}{2}+t_2\cos k_x\cos k_y  
+t_3\frac{\cos 2k_x + \cos 2k_y}{2}
\nonumber \\ &&
+t_4\frac{\cos 2k_x \cos k_y+ \cos k_x \cos 2k_y}{2} 
+t_5 \cos 2k_x \cos 2k_y \; .
\end{eqnarray}
The six parameters $t_0-t_5$ are conveniently determined by using 
characteristic 
features in the experimentally measured electronic normal-state dispersion.
\begin{figure}
\centerline{
\epsfxsize=0.6\textwidth{\epsfbox{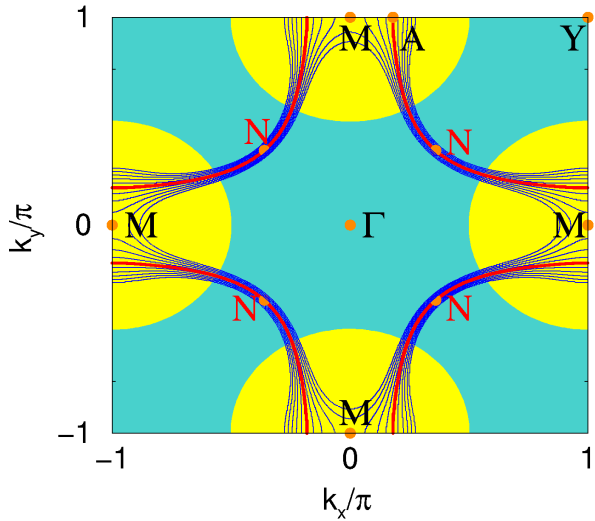}}
}
\caption{
\label{brill}
Equal energy contours around the Fermi surface shown 
as thin curves for energies between $\pm 50 $meV. 
The Fermi surface is shown as a thick curve.  
The dispersion used here
was obtained by a 6-parameter tight binding fit to angle resolved photoemission 
dispersions in optimally doped 
Bi$_2$Sr$_2$CaCu$_2$O$_{8-\delta }$ \cite{Eschrig00}.  The dispersion has a
saddle point at the $M$ point. The $N$ point corresponds to the
node of the $d$-wave order parameter in the superconducting state.
Two characteristic regions can be defined: the regions around the $M$ points
(light) as opposed to the remaining region which includes the $N$ points.
}
\end{figure}

The six features employed in Ref. \cite{Eschrig03} are
the positions of the $N$ (node) and $A$ (antinode) points in Fig.~\ref{brill}, 
parameterized by $k_{\Gamma N}=|\vec{k}_N-\vec{k}_{\Gamma }|$ 
and $k_{MA}=|\vec{k}_A-\vec{k}_M|$, the
band energies at the $M$ and $Y$ points, $\xi_M$ and $\xi_Y$, the
Fermi velocity at the $N$ point, $v_N=|\vec{v}_N|$, and the
inverse effective mass along direction 
$M-\Gamma$ at the $M$ point, $m_{M}^{-1}$. 
Table \ref{tab1} shows as example the parameters appropriate for
optimally doped 
Bi${_2}$Sr${_2}$CaCu${_2}$O${_{8+\delta}}$ (OP Bi2212) \cite{Eschrig03}.
The Fermi surface for such a tight binding fit
and corresponding equal-energy contours in the range between $\pm $50 meV
are shown in Fig. \ref{brill}. 
In this Figure we indicate a separation of the Brillouin zone into two
types of regions. The areas around the $M$ points are dominated by the
flat dispersion near the saddle point, which introduces a new energy scale
given by the distance of the saddle-point singularity from the chemical potential.
In the remaining region, which includes the nodes of the order parameter, the
dispersion is steep in the direction perpendicular to the Fermi surface, 
and excitations are restricted to the close vicinity of the Fermi surface.

The parameter $\xi_Y$ in the tight binding dispersions
is in principle not known from experiment. However, as the high-energy
dispersion is not important for the low-energy physics 
(except for an energy independent
contribution to the quasiparticle renormalization factor), 
it can be set to a reasonable value to preserve the overall dispersion shape 
as observed from experiment.
On a formal level, it is always possible to introduce a
high-energy contribution to the quasiparticle renormalization factor in such a 
way that it accounts for the true high-energy dispersion.

The inverse mass at the $M$ point is known to be negative and
small in the $M-\Gamma$ direction, and it was suggested that it could be
zero, giving rise to an extended van Hove singularity \cite{Gofron94,Abrikosov93}.
The inverse effective
mass decreases when coupling to the spin fluctuation mode is taken
into account, and it is this renormalized inverse mass which is
experimentally observed. 

Similarly, the value of the Fermi velocity at the node is renormalized by self-energy
effects to moderately smaller values. Again, it is these renormalized value of
the Fermi  velocity which is observed in experiment.
The parameters most sensitive to doping variations are $\xi_M$
and $\vec{k}_A$.

For the case of a bilayer splitting, we parameterize the Fermi surface by the
same six parameters plus the bilayer splitting between
bonding ($b$) and antibonding ($a$) band. If the bonding and antibonding
dispersions are written as $\xi_{k}^{(b)} $ and $\xi_{k}^{(a)} $, 
the resulting energy splitting is anisotropic \cite{Chakravarty93,Feng02}:
\begin{equation}
\label{BilSpl}
\xi_{k}^{(a)}-\xi_{k}^{(b)}=
\frac{1}{2}t_{\perp } (\cos k_x  - \cos k_y )^2.
\end{equation}
The normal state dispersion for both bands is then determined by a total
of seven parameters. An example, appropriate for overdoped
Bi${_2}$Sr${_2}$CaCu${_2}$O${_{8+\delta}}$ (OD Bi2212),
and used in Ref.~\cite{Eschrig02}, is
presented in Table~\ref{tab1}. 
\begin{table}
\begin{center}
\tbl{
\label{tab1} 
Parameters for the effective dispersion $\xi_k$ (\cite{Eschrig03,Eschrig02}).
BB and AB in the second line refers to bonding band and antibonding band.
OP and OD Bi2212 stand for optimally doped and overdoped
Bi${_2}$Sr${_2}$CaCu${_2}$O${_{8+\delta}}$.
}
{\begin{tabular}{|c|c|c|c|c|c|c|c|}
\hline
\hline
&&
$k_{\Gamma N}a$ & $k_{MA}a$ & $\xi_M$ & $\xi_Y$ & $\hbar v_N/a$ & 
$\hbar^2/m_{M}a^2$  \\
\hline
OP Bi2212 \cite{Eschrig03}& $-$&
0.36$\sqrt{2}\pi$ & 0.18$\pi$ & $-34$ meV & 0.8 eV & 0.6 eV & $-0.2376$ eV
\\
\hline
OD Bi2212 \cite{Eschrig02}& BB&
0.37$\sqrt{2}\pi$ & 0.217$\pi$ & $-105$ meV & 0.8 eV & 0.6 eV & $-$ 
\\
&AB&
 & 0.135$\pi$ & $-18$ meV & & & $-$ 
\\
\hline
\hline
\end{tabular}}
\end{center}
\end{table}

As mentioned above, the normal state dispersion $\xi_k$, 
and also the order parameter $\Delta_k$, are phenomenological
quantities, which are already renormalized by other effects which we do 
not need to specify, but which are assumed to influence the physics only
on an energy scale large compared to the scale of interest
(50-100 meV). The self energies due to spin fluctuations will have 
a part due to the particle-hole continuum, and another part due to the
resonance. 
In general it is necessary to include both parts of the spectrum.
However, for the low-energy region below 100 meV it is possible to
study a simplified model, in which
the effect of the continuum part of the
spin fluctuation spectrum is included by a constant renormalization of the normal
state dispersion and the order parameter \cite{Eschrig03}. 
In this case the main physics is dominated by
the coupling of the electrons to the resonant spin fluctuations.
The `normal state' reference is here
defined as the state with zero order parameter, interacting with 
a spin fluctuation
spectrum having no resonance part and a continuum part identical to
that in the superconducting state. 
This is different from the physical normal state, because
the spin fluctuation continuum changes when going
from the normal to the superconducting state, leading to an additional
renormalization of the dispersion. 

At higher energies, the spin fluctuation continuum can be excited, and
this leads to an additional strong fermionic damping.
The appropriate model to study this effect is
the extended model which explicitly includes
the gapped spin fluctuation continuum.
For this extended model, the `normal state' dispersion
has a different renormalization factor as compared to the simple
model above. The dispersions in Table~\ref{tab1}
are the appropriate dispersions for the simplified model. For the
extended model it is to be
scaled and shifted back in energy (so that the energy of the
van Hove singularity closest to the $M$ point stays at its original value)
as described below in more detail \cite{Eschrig02,Eschrig03}.

\subsubsection{Model spectrum and basic equations}
\label{BEQ}
\begin{figure}
\centerline{
\epsfxsize=0.44\textwidth{\epsfbox{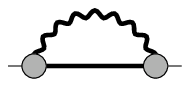}}
\epsfxsize=0.44\textwidth{\epsfbox{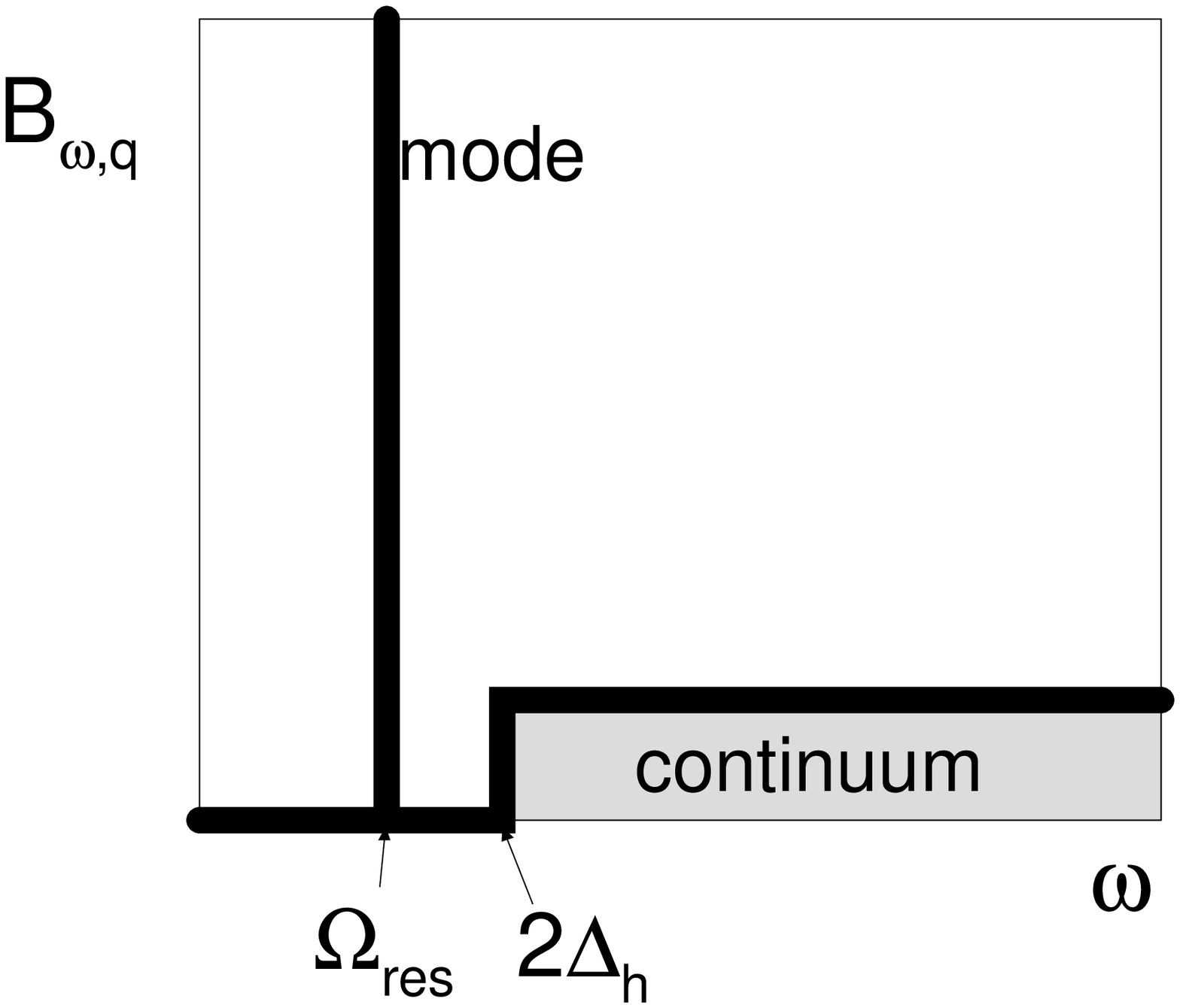}}
}
\caption{
\label{diagram}
Left: Self energy for electrons (full lines). 
The wavy line denotes a spin fluctuation.
Right: the model spin fluctuation spectrum we used for the wavy line
in the Feynman diagram.
The mode affects the low energy fermionic properties.
The continuum part only couples to electrons with higher energies, and
is neglected in the simple form of the model \cite{Eschrig00}.
Damping of electrons at energies above 100 meV is caused by the
continuum part, and is included in the extended model \cite{Eschrig03}.
(From Ref. \cite{Eschrig03},
Copyright \copyright 2003 APS).
}
\end{figure}
It is found that all essential features of the self-energy effects
in the superconducting state are obtained using
a minimal model with a spin fluctuation spectrum shown in
Fig.~\ref{diagram} \cite{Eschrig03}.
The self energy is given by the diagram in Fig.~\ref{diagram} on the left,
and its retarded part is written within standard Keldysh technique as,
\begin{equation}
\label{keld}
\Sigma^R_{\epsilon,k}=
\frac{i}{2} \sum_{q,\omega } \left( G^R_{\epsilon-\omega,k-q} g^2 D^K_{\omega,q}
+G^K_{\epsilon-\omega,k-q} g^2 D^R_{\omega,q} \right),
\end{equation}
where $D=-\chi$ is the bosonic propagator, $G$ the fermionic propagator,
and $g$ the coupling constant between the two.
Analogously, in the superconducting state 
the anomalous (off-diagonal) self energy is obtained from
\begin{equation}
\label{keld_od}
\Phi^R_{\epsilon,k}=
\frac{i}{2} \sum_{q,\omega } \left( F^R_{\epsilon-\omega,k-q} g^2 D^K_{\omega,q}
+F^K_{\epsilon-\omega,k-q} g^2 D^R_{\omega,q} \right),
\end{equation}
with the anomalous fermionic propagator $F$.
In equilibrium, the Keldysh components are given by the expressions
\begin{eqnarray}
D^K_{\omega,q}&=&\left(D^R_{\omega,q}-D^A_{\omega,q}\right)\coth 
\frac{\omega}{2T} 
=-iB_{\omega,q}(1+2b_{\omega}),\quad \\
G^K_{\epsilon,k}&=&\left(G^R_{\epsilon,k}-G^A_{\epsilon,k}\right) 
\tanh\frac{\epsilon}{2T}
=-iA_{\epsilon,k}(1-2f_{\epsilon}), \quad \\
F^K_{\epsilon,k}&=&\left(F^R_{\epsilon,k}-F^A_{\epsilon,k}\right) 
\tanh\frac{\epsilon}{2T}
=-iC_{\epsilon,k}(1-2f_{\epsilon}), \quad 
\end{eqnarray}
where $B_{\omega,q}=-2$Im$D^R_{\omega,q}$ and $A_{\epsilon,k}=
-2$Im$G^R_{\epsilon,k}$ are the bosonic and fermionic 
spectral functions, 
and $b_{\omega }$, $f_{\epsilon }$ their corresponding distribution functions,
respectively.
Note, that the Keldysh components $G^K$ and $D^K$ are purely imaginary.
For the anomalous propagator, the function 
$C_{\epsilon,k}=i(F^R_{\epsilon,k}-F^R_{-\epsilon,-k})=-C_{-\epsilon,-k}$ is only real in the case of a real 
gauge (real order parameter). 

The spectral function for the spin-fluctuation spectrum in Fig.~\ref{diagram}
is the sum between
the resonance part and the continuum part,
\begin{equation}
\label{Btot}
g^2B_{\omega,q} = g^2B^r_{\omega,q} + g^2B^c_{\omega,q}.
\end{equation}
with a resonance part sharp in energy,
\begin{equation}
\label{Bres0}
g^2B^r_{\omega,q} = 2 g^2w_q \left[ \delta (\omega- \Omega_{res} ) -
\delta (\omega + \Omega_{res}) \right]
\end{equation}
and a continuum with onset at $2\Delta_h$,
\begin{equation}
\label{Bcont0}
g^2B^c_{\omega,q} = 2 g^2c_q \left[ \Theta (\omega- 2\Delta_h ) -
\Theta (-\omega - 2\Delta_h ) \right] \; .
\end{equation}
This form for the gapped continuum is similar to the gapped marginal
Fermi liquid spectrum considered in Refs. \cite{Littlewood92,Norman98}.

In order for the real part of the self energy to converge,
the continuum has to be cut-off at high energies. 
The precise form of this high-energy cut-off is irrelevant for the
low-energy part of the fermionic Green function, and variation of the
cut-off leads to only a weakly energy-dependent contribution to the 
renormalization factor 
which can be absorbed in the dispersion $\xi_k$ as described above.
Thus, any change in the high-energy behavior of the 
continuum spectrum is accounted for by a redefinition of the quasiparticle
renormalization factor. 
The energy width of the resonance can be assumed in very good approximation
as zero, as was shown in Ref. \cite{Eschrig03}.

We assume in Eq.~(\ref{Bres0}) that the resonance energy is independent
on momentum. Strictly speaking, this is not true, as the resonance is part of
a dispersive incommensurate response as shown in Fig.~\ref{Fig5_Keimer04}.
However, as can be seen from Fig.~\ref{Fig3_Chubukov01}, in particular on
the left, which is for Bi${_2}$Sr${_2}$CaCu${_2}$O${_{8+\delta}}$,
the weight of the incommensurate response is very quickly reduced
when moving away from the resonance frequency $\Omega_{res}$.
Thus, the approximation of a dispersionless mode is a very good one.

\begin{figure}
\centerline{
\epsfxsize=1.7in{\epsfbox{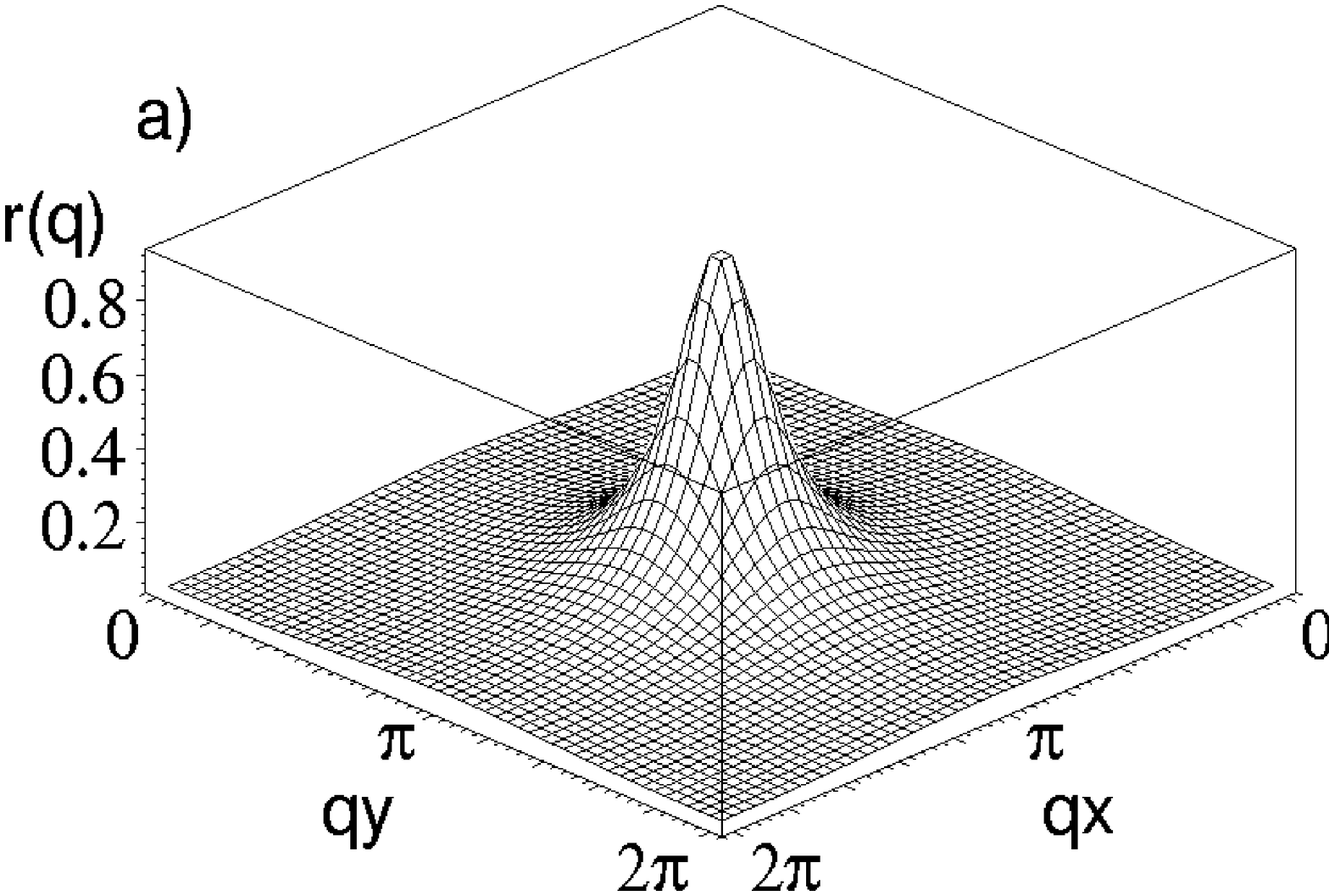}}
\epsfxsize=1.7in{\epsfbox{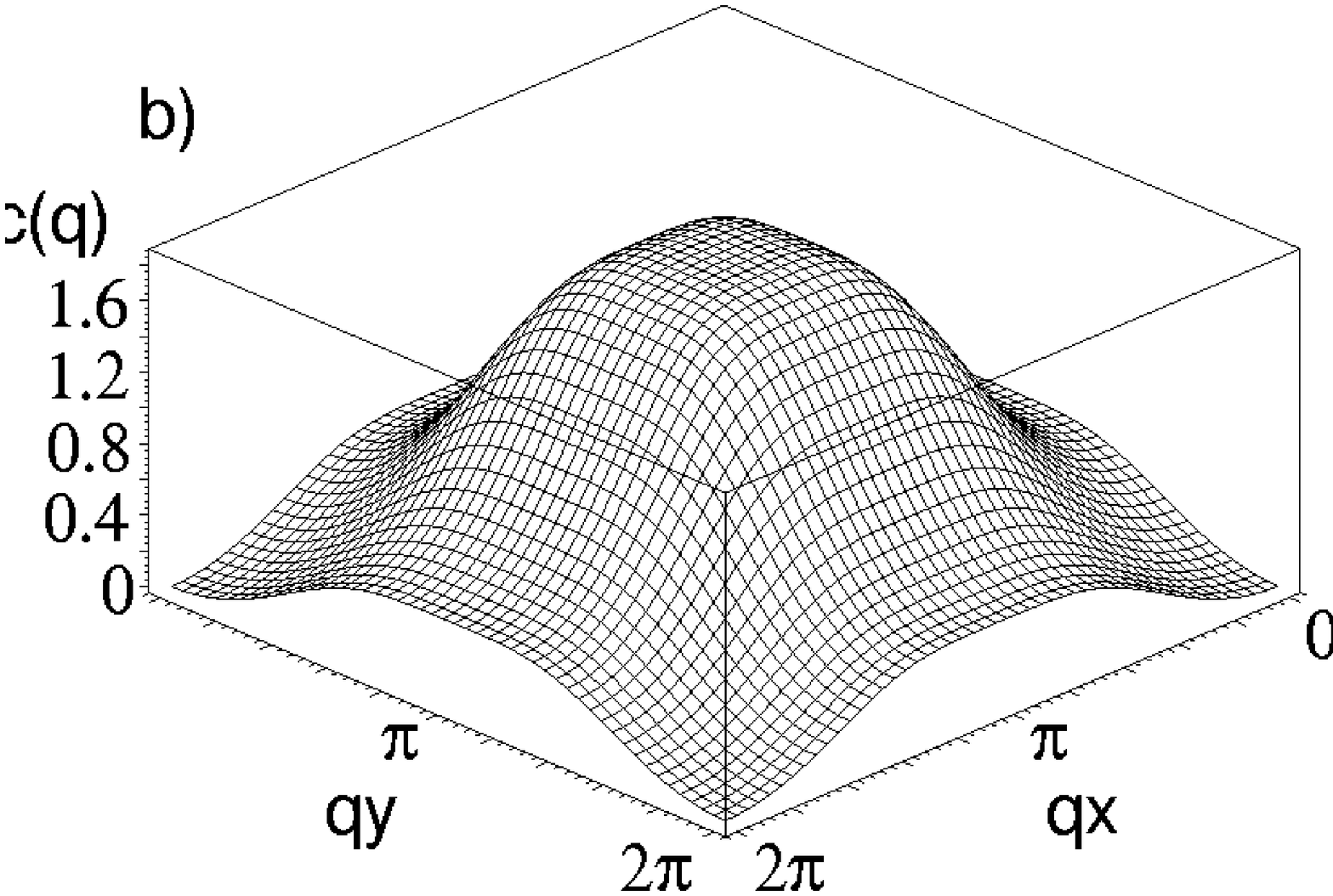}}
}
\caption{ \label{Fig0_Eschrig03}
Momentum dependence of a) the resonance mode
and b) the gapped spin fluctuation continuum.
The resonance mode is peaked at $\vec{Q}=(\pi,\pi)$ with a 
correlation length equal to twice the lattice constant, $\xi_{res} = 2a$.
The continuum spectrum, in contrast, is rather broad around $\vec{Q}$.
(From Ref. \cite{Eschrig02},
Copyright \copyright 2002 APS).
}
\end{figure}

The momentum dependences of the resonance mode and the continuum,
are given by the functions $w_q$ and $c_q$, 
\begin{eqnarray}
\label{wq}
w_q&=&\frac{w_Q}
{1+4\xi_{sfl}^2 (\cos^2\frac{q_x}{2}+\cos^2\frac{q_y}{2})},\\
\label{cq}
c_q&=& \frac{c_Q}
{1+16\xi_{c}^4 (\cos^4\frac{q_x}{2}+\cos^4\frac{q_y}{2})}(1+ \beta) -
c_Q\beta ,
\end{eqnarray}
where $\xi_{sfl}$ is the correlation length of the resonance and
$\xi_{c}$ that of the gapped continuum.
The momentum dependence of the continuum 
takes into account the experimentally observed
flatter behavior around the $(\pi,\pi)$ wavevector at higher energies.
The parameter $\beta $ is chosen in such a way that
$c_q$ is zero at $\vec{q}=0$, and is introduced such
that the response far away from
the $(\pi,\pi)$ wavevector is small, as experimentally observed.
The functions $w_q$ and $c_q$ are plotted in Fig.~\ref{Fig0_Eschrig03}.
The resonance mode, shown in 
Fig.~\ref{Fig0_Eschrig03}a, is peaked at $\vec{Q}$, with a correlation length of
$\xi_{res}=2a $, where $a$ is the lattice constant. The gapped continuum,
shown in Fig.~\ref{Fig0_Eschrig03}b, is much broader.
This is motivated by the experimental data \cite{Rossat91}, which show that the
continuum is enhanced around $\vec{Q}$ with a correlation length of only
0.5 lattice constants. Also, the momentum dependence
of the continuum excitations exhibit experimentally 
a flat behavior around $\vec{Q}$, as in Fig.~\ref{Fig0_Eschrig03}b.
To simplify the model, we use the same functional form for the
gapped continuum in the even and odd scattering channels. This is consistent
with the superconducting state data,
where the gap in the odd channel (about twice the maximal superconducting gap), 
is close to the optical gap in the even channel (see Fig. \ref{Fig1_Dai99}).

The retarded Green function 
as a function of the self energies 
is given as usually by,
\begin{eqnarray}
\label{green}
G^R_{\epsilon,k} [\Sigma^R_{\epsilon,k},\Phi^R_{\epsilon,k}] &=& 
\frac{Z_{\epsilon,k}\epsilon + \bar \xi_{\epsilon,k} }{
(Z_{\epsilon,k}\epsilon + i 0^{+})^2 -E_{\epsilon,k}^2},
\\
F^R_{\epsilon,k} [\Sigma^R_{\epsilon,k},\Phi^R_{\epsilon,k}]&= &
\frac{\bar \Delta_{\epsilon,k} }{
(Z_{\epsilon,k}\epsilon + i 0^{+})^2 -E_{\epsilon,k}^2},
\end{eqnarray}
with excitation energies 
\begin{eqnarray}
E_{\epsilon,k}=\sqrt{\bar \xi_{\epsilon,k}^2+\bar \Delta_{\epsilon,k}
\bar \Delta_{-\epsilon,-k}^{\ast }} \; .
\end{eqnarray}
The renormalized dispersion and gap function are given in terms of
the diagonal 
($\Sigma^R_{\epsilon,k}$) and off-diagonal ($\Phi^R_{\epsilon,k} $)
in particle-hole space self energies, as
\begin{eqnarray}
\bar\xi_{\epsilon,k}&=&\xi_k + 
\frac{\Sigma^R_{\epsilon,k}+\Sigma^{R\ast }_{-\epsilon,-k}}{2} \qquad
\bar \Delta_{\epsilon,k} =  \Delta_k+
\Phi^R_{\epsilon,k} ,
\label{dSig}
\end{eqnarray}
and the renormalization function as
\begin{eqnarray}
Z_{\epsilon,k}&=& 1 -
\frac{\Sigma^R_{\epsilon,k}-\Sigma^{R\ast }_{-\epsilon,-k}}{2\epsilon }
\end{eqnarray}
The self energies $\Sigma^R_{\epsilon,k}$ and $\Phi^R_{\epsilon,k}$ are
even in momentum, and for $\Phi^R_{\epsilon,k}$ the additional symmetry
$\Phi^R_{\epsilon,k}=\Phi^{R\ast }_{-\epsilon,-k}$ holds (we use a
real gauge for the order parameter $\Delta_k$).
The coupling constant $g$
between electrons and the spin fluctuation spectrum 
is assumed to be independent of energy and momentum.

Although the self-energy  Eq.~(\ref{keld}) is generally correct also in
non-equilibrium, it is advantageous in the case of equilibrium to rewrite
the expression.  In equilibrium it is possible to make use of the identity
\begin{eqnarray}
\label{id1}
\sum_{\omega} D^R_{\omega} G^K_{\epsilon-\omega} &=&
-i\sum_{\omega} \tanh \frac{\epsilon - \omega}{2T}
B_{\omega} G^R_{\epsilon-\omega} 
\nonumber \\ &&
+\sum_{\omega} \left( D^A_{\omega} G^R_{\epsilon-\omega} -
D^R_{\omega} G^A_{\epsilon-\omega} \right) \tanh \frac{\epsilon-\omega}{2T} \; ,
\end{eqnarray}
and to convert the second line of Eq.~(\ref{id1})
into a Matsubara sum by noting that $D^A_{\omega }G^R_{\epsilon-\omega} $ is an analytic
function in the lower $\omega $ half plane, and analogously 
$D^R_{\omega }G^A_{\epsilon-\omega} $ analytic in the upper half plane.
An analogous formula holds for 
$\sum_{\omega} D^R_{\omega} F^K_{\epsilon-\omega} $.
The self energies are then given in terms of the spectral function
of the spin fluctuations with energy $\omega $ and momentum $\vec{q}$, 
$B_{\omega,q}$, by the expressions \cite{Marsiglio88}
\begin{eqnarray}
\label{self1}
\Sigma^R_{\epsilon,k}&=&\sum_{\omega,q} 
(b_{\omega} + f_{\omega-\epsilon})
g^2B_{\omega,q}
G^R_{\epsilon-\omega,k-q} 
- T \sum_{\epsilon_n,q} G^M_{k-q}(i\epsilon_n)g^2D^M_q(\epsilon-i\epsilon_n)
\quad
\\
\label{self2}
\Phi^R_{\epsilon,k}&=&\sum_{\omega,q} 
(b_{\omega} + f_{\omega-\epsilon})
g^2B_{\omega,q}
F^R_{\epsilon-\omega,k-q} 
- T \sum_{\epsilon_n,q} F^M_{k-q}(i\epsilon_n)g^2D^M_q(\epsilon-i\epsilon_n),
\quad
\end{eqnarray}
where $G^M$, and $D^M$ are the fermionic and bosonic
Matsubara Green functions.
Analogously, $F^M$ is the anomalous Matsubara Green function.
The Matsubara sums in Eqs.~(\ref{self1}) and (\ref{self2})
only contribute to the real part of the self energies.
The representations (\ref{self1}) and (\ref{self2}) are a bit unconventional,
however for finite temperatures very convenient.
Note that for optimally and overdoped materials
unrenormalized Green functions can be used in
Eqs.~(\ref{self1})-(\ref{self2}) \cite{Eschrig02,Eschrig03}.
This approximation is sufficient to explain a large variety 
of data, and can be justified by considerations discussed in Ref. \cite{Vilk97}.

The function $b_{\omega} + f_{\omega-\epsilon}$ as a function
of $\omega $ for fixed $\epsilon $ is at zero temperature
nonzero only between $\omega =0$ and $\omega = \epsilon $, and is equal to 
sign$(\epsilon )$ in this range.  
This 'box function' is smeared out at finite temperatures by the amount of
the thermal energy $k_BT$. However, because
the spin-fluctuation spectrum is gapped by much more than the thermal energy
in the superconducting state, 
for all practical reasons $b_{\Omega_{res}}=0$ holds.
Thus, thermally excited modes can be safely neglected, and only
emission processes at the resonant mode energy are relevant.
Furthermore, for any gapped spin-fluctuation spectrum with gap 
$\Omega$, the first terms
in Eq.~(\ref{self1}) and (\ref{self2}) are negligible in the range
$-\Omega< \epsilon < \Omega$ (apart from temperature smearing near the
value $\pm \Omega$). Thus, assuming that the spin fluctuation spectrum
is gapped below the resonance energy, at zero temperature scattering of 
electronic excitations is disallowed in the
interval $-\Omega_{res}< \epsilon < \Omega_{res}$. This is an expression of
the fact that at least an energy $\Omega_{res}$ must be spent in order
to emit one spin fluctuation mode. This is the case for optimally
and overdoped cuprates.
For strongly underdoped cuprates, scattering is disallowed
only in the range 
$-E_g < \epsilon < E_g$, where $E_g$ is the spin gap which is smaller
than $\Omega_{res}$.
Also, as an implication, the renormalization function, determined by the
real part of the self energy, is given in the low energy range by the
second terms of Eqs.~(\ref{self1}) and (\ref{self2}) only (but also has
contributions from the first terms for higher energies).

\subsection{Contribution from the spin fluctuation mode}
\label{CSFM}
For a sharp bosonic mode the spectral function is given by Eq.~(\ref{Bres0}) with
the energy integrated weight given by Eq.~(\ref{wq}).
The mode is enhanced at the $\vec{Q}=(\pi,\pi)$ point with a
correlation length $\xi_{sfl}$. It is a good approximation to assume the
mode as perfectly sharp in energy \cite{Eschrig03}.
From neutron scattering data obtained on
Bi${_2}$Sr${_2}$CaCu${_2}$O${_{8+\delta}}$,
the energy integrated weight of the resonance mode was determined 
as 1.9 $\mu_B^2$ \cite{Fong99},
leading (after dividing out the matrix element $2\mu_B^2$) to $w_Q=0.95$.
We fit ARPES data near optimal doping \cite{Eschrig00},
giving $g^2w_Q=0.4 $eV$^2$. This implies that the coupling constant
is equal to $g=0.65$ eV. This is a reliable value as discussed
in \cite{Abanov02a}.
In Table \ref{tab2}, 
the minimal parameter set for optimally doped compounds is presented
(from the band structure tight binding fit, only the parameter $\xi_M$ is listed
as the results are insensitive to reasonable variations of the other parameters)

\begin{table}
\begin{center}
\tbl{
\label{tab2} 
Minimal parameter set appropriate for optimally doped
Bi$_2$Sr$_2$CaCu$_2$O$_{8+\delta}$.}
{\begin{tabular}{|c|c|c|c|c|}
\hline
\hline
$\Delta_{M}$ & $\Omega_{res}$ & $\xi_M$ & $\xi_{sfl}$ 
& $g^2w_Q$ \\
\hline
35 meV & 39 meV & $-34$ meV & 2$a$  & 0.4 eV$^2$
\\
\hline
\hline
\end{tabular}}
\end{center}
\end{table}

In the following sections we review results for numerical calculations
at finite temperatures obtained from solving
Eqs.~(\ref{self1})-(\ref{self2}) using bare Green functions on the right
hand side.
However, for better understanding of the results it will be convenient to
also discuss the zero-temperature limit. 
Using bare Green functions, the  
self energy at zero temperature can be written as
\begin{eqnarray}
\label{self11}
\mbox{Im} \Sigma^R_{\epsilon, k}&=&
-\sum_{q}  g^2w_q A^{-}_{k-q} 
\delta (\epsilon+\Omega_{res}+E_{k-q}) 
\nonumber \\ && 
-\sum_{q}  g^2w_q A^{+}_{k-q} 
\delta (\epsilon-\Omega_{res}-E_{k-q})
\\
\label{self12}
\mbox{Re} \Sigma^R_{\epsilon, k}&=&
-\sum_{q}  \frac{g^2w_q}{\pi} \cdot
\frac{\epsilon + \left( 1+\frac{\Omega_{res}}{E_{k-q}}\right)\xi_{k-q}}{
(\Omega_{res}+E_{k-q})^2 - \epsilon^2}
\end{eqnarray}
where $E_k=\sqrt{\xi_k^2+|\Delta_k|^2}$, $A^{\pm}_k=(1\pm\xi_k/E_k)/2$.
For negative energies, only the first
sum in Eq.~(\ref{self11}) is nonzero. The sum is a weighted average
of the expression $A^{-}_{k-q}  \delta (\epsilon+\Omega_{res}+E_{k-q})$
with weight factors $w_q$. For given fermion energies, $\epsilon $, and
momenta, $\vec{k}$, the delta function restricts the
allowed spin fluctuation momenta $\vec{q}$.
Similar zero temperature formulas hold for the off diagonal self energy,
\begin{eqnarray}
\label{self21}
\mbox{Im} \Phi^R_{\epsilon, k}&=&
-\sum_{q}  g^2w_q \frac{\Delta_{k-q}}{2E_{k-q}}
\Big[ \delta (\epsilon-\Omega_{res}-E_{k-q}) 
- \delta (\epsilon+\Omega_{res}+E_{k-q})  \Big] \quad
\\
\label{self22}
\mbox{Re} \Phi^R_{\epsilon, k}&=&
-\sum_{q}  \frac{g^2w_q}{\pi} \cdot
\frac{\left( 1+\frac{\Omega_{res}}{E_{k-q}}\right)\Delta_{k-q}}{(\Omega_{res}
+E_{k-q})^2-\epsilon^2}.
\end{eqnarray}

\subsubsection{Characteristic electronic scattering processes}
\label{ES}

Scattering of electrons via emission of a magnetic resonance excitation
is characterized by several important peculiarities.
They are a result of the interplay between kinetic scattering restrictions
due to the Pauli principle for electrons and the restriction of the 
resonance momentum to the near vicinity of the $\vec{Q}=(\pi,\pi)$ wavevector.
The kinetic restrictions for electronic scattering are given by the fact
that low-energy scattering is restricted to a close vicinity of the
Fermi surface. In the present case
the electron scattering processes can involve the emission
of a resonance mode with energy $\Omega_{res}$.
However, the electrons also acquire a large momentum $\vec{q}\sim 
\pm (\pi,\pm \pi)$
from the scattering event, which leads to a strongly anisotropic 
scattering rate. 

The scattering rate of electrons is determined by the imaginary part
of the self energy, which can be extracted from photoemission spectra.
In a photoemission experiment a photon with energy $h\nu$ creates a
photo-hole below the chemical potential and the
energy $\epsilon $ and momentum $\vec{k}$ 
of the electron emitted from the sample is detected. The resulting spectra
give information about the spectrum of a hole interacting with
the collective excitations present in the solid.
The creation of such a photohole can be either direct or can be accompanied
by the simultaneous creation of collective excitations, which for the case of
the spin-1 resonance mode we consider have a sharp energy $\Omega_{res}$ and
a momentum not too far away from $\vec{Q}=(\pi,\pi)$.
The detection of the emitted electron 
at a certain momentum $\vec{k}$ will lead to a `coherent peak'
from the creation of photoholes without additional collective excitations, and
a broad incoherent 
continuum from the creation of photoholes accompanied by the creation
of additional collective excitations. The latter has a broad distribution due to
the momentum spread of the resonant magnetic excitations, and this distribution
will be gapped by at least the mode energy, because the highest possible
energy the emitted electron can have in this case is that of the chemical potential
minus the mode energy. The peak determines the binding energy of the well defined
fermionic excitations in the system (quasiholes).

The probability to create a photohole without creating magnetic excitations
is small for a hole near the $M$ point of the Brillouin zone, because of the
large number of states present near the chemical potential at momenta
$(\pi,\pi)$ away (again corresponding to $M$-symmetry points of the Brillouin zone).
In fact, the area of the flat
dispersion region is large enough to exhaust almost the entire weight of
the magnetic resonance during scattering events.
Thus, the peak intensity will be reduced there, and the broad loss spectrum at 
higher energies due to emission of collective spin-1 modes is strong.
Near the nodal points, the probability to create a photohole without additionally
exciting magnetic collective excitations is high, thus the peak intensity is strong
here and the incoherent loss spectrum due to collective spin-1 modes is small
(there will be a considerable incoherent part due to the spin-fluctuation continuum though; this will be discussed later).

From the above it is clear that three effects are contributing to the
physics of the low-energy scattering events: the peaked behavior in
momentum of the magnetic resonance, the fact that scattering events take place
only between points near the chemical potential, and the presence of a large
number of states near the $M$ point of the Brillouin zone.

\begin{figure}
\centerline{
\epsfxsize=0.32\textwidth{\epsfbox{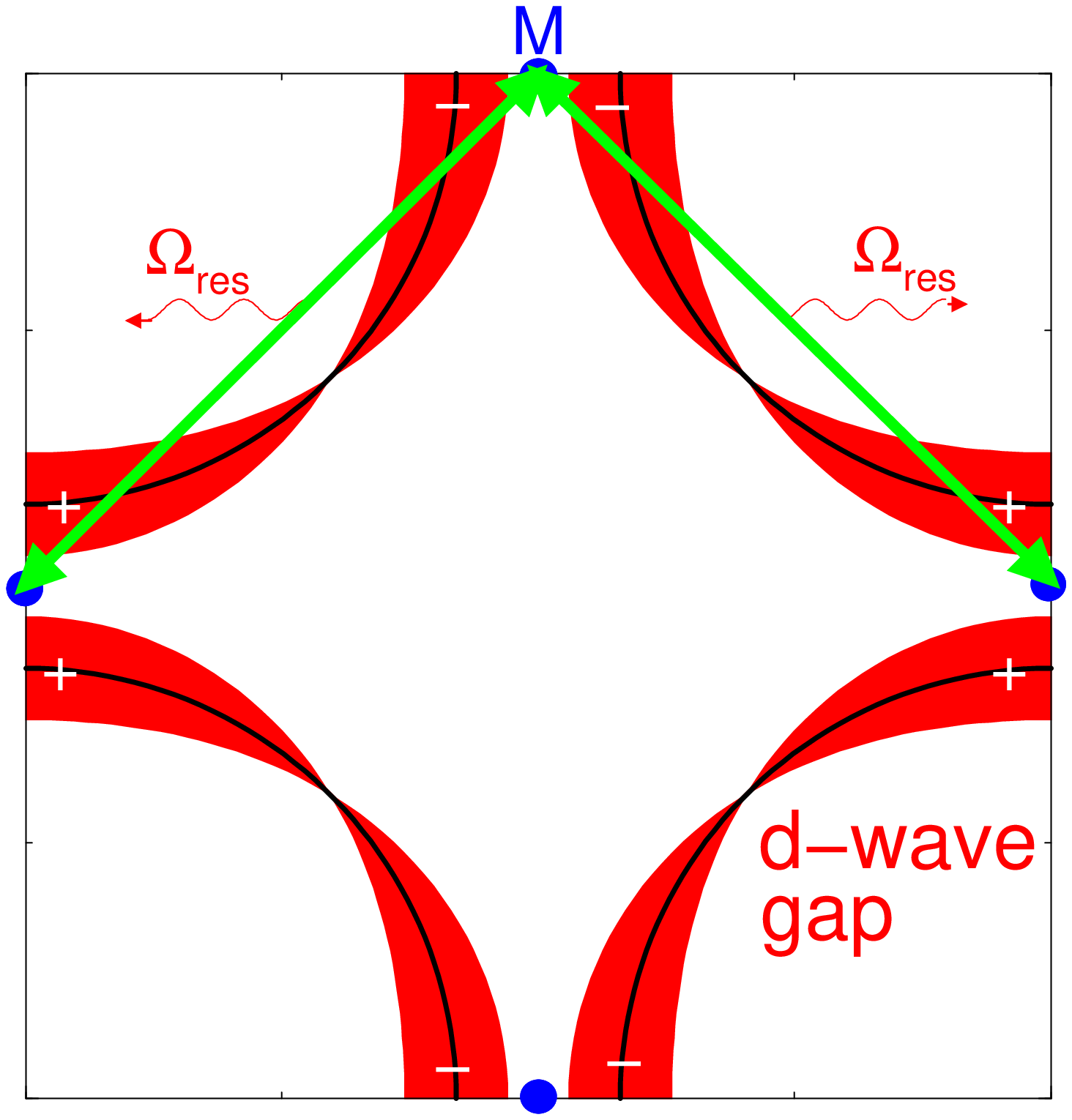}}
\epsfxsize=0.32\textwidth{\epsfbox{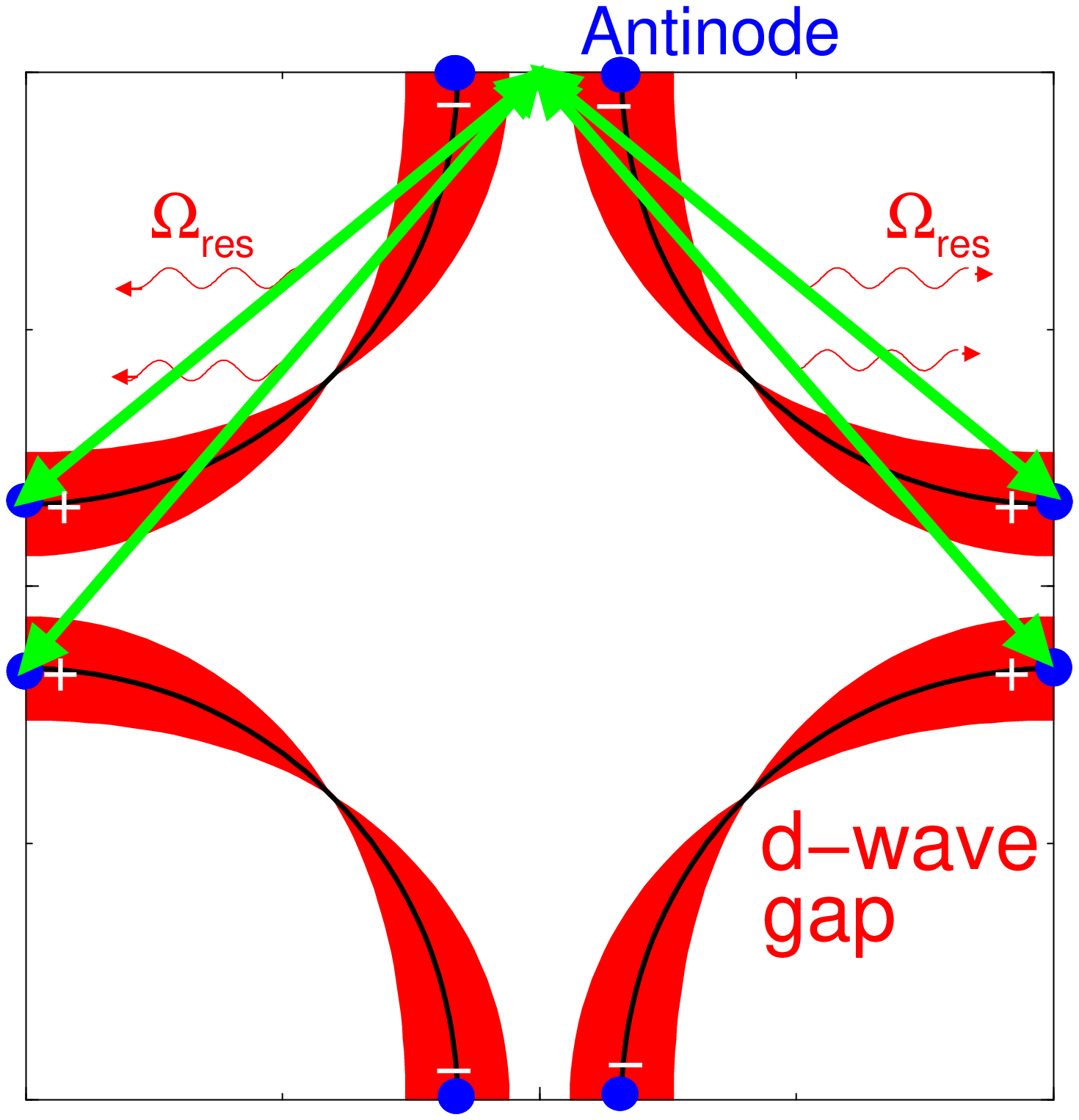}}
\epsfxsize=0.32\textwidth{\epsfbox{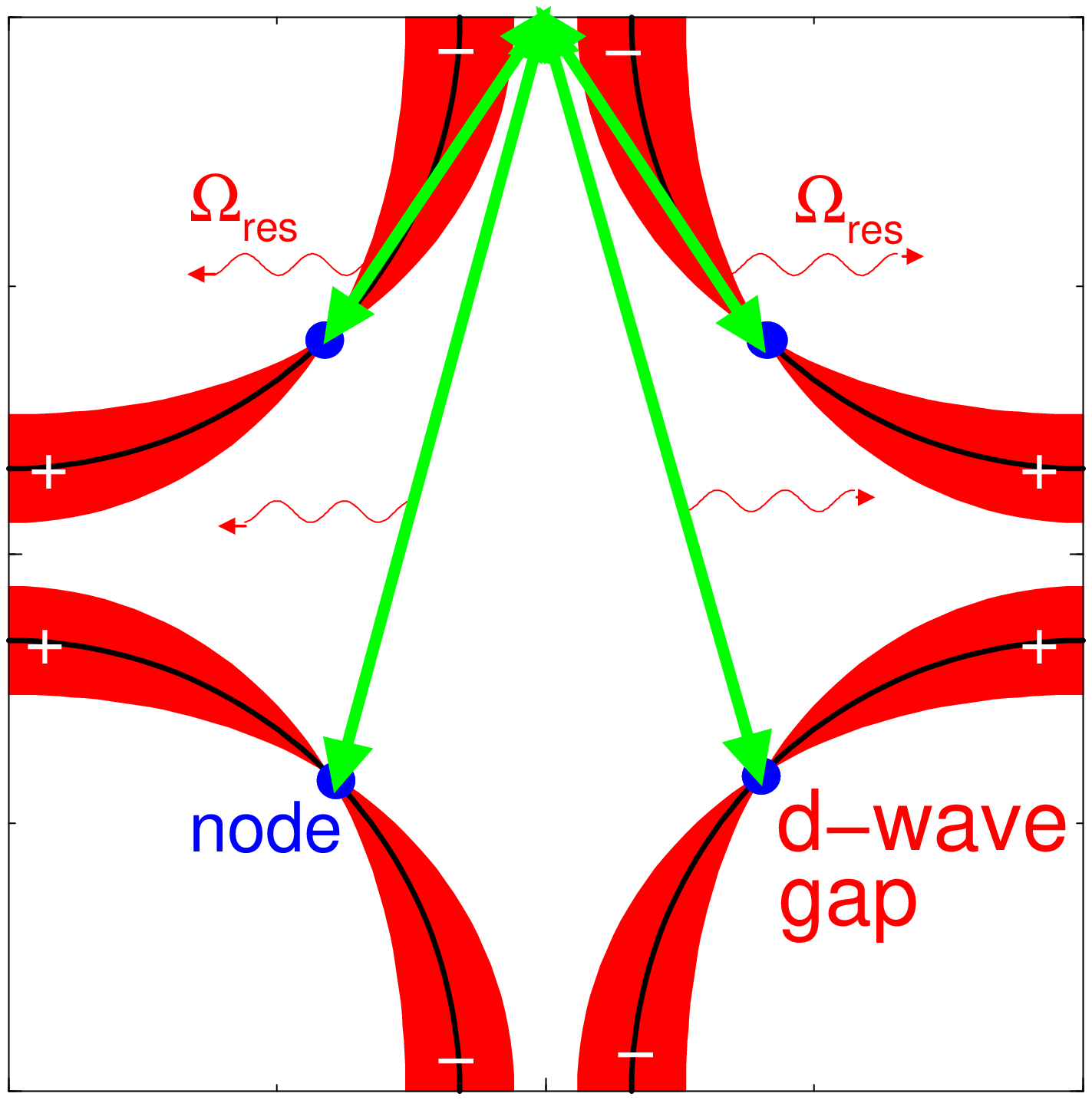}}
}
\begin{minipage}{0.32\textwidth}
\centerline{$-\Omega_{res}-E_M $ }
\end{minipage}
\begin{minipage}{0.32\textwidth}
\centerline{$-\Omega_{res}-\Delta_A $ }
\end{minipage}
\begin{minipage}{0.32\textwidth}
\centerline{$-\Omega_{res} $ }
\end{minipage}
\caption{
\label{scatt}
The relevant scattering processes for an electron removed from the
$M$ point. They correspond to the
characteristic energies: $- (\Omega_{res}+E_M)$  (left),
$- (\Omega_{res}+\Delta_{A})$ (middle),
and $-\Omega_{res}$ (right). The corresponding energies for
an injected electron have opposite sign.
The magnitude of the $d$-wave order parameter is indicated as shading
around the Fermi surface.
}
\end{figure}

There are three characteristic frequencies which determine the onset and
maxima of scattering, and which will show up in the imaginary part of the
self energy, discussed in the following section.
They are shown for the case of an injected hole (removed electron) at the
$M$ point of the Brillouin zone in Fig.~\ref{scatt}.
The first, shown in Fig.~\ref{scatt} on the right, determines the onset
of scattering in the superconducting state
at energy $\epsilon=-\Omega_{res}$, which is connected with the fact that  it costs
a minimal energy $\Omega_{res}$ to emit a resonant excitation.
The important scattering vectors are here the wavevectors connecting the nodes
with the $M$ points of the Brillouin zone,
$\vec{q}=(\vec{k}_M-\vec{k}_N)$ mod $(\vec{G})$
($\vec{G} $ denotes a reciprocal lattice vector).
Below this onset energy, the imaginary part of
the considered electronic self-energy is zero at $T=0$ (at finite temperatures
this onset is smoothened on the temperature scale).
When taking into account the incommensurate spin-fluctuations at lower
energies, a small finite contribution to the imaginary part of the self energy
will be present also below this onset energy.
As the phase space for final scattering states near the nodes is very small, and in
addition the wavevector for scattering from $(\pi,0)$ to the nodal point does not
match the antiferromagnetic wavevector, the
magnitude of scattering near the onset is small as well.

The first maximal effect of scattering with increasing binding energy is reached
when the final scattering states are at the antinodal points, where the 
dispersion in the superconducting state has an extremal point.
This corresponds to Fig. \ref{scatt} middle. The relevant scattering
wavevectors are $\vec{q}=(\vec{k}_M-\vec{k}_A)$ mod $(\vec{G})$,
and electrons are scattered strongly between the $M$ point and the $A$ points,
and the characteristic energy of these scattering events is the sum of the
energy of the emitted mode and the binding energy at the $A$ point,
$\epsilon = -(\Omega_{res}+\Delta_A)$.

A third special point is reached when the binding energy increases even more
to the value $\epsilon = -(\Omega_{res}+E_M)$
(with $E_M=\sqrt{\xi_M^2+\Delta_M^2}$), at which
scattering events between the $M$ points involving spin fluctuations
with momentum $\vec{q}=\vec{Q}$ (and with $\vec{q}=\vec{0}$) are
allowed. This corresponds to Fig. \ref{scatt} left.
At this wavevector the intensity of the magnetic mode is maximal, and the
number of final scattering states is strongly increased due to the flat
dispersion near the $M$ points. Both facts lead to a very strong
scattering at this particular binding energy for quasiparticles 
near the $M$ point of the Brillouin zone.

At even higher binding energies the gapped continuum part of the bosonic spectrum
becomes involved, and this region will be discussed later.
For the parameter set of Table \ref{tab2}, corresponding to optimal doping,
the characteristic energies 
are: $\Omega_{res} = 39 $ meV, $\Omega_{res}+\Delta_A =71.2$ meV, and
$\Omega_{res}+E_M=87.8 $ meV.
The energy range in which the scattering is
maximal is between 70 and 90 meV. In this range also the
strongest renormalization effects are expected.

\subsubsection{Electronic self energy}
\label{RFEL}

The self energy has a characteristic shape as a function of energy,
which is conserved qualitatively for all points in the Brillouin zone.
This is a consequence of the fact that all points are coupled via
the spin fluctuation mode, which has
a finite width in momentum, to
all special points in the Brillouin zone with their corresponding
characteristic energies. These special points are the nodal $N$ points, and
the van Hove singularities at the $M$ points and the $A$ points (the latter
is a dispersion maximum in the superconducting state).
Because the general shape of the energy dependence of the self energy
does not vary much with momentum (although the overall intensity does),
it is sufficient to discuss the important features in 
the energy dependence of the self energy at the $M$ point.

\begin{figure}
\centerline{
\epsfxsize=0.85\textwidth{\epsfbox{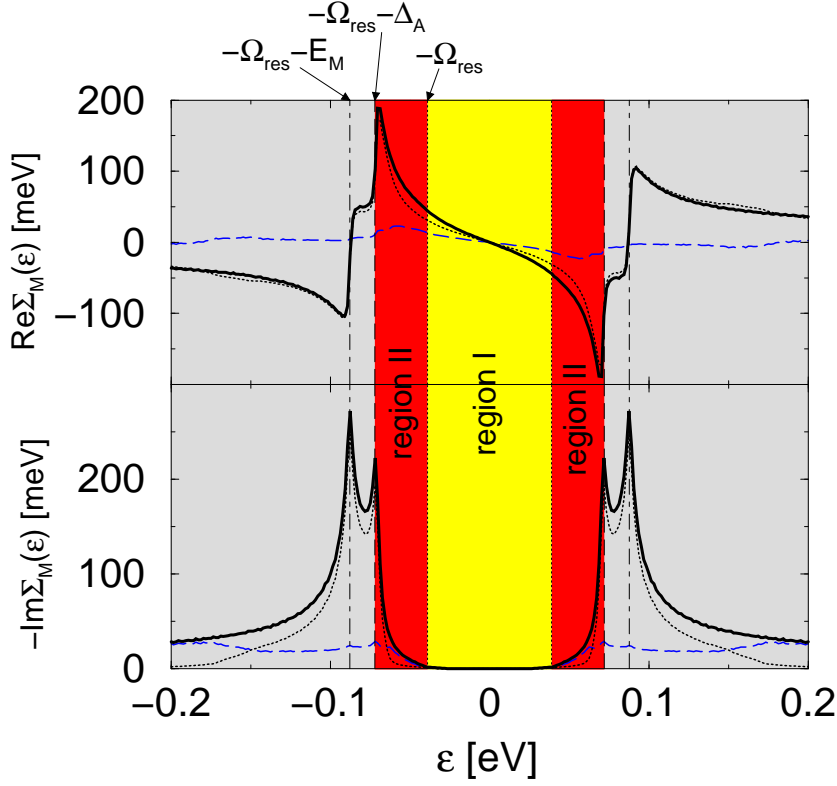}}
}
\caption{
\label{Zfac}
Mode contribution to the
real and imaginary part of the electronic self energy
at the $M$ point, numerically evaluated using a
broadening parameter $\delta=$ 1 meV. Characteristic energies
are indicated.
Electrons at low temperatures are scattered only if their
energy is larger than $\Omega_{res}$, so that they are able to
emit a collective-mode excitation. 
As dashed curves the nodal contributions,
when restricting the quasiparticle momenta to the regions {\it outside} the
area around  the $M$ points discussed in the text, is shown.
The thin dotted curves are the contribution  when
restricting the quasiparticle momenta to the regions {\it inside} the
area around  the $M$ points.
The parameters used are: $T=40$K, $\Omega_{res}=$39 meV, $\Delta_{M}=$35 meV.
(After Ref. \cite{Eschrig03},
Copyright \copyright 2003 APS).
}
\end{figure}

In Fig.~\ref{Zfac}, the results for the electronic self energy at the
$M$ point of the Brillouin zone are shown as a function of energy.
The three characteristic energies discussed in section \ref{ES} can
be readily read off from Fig.~\ref{Zfac}. They define two characteristic
regions.

In region I the scattering rate for scattering of electrons or holes
with the magnetic mode is zero at zero temperature (this statement is true
for electrons at any point in the Brillouin zone).
A finite temperature affects only the onset region, as $k_BT<<\Omega_{res}$.
Concentrating in the following on the hole spectrum (negative energies),
the onset at $\epsilon=-\Omega_{res}$ is determined by
the coupling to nodal electrons via emission of a spin fluctuation mode.
The real part of the self energy is rather linear in region I.

In region II, a larger and larger area around the nodes participates
in scattering events, 
until finally the point at the zone boundary with
maximal gap, $\pm \Delta_{A}$, is reached.
In Fig.~\ref{Zfac} we also show as dotted curves 
the contribution to the electron scattering rate coming from restricting the
final states $\vec{k}-\vec{q}$ in Eq.~(\ref{self1}) to
a region around the $M$ point
that is deliminated by $\pm 0.35 \pi$ in $M-Y$ direction and
by about 0.3 $\pi$ in $M-\Gamma $ direction.
As dashed curves we show the remaining contribution, when the
final states are not too close to the $M$ points.
Clearly, it is the nodal contribution that dominates the onset of scattering
near $\pm \Omega_{res}$.

The scattering near the other two characteristic frequencies discussed
in the last section, $-(\Omega_{res} + \Delta_{A})$ and
$-(\Omega_{res} + E_M)$, is dominated by scattering between the regions
close to the $M$ points of the Brillouin zone.
The proximity of this van Hove singularity leads to a stronger peaked
feature in the scattering rate near $\pm(\Omega_{res} + \Delta_{A})$
compared to the case where this van Hove singularity at the $M$ point is
absent. 
Taking into account a finite intrinsic spectral width of the electrons 
involved in scattering events, the strong peaks in Im$\Sigma $ are smeared out,
leaving a cusp at $-(\Omega_{res} + \Delta_{A})$ and a weak maximum at
$-(\Omega_{res} + E_M)$, and the onset
of scattering at the emission edge for the spin fluctuation mode
occurs as a jump.

At even higher energies, 
the scattering due to the spin fluctuation mode becomes
less effective. 
This region, however, will be dominated by scattering
processes involving the spin-fluctuation continuum, as discussed
later.

\subsubsection{Renormalization function and quasiparticle scattering rate}
\label{QPSR}

An approximate analytical expression for the renormalization function at
the $M$ point, $Z_M(\epsilon )= 1-$Re$\Sigma_M(\epsilon)/\epsilon $,
can be obtained by neglecting the dispersion in the relevant
$M$-point regions between the two Fermi crossings nearby,
and restricting the momentum integral to roughly a square in that regions.
We denote $\sum_q w_q$ over this area by $I_0$. 
For our model we have
$I_0=0.035 $.
Using this approximation \cite{Eschrig03},
\begin{equation}
\label{ZM}
Z_M(\epsilon ) \approx
1 + \frac{g^2I_0}{\pi} 
\frac{1}{(\Omega_{res}+E_M)^2-\epsilon^2}+\lambda_{M}^{(N)}(\epsilon)
\end{equation}
Here, $\lambda_{M}^{(N)}(\epsilon )$ denotes the contributions coming from
the nodal regions discussed above.
The contribution $\lambda_{M}^{(N)}$ is smaller than the first term
in Eq.~(\ref{ZM}), but not negligible.
Because Eq.~(\ref{ZM}) neglects the dispersion
between $(\Omega_{res}+\Delta_A)$ and $(\Omega_{res}+E_M)$
near the $M$ point, it should be used for energies not too close
to the region between these two values. 

For optimally doped and 
overdoped materials the quasiparticle peak at the $M$ point is
situated below the onset of scattering due to emission of spin fluctuations.
In this case the width is determined by other processes, and 
this residual quasiparticle width is modelled by a parameter $\delta $.
In Fig.~\ref{Wwidth} the two top pictures show the numerically evaluated renormalization
function for $\delta=5 $ meV. The renormalization function is
rather constant in region I and shows a peaked behavior in region II.
At higher binding energies the mode contribution to the
renormalization function drops {\it below} one, and goes toward 1
for high energies. 
Later we will discuss the contribution to the renormalization function
of the spin-fluctuation continuum, 
which dominates this high-energy region. This contribution will approach
its high-energy asymptotics from {\it above}.

The behavior of the imaginary part of the self energy near the
onset points, $\pm \Omega_{res}$, in Fig.~\ref{Zfac} is determined by the nodal
electrons. 
For larger residual quasiparticle widths ($\delta = 5 $ meV,
see Fig. \ref{Wwidth}) there are
states available at the chemical potential (coming e.g. from impurity
scattering), which increase the number of final states for scattering
events. Thus, the onset for the electron scattering
rate is stronger in this case than for $\delta = 1$ meV. For zero temperature
there will be a jump at energy $\pm \Omega_{res}$ in the imaginary part
of the self energy. For $\delta =0 $ the onset is linear in energy.

An analytical expression for this onset for $\delta=0$ 
has been derived in Ref.~\cite{Eschrig03}, and is given by,
\begin{eqnarray}
\label{ImSnode}
\Gamma_M^{(N)} (\epsilon ) &=& 
\left\{
\begin{array}{lcc}
\frac{\displaystyle g^2w_{MN}} {\displaystyle \pi v_Nv_{\Delta}} (|\epsilon | - \Omega_{res})  
&\qquad \qquad &|\epsilon |>\Omega_{res}\\
& \mbox{for}&\\
0
&\qquad \qquad &|\epsilon |<\Omega_{res}
\end{array}
\right.
\end{eqnarray}
Here, $w_{MN}=w_{\vec{k}_M-\vec{k}_N}$, $\vec{v}_{\Delta}=\partial_{k} \Delta_k $
and $\vec{v}_N= \partial_{k} \xi_k$ taken at the $N$ point. 
For the parameters in Tables \ref{tab1} and \ref{tab2},
the magnitude of slope 
of the scattering rate at $\epsilon=\pm \Omega_{res}$ 
is equal to 9.5 $w_{MN}/w_Q \approx 0.56$.
Note that Eq.~(\ref{ImSnode})
gives a good approximation of the scattering rate in
the interval $\Omega_{res}<|\epsilon|< \Omega_{res}+\Delta_A/2$ \cite{Eschrig03}.

Finally, for underdoped cuprates the excitation energy at the $M$ point,
$E_M$, is larger than $\Omega_{res}$. Then, the quasiparticle linewidth
at the $M$ point is given by $\tilde \Gamma_M=\Gamma_M^{(N)} 
(-E_M ) /Z_M(-E_M)$.
Thus, for underdoped cuprates it is given by,
\begin{eqnarray}
\label{Gamma_u}
\tilde\Gamma_{M} &=& \frac{g^2w_{MN}}{\pi v_Nv_{\Delta}}
\frac{E_M - \Omega_{res}}{ Z_M }
\end{eqnarray}
with $Z_M \equiv  Z_M (-E_M)$. 
Near the nodes, on the contrary, the quasiparticles will stay relatively 
sharp even in underdoped compounds because
the peak positions are then below the onset energy $\pm \Omega_{res}$.

\subsubsection{Spectral functions at the $M$ point}
\label{SFMP}

In this section, we discuss the spectral lineshape due to
coupling of electrons to a sharp magnetic mode.
The main features of the spectral
lineshape are captured in the simple model neglecting the continuum
part of the bosonic spectrum. We discuss in the following the influence of
the different parameters of the theory on the spectral function, 
\begin{equation}
A (\epsilon , \vec{k}_M) = -2 \mbox{Im} G^R(\epsilon, \vec{k}_M)
\end{equation}
and will discuss changes due to the continuum part of the spin fluctuation
spectrum later. 

\begin{figure}
\begin{minipage}{0.015\textwidth}
$\; $
\end{minipage}
\begin{minipage}{0.48\textwidth}
\epsfxsize=0.88\textwidth{\epsfbox{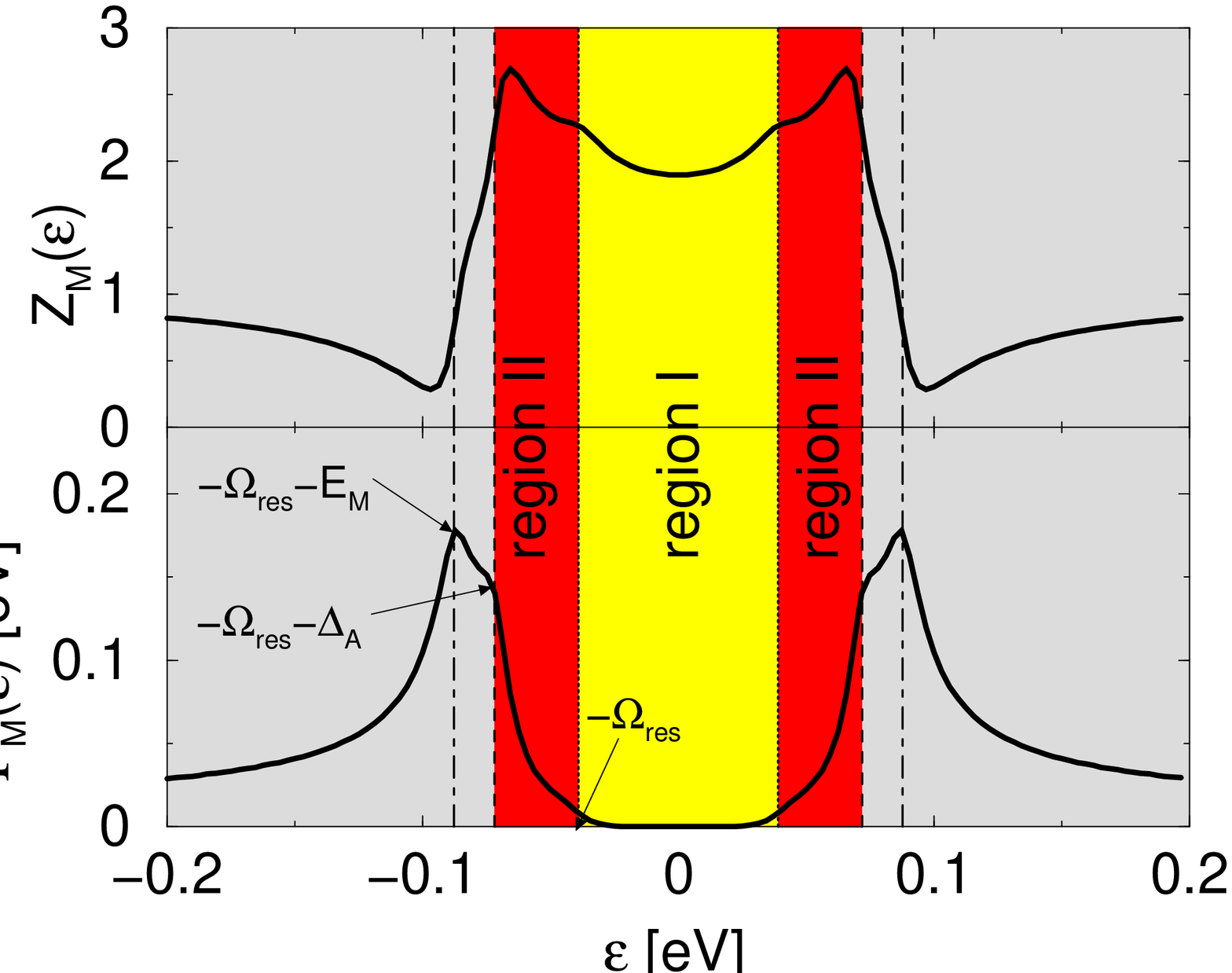}}
\end{minipage}
\hfill
\begin{minipage}{0.48\textwidth}
\epsfxsize=0.88\textwidth{\epsfbox{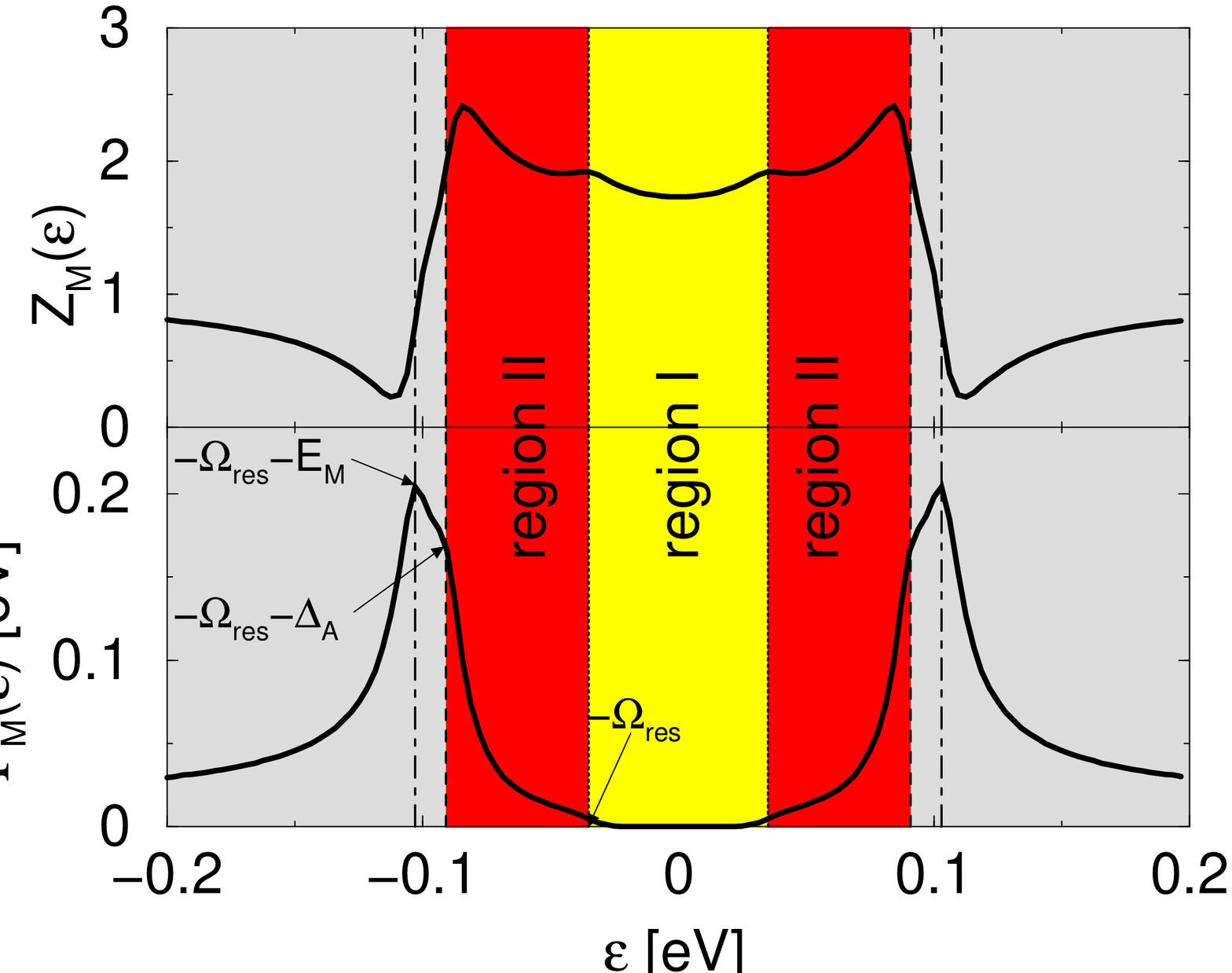}}
\end{minipage}\\
\begin{minipage}{0.48\textwidth}
\epsfxsize=0.96\textwidth{\epsfbox{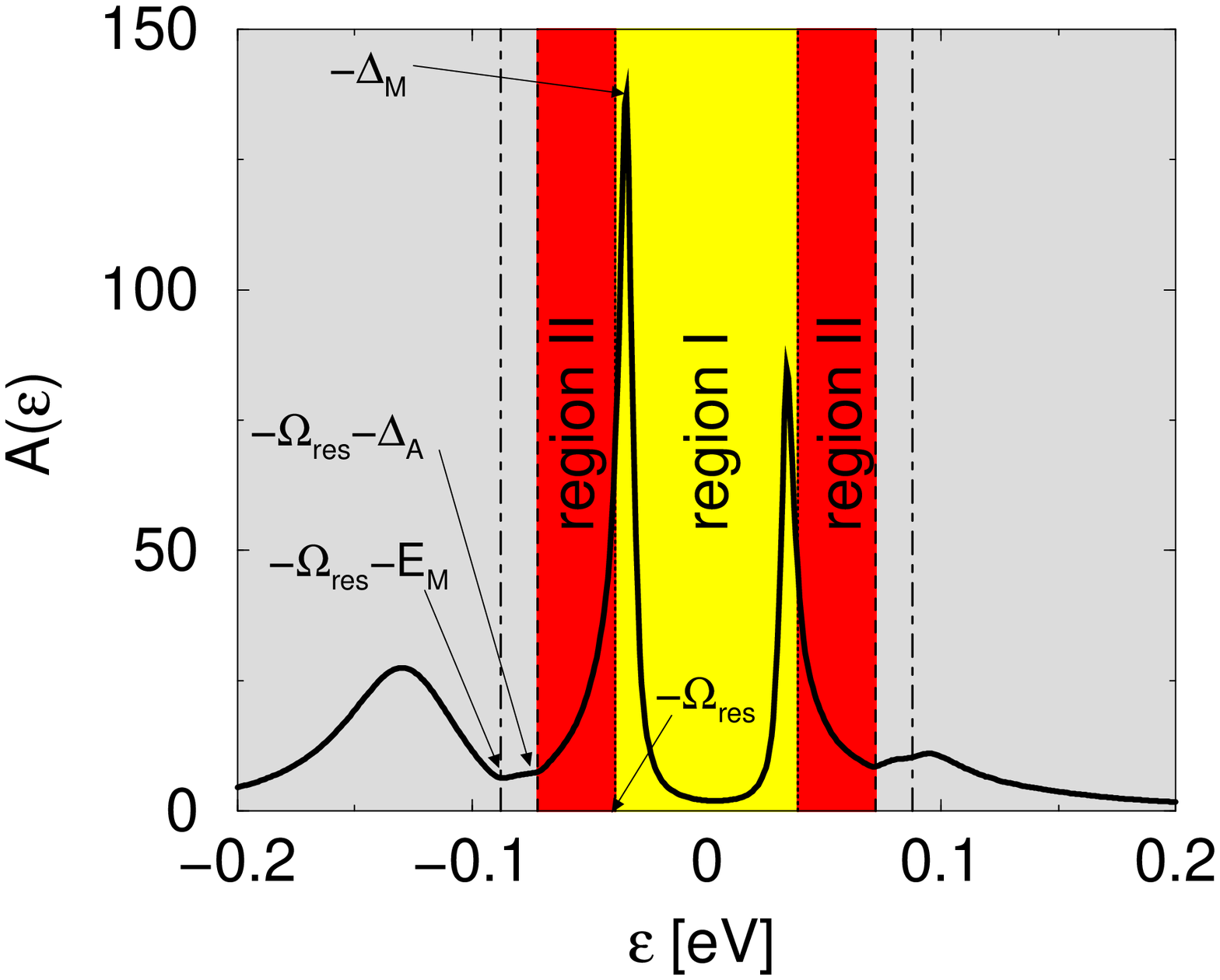}}
\end{minipage}
\hfill
\begin{minipage}{0.48\textwidth}
\epsfxsize=0.96\textwidth{\epsfbox{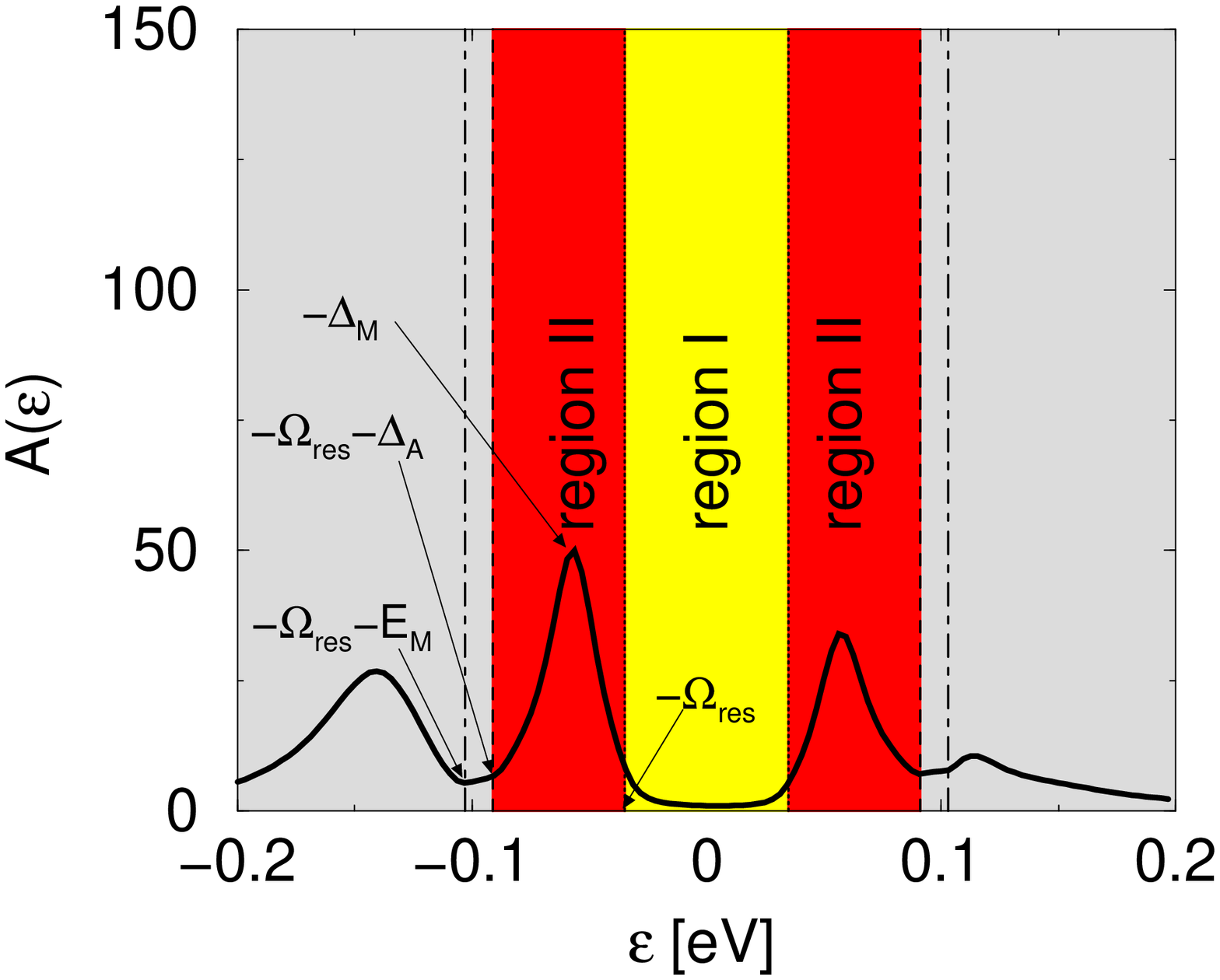}}
\end{minipage}
\caption{
\label{Wwidth}
Spectral functions at $M$ for the self energies shown in the upper panels.
The left two pictures are for optimally doped materials ($\Delta_{max}=35 $meV,
$\Omega_{res}=39$ meV) and the right two pictures for underdoped materials
($\Delta_{max}=60 $ meV, $\Omega_{res}=35 $ meV). A residual quasiparticle
width of $\delta =5$ meV was assumed.
(After Ref. \cite{Eschrig03},
Copyright \copyright 2003 APS).
}
\end{figure}

In Fig.~\ref{Wwidth}, we present the results for the spectral function
at the $M$ point of the Brillouin zone for a perfectly sharp resonance
(a finite width of the resonance of 10 meV does not change the results
significantly, except a slight reduction of the spectral peak height \cite{Eschrig03}). A residual quasiparticle broadening parameter of
$\delta = 5 $ meV was used. 
The main features of the spectral function is the dip feature at an energy
of about the resonance energy relative to the peak \cite{Norman97,Shen97,Norman98,Shen98,Abanov99}. 
The peak position at $-\tilde E_M$ is renormalized by self-energy effects
discussed above,
and is shifted from the bare $-E_M$ to be near $-\Delta_M$.
The dip feature is actually
spread out over a range of size $E_M-\Delta_A$, and it is the onset of
this dip feature which defines the resonance energy, $\Omega_{res}$.
The dip feature is followed by a hump at higher binding energies, and the
position of the hump maximum is very sensitive to the 
coupling constant and to damping due to the spin fluctuation continuum, as
we show later. Thus, we concentrate in the following on the peak-dip structure.
Another feature worth mentioning is the asymmetry of the {\it lineshape}
at positive and negative binding energies, with a relatively weak 
dip feature on the unoccupied side compared to the occupied side.

The important difference between the left and the right pictures in
Fig.~ref{Wwidth} points out the role of the two regions, region I and II,
for the spectral functions. In the left picture, the maximal gap $\Delta_{max}$
is less than the  resonance frequency $\Omega_{res}$. Consequently,
the quasiparticle peaks are situated in region I, slightly below the
onset of damping due to scattering with the resonance mode.
For overdoped materials it moves even further away from region II, becoming
sharper, and the peak width is set in this case
by the residual broadening due to other processes.
In contrast, in the right picture of Fig.~\ref{Wwidth} the quasiparticle
peak is situated in region II. Here, the maximal gap $\Delta_{max}$ is
larger than the resonance frequency, and strong quasiparticle damping 
reduces the heights of the quasiparticle peaks.
The width of the peak is given in this case by Eq.~(\ref{Gamma_u}).
For decreasing resonance mode energy, 
the peak weight is reduced and the incoherent
part of the spectral function grows, taking weight from the
quasiparticle peak. 
The hump energy is moving to higher binding energy with increasing coupling 
constant $g$ and increasing $\xi_M$. 
The weight of the peak is strongly reduced with increasing coupling constant.
This is not the case with varying $\xi_M$. 

The coherent weight of the quasiparticle peak, $z_M$,
is only weakly dependent on the gap and the band structure in the relevant
parameter range. 
It is proportional to the mode energy $\Omega_{res}$;
together with the experimental finding $\Omega_{res}\propto k_BT_c$, this
means $z_M\propto k_BT_c$. 
For coupling constants of order the band width 
or larger, $z_M\propto 1/(g^2w_Q)$;
for smaller coupling constants, $1/z_M \sim A+Bg^2w_Q$ with 
$A$ and $B$ constants. Finally,
$z_M$ weakly decreases with increasing antiferromagnetic correlation length 
$\xi_{sfl}$.

We can understand some of these features using the approximate expression
of Eq.~(\ref{ZM}). Evaluating $Z_M(\epsilon )$ at $\epsilon=-E_M$, and
taking into account the coherence factor at the $M$ point,
$A^{-}_M \equiv A^{-}_M(-E_M)$, and the nodal renormalization factor
$Z_M^{(N)}\equiv 1+\lambda_M^{(N)}(-E_M)$,
gives \cite{Eschrig03}
\begin{eqnarray}
\label{renfactor}
z_M &\approx & \frac{\Omega_{res}A^{-}_M}{
Z_M^{(N)}\Omega_{res}+\frac{g^2I_0}{\pi(\Omega_{res}+2E_M)} }
\end{eqnarray}
which defines the constants $A$ and $B$.

\subsection{Contribution of the spin fluctuation continuum}
\label{CSFC}

For an understanding of the high-energy part of the electronic
spectra and dispersions it is necessary to include also
the continuum part of the spin fluctuation spectrum.
The onset energy for the spin fluctuation spectrum determines
the onset of strong scattering of electrons and the complete
destruction of quasiparticle excitations.
The spin fluctuation continuum extends to
electronic energies ($\sim $ eV), and as a consequence
the electronic scattering rate will increase continuously with energy up to
electronic energies as well. In this section we discuss the implications of
the additional scattering due to the spin fluctuation continuum in an
extended model including the full model spectrum displayed in Fig.~\ref{diagram},
and described by Eqs.~(\ref{Bcont0}) and (\ref{cq}).
The low-energy gap of $2\Delta_h$ is determined by the superconducting gap at
the `hot spot' wavevectors as discussed in section \ref{CE}.
The momentum dependence of the continuum part of the
spectrum is described well by Eq.~(\ref{cq}),
with a correlation length $\xi_{c}=0.5a$ compatible with experimental
findings for near optimal doping. 
For this correlation length, the momentum average of $c_q$ gives
$0.5 c_Q$.
Comparing with the experimental data for the
momentum averaged susceptibility at 65 meV, which was found
to be $6 \mu_B^2/$eV for underdoped YBa$_2$Cu$_3$O$_{7-\delta}$
in the odd channel, and about $3 \mu_B^2/$eV in the even channel \cite{Fong00},
gives $c_Q \approx 6/ $eV and $3/$eV respectively.
The values for optimal doping should be smaller, and in Ref. \cite{Eschrig03}
$c_Q=5.6/$eV and $g=0.65$ eV was found to reproduce well
the experimental high energy (linear in excitation energy) part of the
momentum linewidth in optimally doped Bi$_2$Sr$_2$CaCu$_2$O$_{8+\delta}$, 
which gives $\Gamma_N=0.75 \epsilon $ \cite{Valla99,Yusof02}.
This coupling includes
both the even and odd (with respect to the bilayer indices)
contributions of the spin fluctuations \cite{Stock05}. 
In contrast, the resonance mode is dominant in the odd channel,
\cite{Rossat93}, and only recently there has been resolved a weaker
resonance mode in the
even channel as well \cite{Pailhes03,Pailhes04}.
It is thus a good approximation to neglect the even channel contribution of
the resonance mode and to only take into account the odd channel mode.

In this section we discuss first theoretical results without taking into
account bilayer splitting. We will turn to the case of bilayer splitting in
the next section.

The continuum eventually will be cut-off at electronic energies. 
However, the precise high-energy behavior of the spin fluctuation continuum
is irrelevant, as it only leads to a contribution to the real part of the
self energy that is almost energy independent (varying on the cut-off energy
scale). This energy independent renormalization can be absorbed into
a change of the bandwidth of the (bare) dispersion $\xi_k$ (and possibly of the
`bare' maximal $d$-wave gap $\Delta_k$). 
The corresponding renormalization factor $Z_{HE}$ has to
be regarded as an additional phenomenological parameter.
The experiments near optimal doping can be reproduced best by
rescaling the dispersion from Table \ref{tab1} in the following way:
\begin{equation}
\label{Zh}
\xi_k^{(new)}=Z_{HE} \xi_k - (Z_{HE}-1) \xi_M
\end{equation}
with $Z_{HE}=1.5$ \cite{Eschrig03}.
With this choice, the van Hove
singularity at the $M$ point has the same distance from the chemical
potential as before. 
The choice of parameters for the extended model 
additional to those in Table~\ref{tab2} is summarized in
Table~\ref{tab4}.
\begin{table}
\begin{center}
\tbl{
\label{tab4} 
Additional parameters for the spin-fluctuation continuum
used in the extended model of Ref. \cite{Eschrig03}.
}
{\begin{tabular}{|c|c|c|c|}
\hline
\hline
$2\Delta_h $& $\xi_c$  & $\frac{1}{2}g^2c_Q$ & $Z_{HE}$ \\
\hline
63 meV & $0.5 a$ & 1.18 eV & 1.5 
\\
\hline
\hline
\end{tabular}}
\end{center}
\end{table}

\begin{figure}
\centerline{
\epsfxsize=0.55\textwidth{\epsfbox{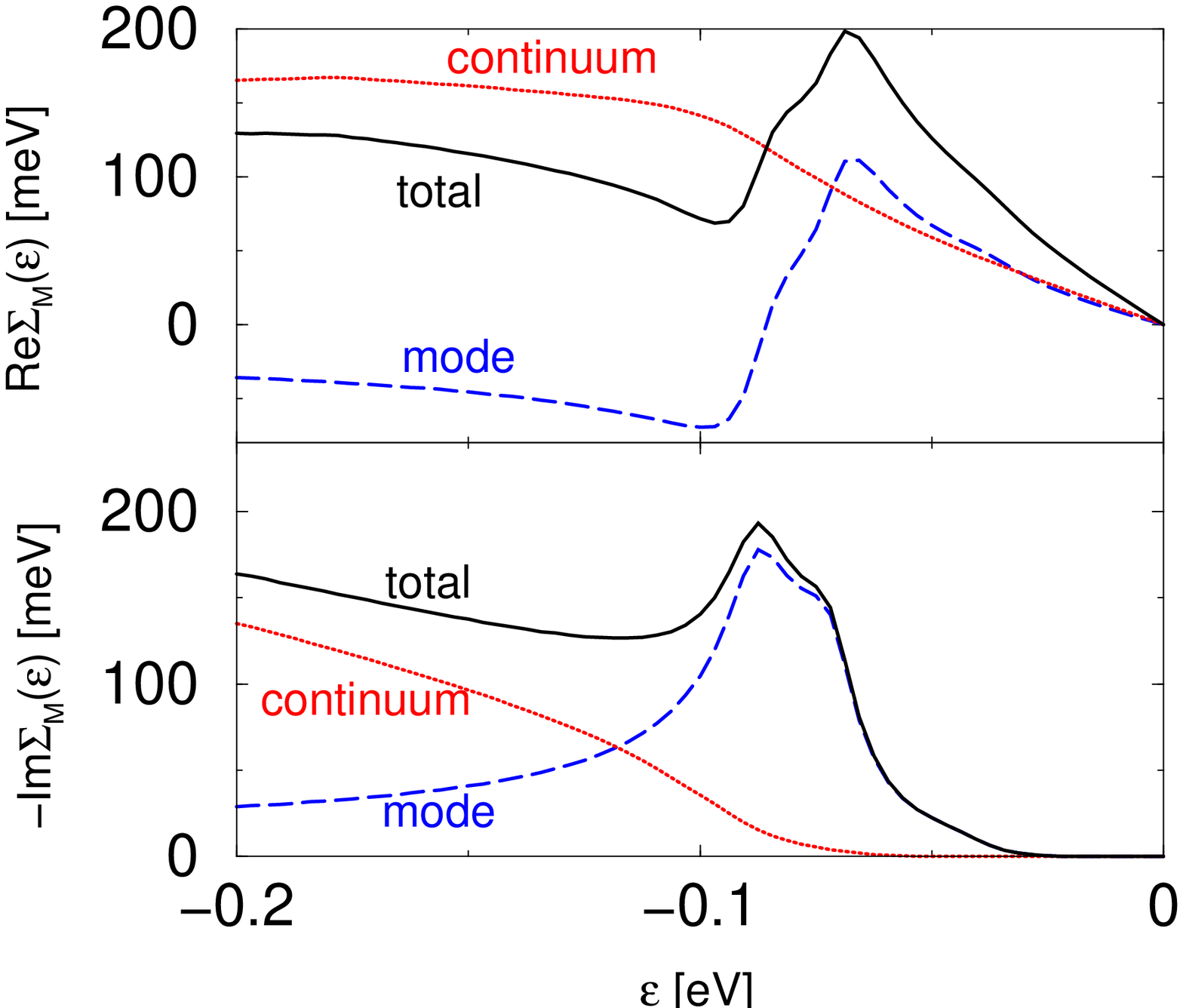}}
\epsfxsize=0.3\textwidth{\epsfbox{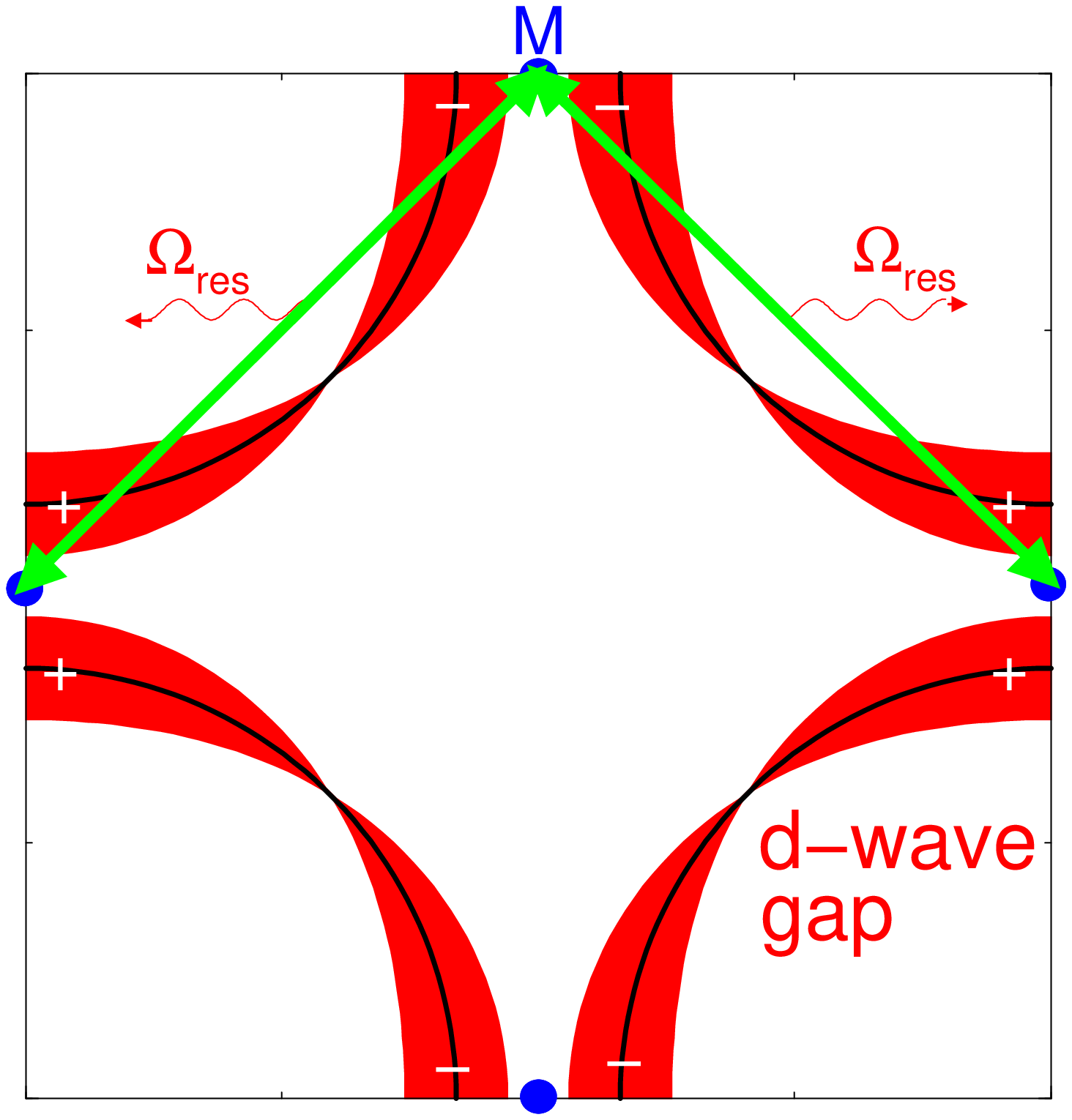}}
}
\caption{
\label{Cont_ModeM}
The different contributions to the real part (top)
and the imaginary part (bottom) of the self energy are shown for the $M$ point.
Dotted curves are the contribution from the
spin fluctuation continuum, dashed the contribution from the spin
fluctuation mode, and full both contributions. 
Calculations were done for the parameters of Tables~\ref{tab2} and \ref{tab4},
except $Z_{HE}=1$. 
The relevant scattering processes are shown on the right.
}
\end{figure}

\begin{figure}
\centerline{
\epsfxsize=0.55\textwidth{\epsfbox{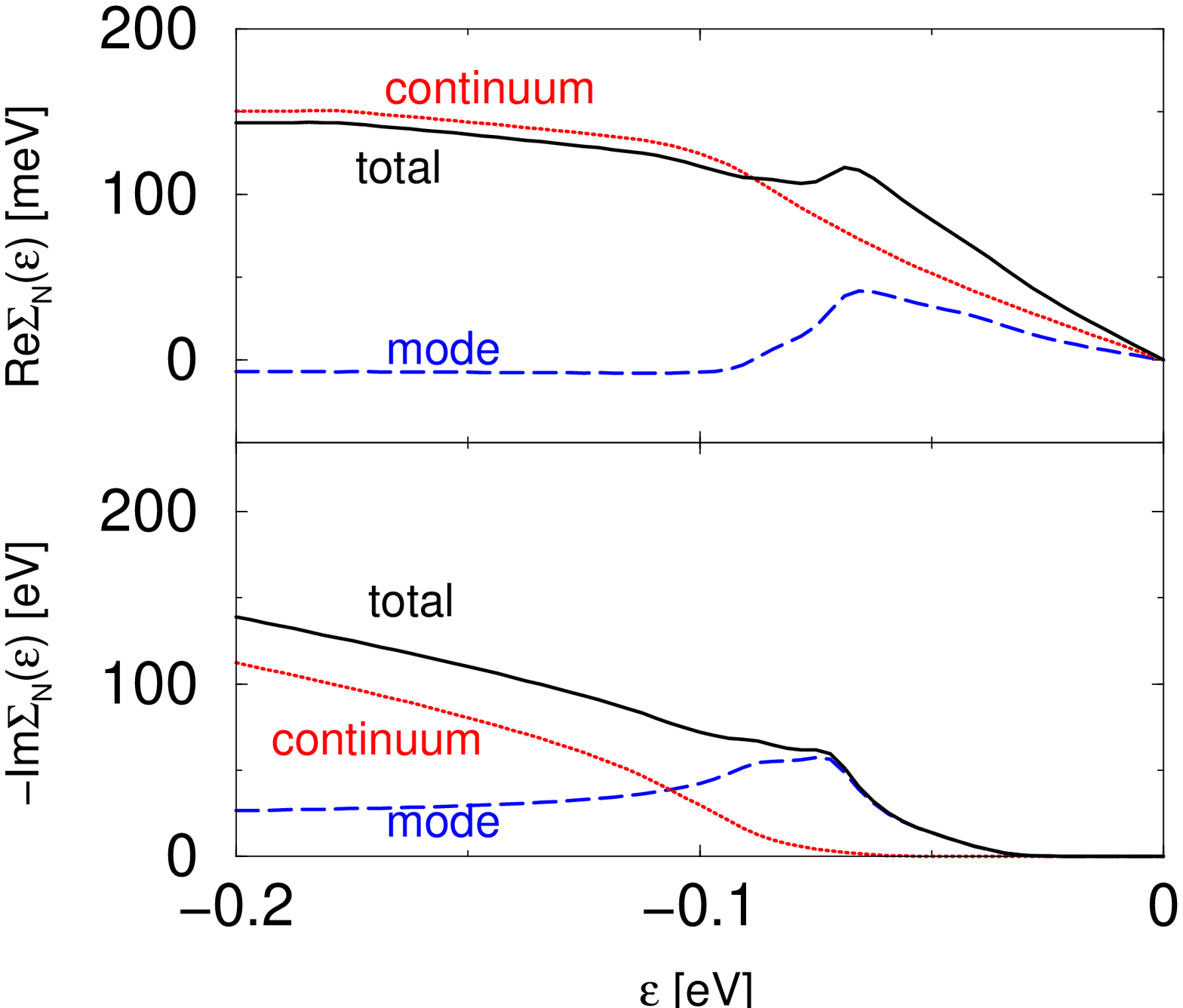}}
\epsfxsize=0.3\textwidth{\epsfbox{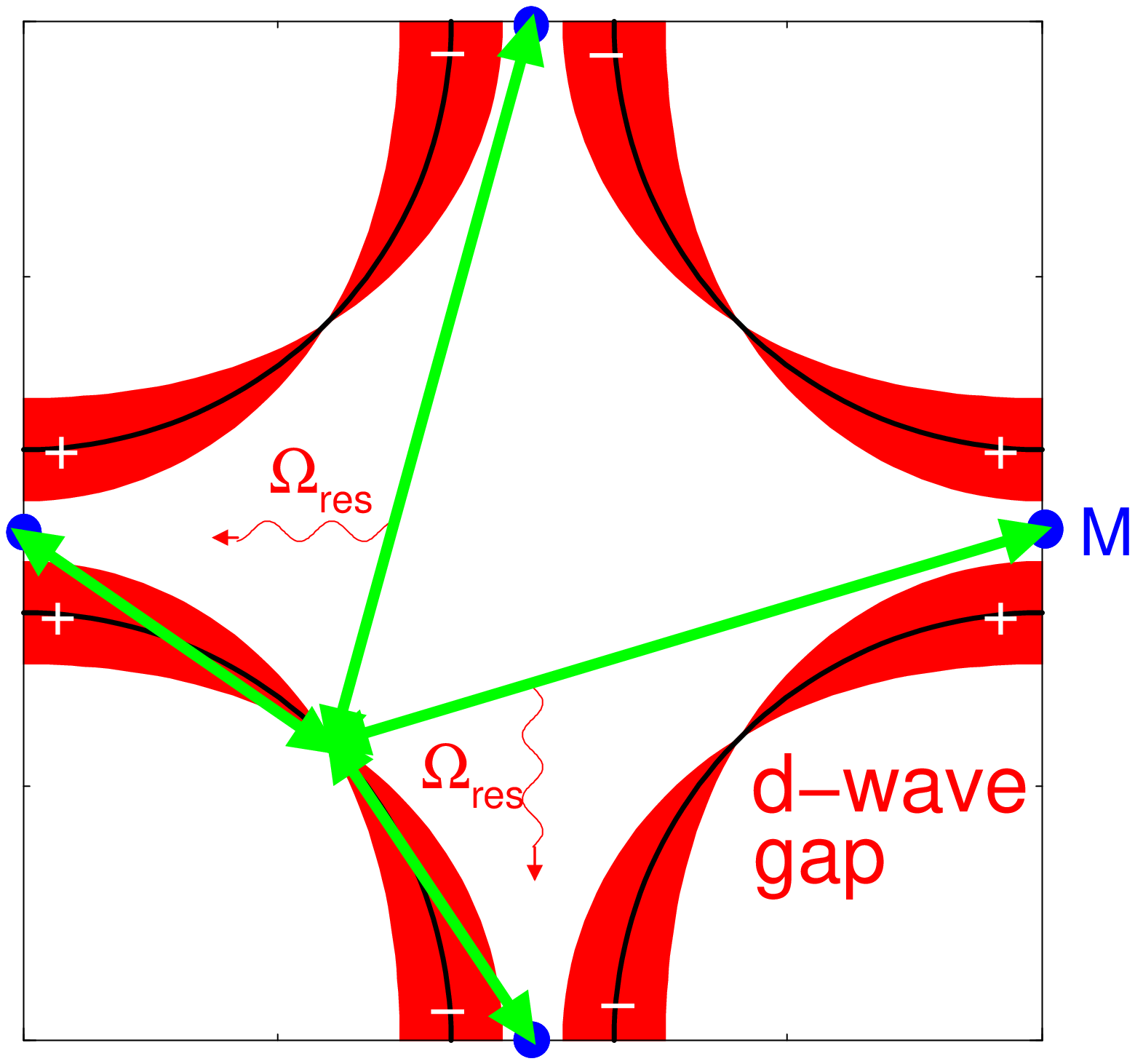}}
}
\caption{
\label{Cont_ModeN}
The different contributions to the real part (top)
and the imaginary part (bottom) of the self energy are shown for the $N$ point.
Dotted curves are the contribution from the
spin fluctuation continuum, dashed the contribution from the spin
fluctuation mode, and full both contributions.
Calculations were done for the parameters of Tables~\ref{tab2} and \ref{tab4},
except $Z_{HE}=1$. 
The relevant scattering processes are shown on the right.
}
\end{figure}

The main results for the self energy effects are summarized for the
$M$ point and the nodal point of the Brillouin zone in Figs.
\ref{Cont_ModeM} and \ref{Cont_ModeN}.
The continuum contribution to the self-energy
is shown as a dotted line, and the mode contribution as dashed line.
As can be seen from the figures, the
continuum contribution to the scattering rate sets in {\it above} the 
structures which are induced by the mode. 
It also contributes considerably
to the renormalization of the low-energy dispersion, as the slope of the
real part of the self energy shows.
The high-energy region of the imaginary part of the self energy is
drastically modified by the continuum contribution, and shows 
a region of linear in energy increase extending to rather high energies
(the high-energy cut-off for the continuum spectrum in the calculations).
In contrast, the mode contribution to the imaginary part of the
self energy is in the high-energy region decaying with increasing energy.
An interesting feature is that the mode contributes to the real part of
the self energy a rather {\it constant negative} contribution.
This feature will have interesting consequences when turning to the
case of bilayer splitting (see next section),
where this constant high-energy contribution 
to the real part of the self energy can differ
for bonding and antibonding bands.

The self energy at the nodal point is very similar to the experimentally
observed ones along the nodal direction \cite{Kaminski00,Kaminski01},
apart from the low-energy scattering gap, which in experiment shows
additional contributions that are not taken into account in the present
theory. These additional contributions most probably stem from the
incommensurate part of the spin-fluctuation spectrum, which couples
most strongly to the nodal quasiparticles.

The difference in the relative magnitude between the mode contribution 
and the continuum contribution in Figs.~\ref{Cont_ModeM} and
\ref{Cont_ModeN} is explained in terms of different scattering geometries
in the right panels of the figures. Because the mode is sharper in momentum
than the continuum is, its contribution in the nodal case is reduced by
the fact that the momentum vectors connecting the nodal points to the 
antinodal regions are far off the $(\pi,\pi)$ wavevector (see
Fig.~\ref{Cont_ModeN} right). Thus, the effect of the mode is reduced at the
nodes, and in strongly overdoped materials in fact not observed in experiment 
(see Fig.~\ref{Fig1_Cuk04}) \cite{Gromko03,Cuk04}.
It is however clearly observed as a kink in Im$\Sigma $
for underdoped and moderately overdoped materials,
as seen from Fig.~\ref{Fig1_Kordyuk04} \cite{Kordyuk04}.
This kink is consistent with the mode contribution
in the theoretical results shown in Fig.~\ref{Cont_ModeN},
but not with the rather smooth onset of the 
continuum contribution only (dotted line).

Finally, note that for underdoped and optimally doped materials,
where in the normal state the even channel stays gapped  \cite{Reznik96},
the corresponding contribution of the even channel spin susceptibility
to the normal state self energy is
given by half the continuum contribution in Fig.~\ref{Cont_ModeN}.
This will induce a weaker kink feature in the normal state at an energy equal 
to the even channel (optical) gap in the spin susceptibility, which
is around 50-60 meV.

\subsection{Renormalization of EDC and MDC dispersions }
\label{RenEDCMDC}

In this section theoretical results for the spectral functions and dispersion
anomalies in a model using both the mode and the gapped continuum of the
spin fluctuation spectrum are summarized. Results are shown
for both EDC and MDC derived dispersions. 

\begin{figure*}
\centerline{
\epsfxsize=0.39\textwidth{\epsfbox{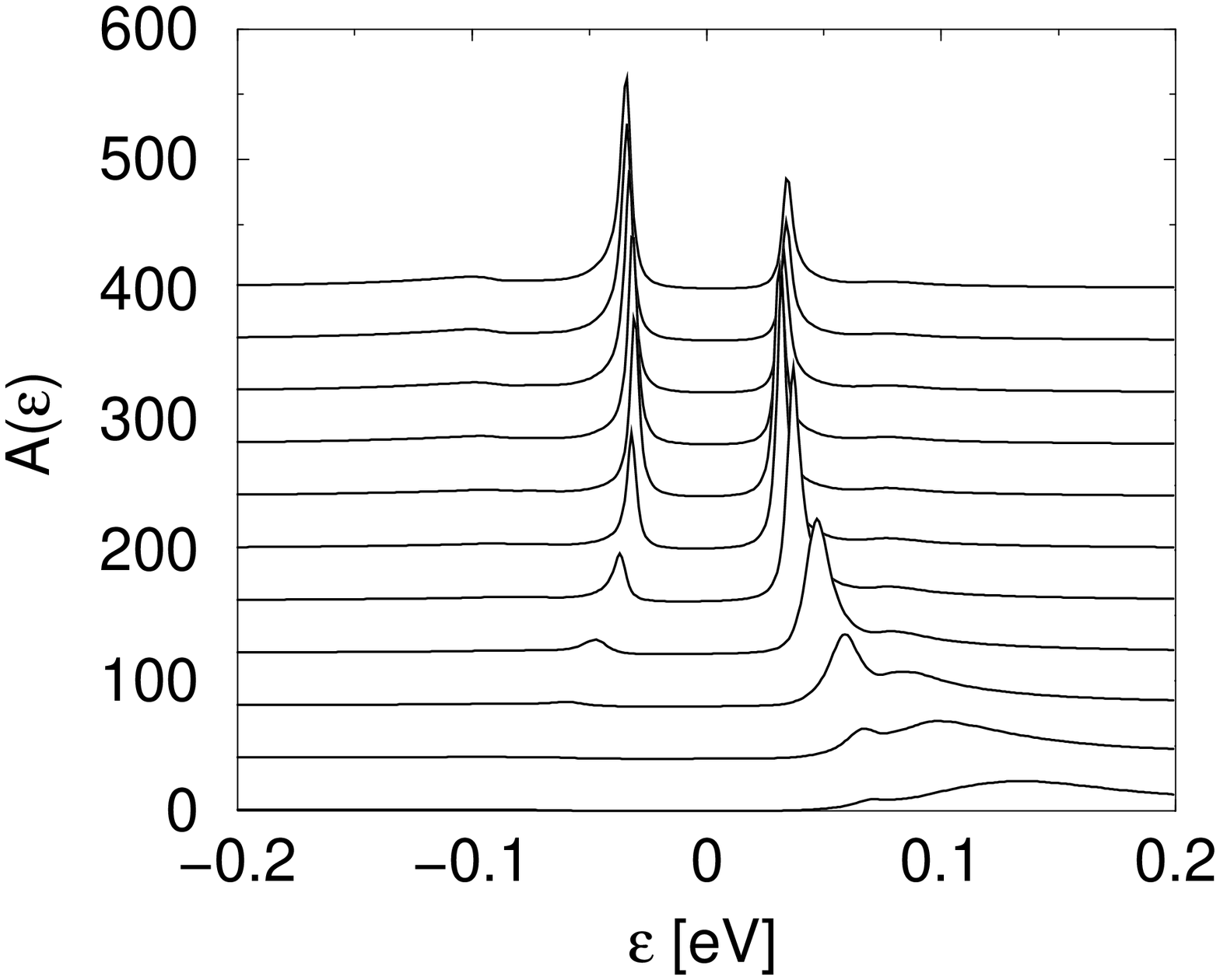}}
\epsfxsize=0.39\textwidth{\epsfbox{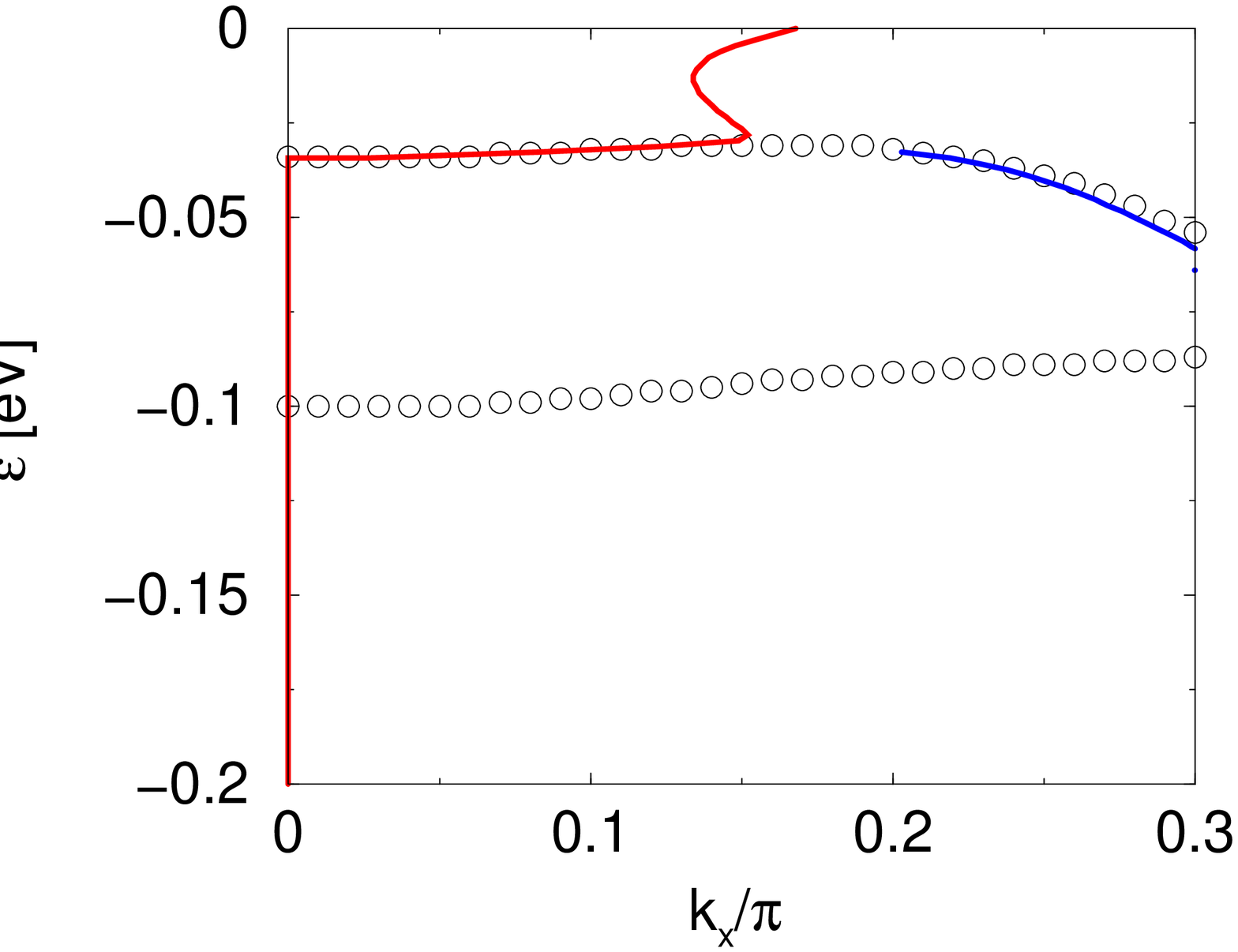}}
\epsfxsize=0.2\textwidth{\epsfbox{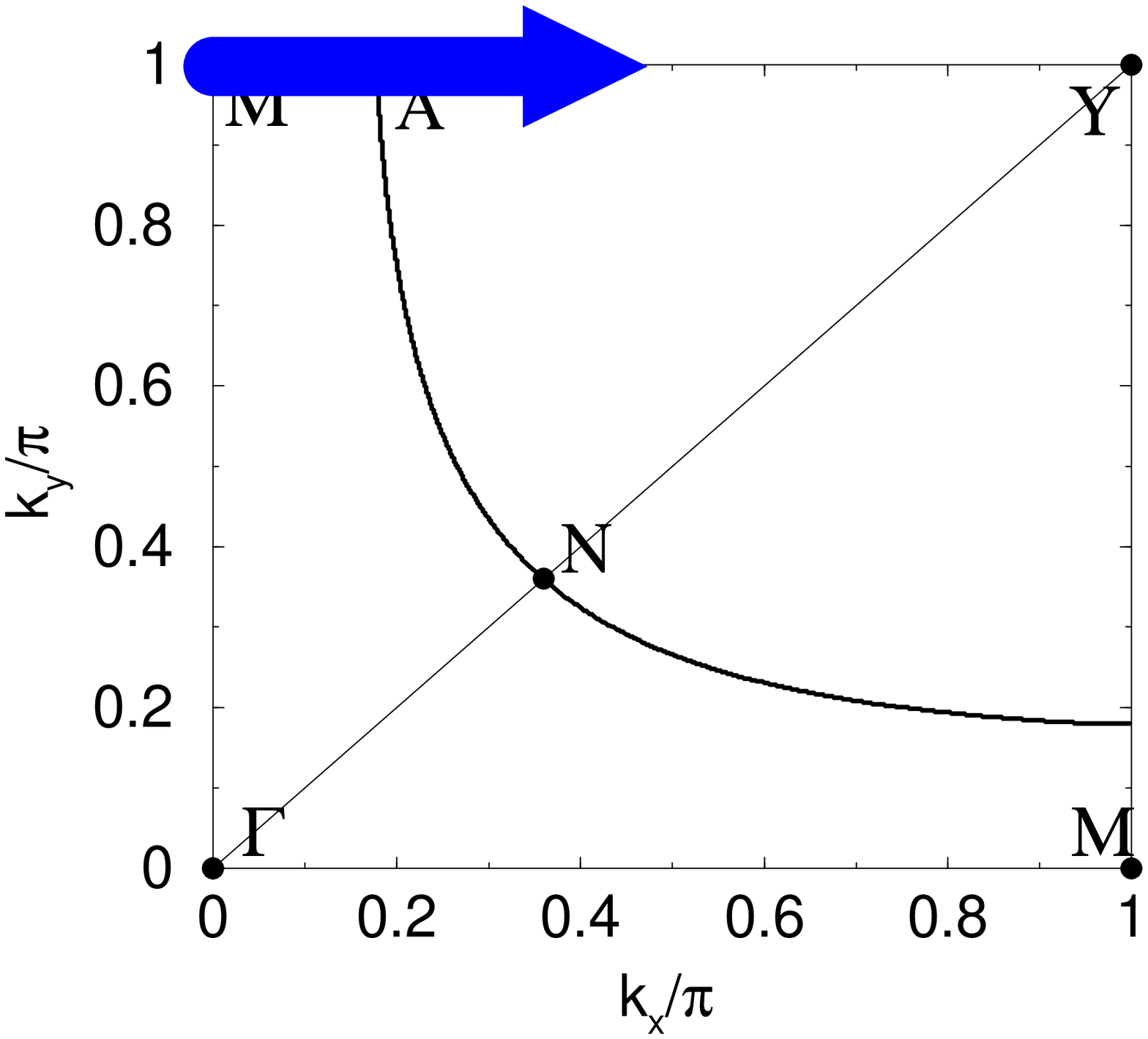}}
}
\caption{
\label{QX1}
Left: Dispersion of the spectral intensity and lineshape 
as a function of momentum along the $M-Y$ cut, ($k_y=\pi $,
$k_x=0...0.4\pi$ in steps of $0.04\pi $ from top to bottom). 
Middle: EDC (circles) and MDC (curve) dispersions from maxima of
the curves shown in the left panel. In the EDC dispersion, the low
energy peak and the high energy hump with the break feature in
between is clearly visible. Because the bottom of the normal
state dispersion is at $\xi_M=-34 $meV, the MDC shows only a
broad maximum at $M$ for high energies.
(After Ref. \cite{Eschrig03},
Copyright \copyright 2003 APS).
}
\end{figure*}

\begin{figure}
\centerline{
\epsfxsize=0.39\textwidth{\epsfbox{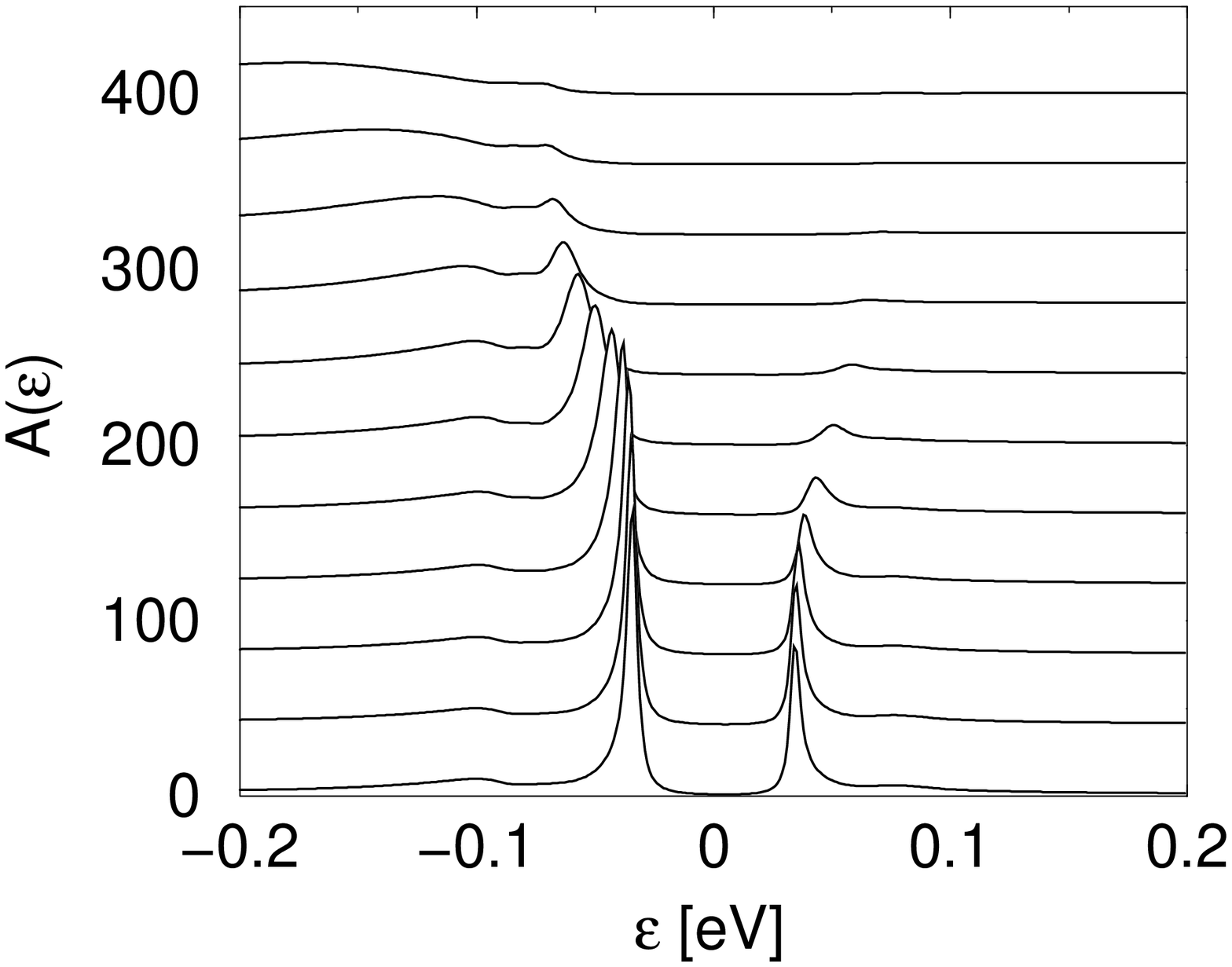}}
\epsfxsize=0.39\textwidth{\epsfbox{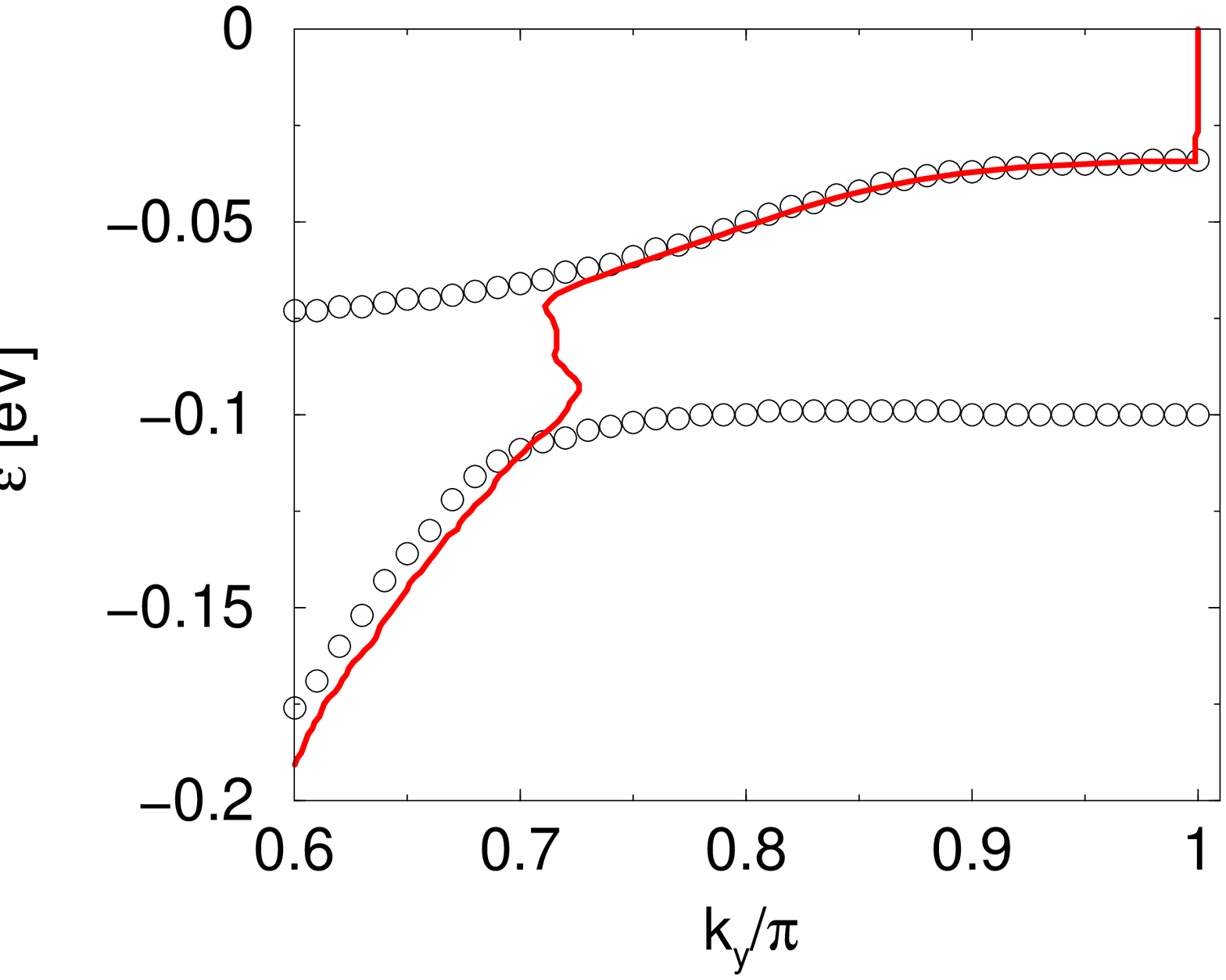}}
\epsfxsize=0.2\textwidth{\epsfbox{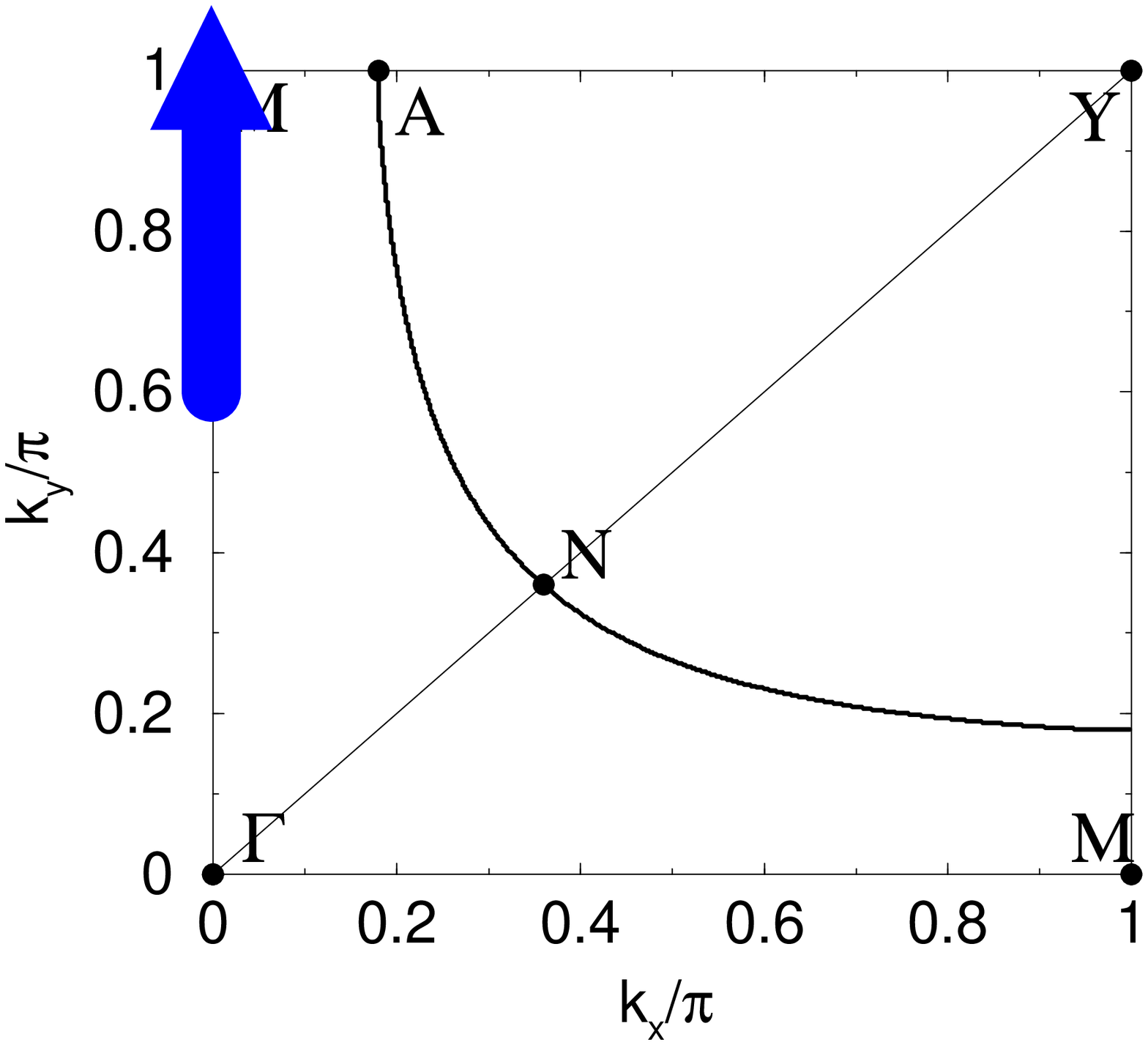}}
}
\caption{
\label{QY2}
Left: Dispersion of the spectral intensity and lineshape 
as a function of momentum along the $M-\Gamma $ cut
($k_x=0 $,
$k_y=0.6\pi..\pi$ in steps of $0.04\pi $ from top to bottom).
Middle:
EDC (circles) and MDC (curve) dispersions from maxima of
the curves shown in the left panel.
(After Ref. \cite{Eschrig03},
Copyright \copyright 2003 APS).
}
\end{figure}

\begin{figure*}
\centerline{
\epsfxsize=0.39\textwidth{\epsfbox{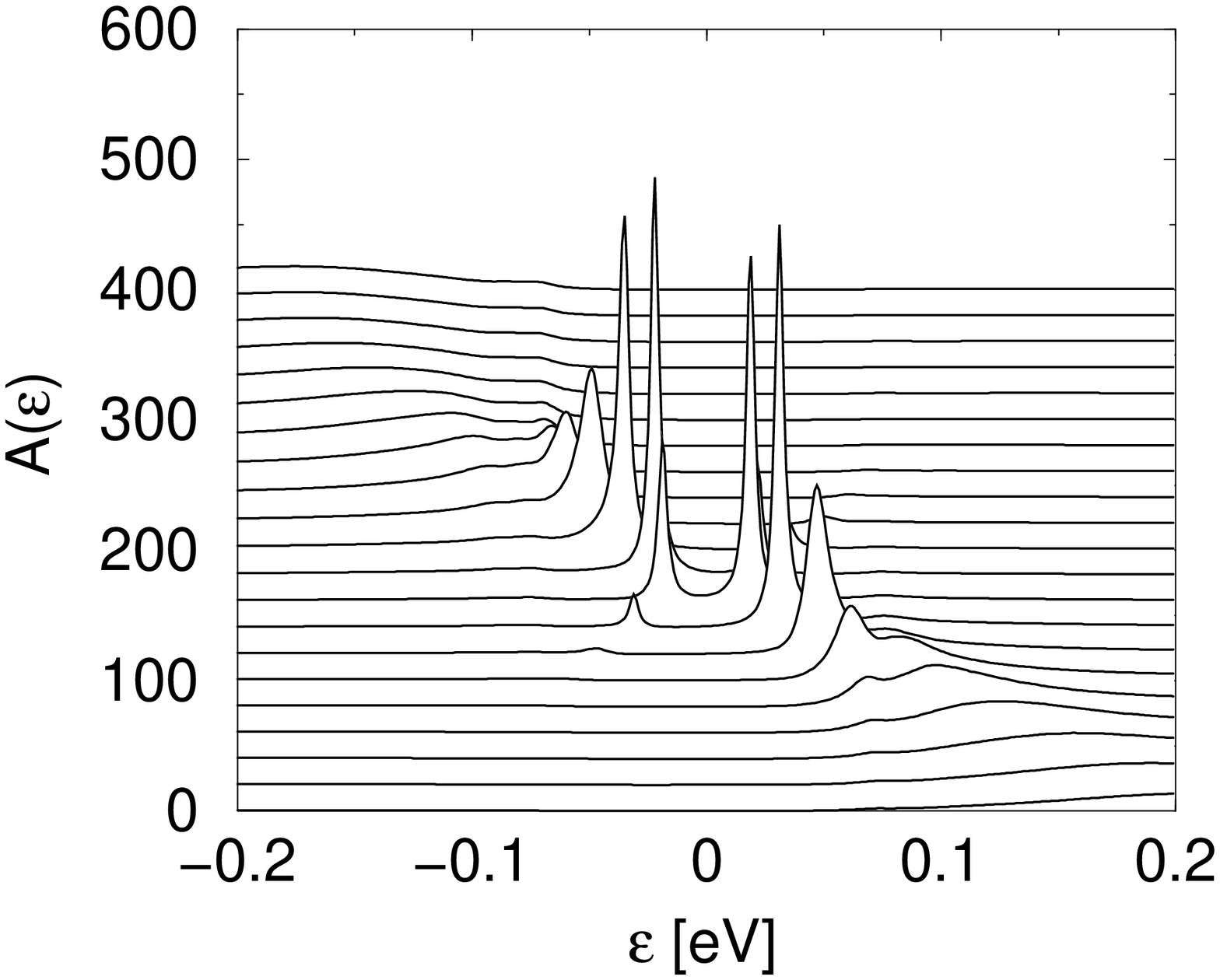}}
\epsfxsize=0.39\textwidth{\epsfbox{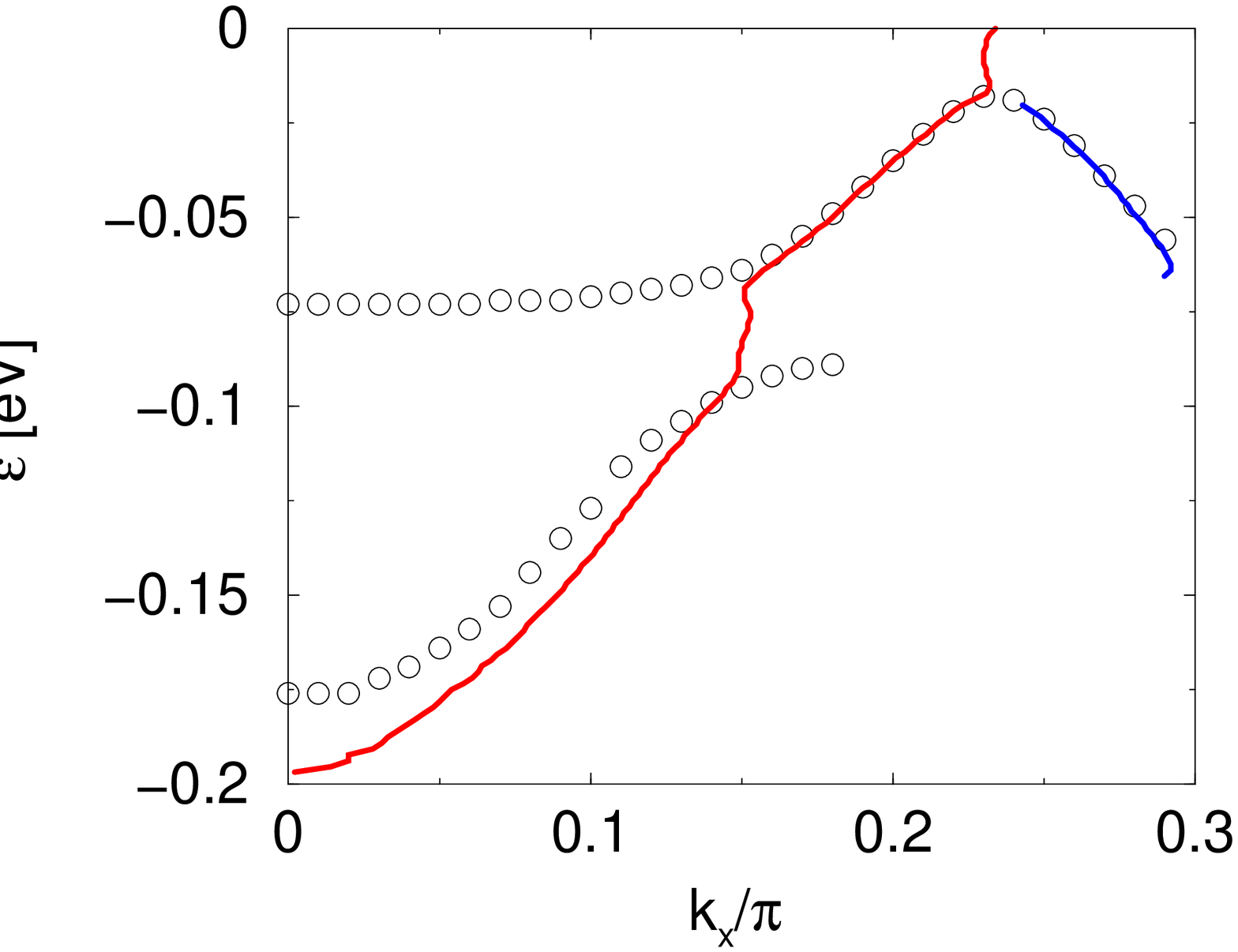}}
\epsfxsize=0.2\textwidth{\epsfbox{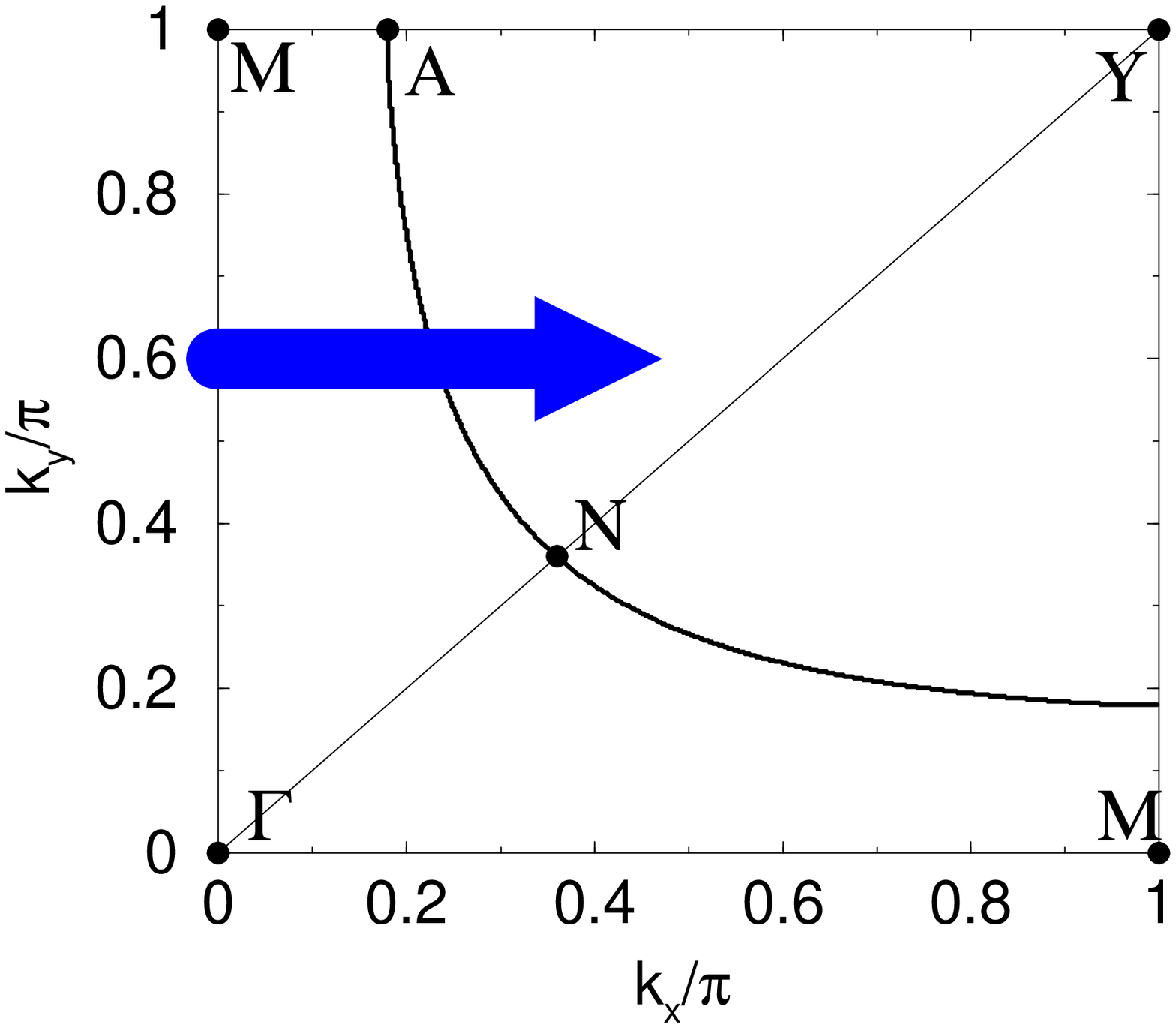}}
}
\caption{
\label{QX2}
Left: Dispersion of the spectral intensity and lineshape 
as a function of momentum $k_y=0.6\pi $,
$k_x=0...0.4\pi$ in steps of $0.02\pi $ from top to bottom. 
Middle: EDC (circles) and MDC (curve) dispersions from maxima of
the curves shown in the left panel. 
(After Ref. \cite{Eschrig03},
Copyright \copyright 2003 APS).
}
\end{figure*}

\begin{figure}
\centerline{
\epsfxsize=0.39\textwidth{\epsfbox{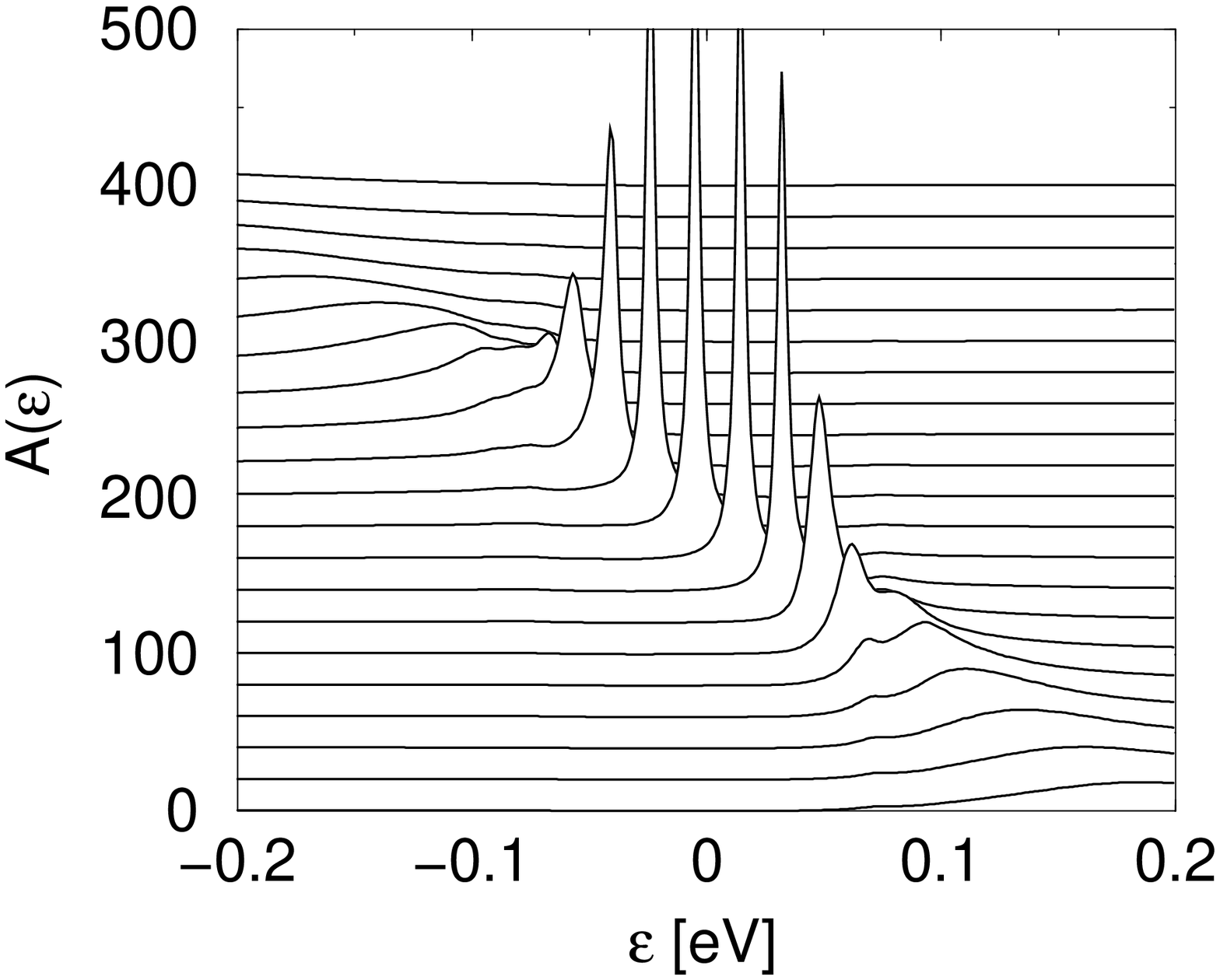}}
\epsfxsize=0.39\textwidth{\epsfbox{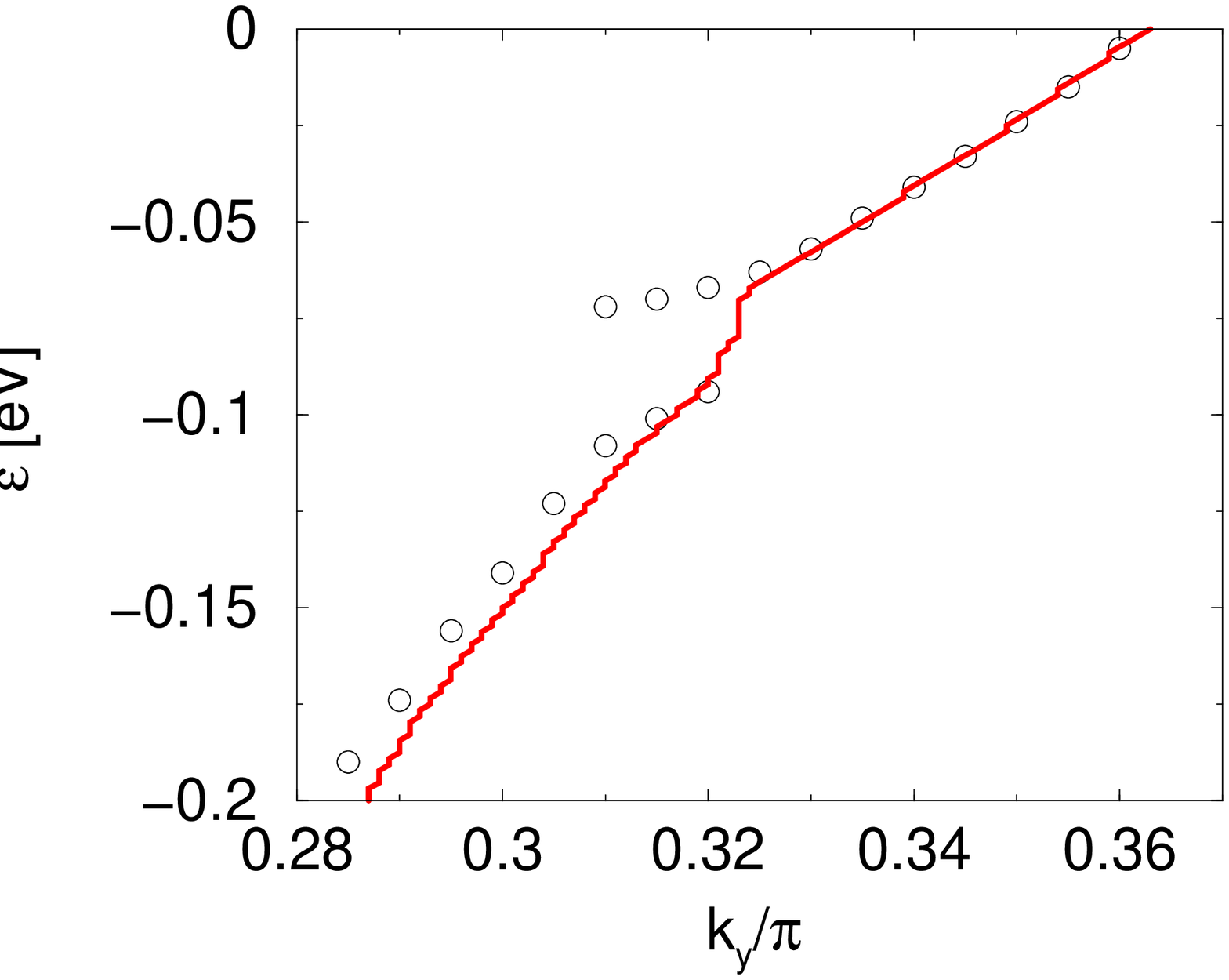}}
\epsfxsize=0.2\textwidth{\epsfbox{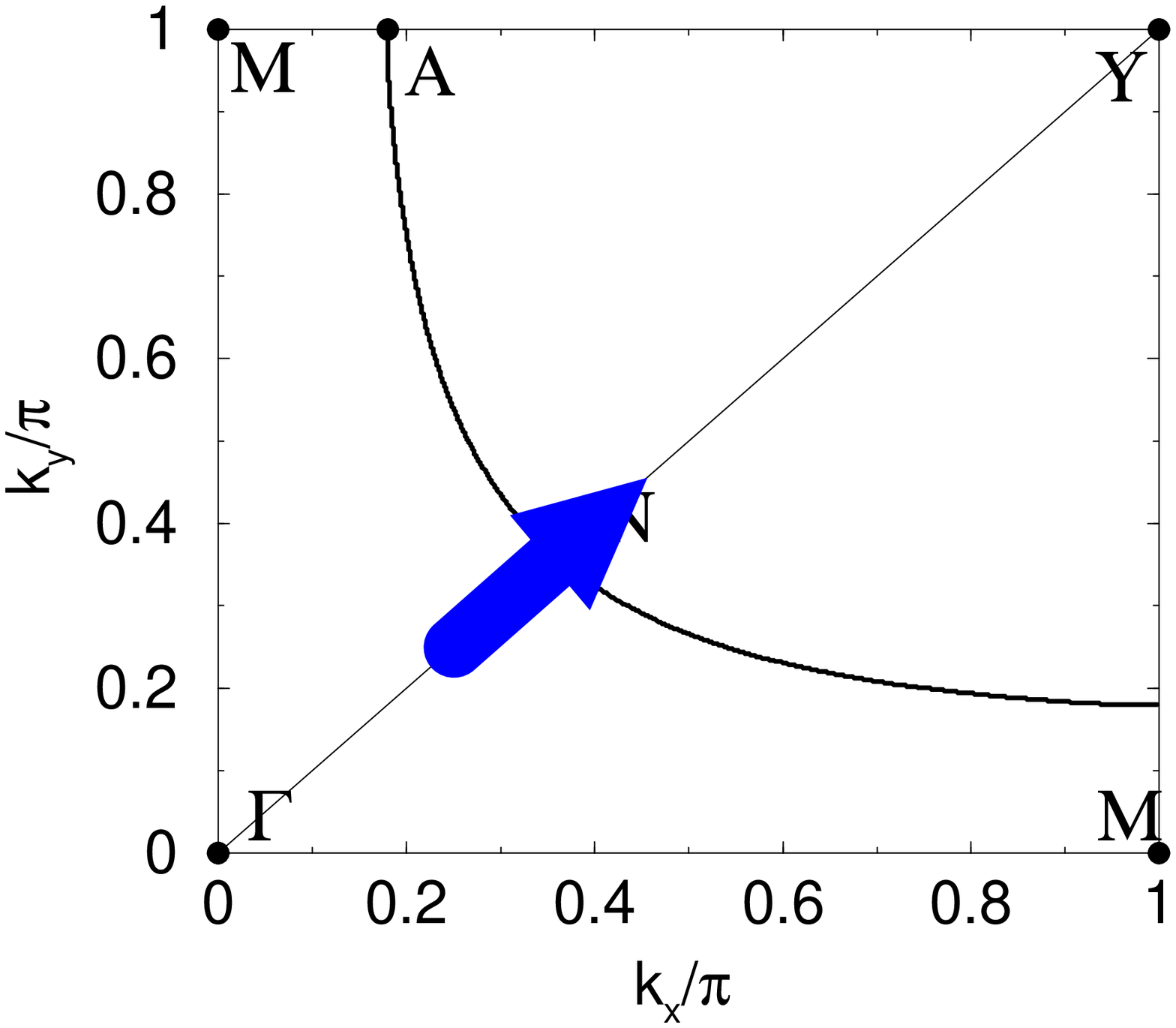}}
}
\caption{
\label{NODE2}
Left: Dispersion of the spectral intensity and lineshape
in the nodal direction ($\Gamma-Y$) as a function
of momentum $k_x=k_y=0.25\pi...0.45\pi$ in steps of $0.01\pi$ from
top to bottom.
Middle: the corresponding EDC (circles) and MDC (curve) 
dispersions.
The kink is most clearly seen in the MDC dispersion. The low energy 
velocity is roughly half the high energy one. The high energy dispersion does 
not extrapolate to the Fermi surface crossing.
(After Ref. \cite{Eschrig03},
Copyright \copyright 2003 APS).
}
\end{figure}

In Figs.~\ref{QX1}-\ref{NODE2}, ARPES spectra and corresponding dispersions
are shown along several selected paths in the
Brillouin zone, indicated in the right panels of the figures. 
In the left panels of the figures, 
the intensities and spectral lineshapes
can be followed, and in the middle panels, the corresponding dispersions of
the peak maxima and hump maxima in the EDC's are shown as circles, and the
maxima in the corresponding MDC dispersions as curves.
As can be seen from these figures, the linewidth of the spectral features
are in general considerably broader in the high-energy region than in
the low-energy region. This is in agreement with the experiments.

In  Fig.~\ref{QX1} the cut is going from the $M$ point of the Brillouin zone toward the $Y$ point. The $A$ point (normal state Fermi crossing)
corresponds to spectra roughly in the middle of the set.
Sharp quasiparticle excitations are present between the $M$ and $A$ points,
and the dispersion (both EDC and MDC derived) is very flat, in agreement
with the experimental findings of Ref.~\cite{Kaminski01} (see Fig.~\ref{Fig4_Kaminski01} top panel).
The MDC variation within the gap edge is observed experimentally
(see section~\ref{S-shape}) and has been discussed in Ref.~\cite{Norman01a}.
At high energies, the MDC is peaked at $M$.

Fig.~\ref{QY2} shows results along a cut in $M$-$\Gamma$ direction.
The main features here are summarized as follows. There is an extremely flat behavior in the
EDC hump-dispersion in the region between the $M$ point and
roughly $0.3\pi$ from there in direction of $\Gamma $,
and in this region the peak disperses only moderately. Quasiparticle peaks are observed in
the entire region where the hump dispersion stays flat.
The MDC shows an $S$-shaped behavior in the break region between the EDC peak and
hump, and roughly at the point where the hump
starts to disperse strongly away from the chemical potential.
There is a weak maximum in the hump dispersion at $q_y\approx 0.85\pi$,
that is due to the coupling of the $(\pi,0)$ and $(0,\pi)$ points by self-energy
effects. This maximum corresponds to the Fermi crossing in the path
displaced by $(\pi,\pi)$ from the one shown in the right panel
of Fig.~\ref{QY2}, and that corresponds to the one shown in the right panel of
Fig.~\ref{QX2}. This effect was observed experimentally \cite{Campuzano99},
as shown in Fig.~\ref{Fig3_Campuzano99}.

Fig.~\ref{QX2} shows the results for a cut parallel to the $M-Y$
between the antinodal and the nodal point, keeping $q_y=0.6\pi $ constant.
This cut compares to the experimental findings in Fig.~\ref{Fig4_Kaminski01}
and Fig.~\ref{Fig1_Norman01a}.
At low energies, the spectral evolution, seen on the left part of
the figure, shows the typical BCS mixing between particle and hole
states. 
The EDC dispersions show a low-energy peak branch and a high-energy
hump branch, separated by a dispersion break. The MDC shows in 
the break region an $S$-shaped dispersion anomaly. 
As was discussed in section \ref{S-shape}, such $S$-shaped
MDC derived dispersions correspond to important many-body
renormalization effects in the superconducting state dispersion
relative to that in the normal state and are observed experimentally.
The MDC dispersion
changes from the low energy peak branch to the high energy hump
branch at roughly the point where the intensity of the peak drops
dramatically. 
Note that the EDC and MDC dispersions are considerably displaced
relative to one another at high energies. Also at low energies, the MDC 
dispersion is stronger
near the break region than the EDC dispersion. This effect increases
when the residual width of the quasiparticle peak increases, and when
convolution with the experimental resolution function is taken into
account \cite{Norman01a}.

It was mentioned in section \ref{S-shape} that
the $S$-shaped dispersion in the MDC spectra is not observed experimentally in nodal
direction, but is replaced by a kink-like feature.
The fact, that an $S$-shaped dispersion is not observed in the nodal MDC spectra
was shown to be inconsistent with an interpretation of the nodal self-energy
effects in terms of electron-phonon coupling only \cite{Chubukov04}. 
In Fig.~\ref{NODE2} we show the theoretical results for the
cut along the nodal direction.
Along the nodal direction the superconducting gap vanishes as a consequence of 
$d$ wave symmetry, and as a result there is a Fermi crossing of the
dispersion. As can be seen in the left panel of Fig.~\ref{NODE2},
The dispersions kink in the MDC derived dispersion is
observed in experiments along the nodal direction. It is reproduced rather well
in the middle panel of Fig.~\ref{NODE2}.
This is a result of the presence of the gapped continuum in addition to the
sharp mode \cite{Eschrig03,Chubukov04}.
Quasiparticle peaks are well defined only in the region below the kink energy,
as seen clearly in the left panel of that figure.
Note that the kink corresponds to the binding energy $\Omega_{res}+\Delta_A$,
where $\Delta_A$ is the gap at the antinodal Fermi surface point.
The damping sets in at binding energy $\Omega_{res}$, 
at slightly {\it lower} energies, due to the onset of
node-node scattering processes, as can be
seen in the left panel of Fig.~\ref{NODE2}.
The velocity renormalization for low energies and high energies differs
by a factor of roughly two, both for EDC's and MDC's, in
agreement with experiment \cite{Kaminski01}. 
The high energy dispersion does not
extrapolate to the Fermi crossing, again in agreement with experiments
\cite{Bogdanov00,Lanzara01}.
Again, note some 
shift between the EDC and MDC dispersions at high energies due to the energy 
variation of the self energy.

Clearly, the velocity break (kink) along the nodal direction and the
break between the peak and hump (dip) near the $M$ point
are occurring in the same energy range between $-\Omega_{res}-\Delta_A$
and $-\Omega_{res}-E_M$ \cite{Eschrig00,Eschrig03}. This is an appealing result
because it explains all features in the dispersion anomalies
in the Brillouin zone seen by ARPES within the same model.

\subsection{Bilayer splitting}
\label{BLS}

ARPES experiments on
bilayer cuprate superconductors have been able
to resolve a bilayer splitting between bonding and antibonding 
bands \cite{Feng02,Gromko02,Kordyuk02,Kordyuk02a}.
The dispersion near the $(\pi,0) $ point of the Brillouin zone
shows an unusual asymmetry between bonding and antibonding 
self-energy effects.  
In particular, Feng {\it et al.} \cite{Feng02} found that the EDC-derived
dispersions
in overdoped Bi$_2$Sr$_2$CaCu$_2$O$_{8+\delta}$ 
($T_c=65$K) consist of three features: an antibonding band (AB)
peak near 20 meV, 
a bonding band (BB) peak near 40 meV, and a
bonding hump near 105 meV.  Gromko {\it et al.} 
\cite{Gromko04,Gromko03}
reported strong self-energy effects in the 
dispersions derived from MDC's in similar samples ($T_c=58$K).
Near momentum $(k_x,k_y)=(1,0.13)\pi/a$, an $S$-shaped
dispersion anomaly, discussed previously in Ref. \cite{Norman01a},
was shown to be present only in the bonding band MDC,
at binding energies between 40 meV and 60 meV.  In both experiments,
a low energy double peak structure in the EDC was only resolvable in the
same momentum region.
In recent experiments by Borisenko {\it et al.} \cite{Borisenko05,Borisenko06} 
it was found that
the asymmetry between bonding and antibonding self-energy effects
leads to a different behavior in the imaginary part of the self energy
as a function of binding energy. This behavior was previously predicted in
Ref.~\cite{Eschrig02}, where it was shown to be a result of the
odd parity of the resonance mode with respect to the exchange of the
bilayer indices. It is argued there, that low energy scattering of electrons 
between the bonding and antibonding bands is strong
compared to scattering within each of those bands.
As scattering events which connect different bilayer bands
are odd with respect to permutation of the layers within a bilayer, this
implies that the corresponding bosonic excitations which mediate
such scattering must be dominant in the odd channel.
The model explains all of the above cited experimental features, and 
the very recent experiments \cite{Borisenko05,Borisenko06} give strong additional support
for the model.

For electrons phase coherent between the two planes of a bilayer,
the spectra will exhibit separate
bonding ($b$) and antibonding ($a$) features with (normal state)
dispersions given by Eq.~(\ref{BilSpl}).
In the superconducting state, the dispersions are modified by the presence
of the $d$-wave order parameter Eq.~(\ref{DWOP}).
In agreement with experiment \cite{Feng02,Kordyuk02,Kordyuk02a},
$\Delta_M$ is assumed to be the same for the bonding and antibonding bands.
Then, the BCS dispersion in the superconducting state takes the form
\begin{equation}
\label{nonint}
E^{(a,b)}_{\vec{k}}= \sqrt{(\xi^{(a,b)}_{\vec{k}})^2+(\Delta_{\vec{k}})^2}.
\end{equation}

It is clear that the observed dispersion features are many-body effects
beyond Eq.~(\ref{nonint}). In the following we review  the theoretical
results from Ref. \cite{Eschrig02} and compare them directly with the
experimental results. The employed model includes coupling of electrons
to the resonant spin-1 mode as well as a gapped spin fluctuation continuum.
From section~\ref{secBil} the spin susceptibility 
in bilayer materials is a matrix in the 
layer indices, having elements diagonal 
($\chi_{aa}$, $\chi_{bb}$) and off-diagonal $(\chi_{ba}$, 
$\chi_{ab}$) in the bonding-antibonding representation. 
The resonance part, $\chi_{res}$, was experimentally found to be dominated by
the odd channel,
whereas the continuum part, $\chi_{c}$, enters in both \cite{Stock05}.
Assuming that the intensity of the even resonance is negligible,
(its intensity is so small that it was found only recently), the
corresponding even and odd susceptibilities are,
\begin{eqnarray}
\chi_{o}(\Omega,\vec{q})&=&\chi_{res}(\Omega,\vec{q})+\chi_{c}(\Omega,\vec{q}), \\
\chi_{e}(\Omega,\vec{q})&=&\chi_{c}(\Omega,\vec{q}).
\end{eqnarray}
This means that the resonance mode can only scatter electrons between
the bonding and antibonding bands. In contrast,
the spin fluctuation continuum scatters both within and between
these bands. As argued in Ref.~\cite{Eschrig02}, 
the odd symmetry of the resonance
is crucial in reproducing the ARPES spectra.

Writing the self energy symbolically as
$\hat \Sigma =g^2\chi \ast \hat G$
(the hat denotes the 2x2 particle-hole space, and $g$ is the coupling
constant),
the self energy $\hat \Sigma^{(a,b)}$
for the antibonding and bonding bands are given as \cite{Eschrig02},
\begin{equation}
\label{bilayer1}
\hat\Sigma^{(a,b)} = \frac{g^2}{2}\left\{ \chi_{res} \ast \hat G^{(b,a)} + 
\chi_{c} \ast \left( \hat G^{(b)}+\hat G^{(a)} \right) \right\}.
\end{equation}
Dispersion anomalies arise mainly from
coupling to the resonance mode. This means that dispersion anomalies
in the bonding band
are determined by the antibonding spectral function and vice versa.
Because the antibonding band is (in contrast to the bonding band)
close to the chemical potential at
$(\pi,0)$ \cite{Feng02}, the associated van Hove singularity
leads to a larger self energy for the bonding band.

The spin-fluctuation spectrum is modeled by Eqs.~(\ref{Btot})-(\ref{Bcont0}).
Parameters are chosen appropriate for the overdoped sample
($T_c=65 $K) studied in experiment \cite{Feng01a,Feng02a}. 
The normal state dispersion is obtained from
a 6 parameter tight-binding fit to experimental data \cite{Norman95},
plus the bilayer splitting discussed earlier.
The seven parameters used for this fit are obtained from the second
line in Table~\ref{tab1}.
The high energy ($|\epsilon| \gsim 200$ meV)
dispersions are not affected strongly
when going from the normal to the superconducting state.
However, even in the normal state, the bare dispersion is
renormalized by the normal state spin fluctuation continuum.
The extra factor $Z_{HE}=1.4$ in Eq.~(\ref{Zh}) accounts for this
extra renormalization.
For the remaining parameters of the model, 
the values shown in Table \ref{tab3} were used.
\begin{table}
\begin{center}
\tbl{
\label{tab3} 
Minimal parameter set used in the calculations for an overdoped 
bilayer material \cite{Eschrig02}.}
{\begin{tabular}{|c|c|c|c|c||c|c|c|c|}
\hline
\hline
$\Delta_{M}$ & $\Omega_{res}$ & $\xi_M^{(a)}/\xi_M^{(b)}$ & 
$\xi_{sfl}$ & $g^2w_Q$ & 
$2\Delta_h $ & $\xi_c$ &  $g^2c_Q$ & $Z_{HE}$ \\
\hline
16 meV & 27 meV & $-18/ -105$ meV & $2a$  & 0.15 eV$^2$ & 
28.8 meV & $0.5a$ & 0.72 eV & 1.4
\\
\hline
\hline
\end{tabular}}
\end{center}
\end{table}
The value for the resonance energy was obtained from the 
relation $\Omega_{res}=4.9 k_BT_c$ found experimentally to hold for
overdoped 
Bi$_2$Sr$_2$CaCu$_2$O$_{8+\delta}$ \cite{Rossat91,Fong99,Zasadzinski01}.
With the parameters of Table~\ref{tab3} the resonance weight is
$0.36 \mu_B^2$ per plane, and the (2D)-momentum averaged continuum contribution
(gotten by summing the even and odd channels
for energies $\omega \lsim 0.2 eV$) amounts to $1.7 \mu_B^2/$eV per plane. 

\begin{figure}
\centerline{
\epsfxsize=0.4\textwidth{\epsfbox{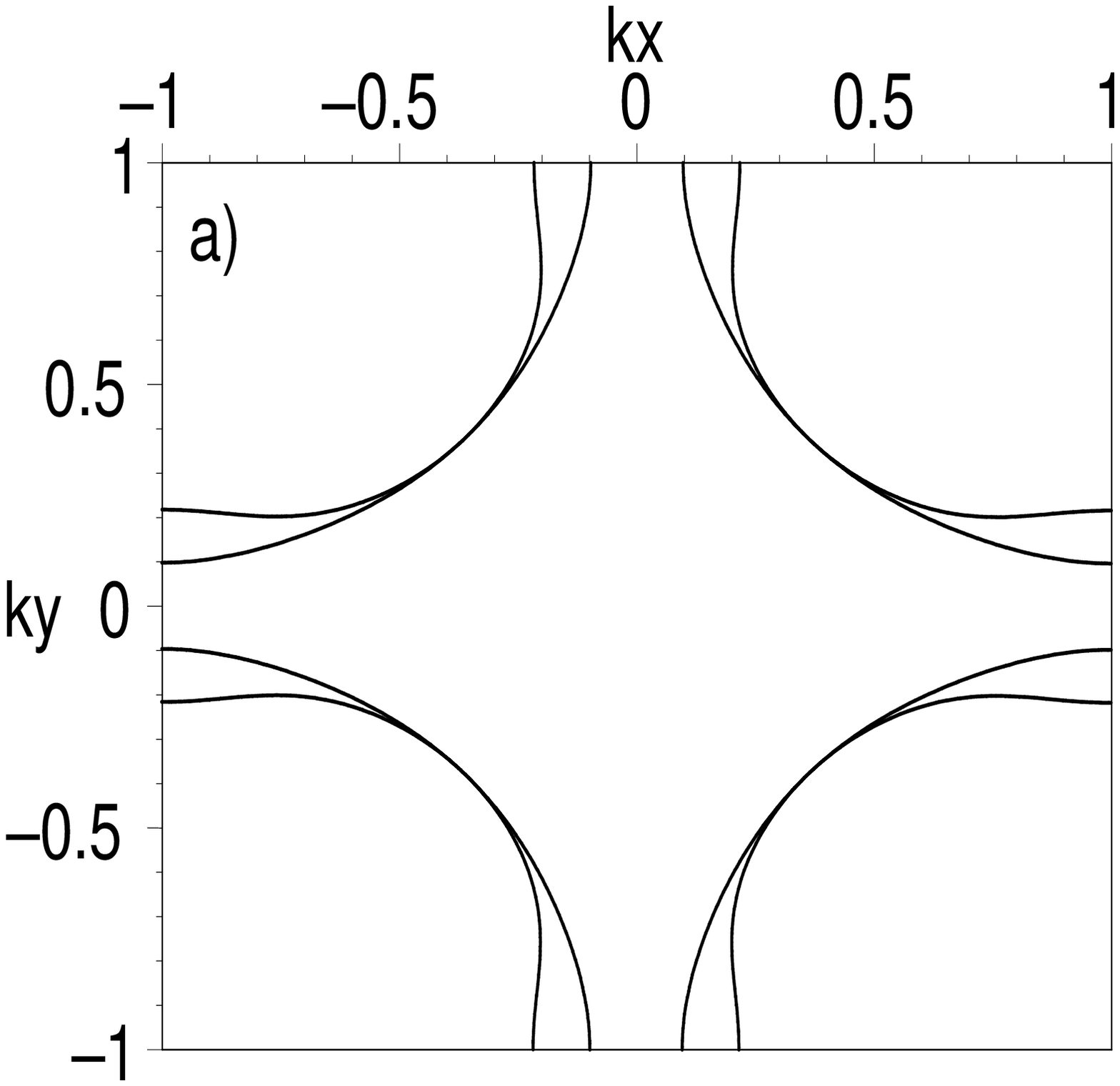}}
\hfill
\epsfxsize=0.5\textwidth{\epsfbox{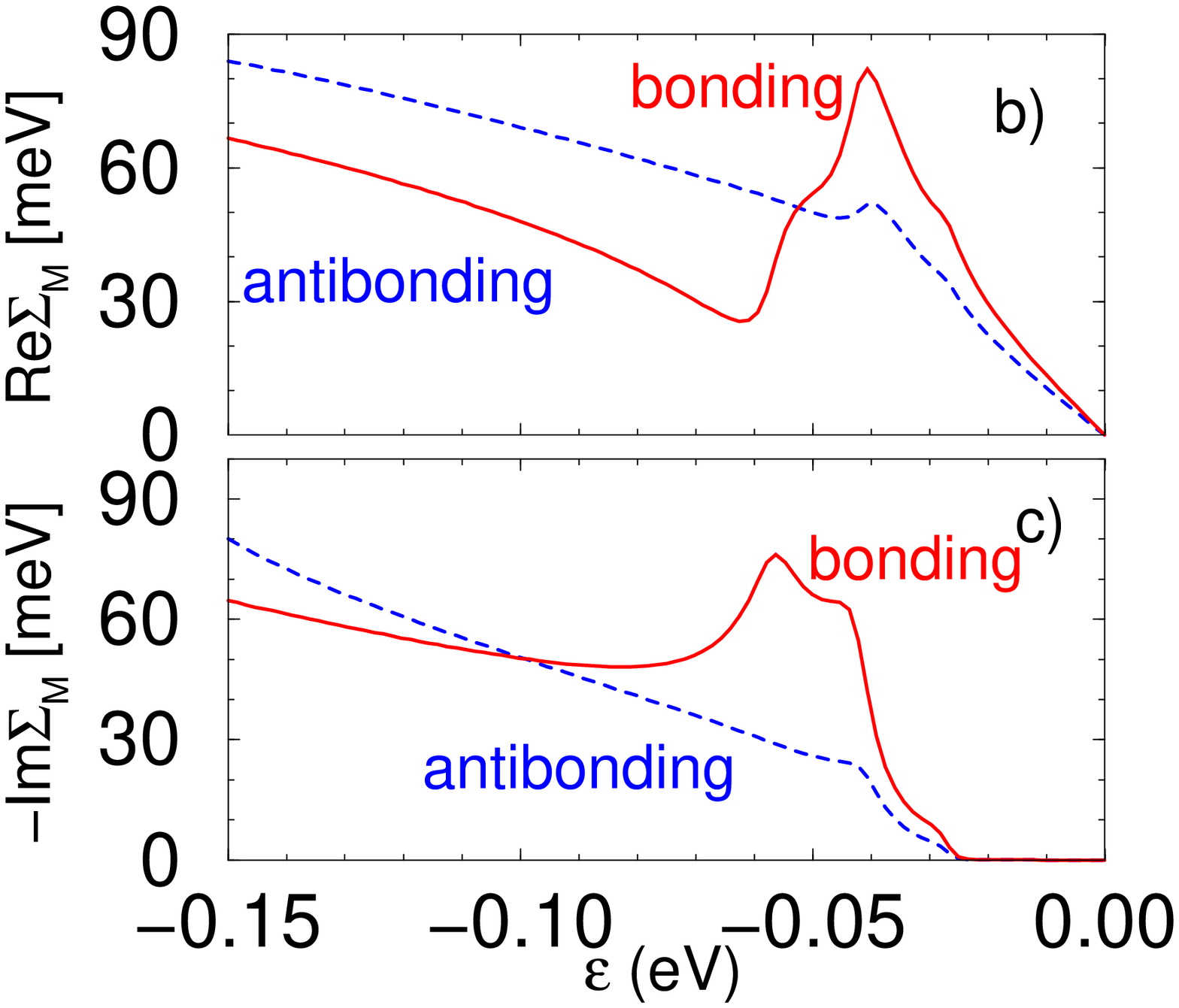}}
}
\caption{\label{Fig1a_Eschrig03}
(a) Tight binding Fermi surfaces for antibonding and bonding bands
in overdoped Bi$_2$Sr$_2$CaCu$_2$O$_{8+\delta}$ ($T_c=65 $K).
(b) Real part and c)
imaginary part of the self energy at
the $(\pi,0)$ point of the zone
for the bonding band (BB) and antibonding band (AB).
(After Ref. \cite{Eschrig02},
Copyright \copyright 2002 APS).
}
\end{figure}

The bonding and antibonding normal
state Fermi surfaces are shown in Fig.~\ref{Fig1a_Eschrig03} a. 
The bilayer splitting is maximal near the $(\pi,0)$ points
of the zone. 
In Fig.~\ref{Fig1a_Eschrig03} b the real part of the self energy 
for bonding and antibonding bands at the $(\pi,0)$ point is shown.
The renormalization effects are stronger for the bonding band
than for the antibonding band. This is a result of the
proximity of the antibonding saddle point singularity to the chemical potential.
As is seen in this figure as well, both bands are renormalized up to high 
energies. 
Note the linear in binding energy high-energy contribution in Re $\Sigma $. 
It has the same slope for bonding and and antibonding self energy, however
there is a shift with respect to each other. The constant 
shift is the consequence of the mode contribution, whereas the term
linear in binding energy comes from the continuum part of the spin fluctuation
spectrum. As the continuum contribution is the same in even and odd channels,
the high-energy slope of Re $\Sigma $ 
in Fig.\ref{Fig1a_Eschrig03} is identical.

The imaginary part of the self energy is shown in Fig.~\ref{Fig1a_Eschrig03} c)
for the bonding and antibonding bands.
As emission processes
are forbidden for $|\epsilon| < \Omega_{res}$, the
imaginary part of the self energy is zero in this range.
Due to scattering events to the antibonding band, electrons
in the bonding band have a large imaginary part of the self energy
in the range between 40 and 60 meV. These events
are dominated by emission of the resonance, and
are enhanced due to the van Hove singularity
in the antibonding band close to the chemical potential.
In contrast,  the imaginary part of the antibonding self energy 
is not enhanced because the bonding band is far from the chemical potential
at $(\pi,0)$. Consequently, it shows linear behavior over a
wide energy range, with a gap at low energies 
($|\epsilon| < \Omega_{res}$).

\begin{figure}
\centerline{
\epsfxsize=0.5\textwidth{\epsfbox{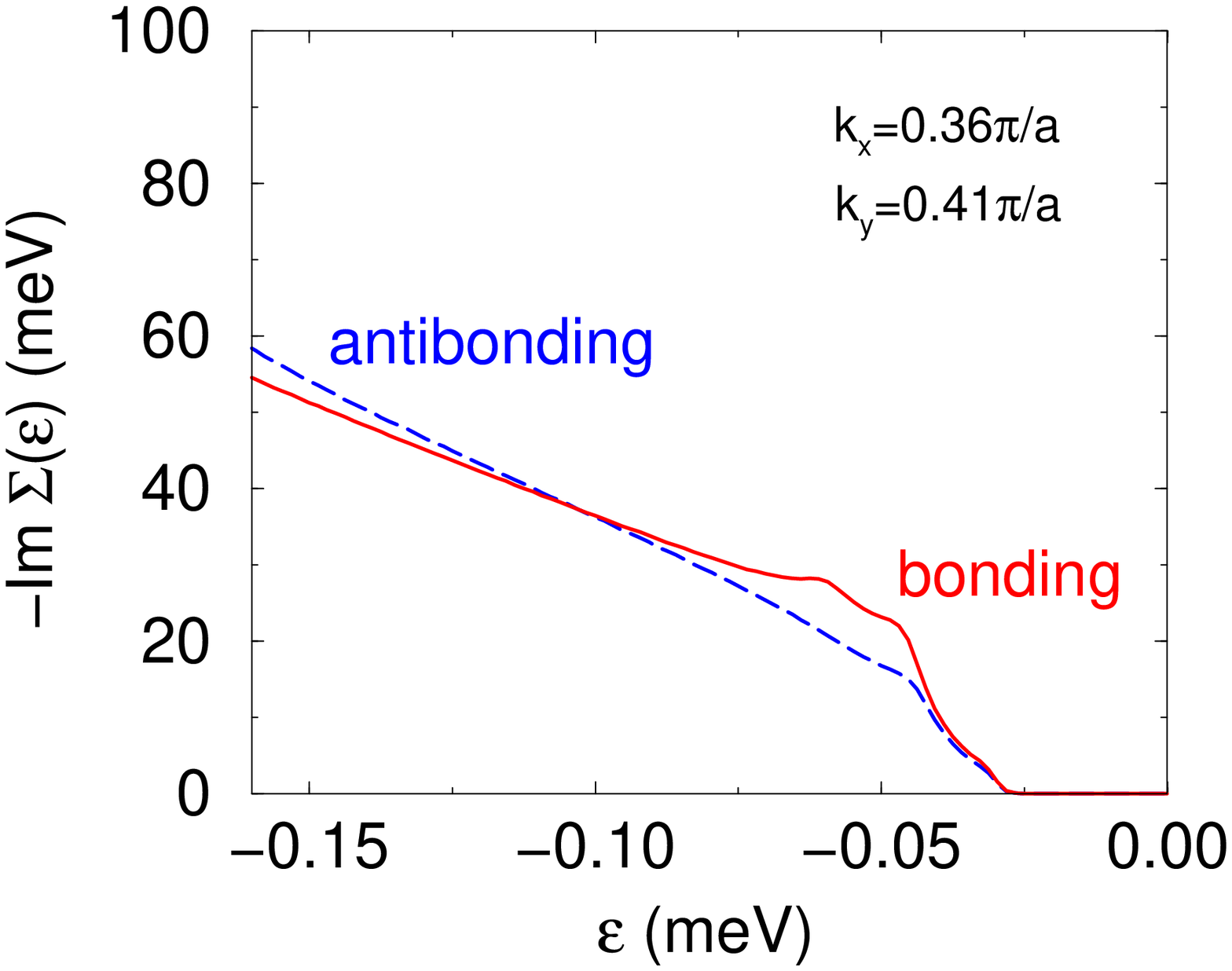}}
\hfill
\epsfxsize=0.5\textwidth{\epsfbox{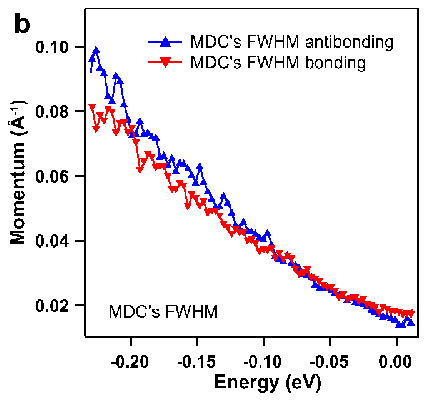}}
}
\caption{\label{Fig1node_Eschrig03}
Left:
Imaginary part of the self energy, calculated within the model in Ref.~\cite{Eschrig02},
at the $\vec{k}$-point $(0.36,0.41)\pi/a$, which is situated at the Fermi surface
close to the nodal point $(0.37,0.37)\pi/a$ of the order parameter.
Calculations are done
for the bonding band (BB) and antibonding band (AB).
Right: Experimental data from Ref.~\cite{Borisenko05}, showing the MDC
full width half maximum for a 
(Pb,Bi)$_{2}$Sr$_{2}$CaCu$_{2}$O$_{8+\delta}$ sample
as a function of energy, separately for the
bonding and antibonding band, at a $\vec{k}$-vector close to the node
of the order parameter.
The MDC-widths are proportional to the imaginary part of the self energy.
(From Ref. \cite{Borisenko05},
Copyright \copyright 2006 APS).
}
\end{figure}

In Fig.~\ref{Fig1node_Eschrig03} we compare theoretical calculations
of the model in Ref.~\cite{Eschrig02} with the experimental results of
Ref.~\cite{Borisenko05}. 
The behavior of the imaginary part of the self
energy, here shown closer to the nodal point, is qualitatively similar
to that of Fig.~\ref{Fig1a_Eschrig03} c). At low energies the bonding
band damping is stronger than the antibonding band damping.
At higher binding energies these roles are switched. The crossover takes
place at binding energy of about 100 meV. This is completely in agreement
with the experiment.

\begin{figure}
\centerline{
\epsfxsize=0.7\textwidth{\epsfbox{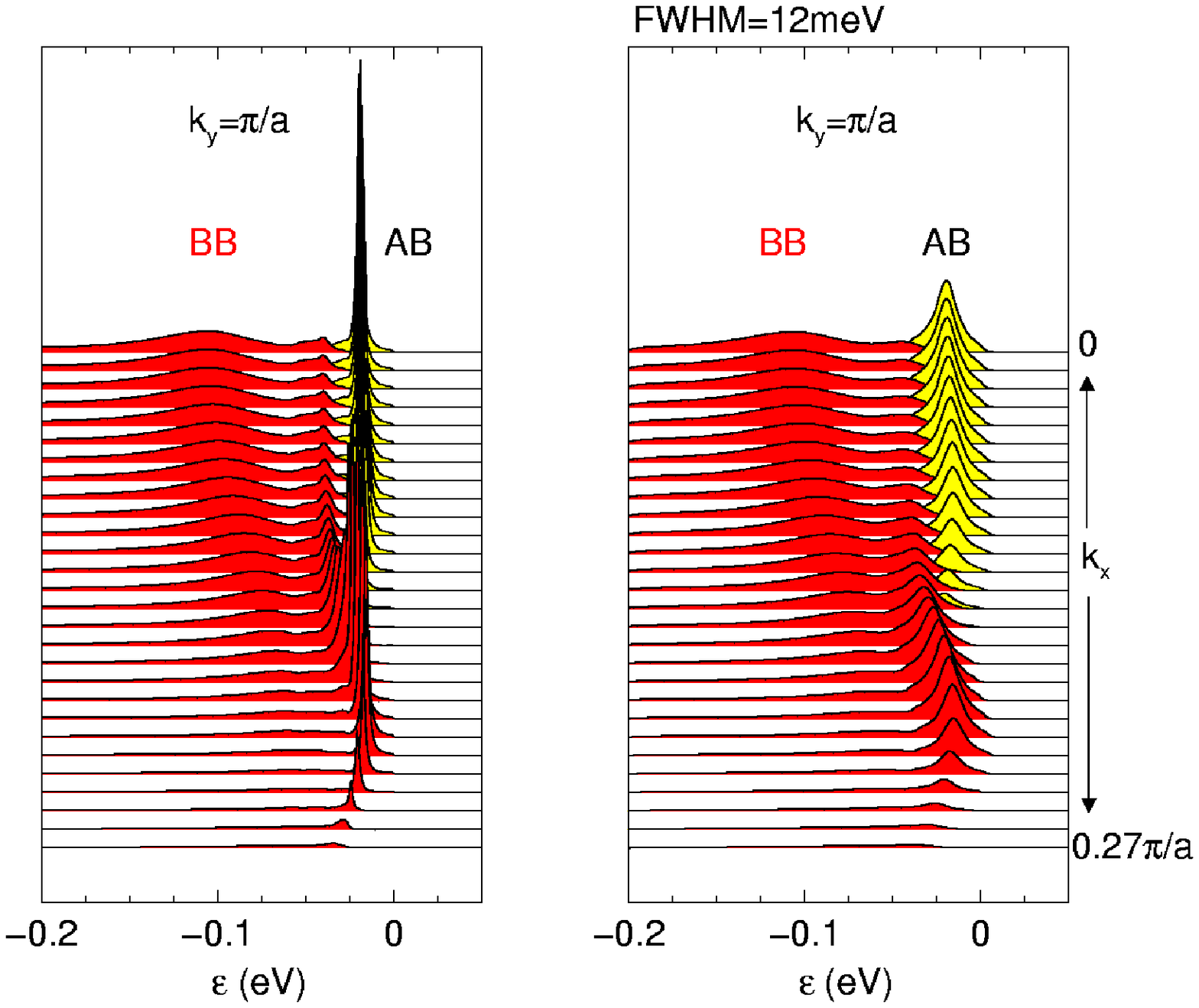}}
\epsfxsize=0.20\textwidth{\epsfbox{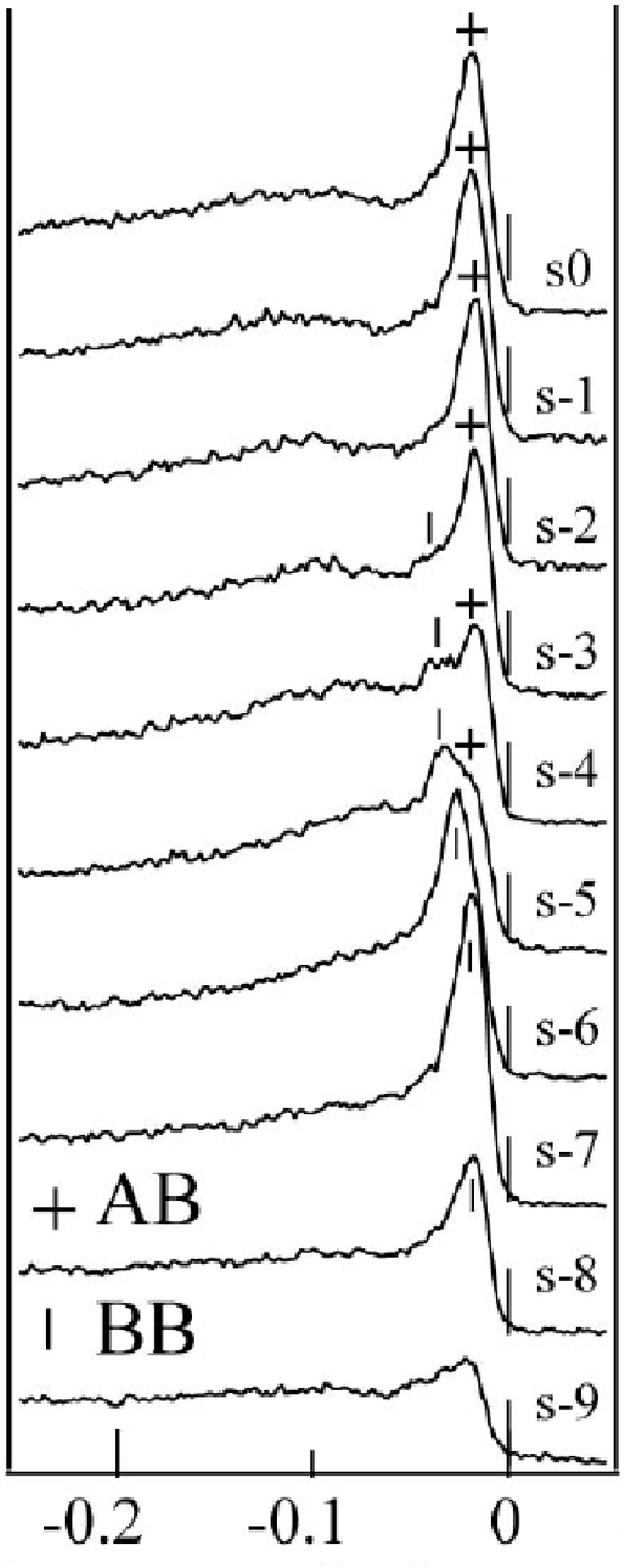}}
}
\caption{\label{Fig1d_Eschrig03}
The left two panels show the calculated ARPES intensity appropriate for
overdoped ($T_c=65 $K) Bi$_2$Sr$_2$CaCu$_2$O$_{8+\delta}$,
for $k_y=\pi /a$, and $k_x $ varying
from $0$ to $0.27 \pi /a$, at $T=10$K. 
For comparison with experiment, in the middle 
the spectra convolved with a
Lorentzian energy resolution function (FWHM 12 meV) are shown
(after Ref. \cite{Eschrig02},
Copyright \copyright 2002 APS).
On the right experimental data from Ref. \cite{Feng01a} are shown for
comparison. The spectrum s0 corresponds to $k_x=0$, and the spectrum
s9 to $k_x=0.24\pi /a$. The crosses and bars denote the antibonding and
bonding peaks, respectively.
(From Ref. \cite{Feng01a},
Copyright \copyright 2001 APS).
}
\end{figure}

Fig.~\ref{Fig1d_Eschrig03} presents the ARPES intensities 
for the bonding and antibonding
spectra and compare with the experimental spectra fro Ref.~\cite{Feng01a}. 
The antibonding spectra consist of a low energy AB peak, and the bonding
spectra have a low energy BB peak and a higher energy BB hump feature.
In agreement with experiment (\cite{Feng01a} and
\cite{Gromko04}), the width of the EDC 
spectrum is large for the BB hump, but not so for the BB and AB peaks.
As can be seen from Fig.~\ref{Fig1d_Eschrig03} (left), the BB peak is
well defined even near $(\pi,0) $, however the finite energy resolution,
(taken into account in middle of Fig.~\ref{Fig1d_Eschrig03})
renders it unobservable in experiment.

\begin{figure}
\centerline{
\epsfxsize=0.36\textwidth
\epsfbox{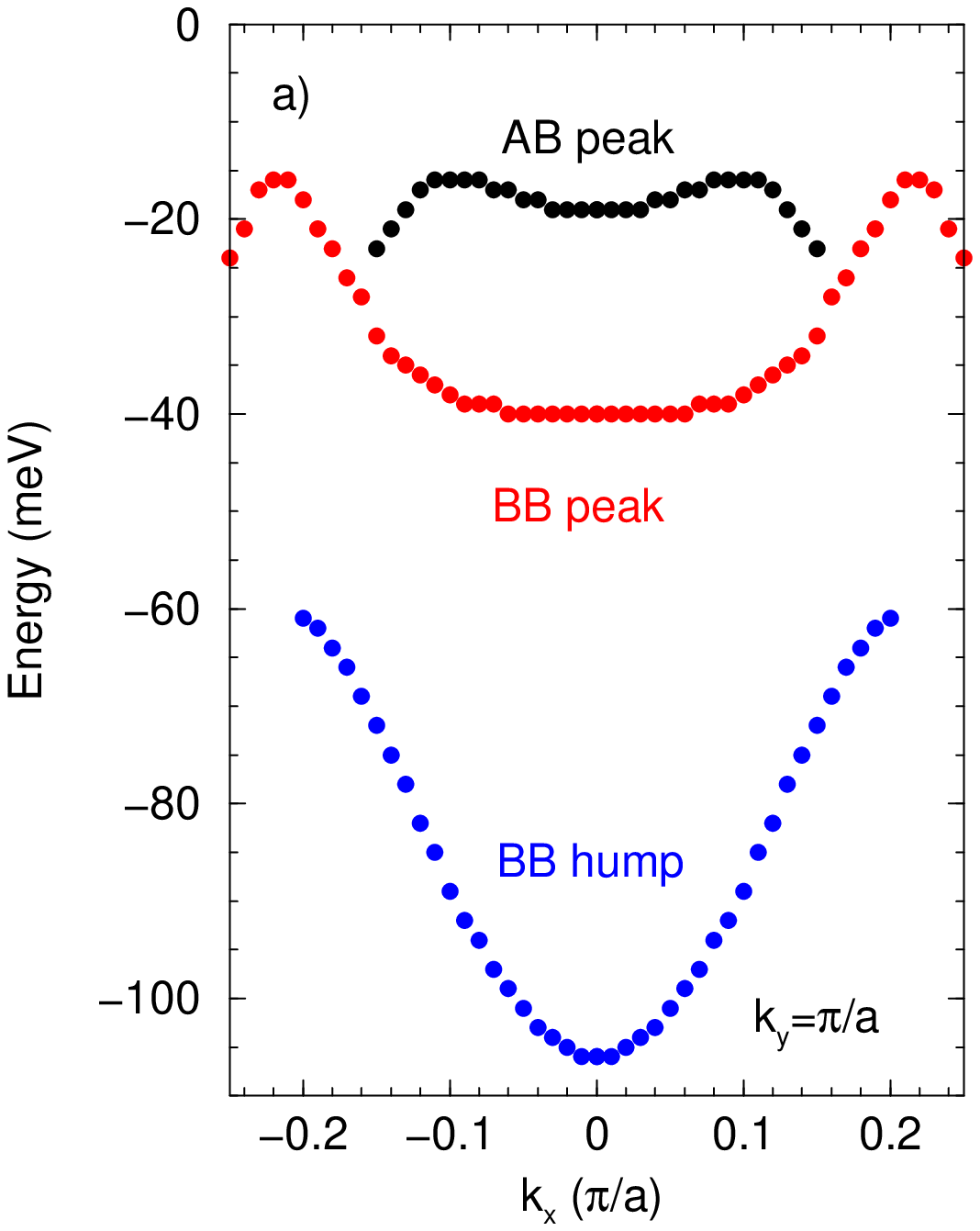}
\hspace{0.1\textwidth}
\epsfxsize=0.46\textwidth
\epsfbox{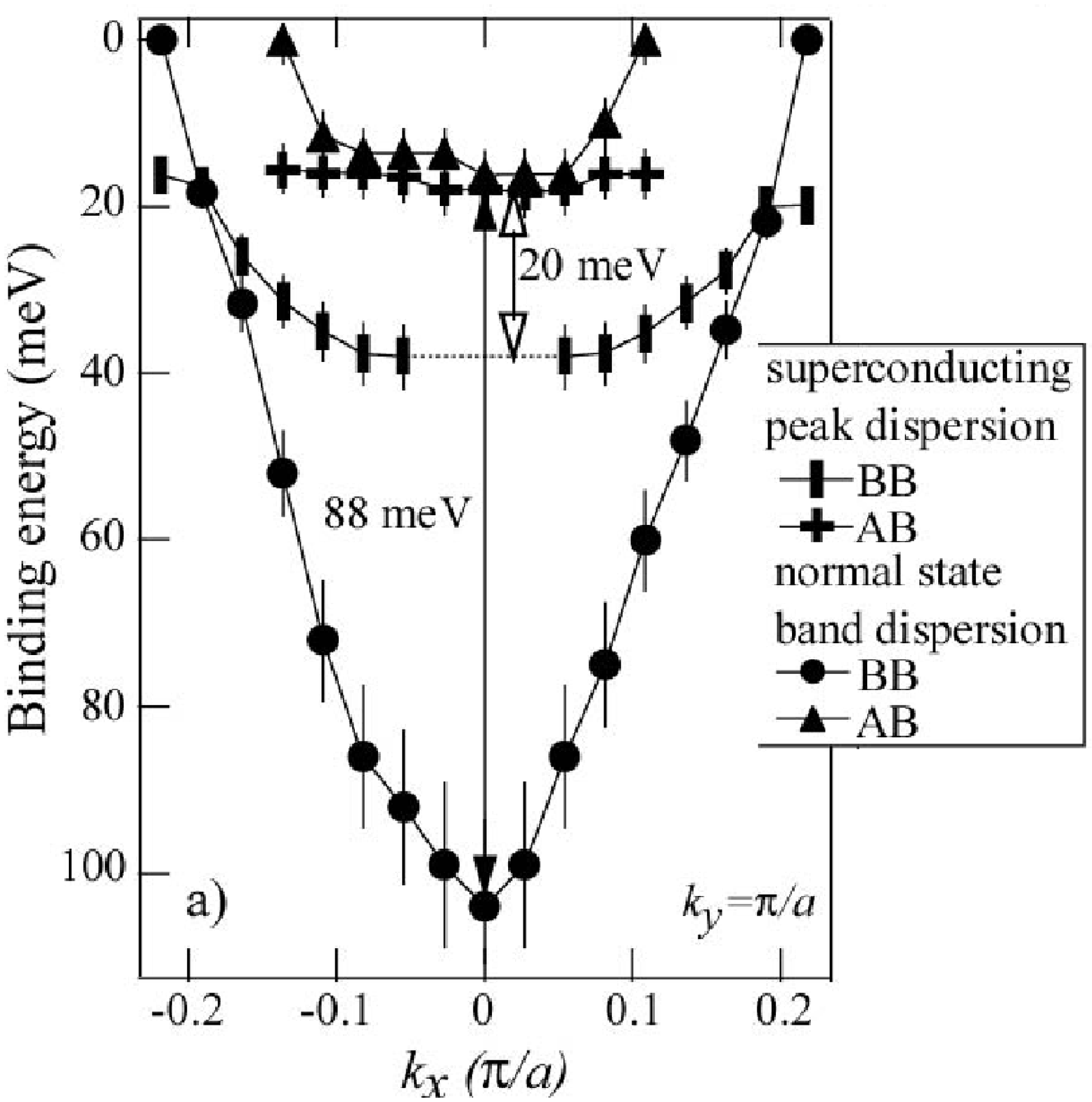}
}
\caption{ \label{Fig2_Eschrig03}
Left: Calculated
dispersion of EDC peak positions 
(appropriate for 
overdoped $Bi_{2}Sr_{2}CaCu_{2}O_{8+\delta}$ with $T_c=$65 K)
near the $(\pi,0)$ point of the Brillouin zone
(after Ref. \cite{Eschrig02},
Copyright \copyright 2002 APS).
The EDC dispersion consists
of three branches, one antibonding peak, one bonding peak and one bonding
hump.  
Right: Experimental
dispersion of the superconducting peaks for the bonding (bars) and antibonding (crosses)
band, compared to the normal state bonding (solid circles) and antibonding (triangles) bands
for an overdoped $Bi_{2}Sr_{2}CaCu_{2}O_{8+\delta}$ sample width $T_c=$65 K
(from Ref. \cite{Feng01a},
Copyright \copyright 2001 APS).
}
\end{figure}

In Fig.~\ref{Fig2_Eschrig03} results for the dispersion of 
the EDC peak positions are shown and compared with the experimental dispersions
of Ref.~\cite{Feng01a}.
The experimentally observed three branches \cite{Feng01a,Gromko04} are
reproduced,
one antibonding peak and two bonding branches, denoted 
`BB peak' and `BB hump'. The BB peak has a very
flat dispersion near $k_x=0$ in agreement with experiment \cite{Feng01a}, as
shown in the right picture in Fig.~\ref{Fig2_Eschrig03}.
The position of that BB peak at 40 meV 
is approximately given by $\Omega_{res} + \Delta_A$,
where $\Delta_A$ is the gap at the antibonding Fermi crossing.
Thus, the energy separation between the AB peak at the AB Fermi crossing 
and the BB peak at $(\pi,0)$
is a measure of the resonance mode energy $\Omega_{res}$ in
overdoped compounds. The BB hump position at high binding
energies (105meV) is determined by the normal state dispersion
of the bonding band.
Because the spin fluctuation continuum changes only at low energies when
going from the normal to the superconducting state, the position
of the BB hump maximum is not very different from the normal state
BB dispersion. This is in agreement with experiment \cite{Feng01a}.
The intensity of the AB peak decreases quickly when it approaches
the BB peak, but is strong at $(\pi,0)$ because of
the proximity of the AB band to the chemical potential in this region.

\begin{figure}
\centerline{
\epsfxsize=0.40\textwidth
\epsfbox{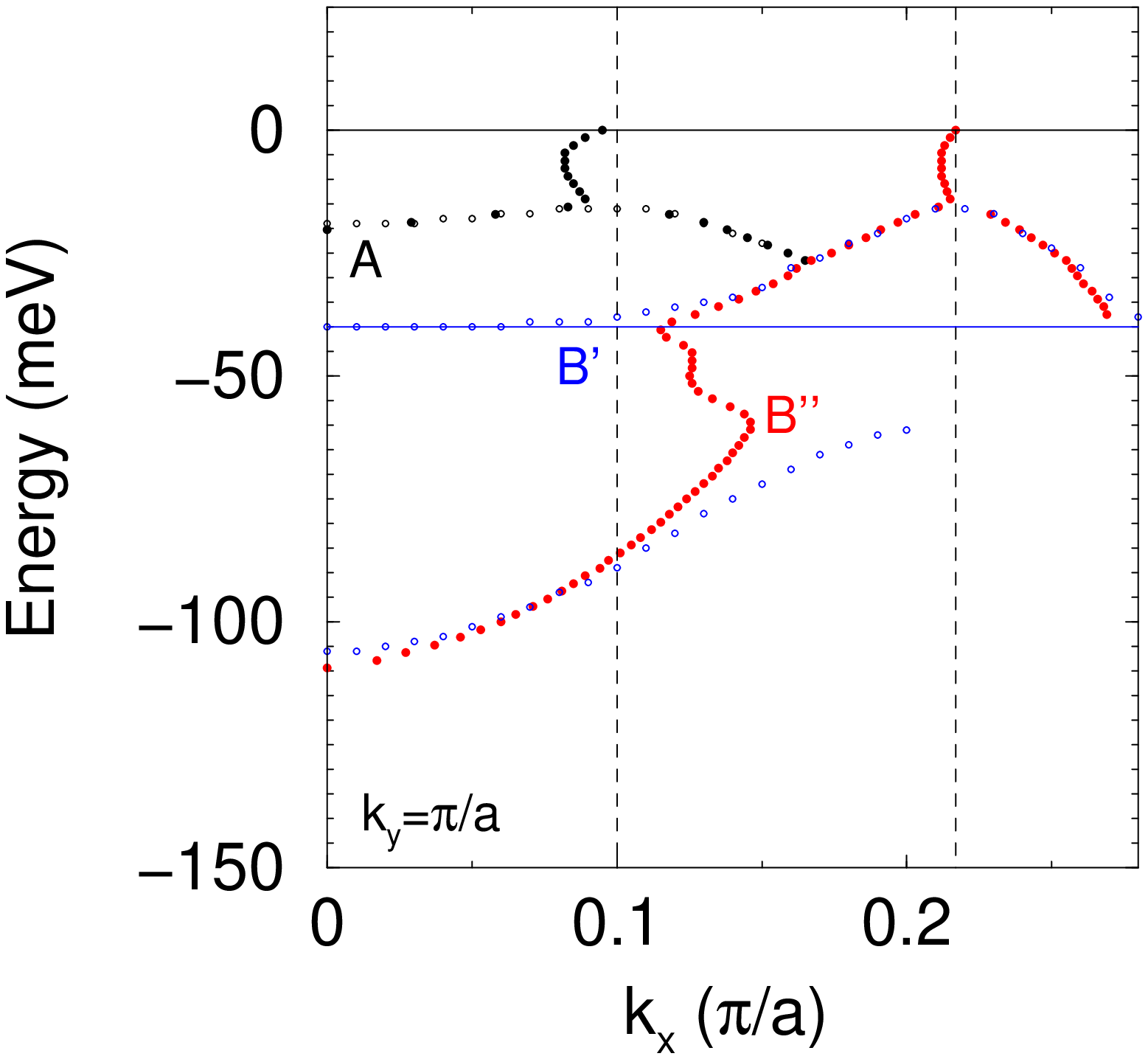}
\epsfxsize=0.46\textwidth
\epsfbox{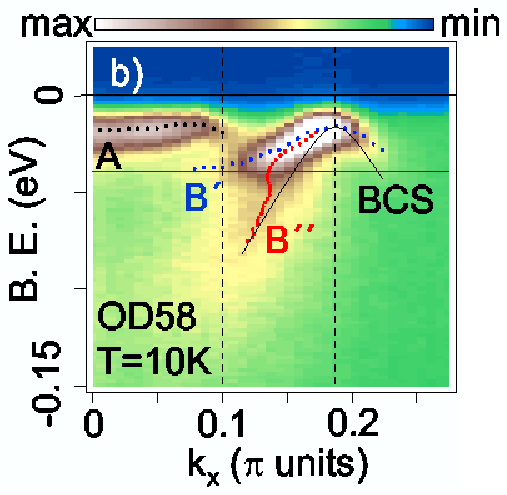}
}
\caption{ \label{Fig2b_Eschrig03}
Left:
Calculated dispersion of MDC peak positions (full symbols) 
(appropriate for 
overdoped $Bi_{2}Sr_{2}CaCu_{2}O_{8+\delta}$ with $T_c=$65 K)
near the $(\pi,0)$ point of the Brillouin zone. 
The MDC bonding dispersion shows a characteristic $S$ shape behavior.
The open symbols show again the EDC peak positions 
from Fig. \ref{Fig2_Eschrig03} for convenience. 
(From Ref. \cite{Eschrig02},
Copyright \copyright 2002 APS).
Right:
Experimental superconducting state ARPES data 
for an overdoped ($T_c=$58 K)
Bi$_{2}$Sr$_{2}$CaCu$_{2}$O$_{8+\delta}$ 
sample near the point $(\pi,0)$. 
(From Ref. \cite{Gromko04},
Copyright \copyright 2002 UC).
In both pictures, $A$ refers to the antibonding
band peak positions, 
$B'$ to the bonding band EDC peak position, and $B''$ to the bonding
band MDC peak positions.
}
\end{figure}

In Fig.~\ref{Fig2b_Eschrig03}, the theoretical results for the
MDC dispersions are shown on the left side
(for comparison also the EDC dispersions is shown as small symbols),
and compared with the experimental ones of Ref.~\cite{Gromko04} (shown
on the right side).
The MDC dispersion consists of two branches, an AB MDC branch and
a BB MDC branch. The self-energy effects are most clearly observable
in the BB MDC branch. In the binding energy range between 40meV and 60meV,
there is an $S$-shaped `break' region, connecting the BB hump EDC branch
with the BB peak EDC branch. This $S$-shaped behavior reproduces the finding
of the experiments \cite{Gromko04}.

\begin{figure}
\centerline{
\begin{minipage}{0.9\textwidth}
\begin{minipage}{0.6\textwidth}
\epsfxsize=1.0\textwidth{\epsfbox{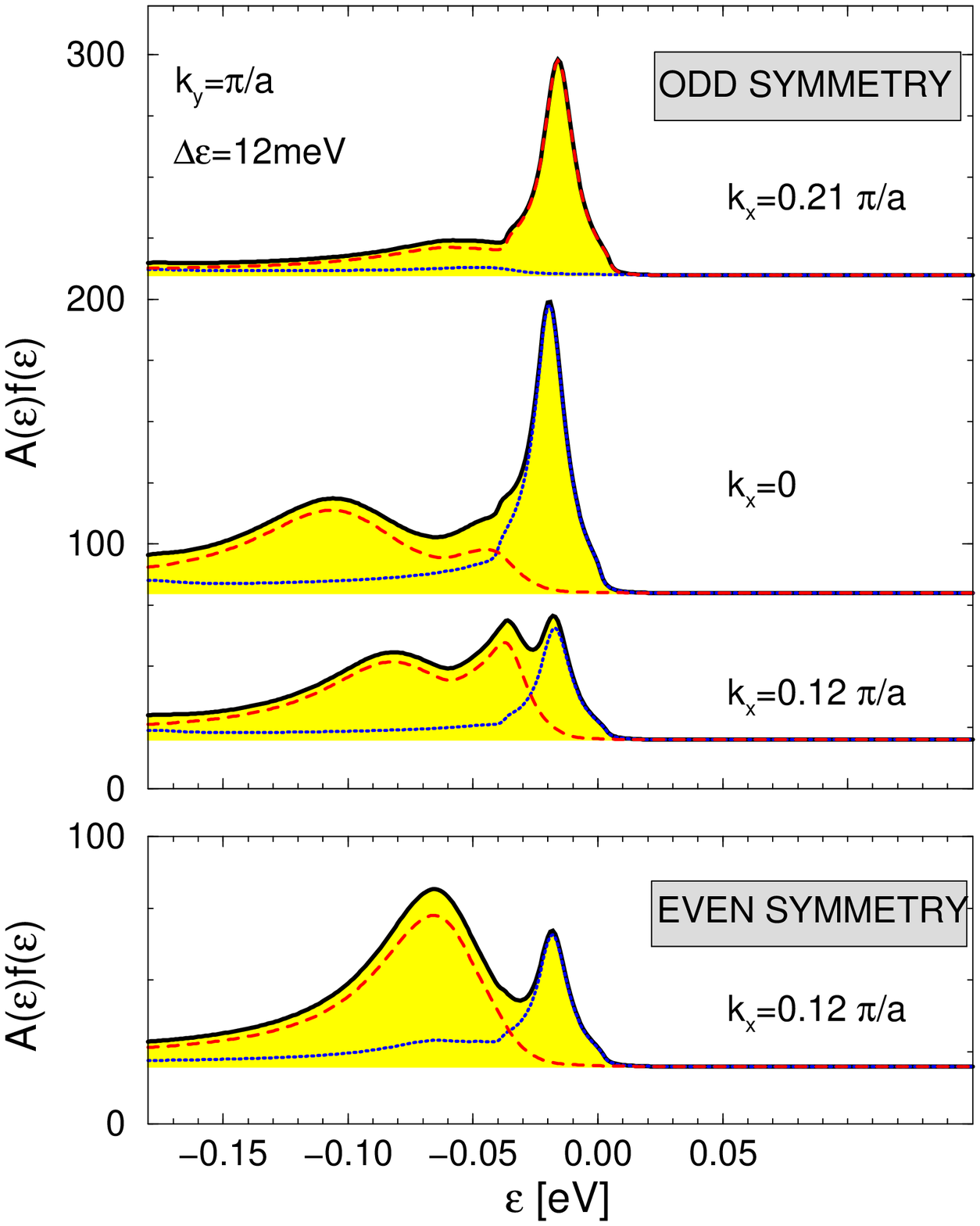}}
\end{minipage}
\begin{minipage}{0.39\textwidth}
\hspace{0.2\textwidth}
\epsfxsize=0.70\textwidth{\epsfbox{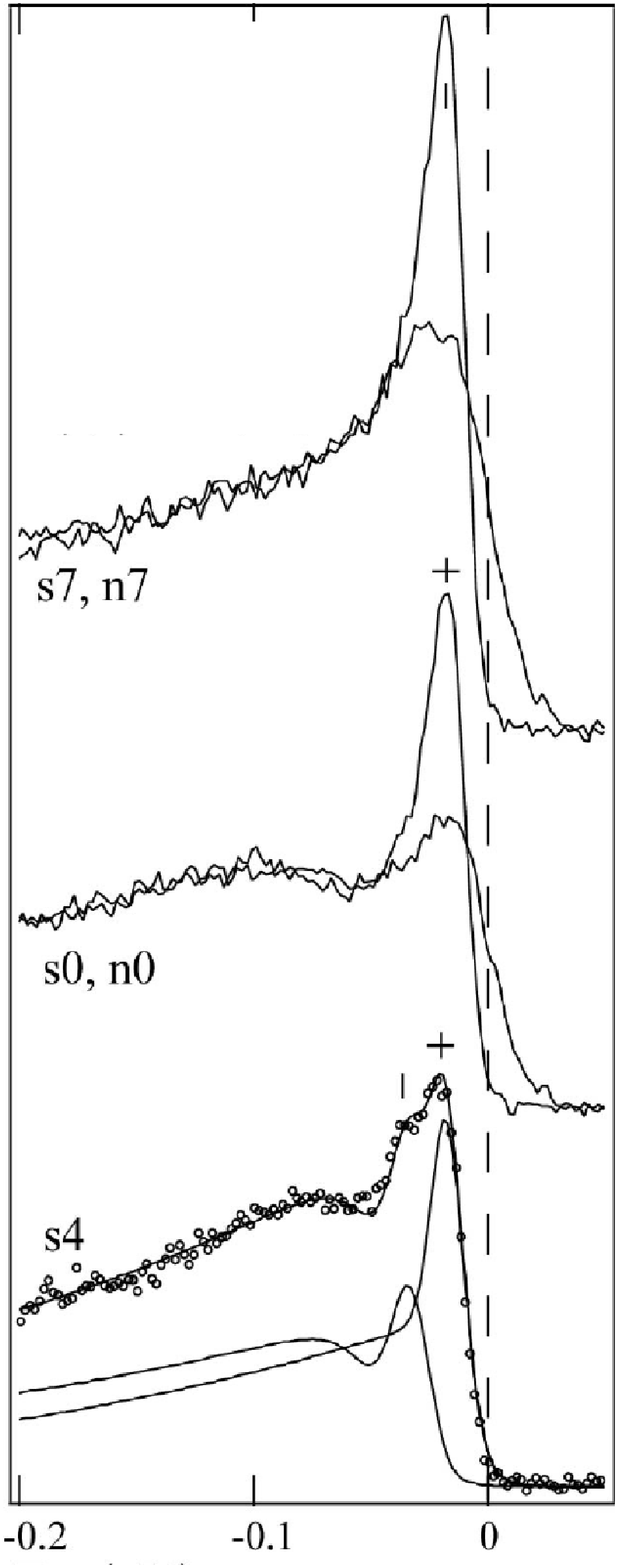}}
\end{minipage}
\end{minipage}
}
\caption{\label{Fig3_Eschrig03}
Left:
Calculated ARPES intensities, appropriate for overdoped ($T_c=65 $K)
Bi$_2$Sr$_2$CaCu$_2$O$_{8+\delta}$,
for $k_y=\pi /a$ at $k_x=0$, $k_x=0.12 \pi /a $ and
$k_x=0.21 \pi /a$. The full lines are the sum of antibonding (dotted)
and bonding (dashed) spectral functions. Calculations are done for $T=10$K.
For comparison with experiment, a
Lorentzian energy resolution function of 12 meV width was assumed.
The low energy double peak structure
is clearly resolved for $k_x=0.12 \pi /a$, as in experiment 
\cite{Feng01a,Gromko04}.
It is not present if the mode is in the even
channel, as demonstrated in the bottom panel.
(After Ref. \cite{Eschrig02},
Copyright \copyright 2002 APS).
Right:
For comparison are shown experimental data 
for an overdoped ($T_c=65 $K) Bi$_2$Sr$_2$CaCu$_2$O$_{8+\delta}$ sample at
$T=10$K from Ref. \cite{Feng01a}
(Copyright \copyright 2002 APS).
The spectral functions s7, s0, and s4 should be compared with the upper three
spectral functions in the left panel 
(the curves n0 and n7, having a broader
peak than s0 and s7, are for the normal state). 
}
\end{figure}

Finally, in Fig.~\ref{Fig3_Eschrig03} 
spectra are shown for three positions in the Brillouin
zone, corresponding to the experimental spectra shown on the right side
\cite{Feng01a}.
For each spectrum, the bonding (dashed) and antibonding (dotted) contributions
are indicated.
The spectra are convolved with a Lorentzian energy resolution function to
allow for direct comparison with experiment.
We reproduce all experimental findings. First, at $(\pi,0)$,
only the BB hump and the AB peak are resolved. This is due to resolution
effects mentioned above.
Second, near the AB Fermi crossing, the spectra show a characteristic
double peak structure, with a relatively sharp
AB peak and a BB peak separated from a broad BB hump. Third, at the
BB Fermi crossing, only the BB peak is observed. The BB hump is so small
in intensity that it only leads to a kink-like feature in the spectrum.

The dispersion anomalies observed in the {\it bonding} band are a mirror of
the large number of states close to the chemical potential 
near $(\pi,0)$ for the {\it antibonding} band.
Scattering events involving a mode with energy $\Omega_{res}$ couple
the bonding band electrons in the energy region between 40 and 60 meV
strongly to those antibonding band electrons.
The corresponding processes are in the odd channel.
These effects would be lost if the resonance were in the even channel.
In this case, the a,b indices in the first part of Eq.~(\ref{bilayer1}) would
be reversed, and thus the assignments listed in Fig.~\ref{Fig1a_Eschrig03}
b and c.
The antibonding van Hove singularity would no longer enter
in the bonding self-energy.  The consequence of this can be seen
in the bottom panel of Fig.~\ref{Fig3_Eschrig03}, where
our calculations were repeated assuming an even symmetry mode.  Only two
spectral features occur now, not three.  Moreover, the resulting MDC
dispersion for the bonding band loses the anomalous $S$-shaped region seen
in Fig.~\ref{Fig2_Eschrig03}b.

As can be seen from Figs.~\ref{Fig1a_Eschrig03}-\ref{Fig3_Eschrig03}
the agreement of the theory \cite{Eschrig02} with
the experimentally
observed self-energy effects in the bilayer split bands in
bilayer high-temperature superconductors is very good. 
The theory reproduces quantitatively
the EDC dispersions, the MDC dispersions, the spectral lineshapes, and
the MDC widths.
Calculations using a more sophisticated
spin-fluctuation spectrum, obtained from a bilayer 
$t-t'-J$ model, have confirmed
this picture \cite{Li05}.
In conclusion, the ARPES data are consistent with the interaction of the
electrons with a sharp bosonic mode which is predominantly {\it odd} in the layer indices,
a property unique to the magnetic resonance observed by 
inelastic neutron scattering.

\subsection{Tunneling spectra}
\label{Tunnel}

Theoretical treatments for the connection between the
dip-features in tunneling spectra and the magnetic resonance mode
have been presented in Refs. \cite{Eschrig00}, \cite{Abanov00b}, 
\cite{Hoogenboom03} and \cite{Zasadzinski03}.
In principle it is straightforward to calculate the tunneling density of states
once the spectral function, $A(\epsilon, \vec{k})$, throughout the zone is known.
The only complication arises from the
tunneling matrix element $T_{\vec{k}\vec{p}}$, which can be very anisotropic
in high-$T_c$ cuprates \cite{Chakravarty93,Andersen95}.

\begin{figure}
\centerline{
\epsfxsize=0.7\textwidth{\epsfbox{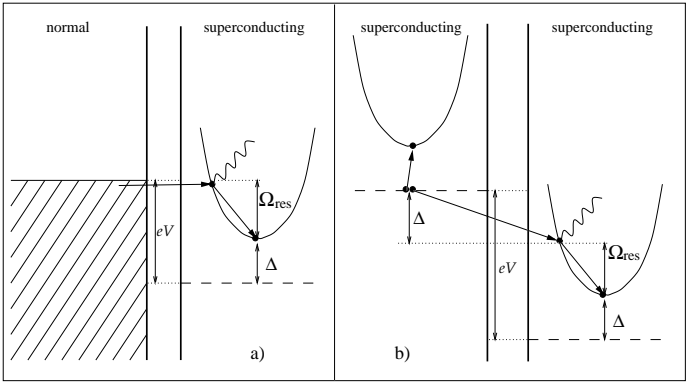}}
}
\caption{
\label{Fig2_Abanov00} 
Schematic diagrams showing the processes responsible for the dip features
in SIN (a) and SIS (b) tunnel junctions. For SIN tunneling, the electron
that tunnels into the superconductor can emit a resonant spin wave with
frequency $\Omega_{res} $ if the
voltage $eV=\Delta + \Omega_{res}$. The electron is left at the bottom of the
band, which leads to a sharp reduction of the current, leading to a drop
in d$I/$d$V$. For SIS tunneling an electron pair must be broken first,
which requires an energy $2\Delta$. In this case the dip in d$I/$d$V$
occurs at $eV=2\Delta + \Omega_{res}$.
(From Ref. \cite{Abanov00},
Copyright \copyright 2000 APS).
}
\end{figure}

The dip-hump structure, which is observed experimentally both in
SIN and SIS junctions, was discussed by Eschrig and Norman \cite{Eschrig00} and
by Abanov and Chubukov \cite{Abanov00b}. The processes, which lead to the
dip in the density of states spectrum, are schematically sketched in
Fig.~\ref{Fig2_Abanov00}.
As can be seen, both for SIN and for SIS junctions the dip is produced
at distance $\Omega_{res}$ from the coherence peaks, which are at $\Delta $ and
$2\Delta $ respectively. 

\begin{figure}
\centerline{
\epsfxsize=0.7\textwidth{\epsfbox{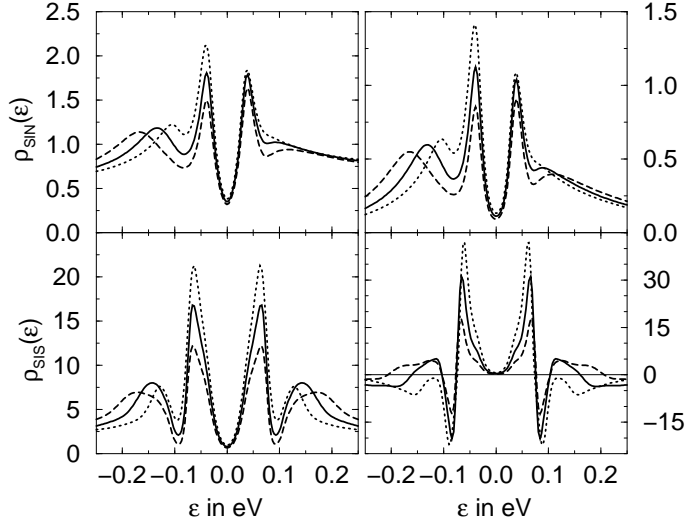}}
}
\caption{
\label{Tunneling} 
Differential tunneling conductance for SIN (top) and SIS (bottom) tunnel
junctions for $T=40 $ K. Units are $eM_i^2$ for SIN and $2e^2T_i^2$ for
SIS. Results for the incoherent (left) and coherent (right)
tunneling limits are shown. 
Curves are for $g$=0.39 eV (dotted), 0.65 eV
(full line), and 0.90 eV (dashed).
The other parameters are given in Table \ref{tab2}.
(From Ref. \cite{Eschrig00},
Copyright \copyright 2000 APS).
}
\end{figure}

In the following we present numerical results from Ref.~\cite{Eschrig00}
which are calculated neglecting for simplicity
the continuum part of the spin fluctuation spectrum.
The numerical results for SIN and SIS junctions are shown in Fig.~\ref{Tunneling}.

From the SIN tunneling current $I(V)$,
\begin{equation}
I(V)=\sum_{\vec{k} } |M_{\vec{k}}|^2
\int_{-\infty}^\infty \frac{d\epsilon}{2\pi}
A(\epsilon,\vec{k})
\left\{f(\epsilon )-f(\epsilon+eV) \right\}
\end{equation}
one obtains the differential conductance, $dI/dV$.
Here, $M_{\vec{k} }$ is the SIN matrix element, assumed to be energy
independent.
The tunneling matrix element can be modelled 
for two extreme cases: for
incoherent tunneling $|M_{\vec{k} }|^2 =M_0^2$,
whereas for coherent tunneling 
$|M_{\vec{k} }|^2=\frac{1}{4}M_1^2( \cos k_x - \cos k_y )^2$
\cite{Chakravarty93}.
Coherent tunneling arises from hopping in c-axis direction via a 
complicated path of intermediate orbitals. The main contribution to
the anisotropy comes from overlap between the 
Cu-$d_{x^2-y^2}$/O-$p_{x,y}$ hybrid orbitals 
with the O-$p_z$ orbitals.
Coherent tunneling in the c-axis direction is
strongly enhanced for the $M$ points
in the Brillouin zone compared to the regions near the zone
diagonal due to the matrix elements \cite{Chakravarty93}.
On the other hand, incoherent tunneling is via inhomogeneities or impurities
that destroy the in-plane momentum conservation during tunneling events.

Results of such a calculation  are shown in the top panels of Fig.~\ref{Tunneling}.
The low energy behavior of the tunneling spectrum in the coherent tunneling
limit does not show the characteristic linear in energy behavior 
for $d$-wave, because the nodal electrons have suppressed tunneling
as a result of the matrix elements. The 
peak-dip-hump features, on the other hand, are not affected by the matrix
elements, as they are dominated by the $M$ point regions which
are probed by both coherent and incoherent tunneling.
The dip position stays roughly at the distance $\Omega_{res}$ from the
peak position, whereas the higher energy features depend strongly 
on the coupling constant.
Note the strong asymmetry of the dip-effect with respect to the 
chemical potential. The self energy effects are much stronger on the
negative side than on the positive side of the spectrum, as in experiments
\cite{Renner95,DeWilde98}.

For an SIS junction, the single particle tunneling current is given in
terms of the spectral functions by
\begin{eqnarray}
I(V)=2e\sum_{\vec{k} \vec{p}} |T_{\vec{k}\vec{p}}|^2
\int_{-\infty}^\infty \frac{d\epsilon }{2\pi }
A(\epsilon, \vec{k}) A(\epsilon+eV, \vec{p}) 
\left[ f(\epsilon ) - f(\epsilon + eV) \right] \quad
\end{eqnarray}
with $|T_{\vec{k}\vec{p}}|^2=T_0^2$ for incoherent tunneling and
$|T_{\vec{k}\vec{p}}|^2=\frac{1}{16 }T_1^2 ( \cos k_x - \cos k_y )^4
\delta_{\vec{k}_{||},\vec{p}_{||}}$ for coherent tunneling
\cite{Chakravarty93}.
Results for d$I$/d$V$ from Ref. \cite{Eschrig00} are shown
in the bottom panels of Fig.~\ref{Tunneling}.
All structures are symmetric around the chemical potential. 
Again, the low energy part of the spectrum is strongly suppressed in the
incoherent tunneling limit.
At higher voltages,
in the coherent tunneling limit, a negative differential 
conductance was predicted \cite{Eschrig00}.  Such an effect has been observed 
in optimally doped Bi$_2$Sr$_2$CaCu$_2$O$_{8-\delta}$
break junctions \cite{Zasadzinski01}.
The negative behavior at higher bias in the (purely) coherent tunneling
limit was explained in Ref.~\cite{Eschrig03}, where it was pointed out that
the continuum contribution will lead to a positive response at high voltages.

The difference between the incoherent tunneling limit and the coherent 
tunneling limit can be most clearly seen by the fact that at zero temperature
the incoherent limit the SIS current is given by,
\begin{equation}
\label{inc}
I^{(incoh)}(V)= \frac{eT_0^2}{\pi}
\int_{-eV}^{0} d\epsilon
N(\epsilon ) N(\epsilon+eV),
\end{equation}
whereas in the coherent limit it can be approximated by \cite{Eschrig03},
\begin{equation}
\label{coh}
I^{(coh)}(V)\approx  \frac{eT_1^2}{\pi}
\int_{-eV}^{0} d\epsilon
A_M(\epsilon ) A_M(\epsilon+eV)
\end{equation}
with $T_1^2=\sum_{\vec{k}\vec{p}} |T_{\vec{k}\vec{p}}|^2$. 
This latter expression results from the fact that tunneling is very effectively
suppressed in the nodal direction for coherent tunneling, and mostly the
$M$ point regions of the Brillouin zone are tested then.
As a consequence, in the coherent limit tunneling tests the
{\it spectral function} at the $M$ point of the Brillouin zone \cite{Eschrig03}. 

Spectra obtained from scanning tunneling spectroscopy (STS) were analyzed
within the model of Eschrig and Norman \cite{Eschrig02} by Hoogenboom {\it et al.} \cite{Hoogenboom03}.
They found that the tunneling data are reproduced only if interaction of
quasiparticles with the collective mode is taken into account. A simple
$d$-wave BCS model as well as a marginal Fermi liquid model fail to
reproduce the tunneling data for reasonable parameters.
\begin{figure}
\centerline{
\epsfxsize=0.7\textwidth{\epsfbox{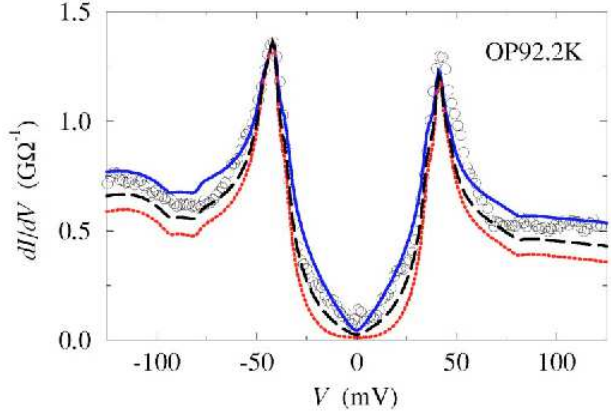}}
}
\caption{
\label{Fig4_Hoogenboom03} 
Comparison of STS spectra including the coupling the the resonance mode
for coherent tunneling limit (dotted line), incoherent tunneling limit 
(full line) and a partially coherent tunneling with
$|T_{\vec{k}}|^2=0.4 |T_{inc}|^2+0.6 |T_{coh}|^2$ (dashed line).
Parameters are $\Delta_M=39 $ meV, $g=0.65 $ eV, $\xi_{sfl}=2a $, and
$\Omega_{res}=5.4 k_{\rm B}T_c$.
(From Ref. \cite{Hoogenboom03},
Copyright \copyright 2003 APS).
}
\end{figure}
In Fig. \ref{Fig4_Hoogenboom03} the comparison of experimental data from
optimally doped Bi$_2$Sr$_2$CaCu$_2$O$_{8+\delta}$ with the calculations
of Ref. \cite{Hoogenboom03} is reproduced, showing that the data
are best accounted for by assuming some partially coherent tunneling
mechanism.

\subsection{Doping dependence}
\label{Doping}
In this section, we deal with the doping dependence of the spectral lineshape
near the $M$ point of the Brillouin zone. 
It was shown in Ref.~\cite{Eschrig03} that there are several general statements
one can draw from the theoretical calculations of the electronic spectra
resulting from an assumed coupling to  the spin-1 magnetic resonance mode.

First, the relevant parameter which determines the behavior of the
spectral functions and separates the underdoped region from the overdoped
region is the parameter $\Omega_{res}/\Delta_M$. This parameter is roughly
1 for optimal doping, larger than  1 in the overdoped region and smaller
than 1 in the underdoped region. 
It means that in the overdoped region the magnetic resonance is closer to
the spin-fluctuation continuum and in the underdoped region closer to zero
energy, the separation at optimal doping being roughly there where the
resonance is halfway between
zero energy and the energy of the continuum onset.

Second, the quasiparticle weight decreases with
decreasing $\Omega_{res}/\Delta_M$ (as it does with increasing coupling constant
$g^2w_Q$).
Third, the quasiparticle scattering rate increases with decreasing 
$\Omega_{res}/\Delta_M$. And fourth, the hump energy disperses to higher binding
energies for increasing coupling constant and increasing $\xi_M$.
Thus, the theoretical predictions for electrons coupled to the spin-1 resonance mode
in cuprates when going from overdoping to underdoping are
a decreasing quasiparticle weight, an increasing 
quasiparticle scattering rate, and an increasing hump binding energy.

\begin{figure}
\centerline{
\epsfxsize=0.75\textwidth{\epsfbox{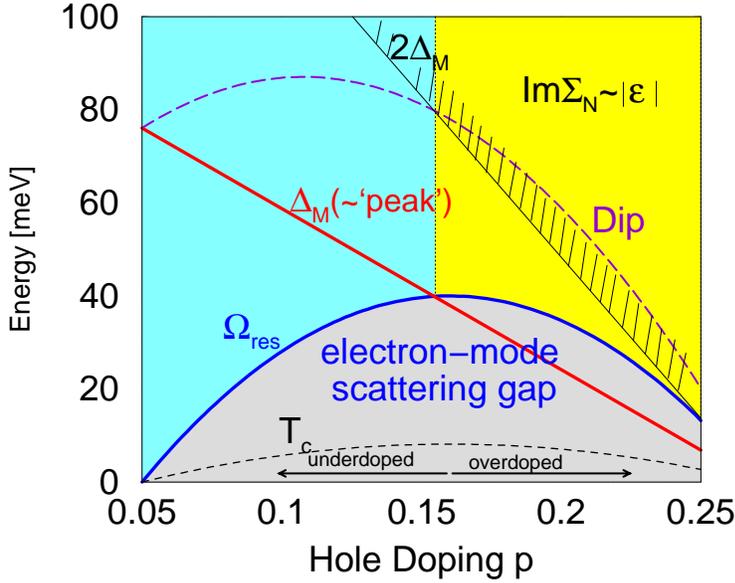}}
}
\caption{
\label{phasediagram}
`Energy phase diagram' for the coupling between electrons and the resonant spin-1
mode in the superconducting state.
The resonance
energy, shown as a thick line, is bounded from above by twice the maximal
gap energy, $\Omega_{res}<2\Delta_{M}$, and approaches it on the 
overdoped side. 
The (antibonding for bilayer cuprates) peak position corresponds roughly to the value
of $\Delta_M$. 
In the region below $\Omega_{res}$ no scattering between electrons and
the resonance mode is possible. Quasiparticle peaks in this region are
sharp. Above this line, strong damping sets in and the peak weight is strongly
reduced on account of an incoherent high-energy background.
The position of optimal doping, at maximal $T_c$ and $\Omega_{res}$,
roughly coincides with the point where $\Delta_M$ as a function of doping
crosses $\Omega_{res}$.
In the overdoped region (toward the right) quasiparticles are well defined
in the antinodal Fermi surface regions. 
With underdoping (toward the left) quasiparticles
in the antinodal Fermi surface regions are progressively destroyed.
(After Ref. \cite{Eschrig03},
Copyright \copyright 2003 APS).
}
\end{figure}

The situation is schematically shown in the phase diagram in 
Fig.~\ref{phasediagram}.
The curves shown are calculated using the formulas
$T_c = 95\mbox{ K}\left(1 - 82.6 (p-0.16)^2\right)$ \cite{Presland91} and
$\Omega_{res}= 4.9 T_c$ \cite{Zasadzinski01}, 
where $p$ is the hole doping level in the Cu-O$_2$ planes.
Optimal doping corresponds to $p=0.16$.
The $\Delta_M$ variation is based on ARPES data \cite{Campuzano99,Mesot99},
and was modelled by
$\Delta_{M} = 38\mbox{ meV}\left(1 - 9.1 (p-0.16)\right)$.
All these quantities approach zero on the overdoped side at $p=0.27$.
The separation between overdoped and
underdoped regions roughly coincides with the regions where
$\Omega_{res}> \Delta_M$ and $\Omega_{res}<\Delta_M$, respectively.
The dip onset is given by $\Omega_{res}+\Delta_A$, where $\Delta_A$ is
the gap at the antinodal point of the Fermi surface. As
$\Delta_A$ is about the same as $\Delta_M$, the line
for $\Omega_{res}+\Delta_{M}$ shown as dashed line in 
Fig.~\ref{phasediagram} determines the position of the dip fairly accurately. 
The continuum in the spin fluctuation spectrum is gapped by $2\Delta_h$,
and consequently 
only affects electrons above $2\Delta_h$, which is near or above
the dip energy. This is the region in which a variation of the
MDC widths linear in the binding energy can be observed (the magnitude of
this linear term drops however towards overdoping, and a quadratic behavior
was suggested in Ref. \cite{Kordyuk04} to take over).
The point of optimal doping for a Cu-O$_2$ 
plane roughly corresponds to the point where $\Omega_{res}/\Delta_{M}=1$. 
Another experimental observation is that this ratio never exceeds the value 
two.  This is expected for an excitonic collective mode below a continuum 
edge \cite{Zasadzinski01}.

The main feature is the `electron-mode scattering gap' shown in 
Fig.~\ref{phasediagram} below
$\Omega_{res}$. As $\Delta_M$ as a function of doping enters this
gapped region at the overdoped side, all quasiparticle excitations along
the Fermi surface (and also in the $M$ point regions) are well defined,
showing up as rather sharp peaks. On the other hand, in the underdoped region,
quasiparticles with binding energy
$\sim \Delta_M$ is strongly scattered by the spin-1 resonance mode and
are progressively destroyed toward underdoping. At the same time
the weight of the quasiparticle peak is reduced on the account of an incoherent 
high-energy background above the dip energy.

\begin{figure}
\centerline{
\epsfxsize=0.49\textwidth{\epsfbox{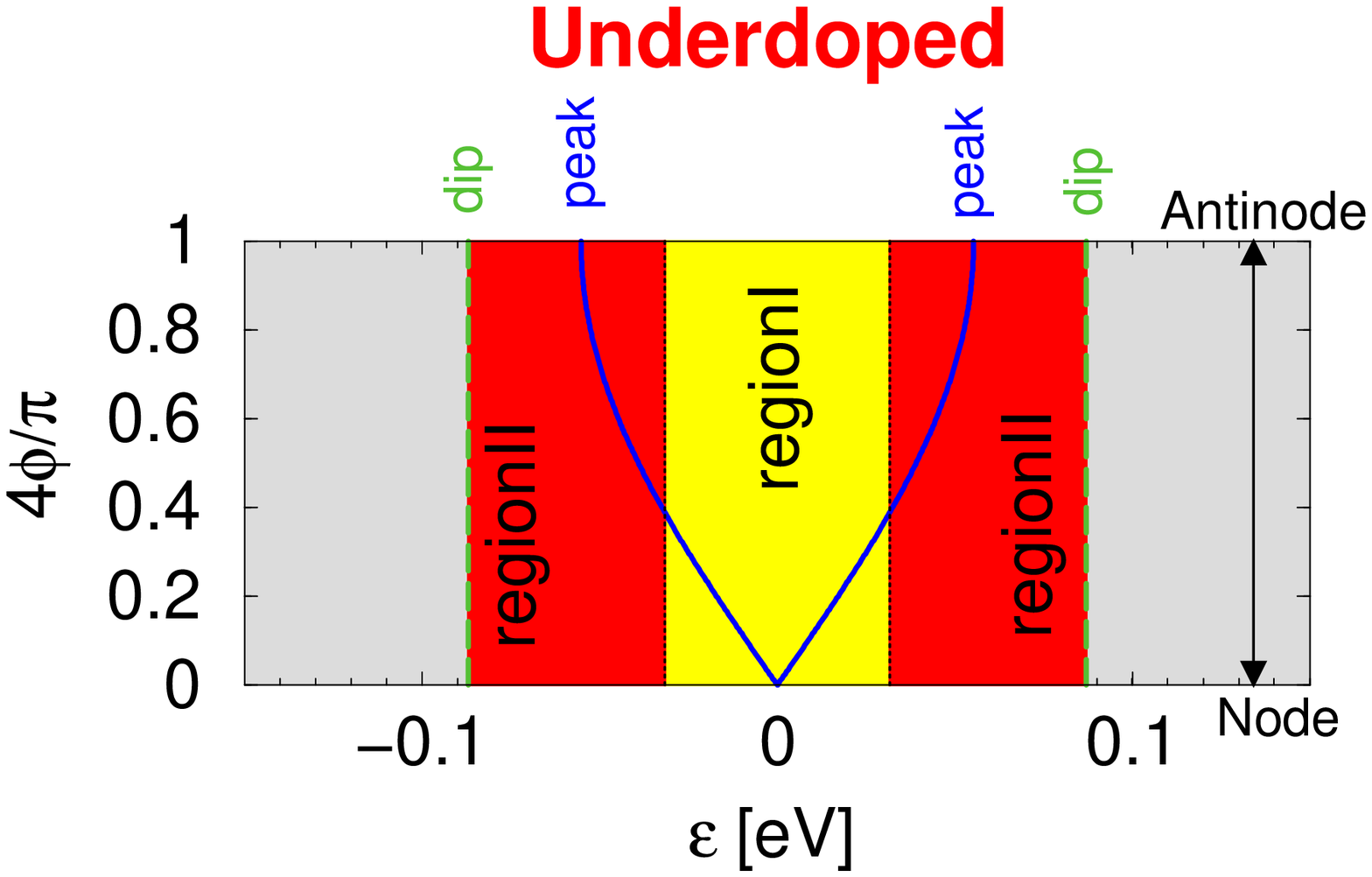}}
\epsfxsize=0.49\textwidth{\epsfbox{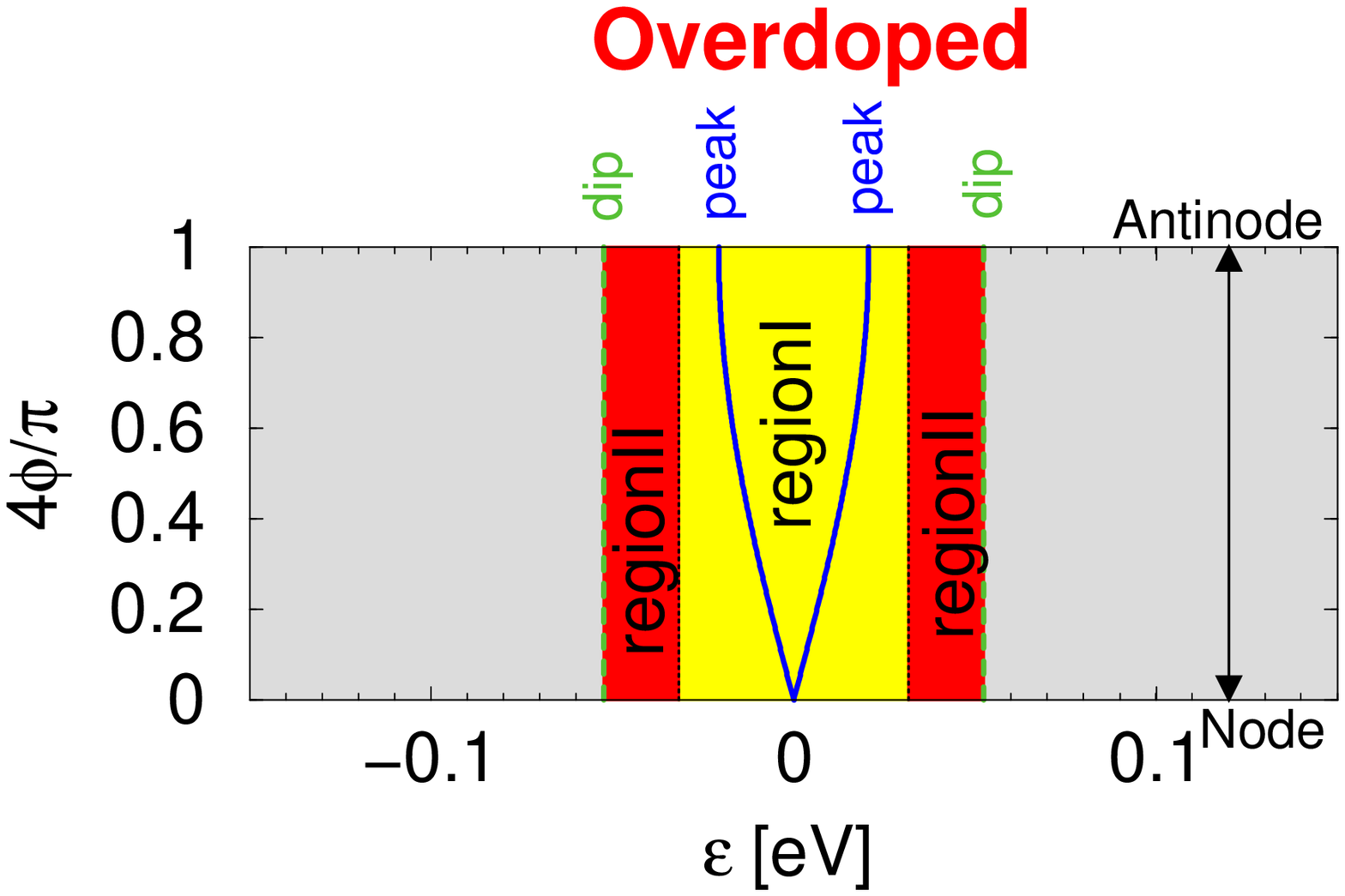}}
}
\caption{
\label{Gapregions}
The full curves determine for each Fermi surface angle $\phi$ 
(which denotes the position along the Fermi surface and is measured 
from the node) the superconducting $d$-wave gap.
The left picture is for the underdoped case, and the right picture
for the overdoped.
Region I is the energy region in which no scattering of quasiparticles
with spin-fluctuations takes place. These regions are bounded by
$\pm \Omega_{res}$.
In region II scattering with the collective spin-1 mode sets in and increases
from $\pm \Omega_{res}$ toward the energies $\pm (\Omega_{res}+\Delta_A)$,
where $\Delta_A$ is the antinodal gap magnitude.
The outer borders of region II (dashed lines) 
denote the position of the dip in the 
EDC ARPES spectrum, the full curves the position of the peak. 
As can be seen, nodal quasiparticles at the Fermi wavevector
are always in region I, and thus not scattered by the resonance mode.
However, when going away from the node toward the antinode, for
underdoped materials (shown on the left) 
scattering by the resonant spin-1 mode sets 
in at a certain critical
Fermi surface angle $\phi$, destroying quasiparticles situated beyond
that critical value. This is not the case for overdoped materials (shown
on the right). Quasiparticles are well defined for all Fermi wavevectors in
this case.
}
\end{figure}

Finally, we discuss an important difference between underdoped and
overdoped materials concerning the anisotropy of the self-energy
effects along the Fermi surface. In Fig.~\ref{Gapregions}
we show for both cases the position of the peak at the gap energy
as a function of the angle $\phi$ which varies from zero at the node
to $\pi/4$ at the antinode.
Within region I in these pictures quasiparticles are sharp, whereas in
region II scattering by the spin-1 resonance mode sets in and increases
toward the borders, indicated by dashed lines, which mark the dip energy
$\pm(\Omega_{res}+\Delta_A)$. Note that nodal quasiparticles are always
well defined, whereas antinodal quasiparticles are only in the
overdoped case. The magnitude of the dip is largest when the peak-dip
separation is smallest, that means near the antinode.
At the nodes, the dip is unobservable in the spectra, and self energy effects
are visible only in the dispersion of quasiparticles away from the Fermi
surface.

Note that the last discussion is appropriate for materials in which
the magnetic resonance mode is the dominant low-energy excitation in
the spin fluctuation spectrum. If the incommensurate spectrum below
the resonance dominates, as it is the case for example in
La$_{2-x}$Sr$_x$CuO$_4$, then the dominant scattering comes from these
incommensurate spin excitations. As incommensurate spin-excitations 
involve scattering of quasiparticles near the nodal points of the
Brillouin zone, in this case the above picture will be reversed:
nodal quasiparticles are strongly damped and antinodal quasiparticles
are sharper. This is in qualitative agreement
with the experimental ARPES results of Refs.~\cite{Ino99,Ino00,Ino02}.
This can also be a potential explanation for the recent findings of a similar
reversed picture in Tl$_2$Ba$_2$CuO$_{6+\delta}$ \cite{Plate05}.

In addition, as these incommensurate spin excitations often 
persist in the normal
state, the self-energy effects near the nodes in these cases 
are also expected to persist in the normal state.

\section{Discussion of phonon effects}
\label{Phonon}
It is to some extent fortunate that the spin fluctuation mode is so sharp
in energy that its effects on the quasiparticle spectra can be separated
from possible additional effects. Such additional effects include those 
due to electron-phonon coupling and 
due to the coupling to the spin-fluctuation continuum.
In the normal state, the spin fluctuation continuum is not gapped but shows
a relaxational behavior with a maximum at a relatively low energy for
underdoped materials. In this case, the quasiparticle dispersion is expected
to be affected
in a similar way as in the superconducting state, however the corresponding
quasiparticle dispersion should not show sharp kinks or sharp $S$-shaped features,
but should be rounded on the scale of the energy where the spin fluctuation continuum
has a maximum.
There are expected additional effects due to electron-phonon coupling, 
which we address in the following.

These phonon effects are stressed in particular by the works of
\cite{Lanzara01,Lanzara02,Cuk04a,Shen02,Zhou04,Devereaux04,Devereaux04b,Gweon04a,Gweon04b}.
There is no doubt that phonons have some contribution to 
modifying the dispersion. 
It has, however, been argued by Chubukov and Norman \cite{Chubukov04} that the 
absence of an $S$-shaped
dispersion anomaly in the nodal region cannot be reconciled with 
the simultaneous presence of a strong Fermi velocity
renormalization if both effects are assigned to 
an optical phonon coupling to electrons. 

In order to obtain such sharp features it is necessary that electrons couple
to one particular optical phonon branch, which is dispersionless to a precision
of 10 meV over an extended region of the phononic Brillouin zone. 
There are mainly two optical phonons which were considered to be responsible for
dispersion anomalies in superconducting cuprates:
the Cu-O buckling mode, which is attractive in the $d$-wave
channel \cite{Song95a,Song95b,Nazarenko96,Bulut96,Dahm96,Sakai97,Nunner99}, and the
Cu-O breathing mode, which is repulsive in the $d$-wave
channel \cite{Nazarenko96,Bulut96,Dahm96,Nunner99}.
Typically, the absolute values of the pairing interactions
in the $B_{1g}$ (`$d$-wave') pairing channel
for both types of vibrations are smaller than 0.1 eV, in the
$A_{1g}$ (`$s$-wave') pairing channel
about 0.5-1 eV; for spin fluctuations, the corresponding numbers
are in the $d$-wave channel 0.5-1 eV and in the $s$-wave channel
1-2 eV \cite{Nunner99}.
The total electron-phonon coupling constant in the $s$-wave channel
amounts to $\lambda_s \approx
0.4-0.6$,\cite{Friedl90,Litvinchuk91,Andersen96,Jepsen98,Nunner99}
and in the $d$-wave channel to
$\lambda_d \approx 0.3$.\cite{Andersen96,Jepsen98}
Thus, in general phonon effects are expected to be moderate.

It was argued that strong coupling of electrons to
the {\it zone boundary half breathing phonon} may be responsible for the
anomalies in the dispersion \cite{Lanzara01,Lanzara02,Shen02}.
It is known for some time that this
phonon shows a dispersion which is strongly renormalized midway between the
zone boundary and the zone center when entering the superconducting state.
These findings show that the zone boundary half breathing phonon is affected
by superconductivity.
It was suggested to be responsible for the renormalizations of the dispersion
observed in ARPES \cite{Lanzara01}.
This zone boundary optical phonon
is a Cu-O bond stretching mode with an energy between 50 and 100 meV.
The characteristics of this mode are well documented
\cite{McQueeney99,Petrov00,McQueeney03}.
In the model of Ref. \cite{Shen02},
the coupling vanishes near the $\vec{q}=(\pi,\pi)$ point,
thus having minimum impact on the electrons
near the $M$ point of their Brillouin zone.
This is in stark contrast to the resonance mode model of
\cite{Eschrig00,Eschrig03}, and can certainly
not explain the effects at the $M$ points. It is, however, possible that
they contribute to the renormalization of the nodal dispersion \cite{Eschrig03}.
The maximal coupling strength was theoretically estimated to
$g_b\approx 0.04 $eV \cite{Nunner99}, but in some models is enhanced by vertex
corrections \cite{Shen02}.
In recent theoretical investigations it was found that this phonon does
only contribute to renormalizations of 
the very near vicinity of the nodal point in the Brillouin zone \cite{Devereaux04}
and cannot describe the experimentally observed features in tunneling spectra
\cite{Zhu05a}.

Another phonon invoked recently in order to explain dispersion anomalies
near the $M$ point of the Brillouin zone is the $B_{1g}$ {\it Cu-O buckling phonon}.
In Refs. \cite{Cuk04,Cuk04a,Devereaux04,Zhu05a} it is 
argued that a highly anisotropic electron-phonon coupling matrix element
can account for the self-energy effects in the antinodal region of
the fermionic Brillouin zone. However, for this phonon the coupling to
linear order in the atomic displacements only arises from a local $c$-axis
oriented crystal field, which breaks the mirror plane symmetry of the
Cu-O plane \cite{Andersen94,Andersen95}. 

In Ref. \cite{Devereaux04} calculations were performed for an assumed 
buckling of the Cu-O planes, that leads to an electric field $eE_z=$
1.85 eV$/\AA $ appropriate for YBa$_2$Cu$_3$O$_{7-\delta}$.
The calculations were able to produce a break effect in the EDC dispersion
near the $(\pi,0)$ point that is comparable in magnitude to the experimentally
in Bi$_2$Sr$_2$(Y$_{x}$Ca$_{1-x})$)Cu$_2$O$_{8+\delta}$ ($x=0.08$) observed. 
It was argued in Ref. \cite{Opel99} that introducing Y for Ca in
Bi$_2$Sr$_2$CaCu$_2$O$_{8}$ leads to the required symmetry breaking,
increasing the intensity of the buckling phonon line. But it was also found
in these experiments that without this Y doping the buckling, if it exists at
all, is at least an order of magnitude smaller as found in 
YBa$_2$Cu$_3$O$_{7-\delta}$ and Bi$_2$Sr$_2$(Y$_{x}$Ca$_{1-x})$Cu$_2$O$_{8}$
with $x=0.38$ \cite{Opel99}. 
Thus, for Bi$_2$Sr$_2$CaCu$_2$O$_{8+\delta}$, the 90 K material that was used for
the vast majority of ARPES experiments, the effect is expected to be small.
However, this is not observed.
In order to reproduce the magnitude of the dip-effect in experimental 
tunneling spectra, in Ref. \cite{Zhu05a} a coupling constant for the
buckling phonon of 10.5 times the gap energy was required, 
which amounts to about 0.28 eV assuming a gap of 27 meV. 

What is missing to date is a careful comparison of both the phonon data
from INS and the corresponding effects in ARPES and tunneling spectroscopy
as a function of doping, temperature, position in the Brillouin zone, 
including bilayer splitting effects, as it has been done for
the spin-1 magnetic resonance.

Thus, although features due to electron-phonon coupling are 
certainly present in the electronic spectra, there is to date no
consistent picture that can describe the dispersion anomalies
in terms of  phonons only. Thus, the most probable scenario is
that of sharp dispersion anomalies due to interaction of electrons
with the sharp spin-1 resonance mode in the antinodal point and
depending on doping also in the nodal point, together with
a superposition of phonon effects that lead to less sharp
features in the electronic dispersion.

\section{Open problems}
\label{OP}

We would like to sketch a collection of open problems which we 
consider as important to be solved in near future.
\begin{itemize}
\item 
One of the main open questions of high-$T_c$ superconductivity is 
nature of the pseudogap phase above $T_c$ in the underdoped state. 
\item 
A related question is if the superconducting state in the underdoped region is
in principle different from that in the overdoped region, and if so whether
the difference is linked to the pseudogap. 
Tunneling experiments indicate that the superconducting states in underdoped and overdoped regions
differ, the underdoped state being inhomogeneous, and the overdoped being
homogeneous \cite{Pan01,Howald01,Lang02,Howald03,Fang04,Mcelroy04a,Mcelroy04b,Fang05}. 
\item 
Can the coupling of electrons to the
spin-fluctuation continuum account for the high transition temperature?
\item 
The fact, that the marginal Fermi liquid hypothesis \cite{Varma89}
works quite well both in the high binding energy
region of the superconducting state and in the normal state, at least near
optimal doping, calls for an theoretical explanation. The main ingredient
seems to be the high-energy inelastic part of 
both the electronic spectrum and the effective
interaction spectrum. It seems that the spin-fluctuation
continuum, extending to very high energies, is able to capture this
physics, although no consensus is reached here yet. 
\item 
A puzzle currently is the observation of the 
unexpected stability of the universal
low-energy renormalization of the Fermi velocity, which does not or only weakly seem
to change with doping or with the chemical composition of the
cuprate superconductor \cite{Zhou03}. 
At the same time the high-energy dispersion is strongly depending both
on doping and material composition. However, instead of a narrowing of
the band with underdoping, a widening of the band is observed,
as the increase of the velocity of the high-energy portion of the
band suggests. The latter effect has been explained in terms of
the development of a Mott-Hubbard pseudogap, that renders the high-energy
dispersion formally equivalent with that of a spin-density wave ordered
state \cite{Chubukov04}.
\item
One open question is also the origin of
the recently observed unusual isotope effect \cite{Gweon04a},
discussed in section \ref{Iso}, that is still unexplained. 
An explanation in terms of a change in the momentum width
of the coupling boson with partial isotope exchange was suggested
in Ref. \cite{Seibold04} (for the case of a charge density wave mode). 
A similar study using the model of Ref. \cite{Eschrig00} has not yet
been performed.
\item 
In La$_{2-x}$Sr$_x$CuO$_4$ the spin excitation spectrum  was recently
found to have a similar low-energy dispersion as in the 90 K 
cuprate superconductors, however with the difference that the
maximum in intensity is situated at $\sim 10$ meV, in a range where
the spectrum is incommensurate. In addition, this incommensurate 
response persists in the normal state, and only sharpens when entering
the superconducting state.
As a result, the corresponding self
energy effects are expected to be 
different from the ones studied in great detail for
Bi$_2$Sr$_2$CaCu$_2$O$_{8+\delta}$. 
A thorough theoretical investigation of this subject
is needed.
\item 
Recent measurements on 
the single layered cuprate Tl$_2$Ba$_2$CuO$_{6+\delta }$ \cite{Plate05}
in the superconducting state 
found a reversal of the linewidths of the quasiparticle peak 
compared to the cases in the bismuth based systems. 
The quasiparticles were found to be sharp near $(\pi, 0)$, i.e. the antinodal
regions, and broad at ($\pi/2,\pi/2$), i.e. the nodal regions of the
Brillouin zone. A similar observation has been made in underdoped
La$_{2-x}$Sr$_x$CuO$_4$, but in this case as a function of doping
the antinodal feature broadens and the nodal feature sharpens \cite{Ino99,Ino00,Ino02}.
This finding poses the problem to explain the fact that the
scattering channels for scattering from the nodes are more effective
than those from the antinodes. Thus clearly, the resonant spin-1 mode
is not the dominating scattering mechanism. The node-node scattering,
however, is dominated by the incommensurate spin-fluctuation spectrum,
as described in section \ref{Incomm}. Thus, an possible explanation is
that similar as in the 
La$_{2-x}$Sr$_x$CuO$_4$ system, the incommensurate part of the spin-fluctuation
spectrum has a higher intensity than the commensurate part. A more 
detailed investigation of
Tl$_2$Ba$_2$CuO$_{6+\delta }$ by inelastic neutron scattering
is needed to support that conjecture.
\end{itemize}
\section{Conclusions}
\label{conclusions}

Experimental investigations of the single particle excitation spectrum
of the superconducting units of high-temperature cuprate superconductors
have revealed a number of dispersion anomalies and unusual lineshapes.
The doping and temperature dependence of these effects as well as
their momentum dispersion within the fermionic Brillouin zone indicate
that there is a contribution to these effects originating from electronic
correlations, in addition to features which can be assigned to 
electron-phonon coupling. Because usually electronic contributions to
self-energy effects in the single-particle spectra are small and smooth
on the low-energy scale, the apparent presence of strong self-energy
effects of electronic origin, which involve a low-energy scale,
are of particular interest. The presence of this common low-energy scale
in spectra taken at various positions in the fermionic Brillouin zone
suggested that all observed dispersion anomalies might have the same
origin. 

This has lead to the development of an explanation in terms of
a peak in the spin excitation spectrum, which couples strongly
to fermionic quasiparticles, is sharp in energy and broadened in momentum
\cite{Norman97,Shen97,Norman98,Shen98,Abanov99}. 
The idea, that the broadening in momentum introduces also dispersion
anomalies in the nodal direction in the fermionic Brillouin zone 
lead to a controversial discussion about the origin of the nodal kink effect
\cite{Eschrig00,Kaminski01,Lanzara01,Johnson01,Shen02,Eschrig03,Sato03,Gromko03,Cuk04a,Cuk04,Kordyuk04,Koitzsch04}.
On one side, an interpretation in terms of electron-phonon coupling was
favored, on the other side in terms of spin-fluctuation exchange.
The situation has been clarified recently by a careful study of the self
energy effects in the nodal direction \cite{Johnson01,Kordyuk04}, 
showing that there are contributions
of both origin, with the spin-fluctuation exchange dominating in
in optimally and underdoped materials whereas they reduce in magnitude
toward overdoping \cite{Johnson01} and are unobservable for highly overdoped 
materials \cite{Gromko03,Hwang04,Cuk04}.

On the other hand, the features in the antinodal regions that originally
lead to the idea that a sharp collective mode might couple to electrons,
had been questioned after the bilayer splitting of the Fermi surface was
experimentally reported \cite{Kordyuk02}. It was claimed that the
anomalous spectral line shape is entirely due to a bilayer split
band, showing a sharper antibonding and a broader bonding quasiparticle peak.
However, further careful experiments revealed that the anomalous lineshape
and the dispersion anomalies
remain present in the bonding band of bilayer materials 
\cite{Gromko03,Feng01a}. 
These dispersion anomalies in antinodal direction are
observed for all doping levels \cite{Gromko03,Cuk04}.
Calculations taking into account bilayer splitting of the quasiparticle bands
accurately describe the overdoped data within a model of electrons
strongly coupled to the sharp spin-1 magnetic resonance \cite{Eschrig02}.
This findings are in agreement with what is expected from the theoretical
treatment in terms of a spin-fluctuation exchange model (in addition to
possible effects due to electron-phonon coupling)
\cite{Eschrig02,Chubukov04,Chubukov04a}. The possibility to separate both
effects allowed to concentrate efforts in particular on the coupling strength between
electronic quasiparticles and the spin excitations.

The theoretical picture that emerged can be summarized as follows \cite{Eschrig03}.
Electronic scattering at low energies is dominated by processes 
which are accompanied by the emission of a sharp spin-fluctuation
mode that is situated below a gapped spin-fluctuation continuum.
The energy required for such scattering
processes is obtained as the sum of the spin fluctuation mode energy and
the binding energy of the quasiparticle at the considered point of the
Brillouin zone.
Accordingly, the coupling between the quasiparticles and the 
sharp spin fluctuation mode
leads to cusps in the energy dependence of the self energy
due to the effect of the van Hove singularities at the $M$ and $A$ points.  
Because of the finite momentum width
of the spin fluctuation mode, there are traces of 
these cusps for electrons at all positions near the Fermi surface.
The {\it position in energy} of these cusps are determined by electrons
near the $M$ and $A$ points only, which explains the observed isotropy of the
involved energy scale around the Fermi surface \cite{Bogdanov00,Kaminski01}.
The {\it intensity} of this self-energy effect is determined by the intensities
of the spin fluctuation mode at such momenta $\vec{q}$ which connect
the electron with momentum $\vec{k} $ 
to electrons near the $M$ point region.
This intensity is large for electrons near the antinodal points, and smaller
for electrons near the nodal points.
This explains the strong anisotropy of the magnitude of the 
effect around the Fermi surface \cite{Eschrig00}.

For overdoped compounds, the intensity of the spin resonance mode for the
wavevector which connects the $M$ point with the $N$ point is too small to
lead to observable effects. This was found in the studies by Johnson {\it et al.} \cite{Johnson01}
and by Gromko {\it et al.} and Cuk {\it et al.} \cite{Gromko03,Cuk04}.
Thus, the self-energy effects at the nodal point
of the Brillouin zone is dominated by other processes, like phonons and
the spin-fluctuation continuum. On the other hand, for optimally and underdoped
compounds the intensity of the spin resonance mode for the
wavevector which connects the $M$ point with the $N$ point is large enough
to modify the dispersion of the quasiparticles even at the nodal points of
the Brillouin zone. This is unambiguously observed in the experiments by
Johnson {\it et al.} \cite{Johnson01} and by Kordyuk {\it et al.} \cite{Kordyuk04}.

The high-energy part of the electronic dispersion is dominated by the
interaction of electrons with a bosonic continuum that extends to
high, electronic ($\sim $1 eV) energies. The scattering linear in 
energy can be well explained by assuming a model with a gapped continuum
that is constant up to a high-energy cut-off. This model also
explains that the high-energy dispersion does not extrapolate to
the normal state Fermi crossing \cite{Eschrig03}.

Finally, a recent aspect is the determination of the parity of the
scattering boson that is responsible for the self energy effects.
It was predicted \cite{Eschrig02}, that for bilayer materials 
the odd symmetry of the resonance-mode under
exchange of the planes within a bilayer has direct consequences for
the strength of the dispersion anomalies of the bonding and antibonding bands.
As the resonance mode scatters predominantly {\it between} bonding and
antibonding bands, in contrast to scattering {\it within} the bands,
corresponding self-energy effects for the bonding band are dominated by
the antibonding van Hove singularity and for the antibonding band by
the bonding van Hove singularity.
Because it is the antibonding band which
is close to the chemical potential in the antinodal region of the Brillouin
zone, the self-energy effects will be stronger for the bonding band \cite{Eschrig02}.
The theoretical results of Ref. \cite{Eschrig02} 
are in full agreement with the experiments
by Feng {\it et al.} \cite{Feng01a}, by Gromko {\it et al.} \cite{Gromko04}
and by Borisenko {\it et  al.} \cite{Borisenko05}.

The success of the theoretical description of the experimentally determined
rather sharp self energy effects allows for the unique possibility to extract
the coupling constant for the interaction between quasiparticles and
resonance mode directly from the measured dispersion anomalies of 
the single particle excitations. 
Assuming that this coupling constant is the same also for the
spin fluctuation continuum, and 
because the 
spin-fluctuation continuum is one of the candidates for providing the pairing
interaction, this gives important information for future studies 
of the origin of the pairing interaction in cuprates.

\section{Acknowledgements}
\label{Ackn}

I would like to thank A.A. Abrikosov, S.V. Borisenko, Ph. Bourges, J. Brinckmann, J.C. Campuzano,
A.V. Chubukov, H. Ding, J. Fink, M. Golden, B. Jank{\'o}, W.P. Halperin,
A. Kaminski, B. Keimer, A.A. Kordyuk, V.F. Mitrovi{\'c}, J. Moreno, 
D. Munzar,
M.R. Norman,
A.I. Posazhennikova, D. Rainer, M. Randeria, A. Rosch, J.A. Sauls, Y. Sidis, O. Tchernyshyov,
I. Vekhter, A. Yurgens, and J. Zasadzinski for useful discussions and 
communications related to the subject.
This work was supported in part by the National Science Foundation under Grant No. PHY99-0794.

\bibliographystyle{prsty1}

\begin{thebibliography}{100}
\label{Refs}

\bibitem{Bednorz86}
J.G. Bednorz and K.A. M\"uller, Z. Phys. B {\bf 64}, 189 (1986).

\bibitem{Tokura89}
Y. Tokura, H. Tagaki, and S. Uchida, Nature (London) {\bf 377}, 345 (1989).

\bibitem{Kastner98}
M.A. Kastner, R.J. Birgeneau, G. Shirane, and Y. Endoh, Rev. Mod. Phys. 
{\bf 70}, 897 (1998).

\bibitem{Dessau91}
D. Dessau, B. Wells, Z. Shen, W. Spicer, A. Arko, R. List, D. Mitzi, and A.
  Kapitulnik, Phys. Rev. Lett. {\bf 66},  2160  (1991).

\bibitem{Hwu91}
Y. Hwu, L. Lozzi, M. Marsi, S.~L. Rosa, M. Winokur, P. Davis, M. Onellion, H.
  Berger, F. Gozzo, F. L{\'e}vy, and G. Margaritondo, Phys. Rev. Lett. {\bf
  67},  2573  (1991).

\bibitem{Matsui03}
H. Matsui, T. Sato, T. Takahashi, S.-C. Wang, H.-B. Yang, H. Ding, T. Fujii, T.
  Watanabe, and A. Matsuda, Phys. Rev. Lett. {\bf 90},  217002  (2003).

\bibitem{Eschrig00}
M. Eschrig and M. Norman, Phys. Rev. Lett. {\bf 85},  3261  (2000).

\bibitem{Kampf90}
A. Kampf and J. Schrieffer, Phys. Rev. B {\bf 41},  6399  (1990).

\bibitem{Dahm96e}
T. Dahm, Phys. Rev. B {\bf 53},  14051  (1996).

\bibitem{Dahm96}
T. Dahm, Phys. Rev. B {\bf 54},  10150  (1996).

\bibitem{Dahm96d}
T. Dahm, D. Manske, and L. Tewordt, Phys. Rev. B {\bf 54},  602  (1996).

\bibitem{Dahm97a}
T. Dahm, D. Manske, and L. Tewordt, Phys. Rev. B {\bf 55},  15274  (1997).

\bibitem{Dahm98}
T. Dahm, D. Manske, and L. Tewordt, Phys. Rev. B {\bf 58},  12454  (1998).

\bibitem{Shen97}
Z. Shen and J. Schrieffer, Phys. Rev. Lett. {\bf 78},  1771  (1997).

\bibitem{Norman97}
M. Norman, H. Ding, J. Campuzano, T. Takeuchi, M. Randeria, T. Yokoya, T.
  Takahashi, T. Mochiku, and K. Kadowaki, Phys. Rev. Lett. {\bf 79},  3506
  (1997).

\bibitem{Norman98}
M. Norman and H. Ding, Phys. Rev. B {\bf 57},  R11089  (1998).

\bibitem{Abanov99}
A. Abanov and A. Chubukov, Phys. Rev. Lett. {\bf 83},  1652  (1999).

\bibitem{Abanov00b}
A. Abanov and A. Chubukov, Phys. Rev. B {\bf 61},  R9241  (2000).

\bibitem{Abanov00c}
A. Abanov and A. Chubukov, Physica B {\bf 280},  189  (2000).

\bibitem{Abanov00d}
A. Abanov and A. Chubukov, Phys. Rev. Lett. {\bf 84},  398  (2000).

\bibitem{Abanov01a}
A. Abanov, A. Chubukov, and J. Schmalian, Phys. Rev. B {\bf 63},  180510
  (2001).

\bibitem{Abanov01b}
A. Abanov, A. Chubukov, and J. Schmalian, J. Electron Spectrosc. Relat. Phenom.
  {\bf 117},  129  (2001).

\bibitem{Abanov02a}
A. Abanov, A. Chubukov, M. Eschrig, M. Norman, and J. Schmalian, Phys. Rev.
  Lett. {\bf 89},  177002  (2002).

\bibitem{Eschrig02}
M. Eschrig and M. Norman, Phys. Rev. Lett. {\bf 89},  277005  (2002).

\bibitem{Eschrig03}
M. Eschrig and M. Norman, Phys. Rev. B {\bf 67},  144503  (2003).

\bibitem{Li00}
J. Li, C. Mou, and T. Lee, Phys. Rev. B {\bf 62},  640  (2000).

\bibitem{Wu01}
C. Wu, C. Mou, and D. Chang, Phys. Rev. B {\bf 63},  172503  (2001).

\bibitem{Manske01}
D. Manske, I. Eremin, and K. Bennemann, Phys. Rev. Lett. {\bf 87},  177005
  (2001).

\bibitem{Manske01a}
D. Manske, I. Eremin, and K. Bennemann, Phys. Rev. B {\bf 63},  054517  (2001).

\bibitem{Norman01a}
M. Norman, M. Eschrig, A. Kaminski, and J. Campuzano, Phys. Rev. B {\bf 64},
  184508  (2001).

\bibitem{Norman03}
M. Norman and C. P{\'e}pin, Rep. Prog. Phys. {\bf 66},  1547  (2003).

\bibitem{Campuzano99}
J. Campuzano, H. Ding, M. Norman, H. Fretwell, M. Randeria, A. Kaminski, J.
  Mesot, T. Takeuchi, T. Sato, T. Yokoya, T. Takahashi, T. Mochiku, K.
  Kadowaki, P. Guptasarma, D. Hinks, Z. Konstantinovi{\'c}, Z. Li, and H. Raffy,
  Phys. Rev. Lett. {\bf 83},  3709  (1999).

\bibitem{Bogdanov00}
P. Bogdanov, A. Lanzara, S. Kellar, X. Zhou, E. Lu, W. Zheng, G. Gu, J.
  Shimoyama, K. Kishio, H. Ikeda, R. Yoshizaki, Z. Hussain, and Z. Shen, Phys.
  Rev. Lett. {\bf 85},  2581  (2000).

\bibitem{Kaminski01}
A. Kaminski, M. Randeria, J. Campuzano, M. Norman, H. Fretwell, J. Mesot, T.
  Sato, T. Takahashi, and K. Kadowaki, Phys. Rev. Lett. {\bf 86},  1070
  (2001).

\bibitem{Gromko02}
A. Gromko, Y. Chuang, A. Fedorov, Y. Aiura, Y. Yamaguchi, K. Oka, Y. Ando, and
  D. Dessau, J. Phys. Chem. Solids {\bf 63},  2299  (2002).

\bibitem{Gromko03}
A. Gromko, A. Fedorov, Y. Chuang, J. Koralek, Y. Aiura, Y. Yamaguchi, K. Oka,
  Y. Ando, and D. Dessau, Phys. Rev. B {\bf 68},  174520  (2003).

\bibitem{Valla00}
T. Valla, A. Fedorov, P. Johnson, Q. Li, G. Gu, and N. Koshizuka, Phys. Rev.
  Lett. {\bf 85},  828  (2000).

\bibitem{Johnson01}
P. Johnson, T. Valla, A. Fedorov, Z. Yusof, B. Wells, Q. Li, A. Moodenbaugh, G.
  Gu, N. Koshizuka, C. Kendziora, S. Jian, and D. Hinks, Phys. Rev. Lett. {\bf
  87},  177007  (2001).

\bibitem{Sato03}
T. Sato, H. Matsui, T. Takahashi, H. Ding, H.-B. Yang, S.-C. Wnag, T. Fujii, T.
  Watanabe, A. Matsuda, T. Terashima, and K. Kadowaki, Phys. Rev. Lett. {\bf
  91},  157003  (2003).

\bibitem{Kordyuk04}
A. Kordyuk, S. Borisenko, A. Koitzsch, J. Fink, M. Knupfer, B. B\"uchner, H.
  Berger, G. Margaritondo, C. Lin, B. Keimer, S. Ono, and Y. Ando, Phys. Rev.
  Lett. {\bf 92},  257006  (2004).

\bibitem{Borisenko03}
S. Borisenko, A. Kordyuk, T. Kim, A. Koitzsch, M. Knupfer, J. Fink, M. Golden,
  M. Eschrig, H. Berger, and R. Follath, Phys. Rev. Lett. {\bf 90},  207001
  (2003).

\bibitem{Kim03a}
T. Kim, A. Kordyuk, S. Borisenko, A. Koitzsch, M. Knupfer, H. Berger, and J.
  Fink, Phys. Rev. Lett. {\bf 91},  167002  (2003).

\bibitem{Koitzsch04}
A. Koitzsch, S. Borisenko, A. Kordyuk, T. Kim, M. Knupfer, J. Fink, H. Berger,
  and R. Follath, Phys. Rev. B {\bf 69},  140507  (2004).

\bibitem{Zasadzinski01}
J. Zasadzinski, L. Ozyuzer, N. Miyakawa, K. Gray, D. Hinks, and C. Kendziora,
  Phys. Rev. Lett. {\bf 87},  067005  (2001).

\bibitem{Hwang04}
J. Hwang, T. Timusk, and G. Gu, Nature {\bf 427},  714  (2004).

\bibitem{McQueeney99}
R. McQueeney, Y. Petrov, T. Egami, M. Yethiraj, G. Shirane, and Y. Endoh, Phys.
  Rev. Lett. {\bf 82},  628  (1999).

\bibitem{Egami02}
T. Egami, J. Chung, R. McQueeney, M. Yethiraj, H. Mook, C. Frost, Y. Petrov, F.
  Do{\u g}an, Y. Inamura, M. Arai, S. Tajima, and Y. Endoh, Physica B {\bf 316},  62
   (2002).

\bibitem{Chung03}
J. Chung, T. Egami, R. McQueeney, M. Yethiraj, M. Arai, T. Yokoo, Y. Petrov, H.
  Mook, Y. Endoh, S. Tajima, C. Frost, and F. Do{\u g}an, Phys. Rev. B {\bf 67},
  014517  (2003).

\bibitem{Lanzara01}
A. Lanzara, P. Bogdanov, X. Zhou, S. Kellar, D. Feng, E. Lu, T. Yoshida, H.
  Eisaki, A. Fujimori, K. Kishio, J. Shimoyama, T. Noda, S. Uchida, Z. Hussain,
  and Z. Shen, Nature {\bf 412},  510  (2001).

\bibitem{Shen02}
Z. Shen, A. Lanzara, S. Ishihara, and N. Nagaosa, Philos. Mag. B-Phys. Condens.
  Matter Stat. Mech. Electron. Opt. Magn. Prop. {\bf 82},  1349  (2002).

\bibitem{Lanzara02}
A. Lanzara, P. Bogdanov, X. Zhou, H. Eisaki, T. Yoshida, A. Fujimori, Z.
  Hussain, and Z. Shen, J. Electron Spectrosc. Relat. Phenom. {\bf 127},  37
  (2002).

\bibitem{Cuk04a}
T. Cuk, F. Baumberger, D. Lu, N. Ingle, X. Zhou, H. Eisaki, N. Kaneko, Z.
  Hussain, T. Devereaux, N. Nagaosa, and Z.-X. Shen, Phys. Rev. Lett. {\bf 93},
   117003  (2004).

\bibitem{Zhou04}
X. Zhou, J. Shi, T. Yoshida, T. Cuk, W. Yang, V. Brouet, J. Nakamura, N.
  Mannella, S. Komiya, Y. Ando, F. Zhou, W. Ti, J.~W. Xiong, Z. Zhao, T.
  Sasagawa, T. Kakeshita, H. Eisaki, S. Uchida, A. Fujimori, Z. Zhang, E.~W.
  Plummer, R. Laughlin, Z. Hussain, and Z.-X. Shen, Phys. Rev. Lett. {\bf 95},
  117001  (2005), cond-mat/0405130.

\bibitem{Gweon04a}
G.-H. Gweon, T. Sasagawa, S. Zhou, J. Graf, H. Takagi, D.-H. Lee, and A.
  Lanzara, Nature {\bf 430},  187  (2004).

\bibitem{Devereaux04}
T. Devereaux, T. Cuk, Z.-X. Shen, and N. Nagaosa, Phys. Rev. Lett. {\bf 93},
  117004  (2004).

\bibitem{Roesch04}
O. R\"osch and O. Gunnarson, Phys. Rev. Lett. {\bf 92},  146403  (2004).

\bibitem{Sandvik04}
A. Sandvik, D. Scalapino, and N. Bickers, Phys. Rev. B {\bf 69},  094523
  (2004).

\bibitem{Lake01}
B. Lake, G. Aeppli, K. Clausen, D. McMorrow, K. Lefmann, N. Hussey, N.
  Mangkorntong, M. Nohara, H. Takagi, T. Mason, and A. Schr{\"o}der, Science {\bf
  291},  1759  (2001).

\bibitem{Lake02c}
B. Lake, H. Ronnow, N. Christensen, G. Aeppli, K. Lefmann, D. McMorrow, P.
  Vorderwisch, P. Smeibidl, N. Mangkorntong, T. Sasagawa, M. Nohara, H. Takagi,
  and T. Mason, Nature {\bf 415},  299  (2002).

\bibitem{Mitrovic01}
V. Mitrovi{\'c}, E. Sigmund, M. Eschrig, H. Bachman, W. Halperin, A. Reyes, P.
  Kuhns, and W.~G. Moulton, Nature {\bf 413},  501  (2001).

\bibitem{Mitrovic03}
V. Mitrovi{\'c}, E. Sigmund, W. Halperin, A. Reyes, P. Kuhns, and W.~G.
  Moulton, Phys. Rev. B {\bf 67},  220503(R)  (2003).

\bibitem{Eschrig94}
M. Eschrig, J. Heym, and D. Rainer, J. Low Temp. Phys. {\bf 95},  323  (1994).

\bibitem{Schrieffer64}
J. Schrieffer, {\em Theory of Superconductivity} (W.A. Benjamin, New York,
  ADDRESS, 1964).

\bibitem{Timusk99}
T. Timusk and B. Statt, Rep. Prog. Phys. {\bf 62},  61  (1999).

\bibitem{Randeria98}
M. Randeria,  in {\em Proceedings of the International School of Physics
  ``Enrico Fermi''}, edited by G. Iadonisi, J. Schrieffer, and M. Chiafalo (IOS
  Press, ADDRESS, 1998), pp.\ 53--75.

\bibitem{Norman05}
M. Norman, D. Pines, and C. Kallin, 
Adv. Phys. {\bf 54}, 715 (2005).

\bibitem{Fauque05}
B. Fauqu{\'e}, Y. Sidis, V. Hinkov, S. Pailh{\`e}s, C. Lin, X. Chaud, and P.
  Bourges, preprint cond-mat/0509201 (unpublished).

\bibitem{Eschrig00b}
M. Eschrig, Applied Magnetic Resonance {\bf 19},  321  (2000).

\bibitem{Chubukov05}
A. Chubukov, D. Maslov, S. Gangadharaiah, and L. Glazman, Phys. Rev. B {\bf
  71},  205112  (2005).

\bibitem{Zaanen89}
J. Zaanen and O. Gunnarsson, Phys. Rev. B {\bf 40},  7391(R)  (1989).

\bibitem{Poilblanc89}
D. Poilblanc and T.M. Rice, Phys. Rev. B {\bf 40}, 7391 (1989).

\bibitem{Machida89}
K. Machida, Physica C {\bf 158},  192  (1989).

\bibitem{Schulz90}
H.J. Schulz, J. Phys. (Paris) {\bf 50}, 2833 (1989); 
Phys. Rev. Lett. {\bf 64},  1445  (1990).

\bibitem{Emery99}
V. Emery, S. Kivelson, and J. Tranquada, Proc. Natl. Acad. Sci. U.S.A. {\bf
  96},  8814  (1999).

\bibitem{Zaanen01}
J. Zaanen, O. Osman, H. Kruis, Z. Nussinov, and J. Tworzyd{\l}o, Phil. Mag. B
  {\bf 81},  1485  (2001).

\bibitem{Kivelson03}
S. Kivelson, I. Bindloss, E. Fradkin, V. Oganesyan, J. Tranquada, A.
  Kapitulnik, and C. Howald, Rev. Mod. Phys. {\bf 75},  1201  (2003).

\bibitem{Tranquada05}
J. Tranquada, preprints cond-mat/0508272 and cond-mat/0512115 (unpublished).

\bibitem{Raczkowski05}
M. Raczkowski, A.M. Ole{\'s}, and R. Fr{\'e}sard, preprint cond-mat/0512420 
(unpublished).

\bibitem{Aeppli89}
G. Aeppli, S. Hayden, H. Mook, Z. Fisk, S. Cheong, D. Rytz, J. Remeika, G.
  Espinosa, and A. Cooper, Phys. Rev. Lett. {\bf 62},  2052  (1989).

\bibitem{Tranquada89}
J. Tranquada, G. Shirane, B. Keimer, S. Shamoto, and M. Sato, Phys. Rev. B {\bf
  40},  4503  (1989).

\bibitem{Rossat93}
J. Rossat-Mignot, L. Regnault, P. Bourges, P. Burlet, C. Vettier, and J. Henry,
   in {\em Selected Topics in Superconductivity}, Vol.~1 of {\em Frontiers in
  Solid State Sciences}, edited by L. Gupta and M. Multani (World Scientific,
  Singapore, ADDRESS, 1993), p.\ 265.

\bibitem{Reznik96}
D. Reznik, P. Bourges, H. Fong, L. Regnault, J. Bossy, C. Vettier, D. Milius,
  I. Aksay, and B. Keimer, Phys. Rev. B {\bf 53},  14741  (1996).

\bibitem{Bourges98}
P. Bourges,  in {\em The Gap Symmetry and Fluctuations in High Temperature
  Superconductors}, edited by J. Bok, G. Deutscher, D. Pavuna, and S.A. Wolf
  (Plenum Press, New York, 1998), p.\ 349.

\bibitem{Regnault98}
L. Regnault, P. Bourges, and P. Burlet,  in {\em Neutron Scattering in Layered
  Copper-Oxide Superconductors}, edited by A. Furrer (Kluwer, Amsterdam,
  ADDRESS, 1998), p.\ 85.

\bibitem{Tranquada92}
J. Tranquada, P. Gehring, G. Shirane, S. Shamoto, and M. Sato, Phys. Rev. B
  {\bf 46},  5561  (1992).

\bibitem{Dai98}
P. Dai, H. Mook, and F. Do{\u g}an, Phys. Rev. Lett. {\bf 80},  1738  (1998).

\bibitem{Mook98a}
H. Mook, P. Dai, R. Hunt, and F. Do{\u g}an, J. Phys. Chem. Solids {\bf 59},  2140
  (1998).

\bibitem{Mook98b}
H. Mook, P. Dai, S. Hayden, G. Aeppli, T. Perring, and F. Do{\u g}an, Nature {\bf
  395},  580  (1998).

\bibitem{Dai01}
P. Dai, H. Mook, R. Hunt, and F. Do{\u g}an, Phys. Rev. B {\bf 63},  054525  (2001).

\bibitem{Aeppli97}
G. Aeppli, T. Mason, S. Hayden, H. Mook, and J. Kulda, Science {\bf 278},  1432
   (1997).

\bibitem{Tranquada04}
J. Tranquada, H. Woo, T. Perring, H. Goka, G. Gu, G. Xu, M. Fujita, and K.
  Yamada, Nature {\bf 429},  534  (2004).

\bibitem{Pailhes04}
S. Pailh\`es, Y. Sidis, P. Bourges, V. Hinkov, A. Ivanov, C. Ulrich, L.
  Regnault, and B. Keimer, Phys. Rev. Lett. {\bf 93},  167001  (2004).

\bibitem{Stock05}
C. Stock, W. Buyers, R. Cowley, P. Clegg, R. Coldea, C. Frost, R. Liang, D.
  Peets, D. Bonn, W. Hardy, and R. Birgeneau, Phys. Rev. B {\bf 71},  24522
  (2005).

\bibitem{Rossat91}
J. Rossat-Mignod, L. Regnault, C. Vettier, P. Bourges, P. Burlet, J. Bossy, J.
  Henry, and G. Lapertot, Physica C {\bf 185-189},  86  (1991).

\bibitem{Mook93}
H. Mook, M. Yethiraj, G. Aeppli, T. Mason, and T. Armstrong, Phys. Rev. Lett.
  {\bf 70},  3490  (1993).

\bibitem{Fong95}
H. Fong, B. Keimer, P. Anderson, D. Reznik, F. Do{\u g}an, and I. Aksay, Phys. Rev.
  Lett. {\bf 75},  316  (1995).

\bibitem{Fong96}
H. Fong, B. Keimer, D. Reznik, D.L. Milius, and I. Aksay, Phys. Rev. B
{\bf 54},  6708  (1996).

\bibitem{Bourges96}
P. Bourges, L. Regnault, Y. Sidis, and C. Vettier, Phys. Rev. B {\bf 53},  876
  (1996).

\bibitem{Fong99}
H. Fong, P. Bourges, Y. Sidis, L. Regnault, A. Ivanov, G. Gu, N. Koshizuka, and
  B. Keimer, Nature {\bf 398},  588  (1999).

\bibitem{Bourges95}
P. Bourges, L. Regnault, J. Henry, C. Vettier, Y. Sidis, and P. Burlet, Physica
  B {\bf 215},  30  (1995).

\bibitem{Balatsky99}
A. Balatsky and P. Bourges, Phys. Rev. Lett. {\bf 82},  5337  (1999).

\bibitem{Dai96}
P. Dai, M. Yethiraj, H. Mook, T. Lindemer, and F. Do{\u g}an, Phys. Rev. Lett. {\bf
  77},  5425  (1996).

\bibitem{Fong97}
H. Fong, B. Keimer, D. Milius, and I. Aksay, Phys. Rev. Lett. {\bf 78},  713
  (1997).

\bibitem{Bourges97}
P. Bourges, H. Fong, L. Regnault, J. Bossy, C. Vettier, D. Milius, I. Aksay,
  and B. Keimer, Phys. Rev. B {\bf 56},  11439  (1997).

\bibitem{Stock04}
C. Stock, W. Buyers, R. Liang, D. Peets, Z. Tun, 
D. Bonn, W. Hardy, and R. Birgeneau, Phys. Rev. B {\bf 69},  014502
  (2004).

\bibitem{Dai99}
P. Dai, H. Mook, S. Hayden, G. Aeppli, T. Perring, R. Hunt, and F. Do{\u g}an,
  Science {\bf 284},  1344  (1999).

\bibitem{He01}
H. He, Y. Sidis, P. Bourges, G. Gu, A. Ivanov, N. Koshizuka, B. Liang, C. Lin,
  L. Regnault, E. Schoenherr, and B. Keimer, Phys. Rev. Lett. {\bf 86},  1610
  (2001).

\bibitem{He02}
H. He, P. Bourges, Y. Sidis, C. Ulrich, L. Regnault, S. Pailh{\`e}s, N.
  Berzigiarova, N. Kolesnikov, and B. Keimer, Science {\bf 295},  1045  (2002).

\bibitem{Fong00}
H. Fong, P. Bourges, Y. Sidis, L. Regnault, J. Bossy, A. Ivanov, D. Milius, I.
  Aksay, and B. Keimer, Phys. Rev. B {\bf 61},  14773  (2000).

\bibitem{Bourges00}
P. Bourges, B. Keimer, L. Regnault, and Y. Sidis, J. Supercond. {\bf 13},  735
  (2000).

\bibitem{Shirane89}
G. Shirane, R.J. Birgeneau, Y. Endoh, P. Gehring, M.A. Kastner, K. Kitazawa,
H. Kojima, I. Tanaka, T.R. Thurston, and Y. Yamada, Phys. Rev. Lett. {\bf 63}
330 (1989).

\bibitem{Mason96}
T.E. Mason, A. Schr{\"o}der, G. Aeppli, H.A. Mook, and S.M. Hayden, 
Phys. Rev. Lett. {\bf 77},  1604  (1996).

\bibitem{Yamada98}
K. Yamada, C.H. Lee, K. Kurahashi, J. Wada, S. Wakimoto, S. Ueki, H. Kimura,
Y. Endoh, S. Hosoya, and G. Shirane, Phys. Rev. B {\bf 57} 6165 (1998).

\bibitem{Christensen04}
N. Christensen, D. McMorrow, H. R{\o}nnow, B. Lake, S. Hayden, G. Aeppli, T.
  Perring, M. Mangkorntong, M. Nohara, and H. Takagi, Phys. Rev. Lett. {\bf
  93},  147002  (2004).

\bibitem{Hayden04}
S. Hayden, H. Mook, P. Dai, T. Perring, and F. Do{\u g}an, Nature {\bf 429},  531
  (2004).

\bibitem{Pailhes03}
S. Pailh{\`e}s, Y. Sidis, P. Bourges, C. Ulrich, V. Hinkov, L. Regnault, A. Ivanov,
  B. Liang, C. Lin, C. Bernhard, and B. Keimer, Phys. Rev. Lett. {\bf 91},
  237002  (2003).

\bibitem{Sidis00}
Y. Sidis, P. Bourges, H. Fong, B. Keimer, L. Regnault, J. Bossy, A. Ivanov, B.
  Hennion, P. Gautier-Picard, G. Collin, D. Millius, and I. Aksay, Phys. Rev.
  Lett. {\bf 84},  5900  (2000).

\bibitem{Pailhes05a}
S. Pailh{\`e}s, C. Ulrich, B. Fauqu{\'e}, V. Hinkov, Y. Sidis, A. Ivanov, C.T. Lin,
B. Keimer, and P. Bourges,
preprint cond-mat/0512634 (unpublished).

\bibitem{Fong99a}
H. Fong, P. Bourges, Y. Sidis, L. Regnault, J. Bossy, A. Ivanov, D. Milius, I.
  Aksay, and B. Keimer, Phys. Rev. Lett. {\bf 82},  1939  (1999).

\bibitem{Millis96}
A. Millis and H. Monien, Phys. Rev. B {\bf 54},  16172  (1996).

\bibitem{Janko99}
B. Jank{\'o}, preprint cond-mat/9912073 (unpublished).

\bibitem{Dai00}
P. Dai, H. Mook, G. Aeppli, S. Hayden, and F. Do{\u g}an, Nature {\bf 406},  965
  (2000).

\bibitem{Mesot99}
J. Mesot, M. Norman, H. Ding, M. Randeria, J. Campuzano, A. Paramekanti, H.
  Fretwell, A. Kaminski, T. Takeuchi, T. Yokoya, T. Sato, T. Takahashi, T.
  Mochiku, and K. Kadowaki, Phys. Rev. Lett. {\bf 83},  840  (1999).

\bibitem{Sidis04}
Y. Sidis, S. Pailh{\`e}s, B. Keimer, P. Bourges, C. Ulrich, and L. Regnault, Phys.
  Status Solidi B-Basic Res. {\bf 241},  1204  (2004).

\bibitem{Sidis96}
Y. Sidis, P. Bourges, B. Hennion, L. Regnault, R. Villeneuve, G. Collin, and J.
  Marucco, Phys. Rev. B {\bf 53},  6811  (1996).

\bibitem{Tarascon88}
J. Tarascon, P. Barboux, P. Miceli, L. Greene, G. Hull, M. Eibschutz, and S.
  Sunshine, Phys. Rev. B {\bf 37},  7458  (1988).

\bibitem{Mendels99}
P. Mendels, J. Bobroff, G. Collin, H. Alloul, M. Gabay, J. Marucco, N.
  Blanchard, and B. Grenier, Europhys. Lett. {\bf 46},  678  (1999).

\bibitem{Pailhes05}
S. Pailh{\`e}s, P. Bourges, Y. Sidis, C. Bernhard, B. Keimer, C. Lin, and J.
  Tallon, Phys. Rev. B {\bf 71},  220507  (2005).

\bibitem{Bourges97a}
P. Bourges, H. Casalta, L.P. Regnault, J. Bossy, P. Burlet, C. Vettier, 
E. Beaugnon, P. Gautier-Picard, and R. Tournier, Physica B {\bf 234}, 830 (1997).

\bibitem{Arai99}
M. Arai, T. Nishijima, Y. Endoh, T. Egami, S. Tajima, K. Tomimoto, Y. Shiohara,
  M. Takahashi, A. Garrett, and S. Bennington, Phys. Rev. Lett. {\bf 83},  608
  (1999).

\bibitem{Bourges00a}
P. Bourges, Y. Sidis, H. Fong, L. Regnault, J. Bossy, A. Ivanov, and B. Keimer,
  Science {\bf 288},  1234  (2000).

\bibitem{Mason93}
T. Mason, G. Aeppli, S. Hayden, A. Ramirez, and H. Mook, Phys. Rev. Lett. {\bf
  71},  919  (1993).

\bibitem{Yamada95}
K. Yamada, S. Wakimoto, G. Shirane, C. Lee, M. Kastner, S. Hosoya, M. Greven,
  Y. Endoh, and R. Birgeneau, Phys. Rev. Lett. {\bf 75},  1626  (1995).

\bibitem{Eremin05}
I. Eremin, D. Morr, A. Chubukov, K. Bennemann, and M. Norman, Phys. Rev. Lett.
  {\bf 94},  147001  (2005).

\bibitem{Lake99}
B. Lake, G. Aeppli, T. Mason, A. Schr{\"o}der, D. McMorrow, K. Lefmann, M. Isshiki,
  M. Nohara, H. Takagi, and S. Hayden, Nature {\bf 400},  43  (1999).

\bibitem{Kakurai93}
K. Kakurai, S. Shamoto, T. Kiyokura, M. Sato, J. Tranquada, and G. Shirane,
  Phys. Rev. B {\bf 48},  3485  (1993).

\bibitem{Regnault95}
L. Regnault, P. Bourges, P. Burlet, J. Henry, J. Rossat-Mignod, Y. Sidis, and
  C. Vettier, Physica B {\bf 213},  48  (1995).

\bibitem{Bourges99}
Ph. Bourges, Y. Sidis, H.F. Fong, B. Keimer, L.P. Regnault, J. Bossy, A.S. Ivanov, D.L. Milius, and I.A. Aksay, in
{\it High Temperature Superconductivity}, 
edited by S.E. Barnes {\it et al} (CP483 American Institute of Physics,
Amsterdam, 1999), pp. 207-212.

\bibitem{Hinkov06}
V. Hinkov, P. Bourges, S. Pailh{\`e}s, Y. Sidis, A. Ivanov, C.T. Lin, D.P. Chen, and B. Keimer,
preprint cond-mat/0601048 (unpublished).

\bibitem{Damascelli03}
A. Damascelli, Z. Hussain, and Z.-X. Shen, Rev. Mod. Phys. {\bf 75},  473
  (2003).

\bibitem{Campuzano04}
J. Campuzano, M. Norman, and M. Randeria,  in {\em Physics of Superconductors},
  edited by K. Bennemann and J. Ketterson (Springer, Berlin, ADDRESS, 2004),
  p.\ 167.

\bibitem{Olson90}
C. Olson, R. Liu, D. Lynch, R. List, A. Arko, B. Veal, Y. Chang, P. Jiang, and
  A. Paulikas, Phys. Rev. B {\bf 42},  381  (1990).

\bibitem{Campuzano90}
J. Campuzano, G. Jennings, M. Faiz, L. Beaulaigue, B. Veal, J. Liu, A.~P.
  Paulikas, K. Vandervoort, and H. Claus, Phys. Rev. Lett. {\bf 64},  2308
  (1990).

\bibitem{Shen95}
Z.-X. Shen and D. Dessau, Physics Reports {\bf 253},  1  (1995).

\bibitem{Ding96b}
H. Ding, A. Bellman, J. Campuzano, M. Randeria, M. Norman, T. Yokoya, T.
  Takahashi, H. KatayamaYoshida, T. Mochiku, K. Kadowaki, G. Jennings, and G.
  Brivio, Phys. Rev. Lett. {\bf 76},  1533  (1996).

\bibitem{Kordyuk02}
A. Kordyuk, S. Borisenko, T. Kim, K. Nenkov, M. Knupfer, J. Fink, M. Golden, H.
  Berger, and R. Follath, Phys. Rev. Lett. {\bf 89},  077003  (2002).

\bibitem{Matsui03a}
H. Matsui, T. Sato, T. Takahashi, H. Ding, H.-B. Yang, S.-C. Wang, T. Fujii, T.
  Watanabe, A. Matsuda, T. Terashima, and K. Kadowaki, Phys. Rev. B {\bf 67},
  060501(R)  (2003).

\bibitem{Sato01}
T. Sato, T. Kamiyama, T. Takahashi, J. Mesot, A. Kaminski, J. Campuzano, H.
  Fretwell, T. Takeuchi, H. Ding, I. Chong, T. Terashima, and M. Takano, Phys.
  Rev. B {\bf 64},  054502  (2001).

\bibitem{Sato02}
T. Sato, H. Matsui, S. Nishina, T. Takahashi, T. Fujii, T. Watanabe, and A.
  Matsuda, Phys. Rev. Lett. {\bf 89},  067005  (2002).

\bibitem{Takeuchi01}
T. Takeuchi, T. Yokoya, S. Shin, K. Jinno, M. Matsuura, T. Kondo, H. Ikuta, and
  U. Mizutani, J. Electron Spectr. Relat. Phenom. {\bf 114-116},  629  (2001).

\bibitem{kondo04}
T. Kondo, T. Takeuchi, T. Yokoya, S. Tsuda, S. Shin, and U. Mizutani, J.
  Electron Spectr. Relat. Phenom. {\bf 137-140},  663  (2004).

\bibitem{Fujimori98}
A. Fujimori, A. Ino, T. Mizokawa, C. Kim, Z.-X. Shen, T. Sasagawa, T. Kimura,
  K. Kishio, M. Takaba, K. Tamasaku, H. Eisaki, and S. Uchida, J. Phys. Chem.
  Solids {\bf 59},  1892  (1998).

\bibitem{Ino99}
A. Ino, C. Kim, T. Mizokawa, Z.-X. Shen, A. Fujimori, M. Takaba, K. Tamasaku,
  H. Eisaki, and S. Uchida, J. Phys. Soc. Jpn. {\bf 68},  1496  (1999).

\bibitem{Ino02}
A. Ino, C. Kim, M. Nakamura, T. Yoshida, T. Mizokawa, A. Fujimori, Z.-X. Shen,
  T. Kakeshita, H. Eisaki, and S. Uchida, Phys. Rev. B {\bf 65},  94504
  (2002).

\bibitem{Kaminski05}
A. Kaminski, H. Fretwell, M. Norman, S.~R. M.~Randeira, U. Chatterjee, J.
  Campuzano, J. Mesot, T. Sato, T. Takahashi, T. Terashima, M. Takano, K.
  Kadowaki, Z. Li, and H. Raffy, Phys. Rev. B {\bf 71},  14517  (2005).

\bibitem{Plate05}
M. Plat{\'e}, J. Mottershead, I. Elfimov, D. Peets, R. Liang, D. Bonn, W. Hardy, S.
  Chiuzb{\u a}ian, M. Falub, M. Shi, L. Patthey, and A. Damascelli, Phys. Rev. Lett.
  {\bf 95},  77001  (2005).

\bibitem{Norman98n}
M. Norman, H. Ding, M. Randeria, J. Campuzano, T. Yokoya, T. Takeuchi, T.
  Takahashi, T. Mochiku, K. Kadowaki, P. Guptasarma, and D. Hinks, Nature {\bf
  392},  157  (1998).

\bibitem{Abrikosov93}
A. Abrikosov, J. Campuzano, and K. Gofron, Physica (Amsterdam) {\bf 214C},  73
  (1993).

\bibitem{Dessau93}
D. Dessau, Z.-X. Shen, D. King, D. Marshall, L. Lombardo, P. Dickinson, A.
  Loeser, J. DiCarlo, C.-H. Park, A. Kapitulnik, and W. Spicer, Phys. Rev.
  Lett. {\bf 71},  2781  (1993).

\bibitem{Gofron94}
K. Gofron, J. Campuzano, A. Abrikosov, M. Lindroos, A. Bansil, H. Ding, D.
  Koelling, and B. Dabrowski, Phys. Rev. Lett. {\bf 73},  3302  (1994).

\bibitem{Massida88}
S. Massida, J. Yu, and A. Freeman, Physica {\bf C 152},  251  (1988).

\bibitem{Andersen94}
O. Andersen, O. Jepsen, A. Liechtenstein, and I. Mazin, Phys. Rev. B {\bf 49},
  4145  (1994).

\bibitem{Feng01a}
D. Feng, N. Armitage, D. Lu, A. Damascelli, J. Hu, P. Bogdanov, A. Lanzara, F.
  Ronning, K. Shen, H. Eisaki, C. Kim, Z. Shen, J. Shimoyama, and K. Kishio,
  Phys. Rev. Lett. {\bf 86},  5550  (2001).

\bibitem{Chuang01}
Y. Chuang, A. Gromko, A. Fedorov, Y. Aiura, K. Oka, Y. Ando, H. Eisaki, S.
  Uchida, and D. Dessau, Phys. Rev. Lett. {\bf 8711},  117002  (2001).

\bibitem{Andersen95}
O. Andersen, A. Liechtenstein, O. Jepsen, and F. Paulsen, J. Phys. Chem. Solids
  {\bf 56},  1573  (1995).

\bibitem{Chakravarty93}
S. Chakravarty, A. Sudb{\o}, P. Anderson, and S. Strong, Science {\bf 261},  337
  (1993).

\bibitem{Feng02a}
D. Feng, C. Kim, H. Eisaki, D. Lu, A. Damascelli, K. Shen, F. Ronning, N.
  Armitage, N. Kaneko, M. Greven, J. Shimoyama, K. Kishio, R. Yoshizaki, G. Gu,
  and Z. Shen, Phys. Rev. B {\bf 65},  220501  (2002).

\bibitem{Chuang04}
Y. Chuang, A. Gromko, A. Fedorov, Y. Aiura, K. Oka, Y. Ando, M. Lindroos, R.
  Markiewicz, A. Bansil, and D. Dessau, Phys. Rev. B {\bf 69},  094515  (2004).

\bibitem{Kordyuk03a}
A. Kordyuk, S. Borisenko, A.~N. Yaresko, S.-L. Drechsler, H. Rosner, T. Kim, A.
  Koitzsch, K. Nenkov, M. Knupfer, J. Fink, R. Follath, H. Berger, B. Keimer,
  S. Ono, and Y. Ando, Phys. Rev. B {\bf 70},  214525  (2004), 
  cond-mat/0311137.

\bibitem{Borisenko06}
S.V. Borisenko, A.A. Kordyuk, V. Zabolotnyy, J. Geck, D. Inosov, A. Koitzsch, J. Fink, M. Knupfer,
B. B\"uchner, V. Hinkov, C.T. Lin, B. Keimer, T. Wolf, S.G. Chiuzb{\u a}ian, 
L. Patthey, and R. Follath, preprint cond-mat/0511596 (unpublished).

\bibitem{Valla05}
T. Valla, T.E. Kidd, J.D. Rameau, H.-J. Noh, G.D. Gu, P.D. Johnson, H.-B. Yang, and H. Ding,
preprint cond-mat/0512685  (unpublished).

\bibitem{Chuang02}
Y. Chuang, A. Gromko, A. Fedorov, Y. Aiura, K. Oka, Y. Ando, and D. Dessau,
  preprint cond-mat/0107002 (unpublished).

\bibitem{Kordyuk02a}
A. Kordyuk, S. Borisenko, M. Golden, S. Legner, K. Nenkov, M. Knupfer, J. Fink,
  H. Berger, L. Forr{\'o}, and R. Follath, Phys. Rev. B {\bf 66},  014502  (2002).

\bibitem{Borisenko02}
S. Borisenko, A. Kordyuk, T. Kim, S. Legner, K. Nenkov, M. Knupfer, M. Golden,
  J. Fink, H. Berger, and R. Follath, Phys. Rev. B {\bf 66},  140509  (2002).

\bibitem{Kaminski03}
A. Kaminski, S. Rosenkranz, H. Fretwell, Z. Li, H. Raffy, M. Randeria, M.
  Norman, and J. Campuzano, Phys. Rev. Lett. {\bf 90},  207003  (2003).

\bibitem{Campuzano96}
J. Campuzano, H. Ding, M. Norman, M. Randeira, A. Bellman, T. Mochiku, and K.
  Kadowaki, Phys. Rev. B {\bf 53},  14737  (1996).

\bibitem{Kaminski00}
A. Kaminski, J. Mesot, H. Fretwell, J. Campuzano, M. Norman, M. Randeria, H.
  Ding, T. Sato, T. Takahashi, T. Mochiku, K. Kadowaki, and H. Hoechst, Phys.
  Rev. Lett. {\bf 84},  1788  (2000).

\bibitem{Gromko04}
A. Gromko, Y.-D. Chuang, A. Fedorov, Y. Aiura, Y. Yamaguchi, K. Oka, Y. Ando,
  and D. Dessau, preprint cond-mat/0205385 (unpublished).

\bibitem{Valla99}
T. Valla, A. Fedorov, P. Johnson, B. Wells, S. Hulbert, Q. Li, G. Gu, and N.
  Koshizuka, Science {\bf 285},  2110  (1999).

\bibitem{Scalapino_Parks}
D. Scalapino,  in {\em Superconductivity}, edited by R. Parks (Marcel Decker,
  New York, ADDRESS, 1969), Vol.~1, p.\ 449.

\bibitem{Randeria04}
M. Randeria, A. Paramekanti, and N. Trivedi, Phys. Rev. B {\bf 69},  144509
  (2004).

\bibitem{Zhou03}
X. Zhou, T. Yoshida, A. Lanzara, P. Bogdanov, S. Kellar, K. Shen, W. Yang, F.
  Ronning, T. Sasagawa, T. Kakeshita, T. Noda, H. Eisaki, S. Uchida, C. Lin, F.
  Zhou, J. Xiong, W. Ti, Z. Zhao, A. Fujimori, Z. Hussain, and Z. Shen, Nature
  {\bf 423},  398  (2003).

\bibitem{Ronning03}
F. Ronning, T. Sasagawa, Y. Kohsaka, K. Shen, A. Damascelli, C. Kim, T.
  Yoshida, N. Armitage, D. Lu, D. Feng, L. Miller, H. Takagi, and Z.-X. Shen,
  Phys. Rev. B {\bf 67},  165101  (2003).

\bibitem{Kordyuk05}
A. Kordyuk, S. Borisenko, A. Koitzsch, J. Fink, M. Knupfer, and H. Berger,
  Phys. Rev. B {\bf 71},  214513  (2005).

\bibitem{Randeria95}
M. Randeria, H. Ding, J. Campuzano, A. Bellman, G. Jennings, T. Yokoya, T.
  Takahashi, H. Katayamayoshida, T. Mochiku, and K. Kadowaki, Phys. Rev. Lett.
  {\bf 74},  4951  (1995).

\bibitem{Ding96}
H. Ding, T. Yokoya, J. Campuzano, T. Takahashi, M. Randeria, M. Norman, T.
  Mochiku, K. Kadowaki, and J. Giapintzakis, Nature {\bf 382},  51  (1996).

\bibitem{Lu01}
D. Lu, D. Feng, N. Armitage, K. Shen, A. Damascelli, C. Kim, F. Ronning, Z.
  Shen, D. Bonn, R. Liang, W. Hardy, A. Rykov, and S. Tajima, Phys. Rev. Lett.
  {\bf 86},  4370  (2001).

\bibitem{Feng02}
D. Feng, A. Damascelli, K. Shen, N. Motoyama, D. Lu, H. Eisaki, K. Shimizu, J.
  Shimoyama, K. Kishio, N. Kaneko, M. Greven, G. Gu, X. Zhou, C. Kim, F.
  Ronning, N. Armitage, and Z. Shen, Phys. Rev. Lett. {\bf 88},  107001
  (2002).

\bibitem{Norman01}
M. Norman, A. Kaminski, J. Mesot, and J. Campuzano, Phys. Rev. B {\bf 63},
  140508  (2001).

\bibitem{Yusof02}
Z. Yusof, B. Wells, T. Valla, A. Fedorov, P. Johnson, Q. Li, C. Kendziora, S.
  Jian, and D. Hinks, Phys. Rev. Lett. {\bf 88},  167006  (2002).

\bibitem{Kuroda90}
Y. Kuroda and C. Varma, Phys. Rev. B {\bf 42},  8619  (1990).

\bibitem{Littlewood92}
P. Littlewood and C. Varma, Phys. Rev. B {\bf 46},  405  (1992).

\bibitem{Borisenko05}
S. Borisenko, A. Kordyuk, A. Koitzsch, J. Fink, J. Geck, V. Zabolotnyy, M.
  Knupfer, B. B\"uchner, H. Berger, M. Falub, M. Shi, J. Krempasky, and L.
  Patthey, Phys. Rev. Lett. {\bf 96}, 67001 (2006), cond-mat/0505192.

\bibitem{Wells92}
B. Wells, Z. Shen, D. Dessau, W. Spicer, D. Mitzi, L. Lombardo, A. Kapitulnik,
  and A. Arko, Phys. Rev. B {\bf 46},  11830  (1992).

\bibitem{Shen93}
Z.-X. Shen, D. Dessau, B. Wells, D. King, W. Spicer, A. Arko, D. Marshall, L.
  Lombardo, A. Kapitulnik, P. Dickinson, S. Doniach, J. DiCarlo, T. Loeser, and
  C. Park, Phys. Rev. Lett. {\bf 70},  1553  (1993).

\bibitem{Tsuei00}
C.~C. Tsuei and J.~R. Kirtley, Rev. Mod. Phys. {\bf 72},  969  (2000).

\bibitem{Sutherland03}
M. Sutherland, D.G. Hawthorn, R.W. Hill, F. Ronning, S. Wakimoto, H. Zhang,
C. Proust, E. Boaknin, C. Lupien, and L. Taillefer,
Phys. Rev. B {\bf 67}, 174520 (2003).

\bibitem{Fedorov99}
A. Fedorov, T. Valla, P. Johnson, Q. Li, G. Gu, and N. Koshizuka, Phys. Rev.
  Lett. {\bf 82},  2179  (1999).

\bibitem{Ding01}
H. Ding, J. Engelbrecht, Z. Wang, J. Campuzano, S. Wang, H. Yang, R. Rogan, T.
  Takahashi, K. Kadowaki, and D. Hinks, Phys. Rev. Lett. {\bf 87},  227001
  (2001).

\bibitem{Feng00}
D. Feng, D. Lu, K. Shen, C. Kim, H. Eisaki, A. Damascelli, R. Yoshizaki, J.
  Shimoyama, K. Kishio, G. Gu, S. Oh, A. Andrus, J. O'Donnell, J. Eckstein, and
  Z. Shen, Science {\bf 289},  277  (2000).

\bibitem{Mook98c}
H. Mook, F. Do{\u g}an, and B.~C. Chakoumakos, preprint cond-mat/9811100
  (unpublished).

\bibitem{Varma89}
C. Varma, P. Littlewood, S. Schmitt-Rink, E. Abrahams, and A. Ruckenstein,
  Phys. Rev. Lett. {\bf 63},  1996  (1989).

\bibitem{Zhu04}
L. Zhu, P.J. Hirschfeld, and D.J. Scalapino, Phys. Rev. B {\bf 70}, 214503 (2004).

\bibitem{Dahm05}
T. Dahm, P.J. Hirschfeld, D.J. Scalapino, and L. Zhu, Phys. Rev. B {\bf 72}, 214512 (2005).

\bibitem{Marshall96}
D. Marshall, D. Dessau, A. Loeser, C. Park, A. Matsuura, J. Eckstein, I.
  Bo{\v z}ovi{\'c}, P. Fournier, A. Kapitulnik, W. Spicer, and Z. Shen, Phys. Rev. Lett.
  {\bf 76},  4841  (1996).

\bibitem{Huang89}
Q. Huang, J. Zasadzinski, K. Gray, J. Liu, and H. Claus, Phys. Rev. B {\bf 40},
   9366  (1989).

\bibitem{Renner95}
C. Renner and {\O}. Fischer, Phys. Rev. B {\bf 51},  9208  (1995).

\bibitem{DeWilde98}
Y. DeWilde, N. Miyakawa, P. Guptasarma, M. Iavarone, L. Ozyuzer, J.
  Zasadzinski, P. Romano, D. Hinks, C. Kendziora, G. Crabtree, and K. Gray,
  Phys. Rev. Lett. {\bf 80},  153  (1998).

\bibitem{Hudson99}
E. Hudson, S. Pan, A. Gupta, K. Ng, and J. Davis, Science {\bf 285},  88
  (1999).

\bibitem{Mandrus91}
D. Mandrus, L. Forr{\'o}, D. Koller, and L. Mih{\'a}ly, Nature {\bf 351},  460  (1991).

\bibitem{Zasadzinski02}
J. Zasadzinski, L. Ozyuzer, N. Miyakawa, K. Gray, D. Hinks, and C. Kendziora,
  J. Phys. Chem. Solids {\bf 63},  2247  (2002).

\bibitem{Yurgens99}
A. Yurgens, D. Winkler, T. Claeson, S. Hwang, and J. Choy, Int. J. Mod. Phys. B
  {\bf 13},  3758  (1999).

\bibitem{Zasadzinski03}
J. Zasadzinski, L. Coffey, P. Romano, and Z. Yusof, Phys. Rev. B {\bf 68},
  180504(R)  (2003).

\bibitem{Zasadzinski00}
J. Zasadzinski, L. Ozyuzer, N. Miyakawa, D. Hinks, and K. Gray, Physica C {\bf
  341},  867  (2000).

\bibitem{Renner96}
C. Renner, B. Revaz, J. Genoud, and {\O}. Fischer, J. Low Temp. Phys. {\bf 105},
  1083  (1996).

\bibitem{Hwang05}
J. Hwang, J. Yang, T. Timusk, S. Shaparov, J. Carbotte, D. Bonn, R. Liang, and
  W. Hardy, 
  Phys. Rev. B {\bf 73}, 24508 (2006), cond-mat/0505302.

\bibitem{Cuk04}
T. Cuk, Z.-X. Shen, A. Gromko, Z. Sun, and D. Dessau, Nature Brief
  Communication {\bf 432} (2004).

\bibitem{Valla04}
T. Valla, P. Johnson, G. Gu, J. Hwang, and T. Timusk, 
preprint cond-mat/0405203v2 (unpublished).

\bibitem{Marsiglio98}
F. Marsiglio, Phys. Lett. A {\bf 245},  172  (1998).

\bibitem{Carbotte99}
J. Carbotte, E. Schachinger, and D. Basov, Nature {\bf 401},  354  (1999).

\bibitem{Casek05}
P. C{\'a}sek, J. Huml{\'i}{\v c}ek, D. Munzar, and Ch. Bernhard, Phys. Rev. B {\bf 72}, 134526 (2005).

\bibitem{Tu02}
J. Tu, C. Homes, G. Gu, D. Basov, and M. Strongin, Phys. Rev. B {\bf 66},
  144514  (2002).

\bibitem{Homes99}
C. Homes, D. Bonn, R. Liang, W. Hardy, D. Basov, T. Timusk, and B. Clayman,
  Phys. Rev. B {\bf 60},  9782  (1999).

\bibitem{Brinckmann98}
J. Brinckmann and P. Lee, J. Phys. Chem. Solids {\bf 59},  1811  (1998).

\bibitem{Brinckmann99}
J. Brinckmann and P. Lee, Phys. Rev. Lett. {\bf 82},  2915  (1999).

\bibitem{Demler98}
E. Demler and S. Zhang, Nature {\bf 396},  733  (1998).

\bibitem{Greiter97}
M. Greiter, Phys. Rev. Lett. {\bf 79},  4898  (1997).

\bibitem{Lavagna94}
M. Lavagna and G. Stemmann, Phys. Rev. B {\bf 49},  4235  (1994).

\bibitem{Abrikosov98}
A. Abrikosov, Phys. Rev. B {\bf 57},  8656  (1998).

\bibitem{Demler95}
E. Demler and S. Zhang, Phys. Rev. Lett. {\bf 75},  4126  (1995).

\bibitem{Zhang97}
Y. Zhang, Science {\bf 275},  1089  (1997).

\bibitem{Tchernyshyov01}
O. Tchernyshyov, M. Norman, and A. Chubukov, Phys. Rev. B {\bf 63},  144507
  (2001).

\bibitem{Vojta00}
M. Vojta, C. Buragohain, and S. Sachdev, Phys. Rev. B {\bf 61},  15152  (2000).

\bibitem{Herbut03}
I. Herbut and D. Lee, Phys. Rev. B {\bf 68},  104518  (2003).

\bibitem{Uemura04}
Y. Uemura, J. Phys.: Condens. Matter {\bf 16},  S4515  (2004).

\bibitem{Uemura05}
Y. Uemura, preprint cond-mat/0512075 (unpublished).

\bibitem{Kruger03}
F. Kr\"uger and S. Scheidl, Phys. Rev. B {\bf 67}, 134512 (2003).

\bibitem{Carlson04}
E.W. Carlson, D.X. Yao, and D.K. Campbell, Phys. Rev. B {\bf 70}, 064505 (2004).

\bibitem{Vojta04}
M. Vojta and T. Ulbricht, Phys. Rev. Lett. {\bf 93}, 127002 (2004).

\bibitem{Uhrig04}
G.S. Uhrig, K.P. Schmidt, and M. Gr\"uninger, Phys. Rev. Lett. {\bf 93}, 267003
(2004).

\bibitem{Uhrig05}
G.S. Uhrig, K.P. Schmidt, and M. Gr\"uninger, preprint cond-mat/0502460 (unpublished).

\bibitem{Andersen05}
B.M. Andersen and P. Hedeg{\aa}rd, Phys. Rev. Lett. {\bf 95}, 037002 (2005).

\bibitem{Seibold05}
G. Seibold and J. Lorenzana, Phys. Rev. Lett. {\bf 94}, 107006 (2005).

\bibitem{Seibold05a}
G. Seibold and J. Lorenzana, preprint cond-mat/0509175 (unpublished).

\bibitem{Tanamoto91}
T. Tanamoto, K. Kuboki, and H. Fukuyama, J. Phys. Soc. Jpn {\bf 60},  3072
  (1991).

\bibitem{Tanamoto94}
T. Tanamoto, H. Kohno, and H. Fukuyama, J. Phys. Soc. Jpn {\bf 63},  2739
  (1994).

\bibitem{Zha93}
Y. Zha, K. Levin, and Q. Si, Phys. Rev. B {\bf 47},  9124  (1993).

\bibitem{Stemmann94}
G. Stemmann, C. P{\'e}pin, and M. Lavagna, Phys. Rev. B {\bf 50},  4075
  (1994).

\bibitem{Liu95}
D. Liu, Y. Zha, and K. Levin, Phys. Rev. Lett. {\bf 75},  4130  (1995).

\bibitem{Brinckmann01}
J. Brinckmann and P. Lee, Phys. Rev. B {\bf 65},  014502  (2001).

\bibitem{Kao00}
Y.-J. Kao, Q. Si, and K. Levin, Phys. Rev. B {\bf 61},  R11898  (2000).

\bibitem{Li02}
J. Li and C.-D. Gong, Phys. Rev. B {\bf 66},  014506  (2002).

\bibitem{Chen05}
W.Q. Chen and Z.Y. Weng, Phys. Rev. B {\bf 71}, 134516 (2005).

\bibitem{Onufrieva95}
F. Onufrieva and J. Rossat-Mignod, Phys. Rev. B {\bf 52},  7572  (1995).

\bibitem{Onufrieva99x}
F. Onufrieva and P. Pfeuty, preprint cond-mat/9903097 (unpublished).

\bibitem{Onufrieva02}
F. Onufrieva and P. Pfeuty, Phys. Rev. B {\bf 65},  054515  (2002).

\bibitem{Sega03}
I. Sega, P. Prelov{\v s}ek, and J. Bon{\v c}a,
Phys. Rev. B {\bf 68}, 054524 (2003).

\bibitem{Sega05}
I. Sega and P. Prelov{\v s}ek, 
preprint cond-mat/0503099 (unpublished).

\bibitem{Prelovsek06}
P. Prelov{\v s}ek, I. Sega, A. Ramsak, and J. Bon{\v c}a,
preprint cond-mat/0601511 (unpublished).

\bibitem{Maki94}
K. Maki and H. Won, Phys. Rev. Lett. {\bf 72},  1758  (1994).

\bibitem{Mazin95}
I. Mazin and V.~M. Yakovenko, Phys. Rev. Lett. {\bf 75},  4134  (1995).

\bibitem{Bulut96}
N. Bulut and D. Scalapino, Phys. Rev. B {\bf 54},  14971  (1996).

\bibitem{Salkola98}
M. Salkola and J. Schrieffer, Phys. Rev. B {\bf 58},  R5944  (1998).

\bibitem{Norman01b}
M. Norman, Phys. Rev. B {\bf 63},  092509  (2001).

\bibitem{Pao95}
C.-H. Pao and N. Bickers, Phys. Rev. B {\bf 51},  16310  (1995).

\bibitem{Takimoto98}
T. Takimoto and T. Moriya, J. Phys. Soc. Jpn {\bf 67},  3570  (1998).

\bibitem{Morr98}
D. Morr and D. Pines, Phys. Rev. Lett. {\bf 81},  1086  (1998).

\bibitem{Chubukov01}
A. Chubukov, B. Jank{\'o}, and O. Tchernyshyov, Phys. Rev. B {\bf 63},  180507
  (2001).

\bibitem{Chubukov04a}
A. Chubukov, D. Pines, and J. Schmalian,  in {\em Physics of Superconductors},
  edited by K. Bennemann and J. Ketterson (Springer, Berlin, ADDRESS, 2004),
  p.\ 495.

\bibitem{Morr00}
D.K. Morr and D. Pines, Phys. Rev. B {\bf 61}, R6483 (2000).

\bibitem{Morr00a}
D.K. Morr and D. Pines, Phys. Rev. B {\bf 62}, 15177 (2000).


\bibitem{Onufrieva99}
F. Onufrieva, P. Pfeuty, and  M. Kiselev,
Phys. Rev. Lett. {\bf 82}, 2370 (1999).

\bibitem{Onufrieva00}
F. Onufrieva and P. Pfeuty, Phys. Rev. B {\bf 61}, 799 (2000).

\bibitem{Ioffe99}
L. Ioffe and A. Millis, Science {\bf 285},  1241  (1999).

\bibitem{Bulut00}
N. Bulut, Phys. Rev. B {\bf 61}, 9051 (2000).

\bibitem{Sachdev99}
S. Sachdev, Ch. Buragohain, and M. Vojta, Science {\bf 286}, 2479 (1999).

\bibitem{Eschrig01}
M. Eschrig, M. Norman, and B. Jank{\'o}, Phys. Rev. B {\bf 64},  134509  (2001).

\bibitem{Matsuda95}
Y. Matsuda, M. Gaifullin, K. Kumagai, K. Kadowaki, and T. Mochiku, Phys. Rev.
  Lett. {\bf 75},  4512  (1995).

\bibitem{Maggio95}
I. Maggio-Aprile, C. Renner, A. Erb, E. Walker, and {\O}. Fischer, Phys. Rev.
  Lett. {\bf 75},  2754  (1995).

\bibitem{Renner98}
C. Renner, B. Revaz, K. Kadowaki, I. Maggio-Aprile, and {\O}. Fischer, Phys. Rev.
  Lett. {\bf 80},  3606  (1998).

\bibitem{Pan00a}
S. Pan, E. Hudson, A. Gupta, K. Ng, H. Eisaki, S. Uchida, and J. Davis, Phys.
  Rev. Lett. {\bf 85},  1536  (2000).

\bibitem{Loeser97}
A. Loeser, Z. Shen, M. Schabel, C. Kim, M. Zhang, A. Kapitulnik, and P.
  Fournier, Phys. Rev. B {\bf 56},  14185  (1997).

\bibitem{Norman98a}
M. Norman, M. Randeira, H. Ding, and J. Campuzano, Phys. Rev. B {\bf 57},
  R11093  (1998).

\bibitem{Norman00}
M. Norman, Phys. Rev. B {\bf 61},  14751  (2000).

\bibitem{Abanov03}
A. Abanov, A. Chubukov, and J. Schmalian, Adv. Phys. {\bf 52},  119  (2003).

\bibitem{Kee02}
H. Kee, S. Kivelson, and G. Aeppli, Phys. Rev. Lett. {\bf 88},  257002  (2002).

\bibitem{Basov96}
D. Basov, R. Liang, B. Dabrowski, D. Bonn, W. Hardy, and T. Timusk, Phys. Rev.
  Lett. {\bf 77},  4090  (1996).

\bibitem{Abanov01c}
A. Abanov, A. Chubukov, and J. Schmalian, Europhys. Lett. {\bf 55},  369
  (2001).

\bibitem{Auerbach00}
A. Auerbach and E. Altman, Phys. Rev. Lett. {\bf 85},  3480  (2000).

\bibitem{Abanov01d}
A. Abanov, A. Chubukov, and A. Finkel'stein, Europhys. Lett. {\bf 54},  488
  (2001).

\bibitem{Monthoux94a}
P. Monthoux and D. Pines, Phys. Rev. B {\bf 49},  4261  (1994).

\bibitem{Zha96}
Y. Zha, V. Barzykin, and D. Pines, Phys. Rev. B {\bf 54},  7561  (1996).

\bibitem{Scalapino95}
D. Scalapino, Phys. Rep.-Rev. Sec. Phys. Lett. {\bf 250},  330  (1995).

\bibitem{Vilk97}
Y. Vilk and A. Tremblay, J. Phys. I {\bf 7},  1309  (1997).

\bibitem{Loram97}
J. Loram, K. Mirza, J. Cooper, and J. Tallon, Physica C {\bf 282},  1405
  (1997).

\bibitem{Shraiman90}
B. Shraiman and E. Siggia, Phys. Rev. B {\bf 42},  2485  (1990).

\bibitem{Schrieffer95}
J. Schrieffer, J. Low Temp. Phys. {\bf 99},  397  (1995).

\bibitem{Sachdev95b}
S. Sachdev, A. Chubukov, and A. Sokol, Phys. Rev. B {\bf 51},  14874  (1995).

\bibitem{Chubukov97}
A. Chubukov and D. Morr, Phys. Rep.-Rev. Sec. Phys. Lett. {\bf 288},  355
  (1997).

\bibitem{Marsiglio88}
F. Marsiglio, M. Schossmann, and J. Carbotte, Phys. Rev. B {\bf 37},  4965
  (1988).

\bibitem{Shen98}
Z. Shen, P. White, D. Feng, C. Kim, G. Gu, H. Ikeda, R. Yoshizaki, and N.
  Koshizuka, Science {\bf 280},  259  (1998).

\bibitem{Chubukov04}
A. Chubukov and M. Norman, Phys. Rev. B {\bf 70},  174505  (2004).

\bibitem{Norman95}
M. Norman, M. Randeria, H. Ding, and J. Campuzano, Phys. Rev. B {\bf 52},  615
  (1995).

\bibitem{Li05}
J.-X. Li, T. Zhou, and Z.D. Wang, Phys. Rev. B {\bf 72}, 94515 (2005).

\bibitem{Hoogenboom03}
B. Hoogenboom, C. Berthod, M. Peter, {\O}. Fischer, and A. Kordyuk, Phys. Rev. B
  {\bf 67},  224502  (2003).

\bibitem{Abanov00}
A. Abanov and A. Chubukov, Phys. Rev. B {\bf 62},  R787  (2000).

\bibitem{Presland91}
M. Presland, J. Tallon, R. Buckley, R. Liu, and N. Flower, Physica C {\bf 176},
   95  (1991).

\bibitem{Ino00}
A. Ino, C. Kim, M. Nakamura, T. Yoshida, T. Mizokawa, Z.-X. Shen, A. Fujimori,
  T. Kakeshita, H. Eisaki, and S. Uchida, Phys. Rev. B {\bf 62},  4137  (2000).

\bibitem{Devereaux04b}
T. Devereaux, T. Cuk, Z.-X. Shen, and N. Nagaosa, preprint cond-mat/0409426
  (unpublished).

\bibitem{Gweon04b}
G.-H. Gweon, S. Zhou, and A. Lanzara, J. Phys. Chem. Solids {\bf 65},  1397
  (2004).

\bibitem{Song95a}
J. Song and J. Annett, Phys. Rev. B {\bf 52},  6930  (1995).

\bibitem{Song95b}
J. Song and J. Annett, Phys. Rev. B {\bf 51},  3840  (1995).

\bibitem{Nazarenko96}
A. Nazarenko and E. Dagotto, Phys. Rev. B {\bf 53},  R2987  (1996).

\bibitem{Sakai97}
T. Sakai, D. Poilblanc, and D. Scalapino, Phys. Rev. B {\bf 55},  8445  (1997).

\bibitem{Nunner99}
T. Nunner, J. Schmalian, and K. Bennemann, Phys. Rev. B {\bf 59},  8859
  (1999).

\bibitem{Friedl90}
B. Friedl, C. Thomsen, and M. Cardona, Phys. Rev. Lett. {\bf 65},  915  (1990).

\bibitem{Litvinchuk91}
A. Litvinchuk, C. Thomsen, and M. Cardona, Solid State Commun. {\bf 80},  257
  (1991).

\bibitem{Andersen96}
O. Andersen, S. Savrasov, O. Jepsen, and A. Liechtenstein, J. Low Temp. Phys.
  {\bf 105},  285  (1996).

\bibitem{Jepsen98}
O. Jepsen, O. Andersen, I. Dasgupta, and S. Savrasov, J. Phys. Chem. Solids
  {\bf 59},  1718  (1998).

\bibitem{Petrov00}
Y. Petrov, T. Egami, R. McQueeney, M. Yethiraj, H. Mook, and F. Do{\u g}an, preprint
  cond-mat/0003414 (unpublished).

\bibitem{McQueeney03}
J.-H. Chung, T. Egami, R. McQueeney, M. Yethiraj, M. Arai, T. Yokoo, Y. Petrov,
  H. Mook, Y. Endoh, S. Tajima, C. Frost, and F. Do{\u g}an, Phys. Rev. B {\bf 67},
  014517  (2003).

\bibitem{Zhu05a}
J.-X. Zhu, A. Balatsky, T. Devereaux, Q. Si, J. Lee, K. McElroy, and J. Davis,
Phys. Rev. B {\bf 73}, 014511 (2006), cond-mat/0507610.

\bibitem{Opel99}
M. Opel, R. Hackl, T. Devereaux, A. Virosztek, A. Zawadowski, A. Erb, E.
  Walker, H. Berger, and L. Forr{\'o}, Phys. Rev. B {\bf 60},  9836  (1999).

\bibitem{Pan01}
S. Pan, J. O'Neal, R. Badzey, C. Chamon, H. Ding, J. Engelbrecht, Z. Wang, H.
  Eisaki, S. Uchida, A. Guptak, K. Ng, E. Hudson, K. Lang, and J. Davis, Nature
  {\bf 413},  282  (2001).

\bibitem{Howald01}
C. Howald, P. Fournier, and A. Kapitulnik, Phys. Rev. B {\bf 64},  100504(R)
  (2001).

\bibitem{Lang02}
K. Lang, V. Madhavan, J. Hoffman, E. Hudson, H. Eisaki, S. Uchida, and J.
  Davis, Nature {\bf 415},  412  (2002).

\bibitem{Howald03}
C. Howald, H. Eisaki, N. Kaneko, M. Greven, and A. Kapitulnik, Phys. Rev. B
  {\bf 67},  014533  (2003).

\bibitem{Fang04}
A. Fang, C. Howald, N. Kaneko, M. Greven, and A. Kapitulnik, Phys. Rev. B {\bf
  70},  214514  (2004).

\bibitem{Mcelroy04a}
K. McElroy, D.-H. Lee, J. Hoffman, K. Lang, E. Hudson, H. Eisaki, S. Uchida, J.
  Lee, and J. Davis, 
preprint cond-mat/0404005 (unpublished).

\bibitem{Mcelroy04b}
K. McElroy, D.-H. Lee, J. Hoffman, K. Lang, J. Lee, E. Hudson, H. Eisaki, S.
  Uchida, and J. Davis, Phys. Rev. Lett. {\bf 94},  197005  (2005), 
  cond-mat/0406491.

\bibitem{Fang05}
A. Fang, L. Capriotti, D. Scalapino, S. Kivelson, N. Kaneko, M. Greven, and A.
  Kapitulnik, Phys. Rev. Lett. {\bf 96}, 017007 (2006),
cond-mat/0508253.

\bibitem{Seibold04}
G. Seibold and M. Grilli, Phys. Rev. B {\bf 72},  104519  (2005).

\end{thebibliography}

\end{document}